LOW-TEMPERATURE COLLECTIVE TRANSPORT AND DYNAMICS IN

CHARGE DENSITY WAVE CONDUCTOR NIOBIUM TRISELENIDE

A Dissertation

Presented to the Faculty of the Graduate School

of Cornell University

In Partial Fulfillment of the Requirements for the Degree of

Doctor of Philosophy

by

Katarina Cicak

December 2020



# LOW-TEMPERATURE COLLECTIVE TRANSPORT AND DYNAMICS IN CHARGE DENSITY WAVE CONDUCTOR NIOBIUM TRISELENIDE


Katarina Cicak, Ph. D.

Cornell University 2020



We investigated low-temperature dynamics in a charge density wave (CDW) conductor NbSe$_3$, a widely studied representative of a class of systems of driven periodic media with quenched disorder and relevant to a wider group of systems exhibiting collective transport behaviors. To date, theoretical efforts have not converged to produce a consistent description of the rich dynamics observed in these systems, especially in the low temperature regime. We developed modern sample preparation techniques and used frequency- and time-domain transport measurements below the second characteristic Peierls CDW transition to investigate the regime of temporally-ordered collective creep in NbSe$_3$ samples in the low temperature regime between 15 K and 32 K. By measuring the frequency of coherent oscillations between two characteristic electric field thresholds, $E_T$ and $E_T^*$, we show that in nine high-quality samples, pure, Ta-, or Ti-doped, the current-field relation for the collective transport in this regime closely follows a modified Anderson-Kim form across five orders of magnitude with thermally- and field-activated behavior above $E_T$ for a range of temperatures. This study, combined with our transport relaxation measurements, provides relevant length, energy, and time scales that set the dynamics in this regime and reveals that the collective dynamics, governed by large length and energy scales, must be reconciled with microscopic local dynamics, with barriers at orders of magnitude smaller scales. The interplay between the


collective and local mechanisms set the dynamics that is responsible for extremely slow (creep-like) collective, yet temporally-ordered behavior. Combined with the existing work, our results paint a consistent picture of a transport phase diagram for charge-density waves, and density-wave systems in general, and provide essential ingredients for a much-needed correct theoretical description of these systems.

# BIOGRAPHICAL SKETCH

Katarina Cicak was born 1974 in the city of Derventa in Bosnia and Hercegovina (then Yugoslavia). She grew up in a small village of Gradina near Derventa and later moved to Derventa where she attended a mathematics/physics/informatics-oriented high school. In 1992 the war in former Yugoslavia forced her to relocate to Zagreb, Croatia. Her childhood home was destroyed and most of her immediate family scattered across Europe. While in Zagreb, she temporarily attended MIOC High School. On November 16, 1992 as an eighteen-year-old refugee with all possessions in one suitcase, she arrived to the U.S. She completed her last semester of high school at Redwood in Visalia, CA and enrolled in a local community college while living with amazing host families who warmly welcomed her in their homes as one of their own children. She received A. S. in Mathematics-Science with Honors in 1995 form College of the Sequoias in Visalia and B. S. in Physics, Magna Cum Laude, in 1997 from University of Southern California in Los Angeles. While at USC she worked in a physics lab that allowed her to spend parts of her summers in NASA's Goddard Space Flight Center in Greenbelt, MD working in payload operations during the space shuttle Discovery mission STS-85 and to launch sounding rockets from White Sands Missile Range in NM. In the summer of 1997, she left California and drove across the country to Ithaca, NY to attend graduate school at Cornell University. She completed her Ph. D. research in 2004 and moved to Boulder, CO to work at National Institute of Standards and Technology. In 2007 she made a trip back to Ithaca and formally defended her dissertation. Finally, in 2020 she officially turned in the dissertation to Cornell to fulfill the requirements for a Ph. D. degree.



To my family Kevin, Tristan, Nalah, and Shea

To my parents Katarina and Marko (Braco) Čičak

To my other parents Bob and Darlene Sartain



ACKNOWLEDGMENTS

I was lucky to be in company of many capable, diverse, and kind people who helped me get to and through my physics Ph. D. journey.  I am grateful to:

Members of my thesis committee: Prof. Dan Ralph who also taught my two favorite classes at Cornell, p=Prof. Piet Brouwer for guiding me to understand CDW state in a broader context of condensed matter, Prof. George Malliaras from Materials Science and Engineering Department for introducing me to optoelectronics and fascinating topics of transport in amorphous materials, and Prof. Jim Sethna who sat in on my Ph. D. defense in place of Prof. Piet Brouwer, and who was always enthusiastic to discuss CDW topics;

My Thorne group lab mates: Serge Lemay for teaching me tricks of the trade in transport measurements on NbSe$_3$, Kevin O'Neill, my "other" Kevin, and a partner in CDW crime in the basement of Clark Hall and during conference trips to France, my countryman Abdel Isakovic, late Zach Stum; my lab mates in "the other Thorne lab" Craig Caylor, Sergei Kriminski, Jan Kmetko, Yevgeniy Kalinin, Slava Berejnov, and Ivan Dobrianov for exposing me to protein crystallography;

Colleagues and collaborators in other labs: Yanping Li with whom I spent many days at CHESS and Brookhaven performing x-ray diffraction experiments on NbSe$_3$ crystals, his advisor, Prof. Joel Brock at Cornell, and Prof. Herre van der Zant and Erwin Slot at Delft University of Technology;

Staff at (then called) Cornell NanoFabrication Facility (CNF): Mike Skvarla, Gary Bordonero, David Spencer, Darren Westely, and Dan Woodie for teaching me the art of modern



microfabrication on which I still heavily relay in my current research and that allowed me to improve on existing $NbSe_3$ sample preparation techniques;

Cornell professors and numerous technical staff with whom I interacted through classes or otherwise, including Prof. Paul McEuen, Prof. Alex Gaeta, Prof. Veit Elser, Prof. David Cassel, Prof. Robert Pohl, Prof David Lee, Prof. Al Sievers, Eric Smith, Prof. Lou Hand, Prof. Carl Frank, and Bob "Sned" Snedeker;

Deb Hatfield, John Minor, and the amazing Physics Department staff of Clark and Rockefeller Halls who made every effort to make our grad student life shielded from distractions of paperwork, and made it easier for us to focus on what we were there to do – physics;

My grad-school physics friends who made life fun at the end of long days in the lab: Harsh Vishwasrao, John Karcz, Frank Albert, Luat Vuong, and many others.

Several people strongly influenced my physics education not necessarily through physics:

The Warren and O'Neill families and the community of Visalia, California who enabled me to stay in the U.S. to continue my education during the war in Bosnia;

Prof. Gene Bickers, my undergraduate advisor at University of Southern California (USC), who introduced me to condensed matter physics, and who charmed me with stories about wonders of Ithaca and Cornell;



My friends and colleagues at National Institute of Standards and Technology: Ray Simmonds, Fabio da Silva, and Joe Aumentado for the support and encouragement to finalize my dissertation and for believing in me;

Bob Sartain, a scary college physics instructor who first made me connect physics in tough homework sets with the real world. Later when I had nowhere to go, he and his wife Darlene took me into their home and made me one of their children. I was very lucky to fall under the good graces of this incredible couple who exemplify goodness in people. My gratitude to, and love for both are beyond words.

I am immensely grateful to my family: my mom and dad who instilled in me a sense of wonder about laws of nature and the idea of education as one of the things that nobody can take away from you, even in a war when you lose everything material; my brother Ivo (Bracan) and sister Sanja who make me laugh until I cry every time I visit; my amazing kids, twins Shea and Nalah, and goofball Tristan, who taught me that hopelessly changing diapers while parenting a teenager shows you what you are made of; and Kevin Moll, a person who held multiple titles through "my lengthy grad-school process", from physics friend, boyfriend, fiancée, and husband, to father of my children…and, during the 2020 pandemic while I was wrapping up my thesis editing until wee hours of the night, a "mother" to my children. I got it done, Kevie! Thank you all for your unconditional love and support.

I am greatly in debt to my research advisor, Prof. Robert Thorne, for ingraining in me a no-nonsense approach to science during my stay in his group between 1998 - 2004 (yes, I have known Rob with the moustache!). It was a privilege to study under a world expert on CDW transport who has an unmatched breath of knowledge on the subject. I admire his versatile,



adaptive, and cross-disciplinary experimentalist approach to physics that exposed me to techniques, methods, and colleagues beyond CDWs.  While I was a teaching assistant in his undergraduate electronics class, I had a chance to watch him lecture and marvel at his innovative teaching approaches.  Most grad students aspire to be like their advisors.  I admired his brilliant technical writing abilities, honed scientific intuition, professional and disciplined work etiquette, standard of excellence, and hard work. Aside from having an advisor who will provide technical guidance, it was important to me to have a high-quality role model as well.  Rob is also a collaborative and approachable scientist and supportive of his students and postdocs.  Rob, thank you for having confidence in me and allowing me independence to explore things in the lab in my own way, and for being an inspiring, supportive, and patient graduate advisor with an unapologetic approach to sound science.



TABLE OF CONTENTS













| | |
|---|---|
| **a** | = unit vector along a crystallographic axis (forming a set with **b**, and **c**) |
| **a**$^*$ | = unit vector of a reciprocal lattice vector (forming a set with **b**$^*$, and **c**$^*$). In NbSe$_3$ **a**$^*$ is parallel to the thickness direction of the whisker. |
| **b** | = unit vector along a crystallographic axis (forming a set with **a**, and **c**) |
| **b**$^*$ | = unit vector of a reciprocal lattice vector (forming a set with **a**$^*$, and **c**$^*$). In NbSe$_3$ **b**$^*$ = **b** and is parallel to the long axis of the whisker. |
| $b_i$ | = impurity coefficient that relates $RRR$ to $n_i$ |
| $b_q^\dagger$, $b_q$ | = phonon creation and annihilation operators at wavevector $q$ |
| $c_{k,\sigma}^\dagger$, $c_{k,\sigma}$ | = electron creation and annihilation operators at wavevector $k$ and spin $\sigma$ |
| **c** | = unit vector along a crystallographic axis (forming a set with **a**, and **b**) |
| **c**$^*$ | = unit vector of a reciprocal lattice vector (forming a set with **a**$^*$, and **b**$^*$) |
| $g_{k,q}$ | = electron-phonon coupling constant |
| $dV/dI$ | = differential resistance |
| $e$ | = electron charge |
| $E$ | = electric field |
| $E_T$ | = threshold field of onset of collective CDW sliding at high temperatures and of collective CDW creep at low temperature. |
| $E_T^*$ | = "switching" threshold at low temperatures (2$^{nd}$ threshold) |
| $F$ | = reduced electric field = $(E-E_T)/E_T$ |
| $f_0$ | = fitting parameter (frequency) in thermally activated form of $f_{NBN}^*$ |
| $f_{NBN}$ | = frequency of narrow band noise (NBN) |
| $f_{NBN}^*$ | = $f_{NBN}(E_T^*)$ |
| $G$ | = conductance |
| $g_{k,q}$ | = electron-phonon coupling constant |
| $H$ | = Hamiltonian |
| $I$ | = electrical current |
| $I_C$ | = CDW current |
| $I_S$ | = single particle current |
| $I_{tot}$ | = total current through a sample crystal |
| $I_T$ | = current at threshold $E_T$ |
| $I_T^*$ | = current at threshold $E_T^*$ |
| $I_{T-}$ | = current where low-field resistance hysteresis loop begins to close in d$V$/d$I$ vs. $I$ plots |
| $I_{T+}$ | = current where low-field resistance hysteresis loop closes in $dV/dI$ vs. $I$ plots |
| $j_C$ | = average CDW current density |
| $K$ | = CDW elastic constant |
| $k_F$ | = Fermi wavevector |
| $L_{barr}$ | = length-scale characterizing local barrier hopping |
| $L_i$ | = length of sample segment between current probes |
| $L_v$ | = length of sample segment between voltage probes |
| $L_{v1}$ | = length of sample segment between 1$^{st}$ current probe and 1$^{st}$ voltage |
| | probe in the standard 4-probe measurement configuration |
| $L_{v2}$ | = length of sample segment between 2st current probe and 2st voltage Probe in the standard 4-probe measurement configuration |
| $L_\phi$ | = length-scale of a phase coherent domain |
| $n$ | = density, concentration |
| $n_{barr}$ | = barrier concentration |
| $n_C$ | = CDW condensate density (i.e. $en_C$ is *charge* density) |
| $n_i$ | = impurity concentration |
| $n_0$ | = electronic density in the metallic state |
| $n_1$ | = amplitude of CDW modulation |
| $Q$ | = quality factor (of a CDW coherent oscillation) |
| $Q_C$ | = CDW wavevector = $2k_F$ |
| $R$ | = resistance |
| $R_i$ | = vector position of impurity i |
| $RRR$ | = residual resistance ratio |
| $R_S$ | = single particle resistance |
| $\Delta R$ | = height of low-field hysteresis loop in $dV/dI$ vs. $I$ measurements at low temperatures |
| $r$ | = position vector |
| $r_o$ | = residual defect coefficient |
| $T$ | = temperature |
| $T_b$ | = fitting parameter (corresponding to energy barrier) in thermally activated form of $f_{NBN}{}^*$ |
| $T_0$ | = fitting parameter in MAK equation |
| $T_P$ | = Peierls transition temperature |
| $T_{P1}$ | = Peierls transition temperature for the upper CDW transition in NbSe$_3$ ($T_{P1}$ = 145 K) |
| $T_{P2}$ | = Peierls transition temperature for the lower CDW transition in NbSe$_3$ ($T_{P2}$ = 59 K) |
| $t$ | = thickness of a NbSe$_3$ sample whisker or time depending on context |
| $u(x)$ | = local displacement of CDW phase with respect to the lattice at position $x$ |
| $V$ | = voltage |
| $V_{barr}$ | = volume per barrier |
| $V_i$ | = impurity pinning potential at site $i$ |
| $v_C$ | = average CDW velocity |
| $v_C{}^*$ | = $v_C(E_T{}^*)$ |
| $v_F$ | = Fermi velocity |
| $w$ | = width of a NbSe$_3$ sample whisker |
| $x$ | = position coordinate |
| $\alpha$ | = fitting parameter in MAK equation proportional to $1/n_{barr}$ |
| $\gamma$ | = intrinsic CDW damping |
| $\Delta_C$ | = ½ of CDW gap energy |
| $\varepsilon_k$ | = energy of electron with wavevector $k$ |
| $\varepsilon_\phi$ | = energy-scale of a phase (FLR) coherent domain |
| $\zeta$ | = critical exponent |
| $\lambda_C$ | = CDW wavelength = $2\pi/Q_C = \pi/k_F$ |
| $\rho_S$ | = single particle resistivity |
| $\sigma_0$ | = fitting parameter in MAK equation (corresponding to conductivity) |
| $\phi(x)$ | = CDW phase at position $x$ |
| $\Psi$ | = CDW order parameter |
| $\omega_q$ | = angular frequency of a phonon with wavevector $q$ |





Charge density waves?

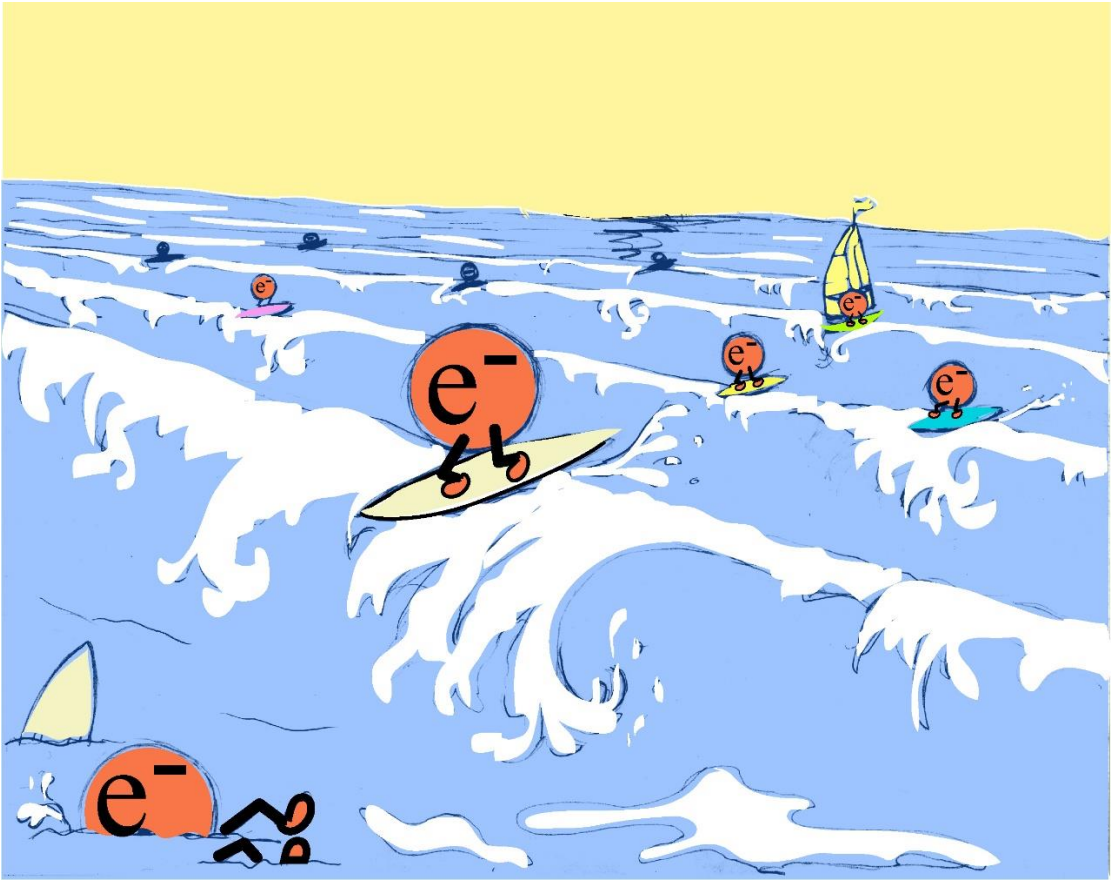



# 1 Charge Density Waves

## 1.1  Introduction

The need to understand progressively more complex dynamic systems has driven the art of modelling them from simple equations describing pendulum motion and coupled harmonic oscillators in the past, to present-day sophisticated models for hydrodynamics, turbulence, and beyond.  A special link in this chain of progression is a class of nonlinear driven systems called driven periodic media with quenched disorder.  The interplay of many internal degrees of freedom in response to an external drive shapes the dynamics of these widely spread systems, from vortex lattices in type-II superconductors and Wigner crystals, to spin- and charge-density waves (SDW and CDW).  Because they exhibit behaviour associated with collective dynamics in the presence of disorder, these systems are also relevant in the understanding of other important problems such as dynamics of liquid interfaces on rough surfaces and motion of tectonic plates.  CDW conductors have been excellent prototypes for study since the collective charge transport they exhibit displays many of the signature characteristics of this class of systems clearly, and they present an ideal experimental platform to probe the interplay of the many degrees of freedom including elasticity, plasticity, pinning by disorder, and thermal fluctuations.  The last forty years had seen a flourish of theoretical activity and significant progress in this field, but ultimately, the transport phase diagrams predicted by the models to date still do not agree with what is experimentally observed in CDW conductors.



The experimental work presented in this dissertation will show that explaining a simple current-voltage (*I-V*) characteristic of these systems remains a challenge within the context of present understanding, thus calling for new theoretical efforts. Our work sheds new light on the dynamics in these systems that we believe will greatly aid future theoretical efforts in producing a picture of microscopic dynamics consistent with experimental observations.

In this chapter we review basic notions of CDW physics, primarily focusing on transport phenomena relevant for understanding the work presented in this dissertation. Due to the complexity of the subject, much of the rich and interesting CDW physics will be presented in a somewhat condensed fashion and the reader will be referred to the abundant literature for a more formal treatment of each topic. Chapter 2 is a summary of our efforts to integrate sample preparation techniques of bulk crystals of niobium triselenide ($NbSe_3$) – the most widely studied CDW conductor – with standard micro- and nano-fabrication technology. In addition to improving the existing techniques to electrically contact samples, our integrated methods add enormous flexibility and customization to sample preparation for studying CDW conductors.

Chapters 3 and 4 investigate CDW transport and relaxation in $NbSe_3$, in the frequency and time domains respectively, in a regime of the transport phase diagram where most of the mysteries remain: below approximately $2T_{P2}/3$ where temporally-ordered collective creep was observed. Here $T_{P2}$ is the Peierls transition temperature for the lower (second) CDW transition in $NbSe_3$. Chapter 5 consolidates our work in a broader context of this field. By incorporating the new findings presented in this dissertation we provide, and attempt to explain, a new phase diagram of CDW transport.



## 1.2    CDW Ground State

Below the Peierls transition temperature $T_P$, a crystal lattice of highly anisotropic conductors can develop a periodic distortion caused by the electron-phonon coupling.  The coupling lowers the energy of the system by creating a gap, $2\Delta_C$, on the Fermi surface.  In the idealized case of a one-dimensional lattice (with Fermi surface at $\pm k_F$), the periodic distortion of the lattice has a wavevector $Q_C = 2k_F$, where $k_F$ is a Fermi wavevector.[*1-4]  The elastic energy cost to distort the lattice competes with the decrease of energy due to formation of a gap on the Fermi surface.  Such a gap pushes the electrons near $k_F$ to lower energies, thus lowering the overall energy of the system.  Electronic density, which is coupled to the ionic lattice, distorts accordingly producing a periodic modulation of charge density, $n(x)$, along a spatial coordinate $x$ (at the same wavevector $Q_C = 2k_F$) and is called a charge density wave (CDW):[5-8]

$$n(x) = n_0 + n_1 \sin[\, Q_C x + \phi(x)]  \qquad \textbf{(1.1)}$$

where $n_0$ is a constant electronic density in the metallic state, $\phi(x)$ is CDW phase related to the local displacement of CDW with respect to the underlying crystal lattice, $u(x)$, given by

$$\phi(x) = -u(x) Q_C  \qquad \textbf{(1.2)}$$

and $n_1$ is the amplitude of the CDW modulation given by

$$n_1 = \frac{n_0 \Delta_C}{\hbar v_F k_F \lambda_C} \qquad (1.3)$$

where $2\Delta_C$ is CDW gap, $\lambda_C = 2\pi/Q_C = \pi/k_F$, $v_F$ is Fermi velocity, $\hbar = h/(2\pi)$ where $h$ is Plank's constant. **Figure 1.1** illustrates the formation of a CDW.

---

[*] Ideal infinite one- and two-dimensional crystals are thermodynamically unstable at finite temperatures because thermal fluctuations destroy long range order.  This is related to Mermin-Wagner theorem, but was also first realized by Peierls and Landau.  In reality, the three-dimensional nature of these materials is very important to producing a phase transition at finite temperature.  Real crystals of these materials have planes of atoms strongly coupled along one (or two) crystallographic axes and much more weakly coupled planes along the two (or one) remaining axes.  This leads to conductivity anisotropy.

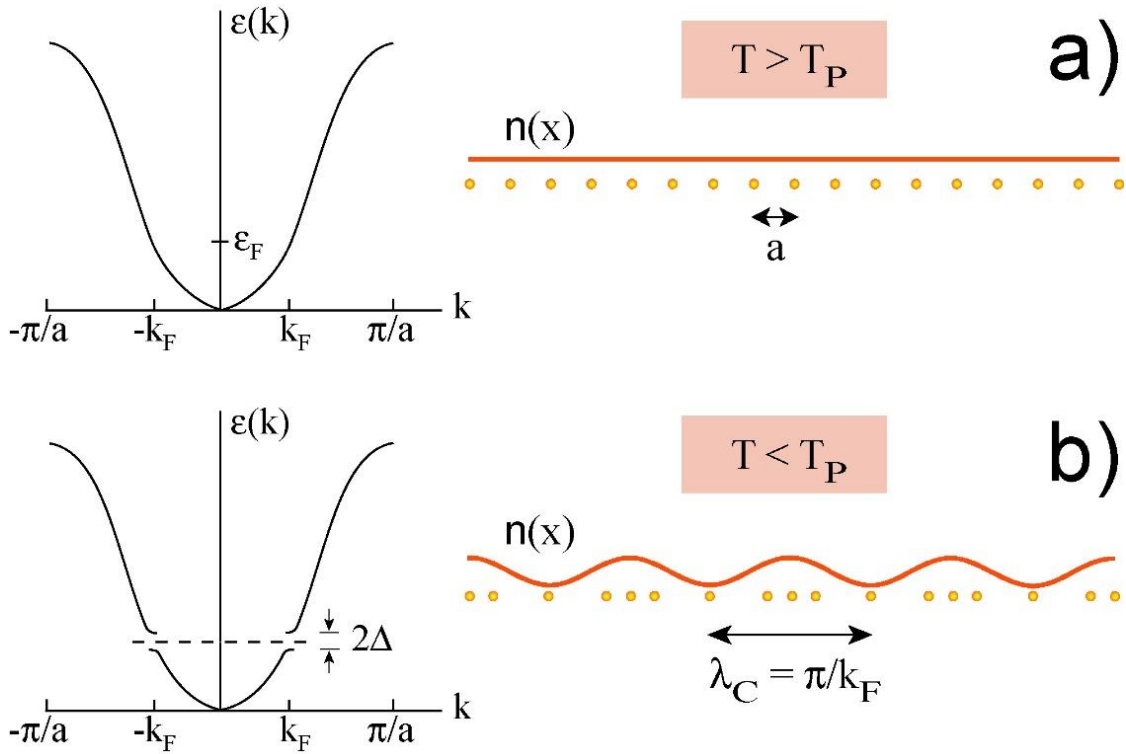

**Figure 1.1**

Electron dispersion relation and the corresponding real space picture of a one-dimensional lattice of ions coupled to electron density $n(x)$ for (a) $T > T_P$ and (b) $T < T_P$ showing charge-density wave.

The formation of a CDW ground state can be discussed in terms of the instability of the free electron gas in one dimension with respect to the formation of a macroscopically occupied phonon mode with a wave vector $q = \pm 2k_F = \pm Q_C$. The electron-phonon coupling causes the



density response function of the electron gas (Linhard response function) to diverge in one dimension.[9]

Formal treatment that leads to CDW ground state begins with a Frohlich Hamiltonian (see for example [9]):

$$H = \sum_{k,\sigma} \varepsilon_k c_{k,\sigma}^+ c_{k,\sigma} + \sum_q \hbar \omega_q b_q^+ b_q + \sum_{k,q,\sigma} g_{k,q} c_{k+q,\sigma}^+ c_{k,\sigma} (b_{-q}^+ + b_q) \tag{1.4}$$

where the first two terms are the usual independent electron and phonon contributions, where $c_{k,\sigma}^+$ ($c_{k,\sigma}$) and $b_q^+$ ($b_q$) are the creation (annihilation) operators for electrons and phonons, respectively. The last term is the electron-phonon coupling term where $g_{k,q}$ is the coupling constant. This term describes electron-phonon scattering: a phonon in mode $q$ ($-q$) is annihilated (created) when an electron is scattered from a state with momentum $k$ into a state with momentum $k+q$. The CDW ground state is a condensate of electron-hole pairs that differ in momentum by $2k_F$ (an equivalent interpretation is that CDW is a condensate of $2k_F$ phonons) just as a superconductor is a condensate of electron (Cooper) pairs of opposite spin and momentum. In fact, a basic theory of CDWs can easily be cast into the equations of the well-known Bardeen-Cooper-Schrieffer (BCS) microscopic theory for superconductors using a mean-field treatment.[9] Appendix A compares the CDW and BCS ground state side by side emphasizing analogous points between the two states.[10]* The CDW ground state can be characterized by an order parameter:

---





$$\Psi = \Delta_C e^{i\phi} \qquad\qquad\qquad \textbf{(1.5)}$$

where $2\Delta_C$ is CDW gap, and $\phi$ is CDW phase. Note that CDW phase has a meaning in real space given by (1.2). The parameter $\phi$ is explicitly used in the CDW phenomenological equations of motion.

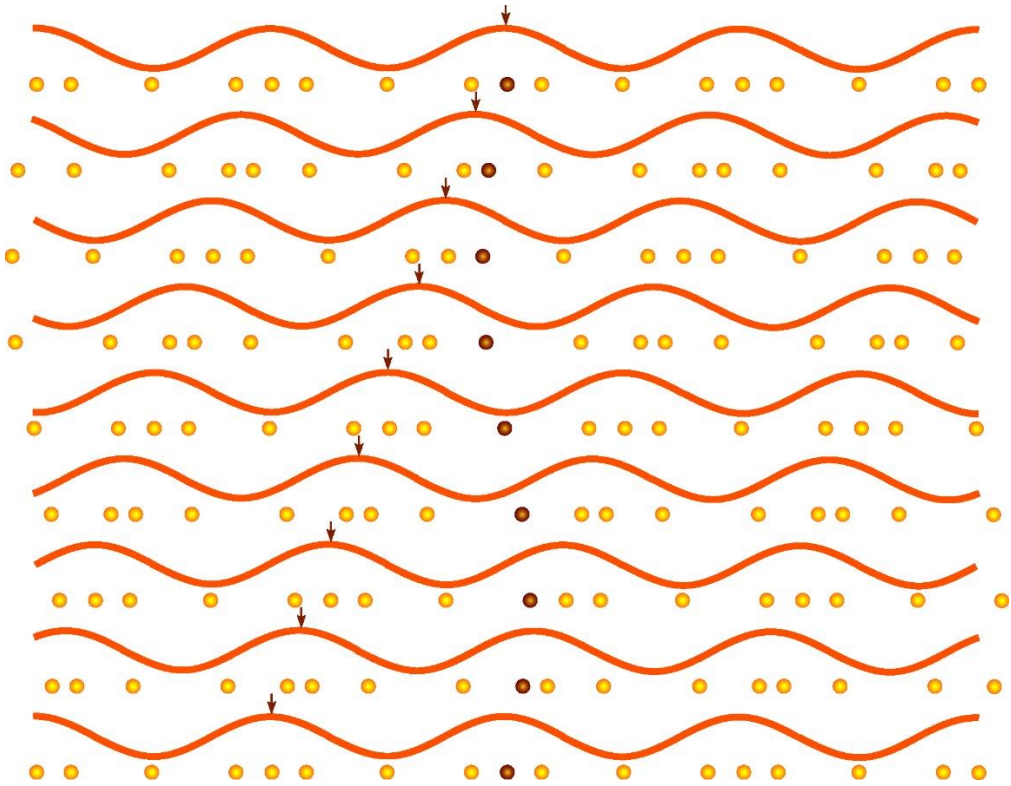

**Figure 1.2**

Frohlich collective transport mode: Each ion oscillates around its equilibrium position while crests of the CDW slide relative to the lattice transporting charge. Each horizontal slice is a snapshot in time, so that the top-down direction shows time progression.

Frohlich predicted a collective transport mode that would allow CDW to slide relative to the lattice and carry charge without friction, i.e. zero-resistance current.[11] **Figure 1.2** illustrates the basic mechanism of the collective transport mode. As mentioned earlier, disorder plays a



major role in collective dynamics of these materials, and in real crystals the CDW is pinned to always-present impurities, defects, and dislocations, and a zero-resistance current is not observed experimentally.  To initiate CDW sliding, a finite electric field, $E$, above a threshold field, $E_T$, must be applied.

## 1.3   Internal Degrees of Freedom in a CDW System

CDW elasticity, interactions with crystal disorder, plasticity, and thermal disorder are important internal degrees of freedom known to be relevant in shaping of CDW dynamics. CDW elasticity is mediated by quasiparticle screening. At finite temperatures quasiparticles excited above the CDW gap couple to the CDW through Coulomb interactions and screen charge fluctuations.[12-16]  CDW elastic constants, thus, become stiffer as the quasiparticles freeze below the gap with decrease of temperature in fully-gapped CDW materials.  In partially-gapped materials this behavior is slightly modified both because the single particle density can remain large, and the single particle mobility increases with decrease of temperature.

Random disorder in a crystal pins the CDW to the lattice.  Strong local pinning centers, like impurities, can cause a CDW to elastically deform locally by adjusting its phase around an impurity site.  The characteristic length scale of the deformation is comparable to the CDW amplitude coherence length $\sim \lambda_C$, and energy scale $\sim \Delta_C$.  A CDW can also deform non-locally or "collectively", by adjusting its phase over much larger length scale $L_\phi$ ($\sim 10^4 \lambda_C$ which in NbSe$_3$ conductor is $\sim 10$ μm) and energy scale $\varepsilon_\phi$ (i.e. $L_\phi \sim 10^4 \lambda_C$ (which in NbSe$_3$ conductor is $\sim 10$ μm) and $\varepsilon_\phi \sim 10^5 \Delta_C$) to find a minimum in a complex potential-energy



landscape caused by a large number of impurities.[17-27]  In both strong local pinning and collective pinning cases, this adjustment of phase costs elastic energy, but the total energy is reduced by a decrease in CDW-impurity interaction energy.  Elasticity and disorder are two very important ingredients in CDW dynamics.

A CDW can also exhibit plastic behavior.  Several models predict thermally assisted incoherent creep and plastic flow at the onset of CDW motion,[28-30] or stick and slip motion with coexisting pinned and sliding regions[31], with the motion becoming more correlated (elastic) at higher velocities.   Experimentally this intrinsic (bulk) plasticity has been observed as rounding of the threshold of motion, $E_T$, only very near $T_P$ and in very thin crystals[19,32-34] where the effects of thermal fluctuations and finite-size effects become imminent (this is further discussed in the next section).  In contrast, plasticity typically observed over a wide range of temperatures and CDW drives is associated with boundary conditions imposed by current injection into samples[35,36] and macroscopic imperfections of crystals.[37-40]  At the current injection contacts, single carriers must condense into a CDW.  The CDW develops phase slips (i.e. discontinuities of the phase in form of CDW dislocations) that facilitate insertion of CDW phase fronts near the current injecting electrode and remove them near the drain electrode.  CDW can also shear and tear near macroscopic imperfections of the crystal lattice like steps on the crystal surface or at boundaries of multiple crystal domains.[40,41]

Together with thermal disorder these factors and internal degrees of freedom can come into play to create rich dynamics that characterize the CDW system.  They can, to a large extent, be probed through transport measurements.



## 1.4 CDW Conductor NbSe₃

Niobium triselenide (NbSe₃) crystals present an excellent platform to study CDW dynamics. Its ribbon-like crystals are grown by a vacuum-transport technique.[42] They have the best crystalline quality of any CDW conductor (more is presented about this in the next chapter) and display the cleanest experimental signatures of collective CDW transport. NbSe₃ undergoes two CDW transitions: one at $T_{P1}$ = 145 K and one at $T_{P2}$ = 59 K (i.e. different parts of the Fermi surface develop gaps at different temperatures).

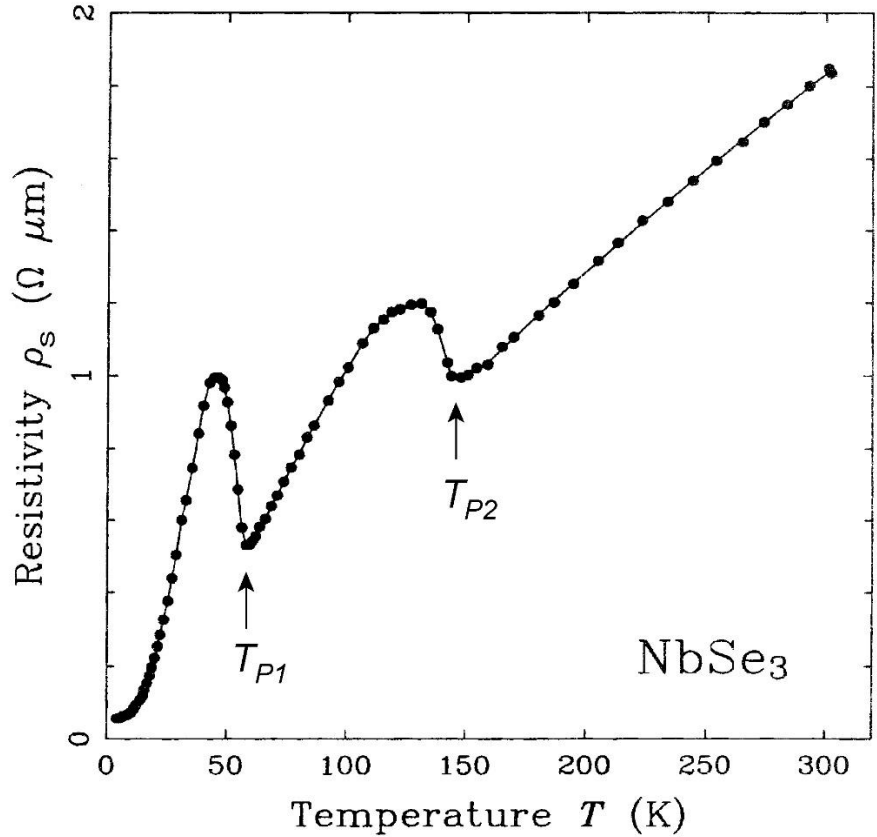

**Figure 1.3**

Single particle resistivity $\rho_S$ vs. temperature in NbSe₃. $\rho_S$ decreases monotonically with decrease of temperature from room temperature, but at $T_{P1}$ and $T_{P2}$ formation of each CDW removes a fraction of carriers from the flowing single particle channel and condenses them into the pinned CDW. This results in increase of $\rho_S$ immediately below $T_{P1}$ and $T_{P2}$. Figure obtained from reference[43] p.20.



It differs from many other studied CDW conductors in that parts of its Fermi surface always remain ungapped, so that a normal-carrier conduction channel, in addition to quasiparticle conduction, shunts the CDW transport for temperatures below the Pierels transition temperatures. Total current, $I_T$, through a biased crystal is thus a sum of collective CDW current, $I_C$, and single particle current, $I_S$. **Figure 1.3** shows signatures of the two CDW condensations in NbSe$_3$. As temperature is decreased through each transition temperature, a fraction of single particles condenses into a pinned CDW, and single particle resistivity $\rho_S$ initially increases. The decrease in resistivity with further lowering of temperature is due to an increase in the mobility of the remaining carriers.

**Table 1.1** summarizes parameter values for the two CDW transitions in NbSe$_3$.

**Table 1.1 Parameters of the two CDW transitions in NbSe$_3$**

| Parameter | Upper CDW ($T_{P1}$ = 144 K) | Lower CDW ($T_{P2}$ = 59 K) | Reference |
|---|---|---|---|
| CDW gap, $2\Delta_C$ (at 1.8 K) | 180 meV | 70 meV | 44,45 |
| CDW wavelength, $\lambda_C$ | 14.4 Å | 13.4 Å | 46 |
| CDW condensate density, $n_C$ | $1.9 \times 10^{21}$ cm$^{-3}$ | $2.05 \times 10^{21}$ cm$^{-3}$ | 47,48 |

## 1.5 CDW Transport: What is Observed and What Can Be Explained

Here we introduce some of the most important features observed in CDW transport experiments with special focus on the velocity-field (or current-voltage) relation. Despite some confusion in this field in its early years, there is consensus now that what we describe in the following has been well established experimentally in virtually all CDW conductors.



## Upper vs. Lower Temperature Regime below $T_P$

Below $T_P$, experimental signatures of CDW transport in the upper temperature range (typically for $2T_P/3 < T < T_P$) differ from what is observed at low temperatures ($T < 2T_P/3$). In the upper temperature range, a CDW depins at a threshold field $E_T$, and the CDW velocity increases smoothly above it. In principle the CDW should exhibit thermally assisted creep below $E_T$ at any finite temperature, rounding the onset of collective conduction. In practice, the pinning energy per collectively pinned domain is so large that thermally assisted motion below $E_T$ is usually negligible. However, very near $T_P$ where the CDW order parameter and pinning energy per volume are smaller and thermal fluctuations are larger, and in very thin samples where finite size effects reduce the pinning energies per unit volume,[19,32-34] threshold rounding produced by incoherent thermal creep can be observed in differential resistance vs. electric field measurements.

$E_T$ depends on bulk impurity concentration $n_i$ and is proportional to $n_i^2$ in thick crystals (three-dimensional case), and to $n_i/t$ in thin crystals where $t$ is crystal thickness (two-dimensional case where finite-size effects come into play).[19,49]

In NbSe$_3$ the average CDW velocity above $E_T$ in this upper temperature regime has a form[50-52]

$$v_C(F) \propto F^\zeta \qquad\qquad (1.6)$$

where $F = (E - E_T)/E_T$ is the reduced electric field, and $\zeta \sim 1.09$. The average velocity $v_C$ increases smoothly with electric field above $E_T$, and at high fields asymptotes to ohmic behavior with $v_C$ proportional to $E$.



The CDW's motion in the presence of impurities is spatially complex and nonuniform in time but (in the absence of phase slip) is periodic in time, with the entire configuration repeating every time the average CDW phase advances by $2\pi$. The dynamic correlation length over which the motion is correlated is predicted to grow as $E_T$ is approached from above, and the effective number of degrees of freedom is expected to decrease. Translation of the elastic CDW through the random potential of disorder in response to applied DC field thus results in temporally-ordered collective motion. The velocity has a periodic component called "narrow-band noise" (NBN) at a frequency $f_{NBN}(E)$ that is proportional to average CDW velocity, $v_C(E)$, and CDW current density $j_C(E)$:

$$j_C(E) = en_C v_C(E) = en_C \lambda_C f_{NBN}(E) \qquad \textbf{(1.7)}$$

where $n_C$ is CDW condensate density. Translation of the CDW by one wavelength $\lambda_C$ corresponds to one period of a coherent oscillation. Coherent oscillations associated with this periodic CDW motion were discovered by Fleming and Grimes in 1979.[53] The most coherent response ever observed in a CDW conductor was measured in high-quality NbSe$_3$ samples where coherent oscillations had narrow spectral widths with quality factors ~30,000 at frequencies of MHz and tens of MHz.[54] **Figure 1.4** shows a voltage spectrum of a DC biased NbSe$_3$ single crystal displaying a sharp fundamental NBN peak at $f_{NBN}$ and large harmonic content visible up to 15$f_{NBN}$. This periodic CDW response can mode-lock onto an external AC drive (applied in combination with a DC bias) producing plateaus in the DC current-vs.-voltage (*I-V*) response analogous to Shapiro steps in superconductors.



Commencement of sliding at $E_T$ is also characterized by a dramatic increase in $1/f$-like or broad-band noise (BBN).

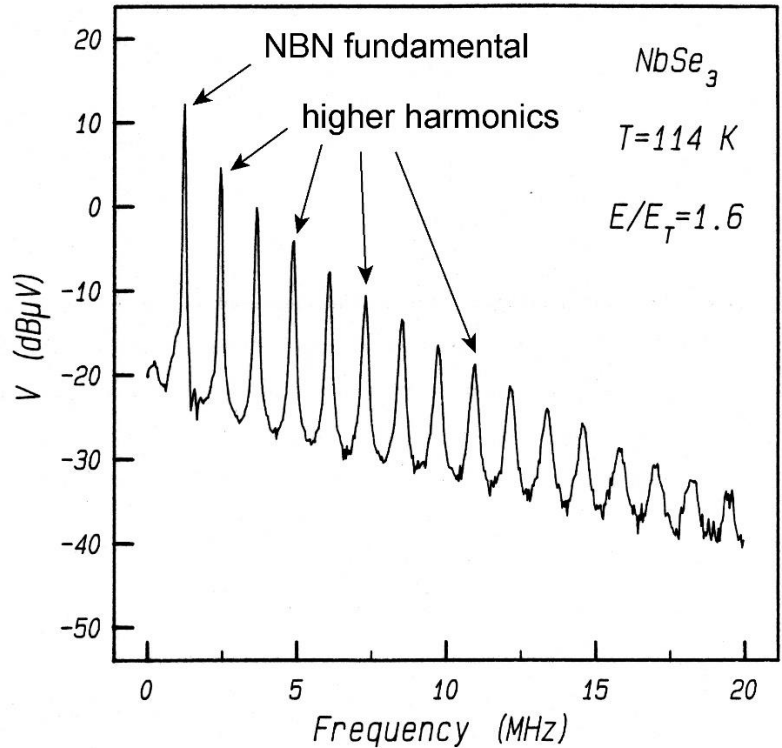

**Figure 1.4**

Signature of CDW collective dynamics: When a CDW is driven by a constant force above the threshold, its motion exhibits coherent oscillations in addition to average phase advance (translation) along the crystal lattice. Periodic components of the motion show up as peaks in the voltage spectrum of a DC driven sample. The figure was obtained from reference [52].

In the lower temperature range, as the temperature is decreased below $2T_P/3$, the observed transport characteristics acquire significant new features signaling changes in internal dynamics. Conduction above $E_T$ begins to freeze out, and CDW velocities become extremely small with creep-like, temperature-activated dependence. At a few times $E_T$, a second characteristic threshold, $E_T^*$, develops at which this slow motion switches into high velocity coherent sliding. This "switch" is highly sample-dependent. In high quality crystals it is



often hysteretic and discontinuous, abruptly increasing the CDW velocity by many orders of magnitude ($10^3$ - $10^6$ in $NbSe_3$[55]) and becoming more pronounced with further decrease of temperature. Above $E_T^*$, CDW conduction has a weak temperature and field dependence with average velocities reaching values typical of the high temperature sliding regime. Coherent oscillations with $f_{NBN}$ given by **(1.7)** are also observed in this regime. A series of experiments on different CDW conductors demonstrate the low-temperature behavior.[56-60] Like $E_T$, the switching threshold $E_T^*$ increases with increasing impurity concentration and decreasing crystal thickness and is independent of design parameters like current-contact separation and position along the sample.[61] The transition to sliding at $E_T^*$ is thus believed to be an intrinsic bulk effect. In this low temperature regime, the part of the current-field relation for $E_T < E < E_T^*$ is called *slow branch*, while the sliding regime for $E > E_T^*$ is called *fast branch*.

**Figure 1.5** illustrates the transport behavior in the upper and lower temperature range observed in partially gapped $NbSe_3$ below the $T_{P2} = 59$ K CDW transition. It shows typical differential resistance *dV/dI* vs. *E* data as well as the corresponding current-field relations measured in the two characteristic temperature regimes. (In addition, the evolution of differential resistance curves from high to low temperature regime can be seen in **Figure 3.5** in section 3.1 where it is presented within the context of the experiments in chapter 3.)



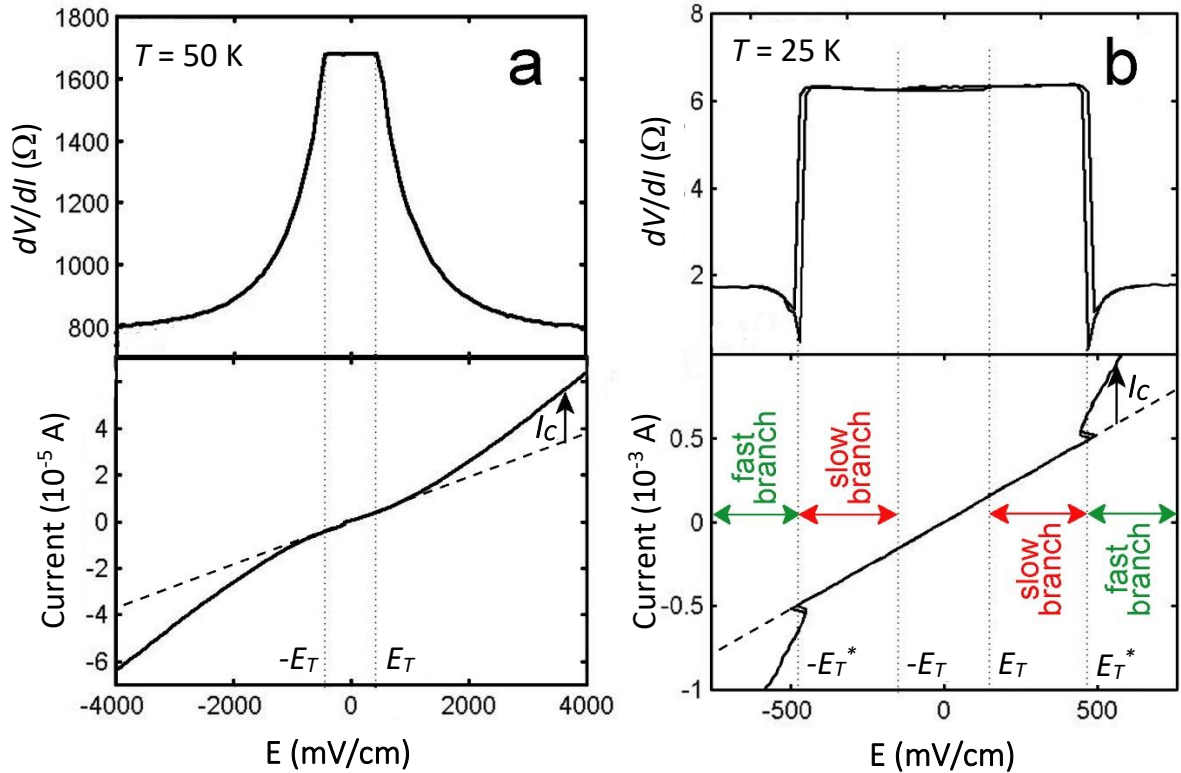

**Figure 1.5**

Differential resistance vs. electric field data and the corresponding current-field data observed in NbSe$_3$ in (a) the high-temperature regime where CDW sliding begins at $E_T$, and in (b) the low-temperature regime where one observes two characteristic thresholds $E_T$ and $E_T^*$ (dotted lines indicate $|E_T|$ and $|E_T^*|$). Current plotted on the $y$-axis in the lower graphs is a total current $I_{tot} = I_S + I_C$, where $I_S$ is the single particle current, and $I_C$ is the CDW current. In NbSe$_3$ at low $T$, threshold $E_T$ is hard to discern since $I_C << I_S$ and $I_{tot} \approx I_S$ in slow branch. $E_T$ can be roughly identified with the field where the low-field resistance hysteresis loop in the upper graph of (b) closes. Here the hysteresis loop is barely discernable for $-E_T < E < E_T$ in the upper plot of (b). This low-field hysteresis as well at the switching hysteresis observed at $E_T^*$ are discussed in chapters 4 and 3 respectively. Dashed lines indicate extrapolated (ohmic) current, $I_S$. Above $E_T$ in a) and above $E_T^*$ in b) $I_C(E)$ can be extracted from $I_C(E) = I_{tot}(E) - I_S(E) = I_{tot}(E) - V(E)/R_S$ where $1/R_S$ is the slope of the $I$-$V$ relation obtained at $|E| < E_T$ where the CDW velocity is zero. Note: Data in (a) and (b) are obtained from two different NbSe$_3$ samples (with different crystal thicknesses and impurity densities which both affect the values of $E_T$ and $E_T^*$. To observe the evolution of $E_T$ and $E_T^*$ with temperature (from a single sample), see **Figure 3.5** in section 3.1.



**Figure 1.6** summarizes the evolution of the CDW transport with temperature and electric field observed in NbSe$_3$ below the lower CDW transition.

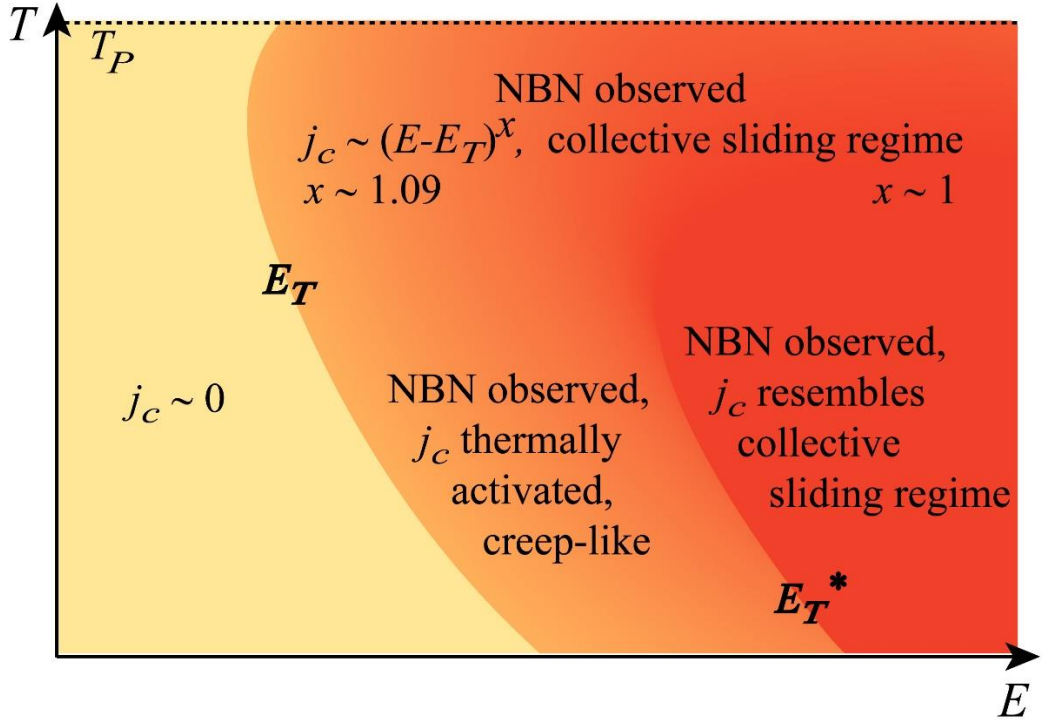

**Figure 1.6**

Evolution of the CDW transport with temperature and electric field observed in NbSe$_3$ for the CDW below $T_{P2}$.

The experimentally observed features discussed so far are common to fully gapped CDW conductors and to partially gapped NbSe$_3$ (an important difference is that in fully gapped materials $E_T$ decreases with decreasing temperature, and the $E_T^*/E_T$ ratio becomes much larger than in NbSe$_3$). Two characteristics threshold fields and temperature activated nonlinear transport between them have been observed in other systems such as spin-density-wave conductor (TMFSF)$_2$PF$_6$ [62-64], and a ladder compound Sr$_{14}$Cu$_{24}$O$_{41}$[65]. Understanding



dynamics in a CDW system thus may have a broad impact on understanding transport in other condensed matter systems.

Until recently, the precise relation between the collective current density and applied field, $j_C(E)$, for slow branch $E_T < E < E_T^*$ in partially gapped NbSe$_3$, was not known. In addition, in fully gapped CDW conductors separating the collective current $I_C$ from the single particle current $I_S$ can be complicated by the fact that the single particle response to applied field in this regime is not ohmic due to CDW–quasiparticle interactions.

Lemay *et al*. showed that in NbSe$_3$ in the low temperature regime, CDW motion immediately below $E_T$ is negligible.[43,66] The CDW velocity just below the threshold is orders of magnitude smaller than above it, i.e. $v_C(E=0.95E_T) < 10^5 v_C(E=1.05E_T)$, and is less than $10^{-4}$ Å/s. This established that $E_T$ is a true threshold for collective motion and that incoherent thermal creep below it is negligible. Lemay *et al*.[55] observed coherent oscillations with relatively large quality factors ($Q$'s up to 130) between $E_T$ and $E_T^*$. By comparing them with measurements of relaxation of CDW polarizations associated with current conversion at contacts when the driving current reversed direction, he showed that the frequency of those oscillations corresponded to the CDW current with the usual proportionality given by **(1.7)** to within the measurement uncertainty. Thus, the remarkable observation that the low temperature creep in the slow branch in NbSe$_3$ is temporally-ordered, exhibiting coherent oscillations[55], provided the means to directly measure $j_C(E)$ via the frequency of coherent oscillations $f_{NBN}(E) \propto v_C(E) \propto j_C(E)$ and by using **(1.7)** in a regime where the CDW current is many orders of magnitude smaller ($10^{-9}$ to $10^{-7}$ at 20 K) than the single particle current.[43,55] Furthermore, the discovery



prompted for a re-examination of the CDW transport phase diagram and its defining underlying mechanisms. This will be the subject of Chapters 3, 4, and 5.

## Models of CDW Transport

The simplest phenomenological model of CDW transport is a "particle in a washboard potential" also used to describe tunneling of Cooper pairs and quasiparticles in a superconducting Josephson junction and known as resistively shunted junction model (RSJ). A CDW is treated as a rigid periodic medium (a "particle") that interacts with random impurities producing a periodic ("washboard") potential. The equation of motion in terms of CDW phase $\phi$ as a variable is essentially the one of overdamped harmonic oscillator with the inertial term being negligible compared to the dissipative one (see for example [67]). This model can account for the very basic experimental signatures of CDW transport such as the characteristic threshold field $E_T$ for onset of conduction and the periodic component of the motion (NBN) with frequency proportional to CDW velocity. **Figure 1.7** illustrates the basic idea of the model.



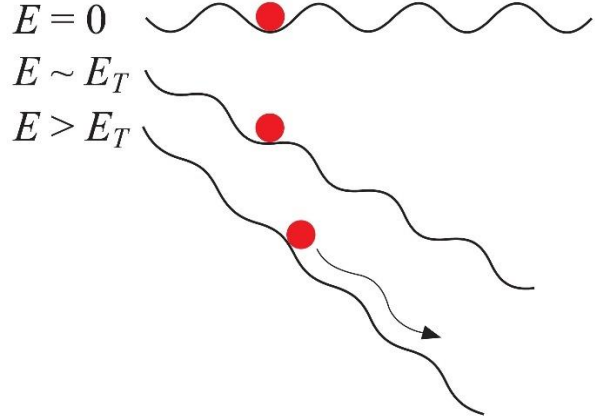

$E = 0$

$E \sim E_T$

$E > E_T$

**Figure 1.7**

CDW particle in a washboard potential. CDW particle is trapped in the potential until a sufficient threshold field $E_T$ is applied to "tilt" the potential. The particle then begins to roll down the washboard. For a constant field when $E > E_T$, CDW velocity exhibits a periodic component of motion with frequency proportional to average CDW velocity.

A more realistic model proposed by Fukuyama, Lee, and Rice (FLR model) takes into account internal elastic degrees of freedom of a CDW and treats a CDW as a classical deformable medium that interacts with randomly distributed impurities[17,18]. Spatial and temporal variations of the CDW phase, $\phi(r,t)$ are described by

$$\gamma \frac{\partial \phi(\vec{r},t)}{\partial t} - K\nabla^2\phi(\vec{r},t) + \sum_{\vec{R}_i} V_i(\vec{r}-\vec{R}_i)en_1\sin\left[\vec{Q}_C \cdot \vec{r} + \phi(\vec{r},t)\right] = \frac{en_C}{Q_C}E \qquad \textbf{(1.8)}$$

where $\gamma$ is intrinsic CDW damping, $K$ is the CDW elastic constant, $V_i$ is impurity pinning potential due to impurity at a site $R_i$, and $E$ is the applied electric field. In this model, impurities can have both local and collective effects. Fluctuations in pinning interactions within volumes containing large numbers of impurities can pin the CDW collectively. A length $L_\phi$ known as the FLR length (also known as the Larkin length for flux line lattices in



superconductors) sets the scale of the resulting deformations of the CDW phase.

Characteristic energy scale $\varepsilon_\phi$ of collective pinning is roughly $(en_C E_T \lambda_C) L_\phi^3$ and is of the order

$\varepsilon_\phi \sim 10^3$-$10^5 k_B T \gg \Delta_C$. For impurities with sufficiently strong interaction with CDW, the

phase may also be pinned at the impurity (locally), producing much stronger deformations

within a length scale of roughly the amplitude coherence length $\sim \lambda_C$ that is typically $\ll L_\phi$.

The local pinning has a characteristic energy scale of the order of the CDW gap, $\Delta_C \ll \varepsilon_\phi$.

These collective and local effects of impurities are associated with vastly different length and

energy scales and are not mutually exclusive. This contrasts with the more common but naive

notion of strictly "weak" versus "strong" pinning scenarios illustrated in **Figure 1.8**.

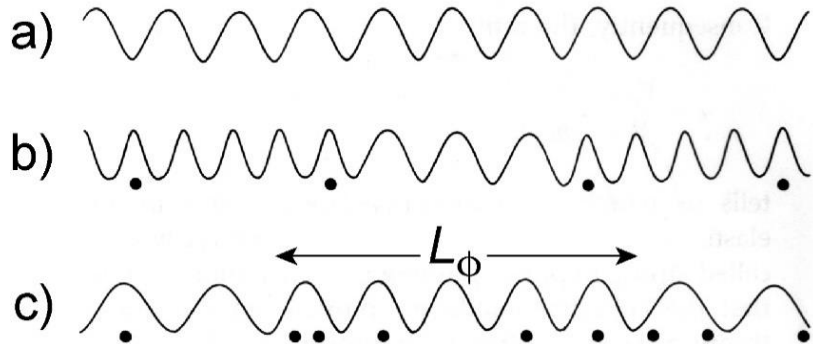

**Figure 1.8**

A common but somewhat oversimplified picture of CDW interaction with impurities (shown as black dots): (a) No impurities – CDW phase is preserved throughout the crystal. (b) Strong pinning – CDW phase is adjusted locally so impurities sit in troughs (or crests) of the CDW. (c) Weak pinning – CDW phase adjusts over a large length scale to "accommodate" impurities. A phase-coherent domain contains many impurities.



**Table 1.2  Relevant length and energy scales**

| Parameter | Parameters characterizing CDW state | Local (strong) pinning | Collective pinning |
|---|---|---|---|
| Energy scale | $\Delta_C$ | $\varepsilon_{strong} \sim \Delta_C$ | $\varepsilon_\phi \sim 10^5 \Delta_C$ |
| Length scale | $\lambda_C$ | $L_{strong} \sim \lambda_C$ | $L_\phi \sim 10^4 \lambda_C$ |

**Table 1.2** summarizes length and energy scales for strong local and collective pinning in terms of relevant CDW parameters.

The collective-pinning theory and its many variants can account for many features of CDW transport, namely the ones observed experimentally in the upper temperature range $2T_P/3 < T < T_P$. These include collective effects such as the onset of CDW conduction at a finite collective threshold $E_T$, $E_T$ dependence on impurity concentration and sample thickness,[19,49] existence of coherent oscillations with frequency proportional to CDW current and their sample volume dependence,[53] mode-locking of CDW coherent oscillations to external AC drive,[68] large dielectric constants, etc. The theory predicts that in the limit of perfectly rigid medium $E_T$ should vanish as $T \to 0$, which indeed is the behavior observed in most fully-gapped CDW conductors with decrease of temperature,[57,60] * since depletion of single particles causes descreening of CDW fluctuations and stiffens CDW elastic constants. In partially gapped NbSe$_3$ this behavior is modified since single particle density remains large with decrease of temperature while single particle conductivity increases. This causes increased screening of CDW thus softening the elastic constants and increasing $E_T$ with decrease of temperature. The FLR model allows for metastable (energy) configurations of CDW phase within FLR domains that result in observable effects such as metastable CDW phase polarization between current contacts during external drive or upon drive termination, and metastable low-field ($-E_T < E < E_T$) resistance hysteresis in differential resistance measurements (metastability will be further discussed in chapter 4).

---

* Temperature dependence of $E_T$ in TaS$_3$ is somewhat more complex, but this CDW conductor is anomalous in other respects

The FLR model and variants of it have been used to make detailed predictions about the form of the CDW current-field relation. The model in reference [69] predicts that in the high-field limit the CDW conductivity $G_C(E) = G_\infty - CE^{-1/2}$ where $C$ is a constant and $G_\infty$ is conductivity at high fields, in agreement with very early measurements in NbSe$_3$ up to $E/E_T \sim 8$ in the upper temperature range,[53] but subsequent detailed experiments[50-52] in high-quality NbSe$_3$ crystals up to $E/E_T \sim 200$ showed that $G_C(E) = G_\infty - C/E$.

D. S. Fisher[70,71] examined CDW dynamics near $E_T$ threshold. Starting from the FLR model, Fisher showed that CDW depinning is a dynamic critical phenomenon with predicted mean field scaling exponent $\zeta$ in **(1.6)** equal to 3/2. A later renormalization-group treatment[72] predicted $\zeta = 5/6, 2/3, 1/2$ in 3, 2, and 1 dimensions respectively. The predicted exponents in references [70-72] have not in general fared well with experiments in different CDW conductors,[50-52,54,73-76] but it turns out that accessing the critical regime experimentally is not trivial, partly due to the fact that many sample-specific, external factors can affect the non-linearity near $E_T$, and partly due to finite-size effects in ordinary NbSe$_3$ samples that become important before the critical regime can be reached, as pointed out by simulations in [77-80].

When plasticity and thermal fluctuations are included, models[22,28,81,82] typically predict a region of incoherent thermal creep below $E_T$, a region of plastic flow near and above $E_T$, and eventually a crossover or transition to nearly elastic flow at high fields and velocities. **Figure**



**1.9** shows an example of a transport phase diagram[28] which predicts an incoherent-flow phase above a first threshold followed by a phase transition into a moving solid phase at a second threshold (here labelled $E_C$).

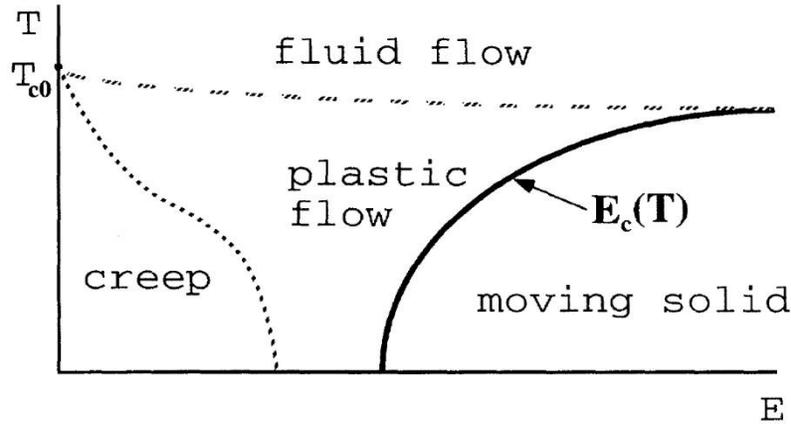

**Figure 1.9**

Transport phase diagram for the three-dimensional CDW predicted by Balents and Fisher.[28] When thermal effects or phase slip are included in the model, the CDW depinning transition is predicted to become a rounded crossover from thermal creep into plastic flow (dotted line). $E_C(T)$ is a true phase transition from plastic flow to a temporally-periodic moving solid phase. The figure was obtained from reference [28].

Incoherent thermal creep produces negligible threshold rounding and negligible motion below threshold except at temperatures very close to $T_P$ as mentioned earlier. Above $E_T$, when plasticity associated with current conversion at current contacts and with shear at grain boundaries are eliminated, there is no evidence of significant CDW plasticity. All extrinsic plasticity broadens the spectral width of the NBN peaks (in some materials making them unobservable) and increases the broadband low frequency noise. But in carefully prepared samples in the high temperature regime above $T_P/2$, the spectral width of the coherent oscillations is extremely small right above $E_T$ and shows little variation with CDW velocity,[54] and any broadening of coherent oscillations that is observed is roughly consistent with



expected broadening due to longitudinal current variation associated with phase slip near contacts.[36,83] Moreover, the functional form of the velocity-field relation shows no kinks or inflections above $E_T$ that would indicate a plastic-to-elastic crossover or transition. Even in the slow branch above $E_T$ in the low temperature regime, the spectral width of the coherent oscillations is small and neither it nor the broadband noise amplitude show any dramatic variation with current or field. Thus, while one cannot rule out plasticity, there is no evidence for plastic-to-elastic transitions. Our measurements presented in chapter 3 also confirm this even for intentionally doped NbSe$_3$ samples for various doping levels. So far, the instances where plasticity matters were associated with boundary conditions and include CDW phase-slip near current contacts,[36,83] or shear near macroscopic defects of the crystal such as steps on crystal surface or near grain boundaries as we discussed earlier.[38,40] To our knowledge, a regime where intrinsic, bulk plasticity dominates CDW transport has not been confirmed experimentally to date.

So far, all the models we discussed address only features observed in the upper temperature range where only one characteristic threshold is observed. A model by Littlewood[14,15] and a latter simulation by Levy _et al._[84] tried to explain the abrupt, hysteretic, switching threshold $E_T^*$ observed at low temperatures. The theory in these references is based on single-particle screening of the CDW which leads to increased dissipation. The models do predict a "switching" threshold for onset of sliding at low temperatures but do not predict _two_ characteristic thresholds observed in all CDW conductors. Furthermore, unlike in fully gapped conductors, in partially gapped NbSe$_3$ single particle density remains large and increases with a decrease of temperature, and yet transport in NbSe$_3$ exhibits a behavior at $E_T^*$ almost identical to what is observed in fully gapped conductors. De-screening of CDW is



thus an unlikely explanation (or at least an incomplete one) for switching at $E_T^*$ observed in both fully and partially gapped conductors. To our knowledge, a complete microscopic model that can account for features observed for $T < 2T_P/3$ including two characteristic thresholds observed at low temperatures, and the collective CDW creep exhibiting coherent oscillations that separates pinned from the coherent sliding regime, does not exist to date.

Surprisingly, even after many decades since Monceau *et al.*[85] first observed nonlinear electrical conduction in NbSe$_3$ in 1978, the basic *I-V* characteristic and its evolution with temperature exhibiting features common to all CDW conductors can still not be explained successfully. Existing models are either valid in very narrow, specific regimes, or address specific features in specific CDW conductors (i.e. different model for each CDW conductor), or simply do not agree with what is experimentally observed. More in-depth critical review of the models we discussed is given in [86-88]. While researching the wealth of existing literature, one finds that despite the ubiquity of transport signatures experimentally encountered across different CDW conductors, a clear picture that underlies *common* microscopic dynamics in these systems is still eluding us today. Pieces of this picture are coming together, however, and we hope that the work presented in this dissertation should clarify some mysteries that remain.



# 2  Processing NbSe₃: A Novel Material in the Cleanroom

NbSe$_3$ sample preparation involves manipulating microscopic crystals of NbSe$_3$ to electrically contact them or modify their shape.  Over the years this has always been a messy and tricky process.  Traditionally, electrical contacts were applied manually as blobs of conductive silver paint (or epoxy), or as thin, extruded indium wires pressed over delicate NbSe$_3$ ribbons.  Ribbons were cleaved by peeling crystal layers using adhesive tape or by using tweezers, scissors, and various "poke tools" while looking and manipulating them under microscope.  More recently, some effort has been made to apply standard microfabrication techniques to make these tasks easier and to obtain novel NbSe$_3$ device structures.[35,36,66,83,89-97]

However, the attempts to adapt NbSe$_3$ material to the cleanroom tools, techniques, and processes have been challenging.  Development of micro and nano fabrication technology was originally driven by semiconductor industry.  Over time, fabrication techniques have been standardized and cleanroom equipment has been commercialized.  Processing typically involves silicon (or other semiconducting materials) and thin films on silicon wafers.  As the industry evolved, the fabrication and sample analysis techniques were applied to materials beyond silicon in both industrial and academic settings. However, the anisotropic nature of NbSe$_3$ crystal that causes it to grow as whisker-like ribbons rather than thin films on standard wafers, made NbSe$_3$ poorly suited for use with these methods.  An additional challenge is that chemistry and reactivity of NbSe$_3$ differ from standard semiconducting materials



complicating things in terms of cross-contamination of equipment, material compatibility, and other unknowns. As a result, historically there likely was some reluctance to introduce NbSe$_3$ into a cleanroom environment, even when cleanrooms were accessible to researches in academia.

The potential of using modern cleanroom fabrication techniques to expand our understanding of CDW conductors like NbSe$_3$ is enormous. To this end we have developed and optimized reliable sample preparation techniques for NbSe$_3$ by using standard micro- and nanofabrication to produce recipes for applying electrical contacts to NbSe$_3$, and to pattern bulk NbSe$_3$ crystals both laterally and vertically. Furthermore, we have successfully integrated this new material with standard cleanroom processing techniques, allowing novel device structures to be created routinely and the physics of CDWs to be explored on much smaller length scales.

The bulk of the developed work presented in this chapter was performed at Cornell Nanofabrication Facility (CNF). We describe a wide variety of tests and experiments aimed at developing reliable protocols for fabricating sample structures from grown NbSe$_3$ ribbons. Because of the extremely large parameter space that can be explored when introducing a new material to the cleanroom environment, in-depth investigations were not always possible due to time constraints, and the results presented here provide valuable guidelines and pragmatic "rules-of-thumb", and not necessarily the last word on the subject. The objective was to quickly determine protocols that could deliver desired sample structures and then to study the electronic transport properties in those structures. It is hoped that the relatively complete discussion presented here will also facilitate future microfabrication work.



To avoid confusion in the following discussion, we define the following terms:

- *whisker* or *ribbon* refers to a strand of bulk, as-grown NbSe₃. A strand contains a single crystal or multiple crystal grains aligned along the long axis of the whisker.

- *crystal* is used when emphasizing the ordered structure of NbSe₃

- *sample* refers to a whisker or a crystal prepared for measurements (typically the crystal is mounted on a substrate and electrical contacts have been fabricated/attached to the whisker).

## 2.1 NbSe₃ Crystals

Bulk NbSe₃ crystals are grown in quartz tubes by a vapor transport method at high temperatures.[42] Our group has developed methods to obtain relatively large, very high purity NbSe₃ crystals. A typical crystal batch is shown in **Figure 2.1**.

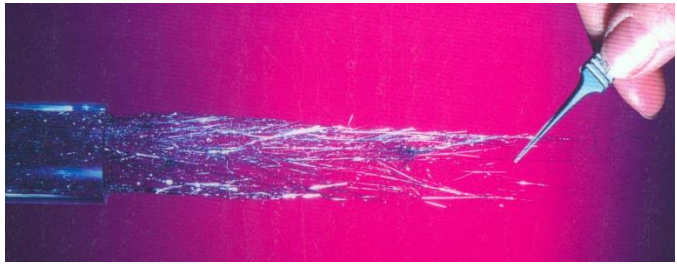

**Figure 2.1**

A batch of NbSe₃ whiskers grown by vapor transport in a quartz tube.

Anisotropic bonding in NbSe₃ gives this material a quasi-one-dimensional (1D) character. Bulk crystals grow to form ribbon-like whiskers with typical dimensions ranging in thickness



$t$ = 0.1 - 1 μm, width $w$ = 1 - 100 μm, and length of a few centimeters. Whisker facets have a shiny un-tinted metallic appearance as seen in **Figure 2.1**. Bulk grown NbSe$_3$ typically exhibits small angle grain boundaries and other defects producing whiskers that are composed of long crystalline grains as shown in **Figure 2.2.**

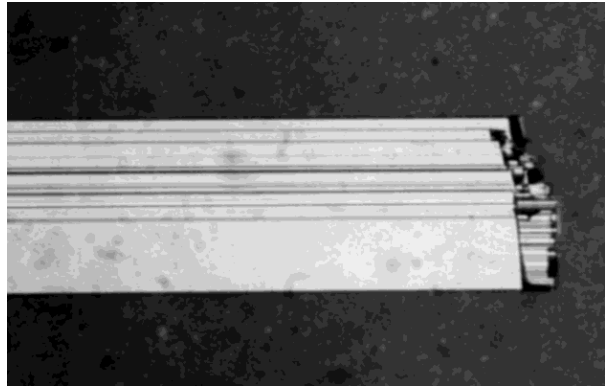

**Figure 2.2**

Micrograph of a NbSe$_3$ whisker end showing multiple crystal domains on and below the surface caused by small angle grain boundaries and other defects.

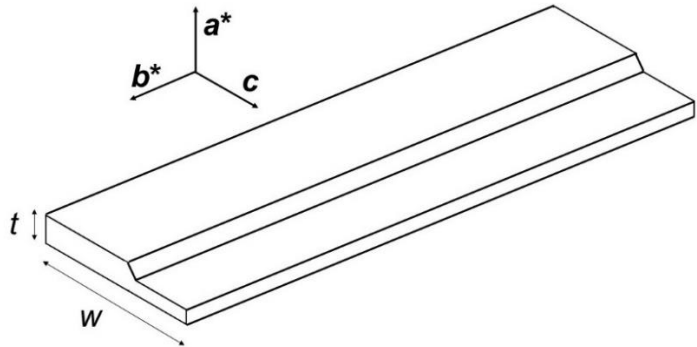

**Figure 2.3**

A sketch of a ribbon-like whisker showing a step edge along the long **b** (= **b**$^*$) axis of the whisker that produces a non-uniform thickness, $t$, of the whisker cross section. Here **a**$^*$, **b**$^*$ and **c**$^*$ are unit vectors along the directions of the reciprocal lattice vectors. **c**$^*$(not show) lays in **b**$^*$-**c** plane.



The grains are oriented with their **b** (= **b**$^*$) crystallographic axis aligned along the long axis of the whisker, but with grains slightly misoriented in the **a**$^*$-**c**$^*$ plane,[38]$^*$ here **a**$^*$, **b**$^*$ and **c**$^*$ define directions of the reciprocal lattice vectors for the NbSe$_3$ crystal (see **Figure 2.3**). These longitudinal grains produce non-uniform whisker cross-sections and often manifest as steps on the whisker surface with step edges running along the whisker's long axis as shown in **Figure 2.3**.

Whiskers easily cleave and crack along **b**$^*$ axis. **Figure 2.4** is a scanning electron microscope (SEM) image of a ribbon end that was handled and damaged by tweezers. Cracks can be seen propagating along **b**$^*$ away from the damaged region.

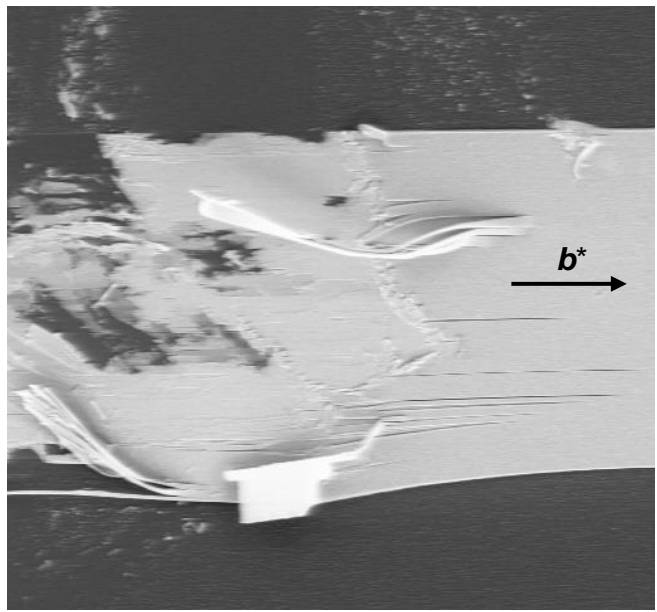

**Figure 2.4**
SEM image of a NbSe$_3$ ribbon handled and damaged by tweezers.

---

$^*$ X-ray diffraction measurements show very small mosaic widths when the crystal is rocked in the **a**$^*$-**b**$^*$ plane, but relatively large (~1 degree) mosaicities when rocked in the **b**$^*$-**c**$^*$ plane.



Larger steps are easily visible under an optical microscope as lines along the whisker. **Figure 2.5** shows images of three whiskers with various degrees of steps dominating the surface: one with many steps, one with few, and one with no visible steps on the surface. Depending on its hidden subsurface structure, the whisker in **Figure 2.5 (c)** with no visible steps may or may not be a single crystal of NbSe$_3$. Single-crystal whiskers are extremely rare. They are the most sought-after since they display the most nearly ideal collective transport properties.

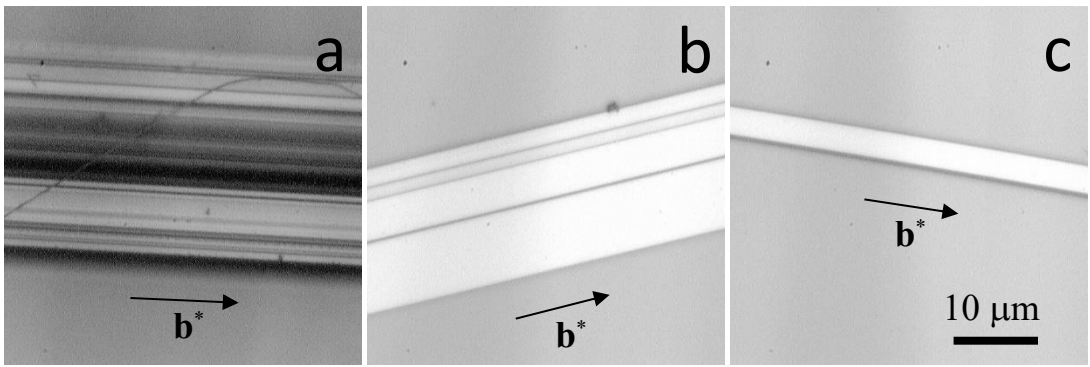

**Figure 2.5**

Micrographs of three NbSe3 whiskers: (a) a whisker with steps dominating the whole surface, (b) a whisker with few steps, and (c) a whisker with no visible steps on the surface.

When searching for a single crystal of NbSe$_3$, one initially relies on visual inspection of the whisker surface under an optical microscope. Methods like X-ray diffraction can reveal subsurface grain boundaries not visible under microscope, but as a sample selection method for transport experiments this is too involved and impractical. Typically, a more feasible procedure is to select the best-looking whisker one can find, electrically contact it, cool it down below $T_P$, and check the quality of its transport characteristics by measurement. Non-ideal whiskers with multiple domains exhibit compromised collective transport signatures such as multiple or rounded thresholds, broad or multiple peaks of a coherent oscillation



fundamental mode, incomplete mode-locking to external AC drive, large BBN level in the sliding regime, etc.[37,98-100] and can be identified based on these checks. Selection of best samples is, thus, most often performed directly by transport measurement characterization. This selection procedure is very inefficient because most prepared samples exhibit multiple crystal domains hidden below the whisker surface resulting in non-ideal CDW transport. This then requires an experimentalist to repeat the tedious process of finding a new sample, cooling it down, and performing transport characterization. As an example, more than forty visibly-step-free crystals were carefully prepared and characterized in the experiments described in Chapter 3, but only nine yielded data with useful information uncomplicated by behavior associated by multidomain samples. As discussed later in this chapter, we have developed a technique involving reactive ion etching (RIE) that can provide a relatively quick and easy method to reveal grain boundaries within NbSe$_3$ whiskers *before* they are prepared for cooldown to avoid wasting measurement time on sub-optimal samples. Another nuisance and source of inefficiency with transport experiments on NbSe$_3$ is frequent failing of electrical contacts mid experiment. Numerous tests with NbSe$_3$ described in the following sections helped us tailor a sample preparation recipe that produces high-yield, reliable, low-noise electrical contacts that can withstand multiple thermal cycling of NbSe$_3$ samples from room temperature to < 20 K.

## 2.2 Heating

Most microfabrication recipes that involve lithographic patterning of resist masks on samples call for moderate heating of samples on a hot plate or in an oven to cure the resists. Many dry



etching processes, such as plasma etching or ion milling, generate heat and elevate sample temperatures. It is thus important to be aware of $NbSe_3$ heating limitations in order to properly design a sample fabrication process that will preserve crystal quality, minimize surface oxidation, and yield good electrical contacts. To characterize how exposure to elevated temperatures affects $NbSe_3$ samples, we heated $NbSe_3$ whiskers in air, vacuum, and in an ambient where oxygen was displaced by an inert gas (argon). The resulting changes in crystal surface (appearance, conductivity, composition, oxidation) were observed.

## Heating in Air

Nine samples were prepared by mounting $NbSe_3$ whiskers on nine chips of oxidized silicon wafer. The two ends of each whisker were affixed to the substrate by Dupont silver paint (often used to electrically contact whiskers). Samples were then placed in an air-ambient oven (available at the Cornell's CCMR TOL facility in Clark Hall) near a thermocouple that monitors temperature of the oven interior. The oven was incrementally heated to nine different temperature plateaus between room temperature and 670° C over a period of a few hours. At each temperature plateau one samples was taken out of the oven for inspection while the others remained in the oven for heating to higher temperatures. Each of the nine samples was thus exposed to a different maximum temperature. The whiskers were then checked under an optical microscope for any visible changes or degradation of the surface. **Figure** 2.6 shows samples before and after heating. The samples exposed up to 185° C (samples 1, 2 and 3) remained visibly unchanged. Sample 4 heated to ~240° C changed color from its natural metallic-gray to tints of dark yellow, green, blue, and reddish-brown as shown on the color micrograph in **Figure 2.7**. Degraded surfaces also exhibited non-uniform, spotted



discoloration.  The degradation becomes more pronounced for samples heated to higher temperatures.  Samples 8 and 9, heated above 555° C, show blob-like clusters of sample material, suggesting that the high temperatures melted $NbSe_3$ forming surface-tension-defined puddles on the surface, which then froze into these non-uniform clusters on cooling to room temperature.  Heat induced Se depletion from the bulk perhaps yielded the more stable $NbSe_2$ as a final product.



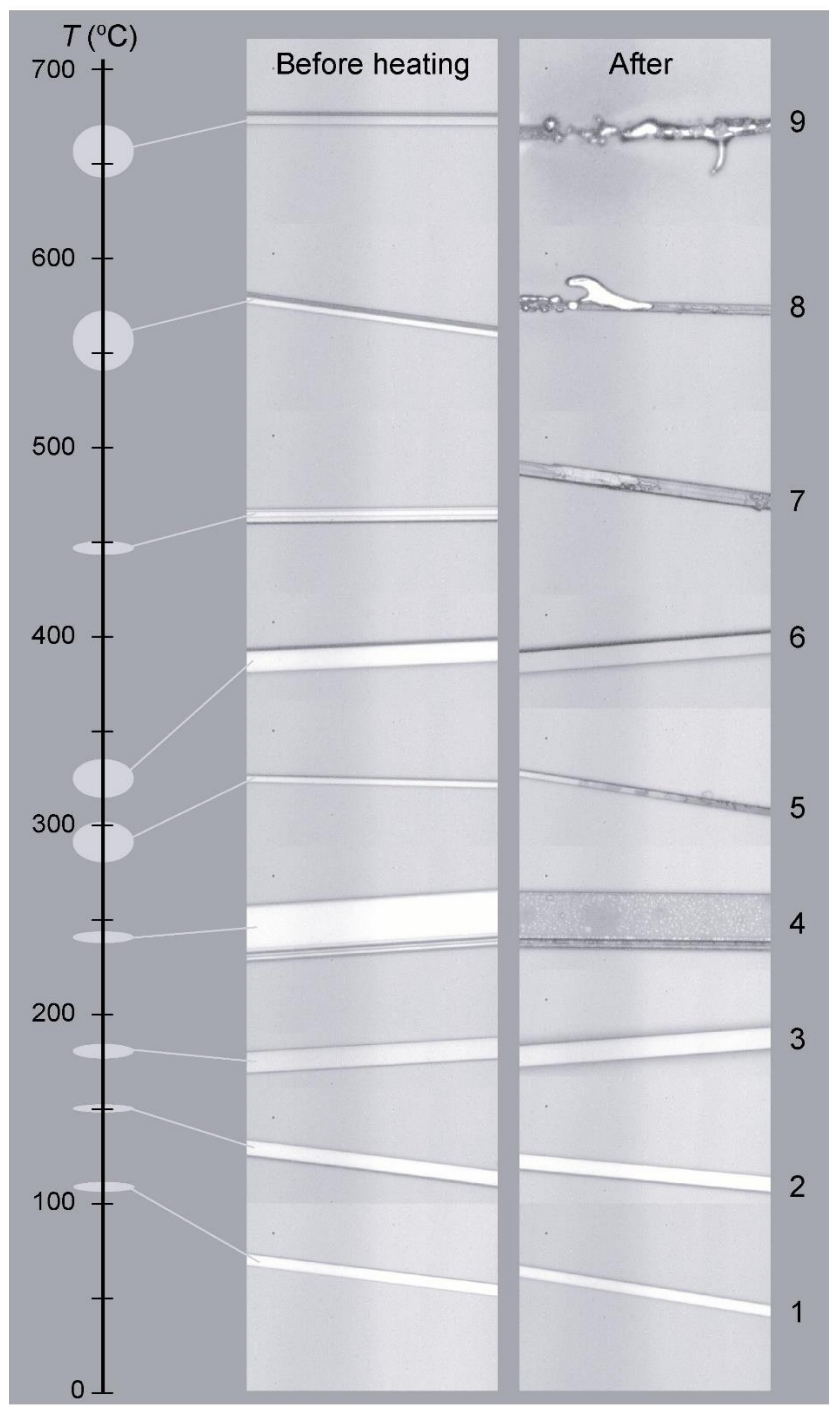

**Figure 2.6**

Black and white micrographs of NbSe₃ samples before and after heating in air ambient. Silver paint was used to mount whiskers to substrates and was present in the oven during heating. Light gray ellipses on the temperature scale indicate the range of temperature variation at each temperature plateau.



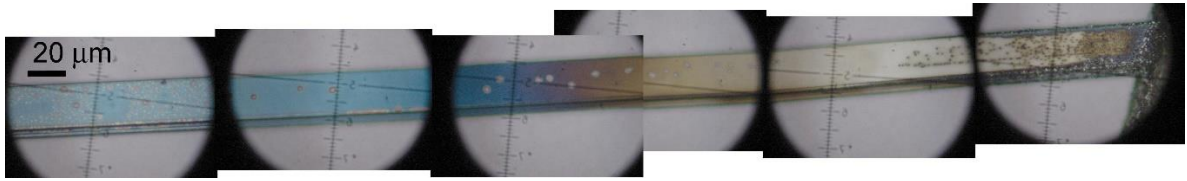

**Figure 2.7**

Composite of color micrographs of sample 4 in **Figure 2.6** after heating in air ambient oven to ~240° C.  Silver paint residue is visible on the right end of the whisker.  Silver paint was used to affix the whisker to the substrate.

**Figure 2.7** shows grainy silver paint residue extending over the whisker away from the silver paint blob bonding the right end of the whisker to the substrate.  This residue was not present on the whisker before heating, suggesting that silver paint becomes unstable when heated.  To clarify whether the surface color change away from the silver paint blob observed on heating above 240° C is due to surface degradation or due to re-condensation of silver paint residue, the tests were repeated on a second set of whiskers using carbon paint (often used to mount samples for scanning electron microscopy) instead of silver paint.  Carbon paint is a suspension of graphite particles in isopropyl alcohol. The alcohol evaporates completely when heated leaving graphite, which is non-volatile and stable when heated to several thousand degrees C.  **Figure 2.8** shows this set of samples before and after heating in air.



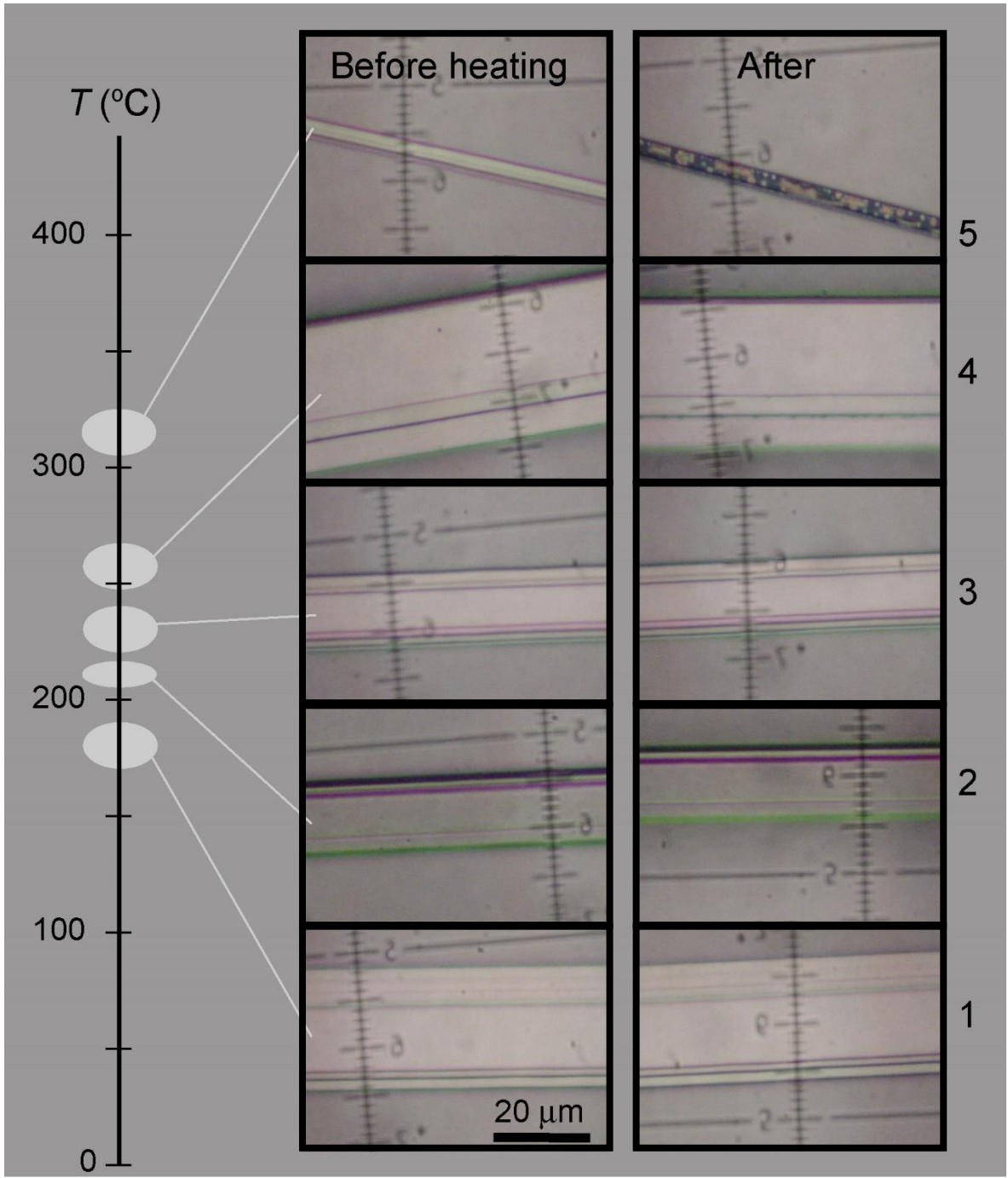

**Figure 2.8**

Color micrographs of NbSe$_3$ samples before and after heating in air ambient oven. Whiskers were affixed to substrates with carbon paint which after drying does not outgas or degrade upon heating. Light gray ellipses on the temperature scale indicate the range of temperature variation at each temperature plateau.



Samples held by carbon paint and heated to 230° C appear unchanged. Degradation first appears on the sample 4 heated to 255° C as faintly visible dots along the lower step on the whisker. Neither sample 3 (heated to 230° C), nor sample 4 (heated to 255° C) exhibit the surface discoloration of silver-paint-mounted sample 4 (240° C) of **Figure 2.6** and **2.7**. This confirms that heat-induced outgassing from silver paint is the likely cause of the drastic color changes observed in **Figure 2.7**. Melting point of silver is 961° C, which is well above the temperatures we investigated, but the organic components in the colloidal silver compound likely decompose and degrade at much lower temperatures and could thus be responsible for discoloration of NbSe$_3$ surfaces observed in **Figure 2.6** and **2.7**. Based on this we advise that one should avoid using silver paint in fabrication recipes that call for sample heating.

Both tests (with silver paint and carbon paint) suggest that it may be safe to heat samples at least up to 200° C in air ambient before degradation of NbSe$_3$ surface can be visually observed.

Next, we tested how heating in air ambient affects surface conductivity of a NbSe$_3$ whisker. A set of three whiskers were mounted on alumina substrates using carbon paint and were heated to 230° C - 246° C in air ambient oven for two hours. Each whisker was then contacted by fresh carbon paint near the two ends. Each whisker was biased by 10 μA current, and a voltage drop across the whisker (including contacts in this 2-probe configuration) was measured. The corresponding resistances were found to be in a 0.1 MΩ - 2 MΩ range. This range is 2-3 orders of magnitude larger than typical resistance values measured across fresh, non-heat-treated whiskers of similar size that were contacted and measured in the same way



suggesting that heating in air ambient can induce changes in surface or bulk (or both) of NbSe$_3$ that affect sample conductivity.

Since most resists used in micro- and nano-fabrication are cured on hotplates, we also wanted to investigate how whiskers were affected when heated on a hotplate in air ambient. A set of whiskers was mounted onto oxidized silicon substrates using carbon paint. A thermocouple was placed in contact with the hotplate surface to measure the temperature. Samples were heated for 12-13 minutes to different temperatures: 120° C, 150° C, 170° C, 190° C, 210° C, 240° C, 270° C. Whiskers heated up to and including 190° C did not show surface changes when inspected under an optical microscope. Whiskers heated to 210° C and above became discolored, acquiring a brownish tint on the surface. The whiskers were allowed to cool to room temperature and were then contacted by silver paint. Each sample was biased with a 10-μA current and a voltage drop was measured across each whisker. The measured voltages were much noisier than the ones measured on fresh whiskers contacted in the same way, and all resistances were above 100 kΩ, two orders of magnitude larger than for bulk NbSe$_3$ samples.

To test if the resistance increase is associated with a surface effect rather than a change in bulk resistance, we plasma etched one of the previously heated whiskers using reactive ion etching (RIE) to remove a thin surface layer. Carbon paint contacts were then applied over the freshly etched surface. The low resistance measured in this sample was consistent with measurements on fresh, unheated whiskers. This implies that the resistance increase observed upon heating samples to moderate temperatures in air ambient is associated with surface



rather than bulk sample degradation and is most likely a result of heat-accelerated oxidation of surfaces in an ambient containing oxygen.

To investigate this hypothesis further, a Scanning Auger Nanoprobe (PHI-670) was used to determine surface atomic composition (typical analysis depth ~ 5-50 Å). Two samples were investigated: a fresh, non-heated $NbSe_3$ whisker, and a $NbSe_3$ whisker that was heated in air ambient at 135° C for 30 minutes[*]. Inspection under optical microscope revealed that the surface of the heated crystal had acquired a slight yellowish-brown tint. (Note that this contrasts with our observation that samples in **Figure 2.6** and **2.8** remained unchanged up to temperatures well above 135° C. Heating in air ambient at temperatures lower than 200° C can produce discoloration of a whisker surface *if heating is prolonged*.) **Figure 2.9** shows concentrations of Nb, Se, O, and C atoms on the surfaces of the two whiskers. The Nb fraction is almost identical in both samples, but the data suggests that heating in air replaces a large fraction of selenium by oxygen. This is consistent with a known fact that niobium oxides have greater stability over niobium selenides[†].

---

[*] This heating procedure is a step in a process to cure polyimide, a polymer compatible with standard fabrication practices that can be used as a glue to affix whiskers to substrates. See Recipe 2 in Appendix C.
[†] Oxygen in the air reacts with both Nb and Se. Niobium oxide is extremely stable and non-volatile, while the $SeO_2$ sublimes readily at few hundred degrees C.



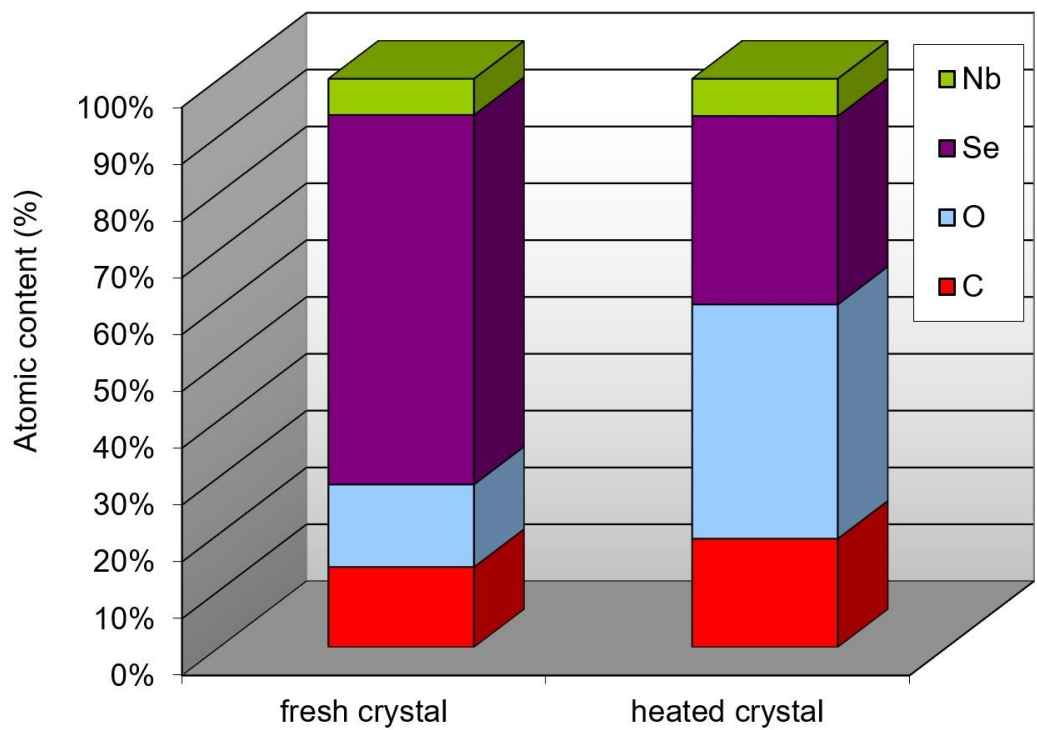

**Figure 2.9**

Results of an Auger analysis performed on surfaces of two NbSe₃ samples: an as-grown fresh crystal, and a crystal heated in air ambient for 30 minutes at 135° C. Heated sample had acquired a brownish–yellow tint.

The ratio of Se to Nb atoms in the fresh crystal is observed to be 9.5:1 as opposed to expected stoichiometric ratio of 3:1. Larger content of Se on the crystal surface is not surprising since the crystals are grown in an ambient with overpressure Se, and at the end of the growth process, when the vapor transport oven is cooled to room temperature, the excess Se will condense and solidify on the crystal surface most often in a form of small dots typically around a hundred nm tall and several hundred nm wide with density of a few dots/$\mu m^2$ (also see section 2.6 on surface dots). If the cross-section of the nanoprobe beam is large enough to encompass many dots while probing the surface (likely), then the measurement can grossly



overestimate the fraction of Se within the whisker surface. Note that even the heated whisker surface was found to have a ratio above 3:1. The loss of Se due to heating is most likely caused by outgassing of adsorbed and surface-localized Se rather than by removal of Se from the bulk of the crystal. This is consistent with our observations of nominal CDW transport signatures in some moderately heated whiskers.

The presence of observed carbon in the analysis was expected due to unavoidable presence of some carbon in the Auger analysis chamber. Since Auger probe is not sensitive to chemical composition, but only to the atomic content of the surface, it is impossible to tell what exact chemical species were formed on the surface of the heated crystal. The species could include resistive oxides (likely) and/or resistive organic compounds including C, H, and O (less likely) which could explain why heated $NbSe_3$ samples exhibited higher resistance contacts in our tests. Whatever the species, the rates of reactions forming them depend both on temperature and on presence of the reactants in the heating environment. This should be taken into consideration when creating new sample preparation recipes that involve heating of $NbSe_3$ in various environments.

## Heating in Argon

We next investigated heating of samples in an inert gas environment. $NbSe_3$ whiskers affixed to substrates by carbon paint were heated in argon gas. The samples were placed in a quartz tube which was then placed in the oven. During heating the tube was purged with an Ar flow of 0.75 $ft^3$/hour to maintain $O_2$-free environment. A thermocouple probe placed inside the



tube recorded the temperature. Samples were heated individually to different temperatures: 214° C, 247° C, 287° C, 313° C, 340° C, 383° C, and 443° C, for 5-10-minute intervals.

In contrast to observations on whiskers heated in air, no visible degradation or discoloration or other change of the surface was observed under optical microscope on any of the whiskers.

## Heating in Vacuum

To further test the idea that the degradation of whiskers heated in air ambient above ~220° C is primarily due to oxidation, we heated a set of three whiskers affixed to an alumina substrate by carbon paint in vacuum at a base pressure of $1.3 \times 10^{-6}$ torr – an environment where most oxygen has been removed. The whiskers were heated for two hours to a temperature of T = 300° C. The oven was then left to cool under vacuum overnight.

There was no visible degradation of the surface on either of the three whiskers (no discoloration observed when the crystals were heated in air ambient to 255° C). In fact, after heating the crystal surfaces appeared clean, and were free of selenium "dots" that were present before heating and that are often observed on crystals post growth (see section 2.6 on surface dots). After cooling, the whiskers were contacted with silver paint. When biased with 10 µA current, the voltage drop across each whisker was stable to roughly one part in 10,000, and the measured resistances below 200 Ω were comparable to those obtained with freshly grown whiskers. Heating crystals in vacuum up to 300° C therefore does not compromise the surface conductivity. Many samples that have undergone heating in vacuum have later been investigated below Peierls transition temperatures, and it was found that their CDW transport properties (including CDW depinning thresholds which are extremely sensitive to crystal



imperfections) do not differ from the nominal transport properties of crystals that have not been heated.

We summarize the results of the heating tests:

- Heating in air ambient (oven or hotplate) appears to oxidize NbSe$_3$ surfaces yielding highly resistive, noisy electrical contacts. While it decreases surface conductivity, heating whiskers to moderate temperatures in air appears to leave bulk conductivity unchanged. Surfaces begin to visibly degrade when whiskers are heated above ~200° C on a timescale of a few minutes, and above ~550° C whiskers disintegrate. Surface discoloration can appear for temperatures below 200° C when the heating is prolonged (i.e. 30 min at 135° C).

- Heating in inert environments such as argon (up to 443° C) and vacuum (up to 300° C) did not produce visible surface changes.

- Whiskers heated in vacuum to 300° C retain conductive whisker surfaces, comparable to surfaces of fresh whiskers, which enables highly stable low-resistance electrical contacts to be produced by applying conductive paints.

- Fresh whiskers have surfaces coated with adsorbed selenium (from growing in selenium overpressure). This surface adsorbed selenium evaporates when samples are moderately heated, but selenium out-gassing from the whisker bulk does not appear to be significant (see section 2.6 on surface dots).



- Use of silver paint in fabrication recipes that call for heating of samples should be avoided.

## 2.3 Resistance to Chemicals

NbSe$_3$ whiskers were soaked in solvents commonly used during processing for cleaning wafer surfaces and for stripping resists, including isopropanol, acetone, and methylene chloride, on multiple occasions for periods over 24 hours. These solvents do not degrade the quality of the crystals as revealed by inspection under optical microscope, contact resistance measurements, and collective transport properties. Contact with polyimide (Probromide 285) and standard resists (Shipley resists used in optical lithography; and PMMA, copolymer, and UV5 used in e-beam lithography) does not appear to affect the whiskers. However, a whisker coated with PMMA resist and heated for extended period of time in air ambient can acquire "processing dots" on its surface (see a discussion in section 2.6 on dots induced by processing). Whisker surfaces were also found to be resistant when exposed to certain aggressive solutions and do not degrade when in contact with chemicals used in gold electroplating including a cleaning solution consisting of hydrofluoric acid highly diluted in buffered oxide etch solution, and a sulfate-based gold-plating solution called Microfab Au.

## 2.4 Mounting to Substrate

Standard microfabrication tools and processes are optimized for processing of thin films on size-standardized substrates like silicon or sapphire wafers. Most often the machines cannot handle non-standard samples like three-dimensional microscopic crystals that come in varying



shapes and sizes. The easiest way to adapt whiskers to standard processing tools and techniques is to affix them to standard substrates and then process them on top of the substrate. Here we describe several methods to mount a whisker on a substrate and discuss how each method is suited to different processing requirements.

When carefully laid down flat on a clean silicon wafer, ribbon-like whiskers of $NbSe_3$ usually adhere well to the wafer by Van der Waal's interactions. This adhesion can be made more reliable by carefully applying a small drop of isopropyl alcohol over the whisker on the substrate and letting it evaporate. The alcohol fills the gaps between the wafer surface and the whisker, and as it slowly evaporates its surface tension pulls the bottom whisker surface into a good contact with the wafer. Isopropyl alcohol is not the only liquid that can be used, but it is a good choice since it evaporates quickly and completely without leaving any residue on either the whisker or the wafer, and it is does not chemically react with crystals.

This technique provides a simple way to attach whiskers to substrates, and samples prepared in this way can withstand usual handling and transporting well. The whisker is not permanently attached, and it can be separated from the wafer by tweezers and repositioned if needed. It is even possible to dispense resist onto a spinning sample (by utilizing standard cleanroom resist spinners) without washing the whisker off the wafer. However, it is difficult to do this reliably, and the whisker must be carefully positioned in the center of the wafer (i.e. in the center of rotation) to minimize the forces it experiences during rotation and decrease chances of being washed off the wafer with excess resist. Another downside is that whiskers mounted on a substrate in this way cannot withstand "wet" processing, i.e. upon immersion in



wet agents like acetone or isopropanol (for stripping resist, or cleaning) the whisker usually detaches from the substrate.

We have developed a process to permanently affix a whisker to a substrate by using polyimide (Probromide PI-285) as a glue. Cured polyimide is chemically resistant to almost all solvents used in standard processing, it is a good dielectric, it withstands cooling to 10 K without cracking or other structural degradation, and it can be spun on a substrate by using a standard resist spinner. Polyimide is first spun on a bare substrate. While still uncured and wet (within two hours or so), whiskers are carefully dropped on it. The samples are then heated in an oven evacuated by a turbo-molecular pump to cure the polyimide. The vacuum environment minimizes oxidation of the whisker surface. Whiskers attached to the substrate in this way can be manipulated during processing without fear of loss or damage.

A downside of this method is that polyimide is highly resistive, so that a silicon wafer coated with it traps charges when illuminated by an electron beam. Charged samples are difficult to characterize by SEM and limit the use of electron beam lithography. The remedy is to dilute the polyimide, allowing it to be spun as a very thin film. We have reliably glued whiskers using 60 nm thick polyimide films. Thinner films are more "transparent" to electron beams and tend to charge less than thicker films. Excess polyimide on the substrate away from the whisker can be removed by a short exposure to oxygen and $CF_4$ plasma easily without notable damage to the whisker when the polyimide is thin. The full process for attaching $NbSe_3$ whiskers to substrates by use of polyimide is given in Recipe 2 in Appendix C.

Whiskers mounted to substrates in this way can withstand a multitude of processing steps. For example, resist can be spun over the sample, baked, developed and stripped multiple



times. A sample can be exposed to etching, and other materials (e.g., evaporated metal contacts) can be added without fear that a whisker would delaminate from the substrate. Polyimide can also be used to fully encapsulate a whisker on top of a substrate. Polyimide applied over the whisker can also be patterned to expose/uncover desired parts of the crystal (this can be easily achieved by use of a photo-exposable polyimide like PWDC-1000). This is an essential step in our recipe that utilizes gold electroplating in order to imbed metal probes into the bulk of the crystal cross-section (see section 2.8. and Recipe 5 in Appendix C).

This method for permanently attaching whiskers to substrates was a first step in integrating NbSe$_3$ whiskers with standard clean-room processing techniques. The ability to perform lift-off and multiple lithography steps on a whisker allows us to develop and apply different processing techniques to produce complex structures with bulk NbSe$_3$ samples.

## 2.5 Dry Etching Techniques

When it comes to control and precision of patterning, dry etching by plasma techniques such as reactive ion etching (RIE) and ion milling (IM) are far superior to techniques that use wet chemical agents to transfer a pattern from a resist mask into a thin film below. We have investigated and characterized dry etching of NbSe$_3$ by various reactive ion plasma species in RIE, and by milling with argon ion beams. These dry etch techniques can be used to shape whiskers into a variety of geometries in order to probe different aspects of CDW physics. These techniques are essential in fabricating heterostructures and devices; they can be used for contacting whiskers with electrical probes, or simply to thin a whisker to a desired thickness.



The first task was to identify useful plasma species and required etch parameters (such as power, pressure, and gas flow) to optimize etching of NbSe$_3$ as well as to determine its etch rates. In characterizing the effects of etching on NbSe$_3$, we have paid special attention to post-etch surface roughness and surface conductivity. Both attributes are relevant when it comes to contacting whiskers with electrical probes as well as in preserving the intrinsic properties of CDW transport in measurements. We found that controlled etching can also be a powerful technique to tap beneath the surface of the whisker and reveal hidden grain boundaries that significantly affect collective transport properties.

## Surface of Unprocessed Whiskers

To establish a baseline for the etch experiments presented in the following sections we have analyzed the surface of a fresh, unprocessed NbSe$_3$ whisker by atomic force microscopy (AFM). **Figure 2.10** are AFM images of a crystal surface, one of a patch area of approximately 6 $\mu m^2$ and the other of 1 $\mu m^2$. The surface is devoid of dots that are sometimes visible on whiskers after crystal growth (see section 2.6 on surface dots). The z-range of the surface (or maximum achieved span of AFM tip amplitude during scan) is 2.4 nm, with 95 % of surface points spanning a height of less than 1 nm. Typical length-scale associated with a lateral variation in surface texture is on the order of 50-150 nm (as one moves across the surface, high areas change to low areas over this length-scale). No discernable crystal terraces or atomic steps were observed. It should be noted that in our measurement no attempt was made to carefully prepare the surfaces, optimize the AFM scanning, and reduce the noise in order to resolve lattice order. We thus expect that the height variations we observe are within the noise of the measurement. These images were taken in a quick fashion



to obtain an estimate of the surface roughness for base comparison with samples exposed to different etch processes, which have much rougher surfaces (see **Table 2.1**).

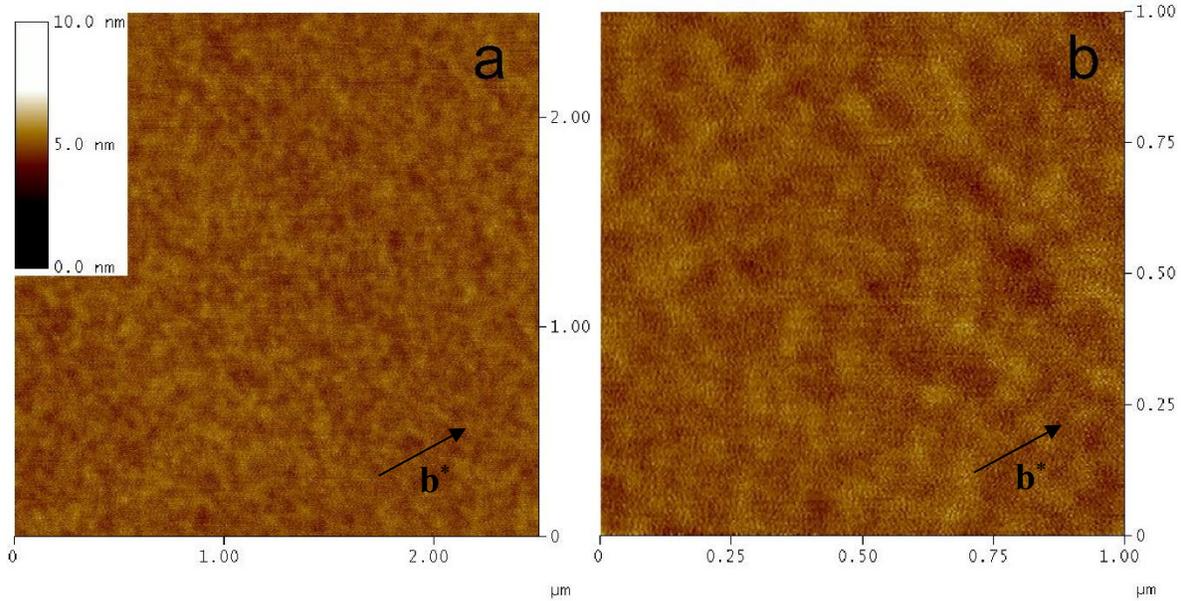

**Figure 2.10**

AFM scans of (a) approximately 6 μm² area, and (b) 1 μm² area of a surface of an as-grown (unetched) NbSe₃ crystal. Here **b**$^*$ long axis of the whisker makes an angle of approximately 30° with the horizontal.

## SF₆ Plasma Etch

Mantel *et al*. [93,94] reported patterning ribbons of NbSe₃ samples using a dry etching technique. A resist mask pattern on top of the ribbon was transferred into the crystal by an SF₆ RIE plasma. A wide crystal ribbon was shaped into a narrower sample with side probes of NbSe₃ that were subsequently contacted by gold, as shown in **Figure 2.11**. The desired crystal pattern was produced by etching away the unmasked parts of the crystal while the masked parts were protected from the SF₆ plasma.



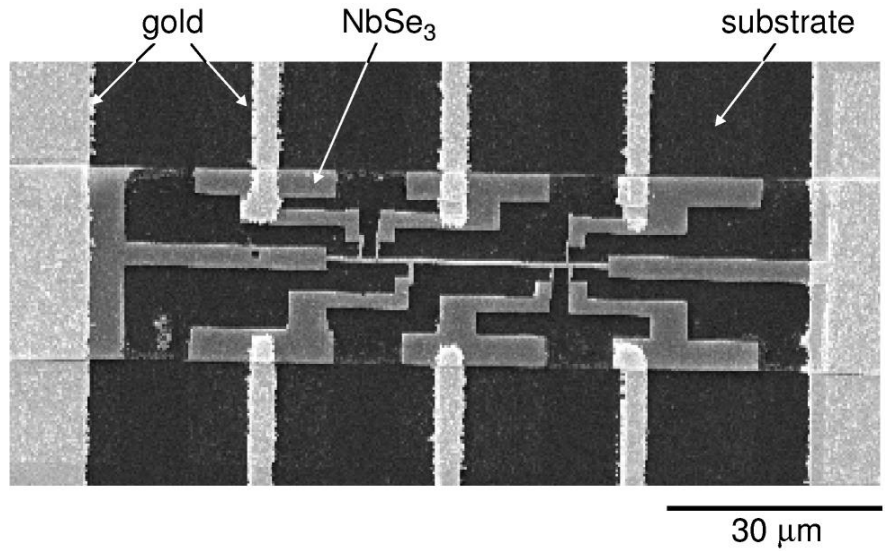

gold    NbSe₃    substrate

30 µm

**Figure 2.11**

NbSe$_3$ ribbon patterned into a narrow crystal segment with side probes shaped from the same ribbon.  Side probes were subsequently contacted by gold.  The crystal was patterned by SF$_6$ plasma etching through a resist mask.  Image obtained from reference [94].

In addition to this lateral patterning of whiskers by masking and completely removing parts of the whisker, the ability to control vertical patterning would also be very useful, i.e. to thin crystals in a controlled way to arbitrary desired thickness.  Crystal thickness is a very important parameter in CDW transport, because the transverse phase-phase correlation length, the pinning energy per phase correlated domain, and the depinning electric field are typically limited by the sample thickness.

Following the work of Mantel, we used SF$_6$ plasma etching to thin and pattern NbSe$_3$ whiskers.  Initial tests showed that prolonged etching even at moderate plasma powers can destroy NbSe$_3$ crystals.  **Figure 2.12** shows a series of micrographs of two NbSe$_3$ crystals before and after RIE etch by SF$_6$ plasma at a moderate RIE power setting of 45 W for a total



of 4.5 minutes. The etching was performed in three 1.5-minute intervals, with short pauses

after each etch step to allow for sample cooling.

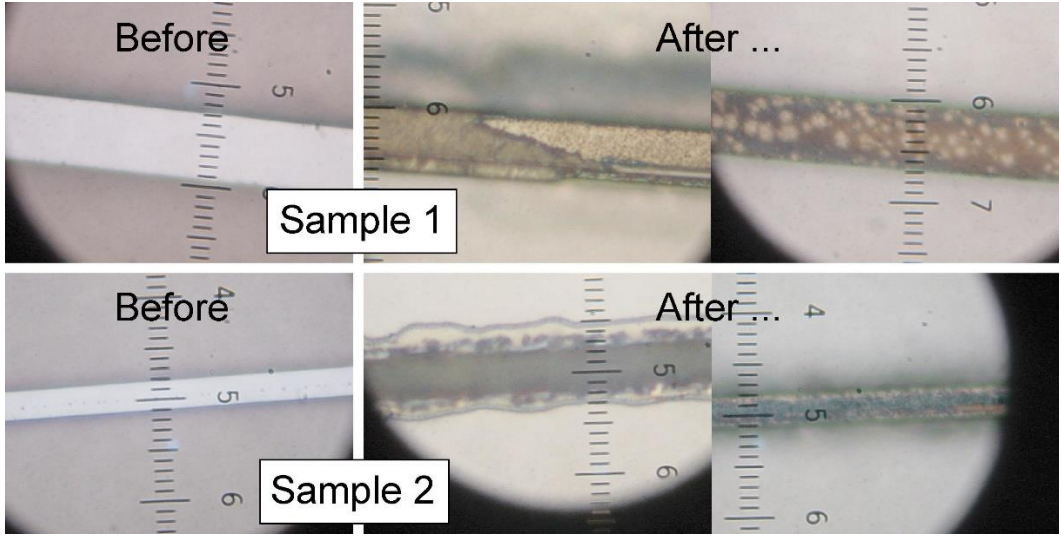

**Figure 2.12**

Two NbSe$_3$ whiskers before and after 4.5 minutes etch by SF$_6$ plasma at RIE power of 45 W, 15 mTorr pressure, and 20 sccm SF$_6$ gas flow. Etching was done in 1.5-minute intervals, with short pauses of several minutes between etching to allow for sample cooling. The two images after the etch (on the right) show different sections of the etched sample.

After etching, these crystals somewhat resemble the ones heated in air ambient described in

section 2.2. The observed damage is most likely caused by excessive heating during the etch

and exposure to prolonged moderate-power plasma despite our attempt to minimize heating

by performing the etch in several short intervals. Note that heating here occurs in the

environment mostly depleted of O$_2$ but in the presence of energetic ions and radicals. The

following tests suggest that etching should be performed at very low powers to further

minimize heating.



After carefully manipulating etching parameters towards a less aggressive etch we determined the optimal parameter space for etching $NbSe_3$ by $SF_6$ plasma. We then characterized the etched whisker surfaces by optical, SEM and AFM microscopy. Appendix B gives the optimal parameters found with corresponding etch rates.

$SF_6$ plasma etching produces topography on the whisker surface that is quite distinct from those produced by use of other etch techniques. This topography is not easily discernable under an optical microscope due to small feature sizes. In fact, most often the etched surfaces do not appear different in any respect from the etched ones (including no change in color) under an optical microscope. Images at higher magnifications, obtained by use of SEM and AFM, reveal pitted topography that somewhat resembles distorted honeycomb-like structure as shown on **Figure 2.13**, **2.14**, and **2.15** for two different $NbSe_3$ samples exposed to $SF_6$ plasma etching. The sample in **Figure 2.13** was first exposed to a short $O_2$ plasma etch[*] (to remove polyimide glue around the whisker in order to prevent charging during SEM imaging), followed by an $SF_6$ etch for 120 s that removed ~200 nm from the surface of $NbSe_3$.

AFM scans in tapping mode confirm that the observed structures are indeed "pits" rather than "bumps" (depressions and protrusions are sometimes hard to distinguish in SEM images). **Figure 2.14** shows an AFM image acquired on the same sample. Typical pit depth was found to be in a range of 10-20 nm (or 5% - 10% of the thickness removed by the etch) with lateral pit sizes between 150-500 nm.

---

[*] $O_2$ plasma was shown not to etch $NbSe_3$ (see **Table 2.1).**



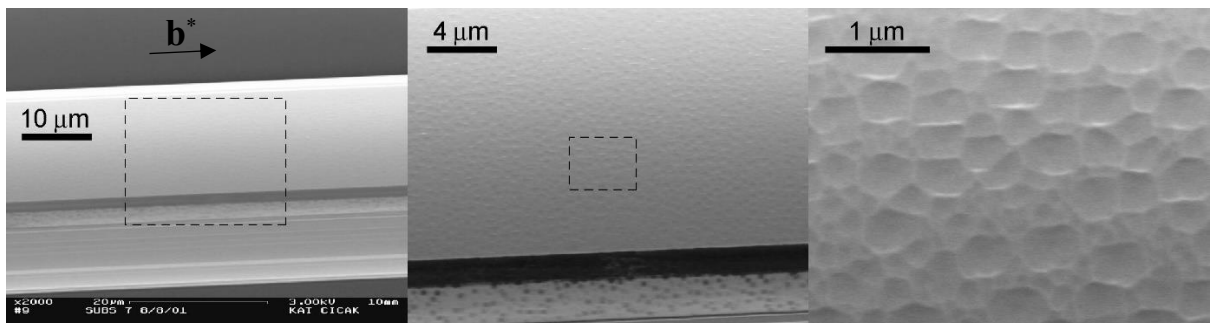

**Figure 2.13**

Three SEM images of a NbSe₃ whisker surface at progressively higher magnifications, i.e. each dashed rectangle is a magnified area showed on the right of the image. The whisker was mounted with polyimide on an oxidized silicon wafer. The sample was first exposed to $O_2$ plasma (30 sccm, 30 mTorr, 90 W) for 20 s to remove excess polyimide from the substrate around the whisker, and then etched by $SF_6$ plasma (20 sccm, 15 mTorr, 20 W) for 120 s which removed ~ 200 nm of NbSe₃ from the surface.

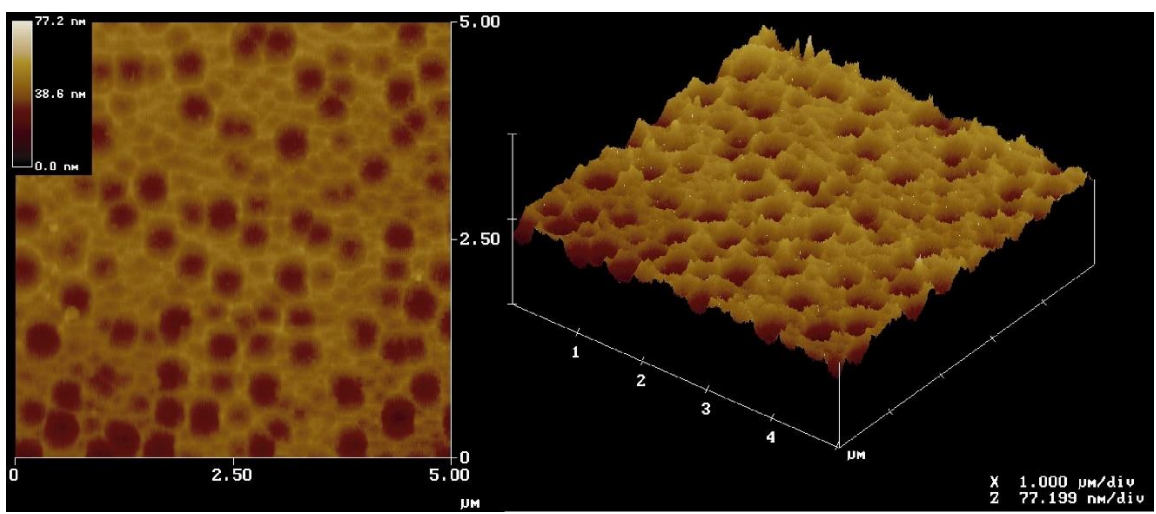

**Figure 2.14**

AFM images of the sample shown in **Figure 2.13**.

**Figure 2.15** shows another crystal etched by $SF_6$ plasma for the same duration and with the same etch parameters that were used on sample in **Figure 2.14** but without a prior exposure to $O_2$ plasma or any other treatment. A similar honeycomb-like pattern is observed, confirming



that the characteristic topography is formed due to $SF_6$ plasma etching, and not due to $O_2$ plasma. Typical pit depths and widths on this sample were found to be 15-30 nm and 100-200 nm, respectively, similar to what was observed on the previous sample.

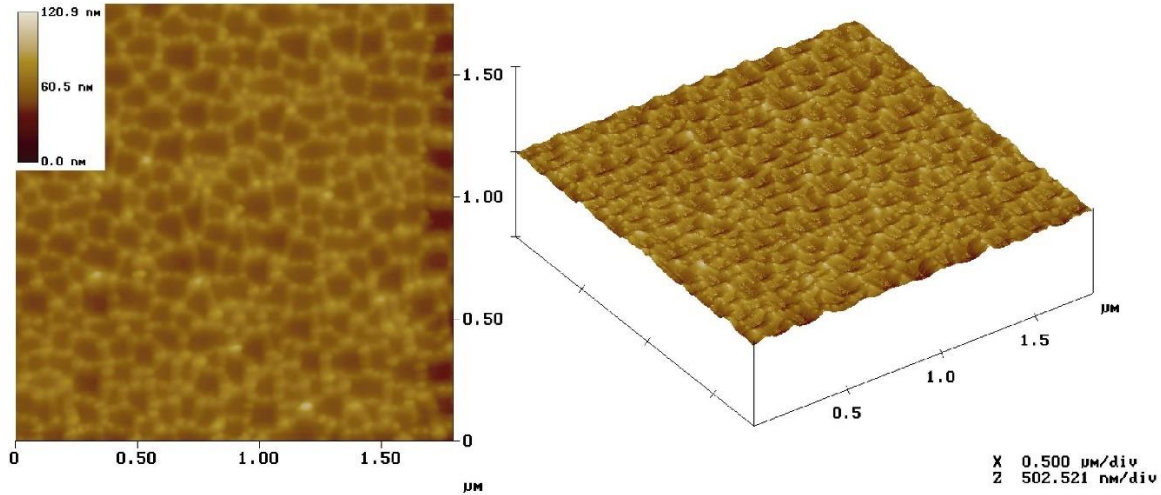

**Figure 2.15**
AFM images of another $NbSe_3$ sample that was etched in $SF_6$ plasma (20 sccm, 15 mTorr, 20 W) for 120 s without prior $O_2$ etching. Approximately 200 nm of whisker was removed. The whisker was mounted on alumina substrate by applying silver paint to whisker ends.

Etching samples at a higher $SF_6$ gas pressure and flow (30 mTorr, 30 sccm instead of 15 mTorr, 20 sccm) but at same powers (20 W) and for the same duration (120 s) removed substantially more $NbSe_3$ (about 600 nm as compared to 200 nm), but did not significantly affect the surface configuration or roughness. The pit depth range increased only slightly to 20-50 nm for the more aggressive etch. Due to time constraints we did not more systematically evaluate how surface roughness depends on the etch length and thickness of material removed.



Whiskers etched by $SF_6$ plasma and subsequently contacted by silver paint revealed contact resistances comparable to those measured on fresh whiskers. On whiskers with the surface conductivity compromised by prior chemical or other (e.g., $O_2$ plasma) treatment, a short $SF_6$ plasma etch can be used to "recover" surface conductivity by removing a thin resistive layer from the surface of the whisker. Even though the $SF_6$ etch produces a relatively rough surface on $NbSe_3$, (compared to a surface of an un-processed fresh whisker shown in **Figure 2.10** with at least an order of magnitude smaller roughness variations in z-direction) the resulting surface is clean, conductive and allows formation of low resistance contacts.

We note that a prolonged $O_2$ plasma etch to strip/remove a thick layer of resist can leave organic scum on the surface. This scum is difficult or impossible to remove and will remain on the surface even after exposure to $SF_6$ etch, as shown in the SEM images in **Figure 2.16**.

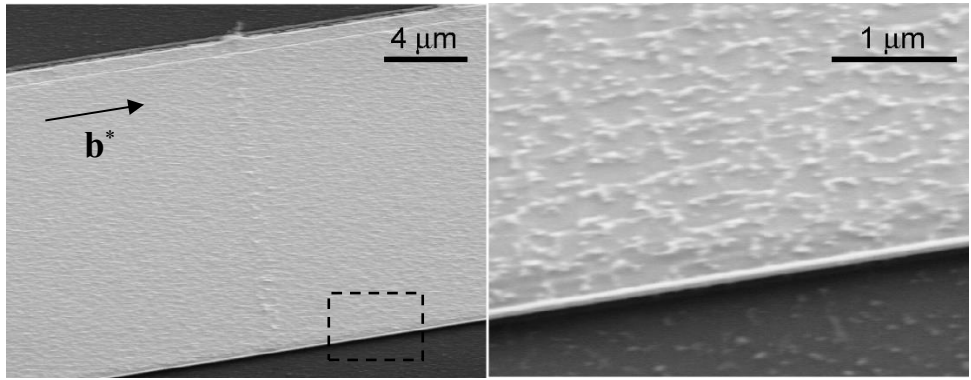

**Figure 2.16**

SEM images of a $NbSe_3$ whisker first exposed to a prolonged $O_2$ plasma etch to strip a thick layer of resist from its surface, and then exposed to $SF_6$ plasma for etching. The scum (most likely cross-linked resist or re-sputtered organic residue from resist) is visible both on the surface of the whisker and on the substrate next to the whisker in the magnified image on the right (indicated by the dashed rectangle in the left image).



We recommend that all resist stripping be performed by using wet chemicals instead of by dry etching (e.g., electron beam lithography resist PMMA with a methylene chloride and acetone mixture in 9:1 proportion, and optical lithography resists with acetone).

$SF_6$ plasma provides relatively high etch rates of $NbSe_3$ even at the lowest available plasma bias settings, i.e. powers of 10-20 W (see Appendix B). $SF_6$ plasma can be useful when parts of a whisker are to be completely removed from the substrate (as in the structure in **Figure 2.11**), but for thinning crystals in a controlled way, especially down to thicknesses below 100 nm, etching by $SF_6$ plasma may not be suitable. At high etch rates, terminating the etch properly to achieve a precise crystal thickness is difficult. In addition, the substantial surface roughness (10-50 nm deep pits) produced by $SF_6$ etching renders a large fraction of the whisker cross-section non-uniform which can produce undesirable transport effects we mentioned earlier. Minimum required power (DC bias) to maintain plasma in the chamber prevented us from exploring the regime with lower etch rates. Another way to decrease the etch rate is to use a plasma specie that is less reactive with $NbSe_3$.

## $CF_4$ Plasma Etch

In search of a less aggressive and more controllable etch process that also yields smoother surfaces, we explored $CF_4$ plasma etching. With fewer available reactive fluorine ions per molecule compared to $SF_6$, $CF_4$ plasma etching indeed produces lower etch rates and appears to yield smoother surfaces. Lower etch rates provide much better control in achieving a desired crystal thickness, particularly when there is a need to remove a very thin layer of $NbSe_3$ or to produce a very thin whisker on a substrate. AFM scan in **Figure 2.17** shows a



whisker surface after it is exposed to a $CF_4$ plasma etch.  Inspection of this sample under an optical microscope prior to the plasma exposure revealed a presence of dots that are sometimes present on whisker surfaces after crystal growth (see section 2.6).  After the etch the AFM images reveal scattered bumps with a density similar to the density of dots observed before the etch.  Most likely these structures are dot remnants and/or patterns transferred into the whisker material due to masking effects of the dots during the etch.

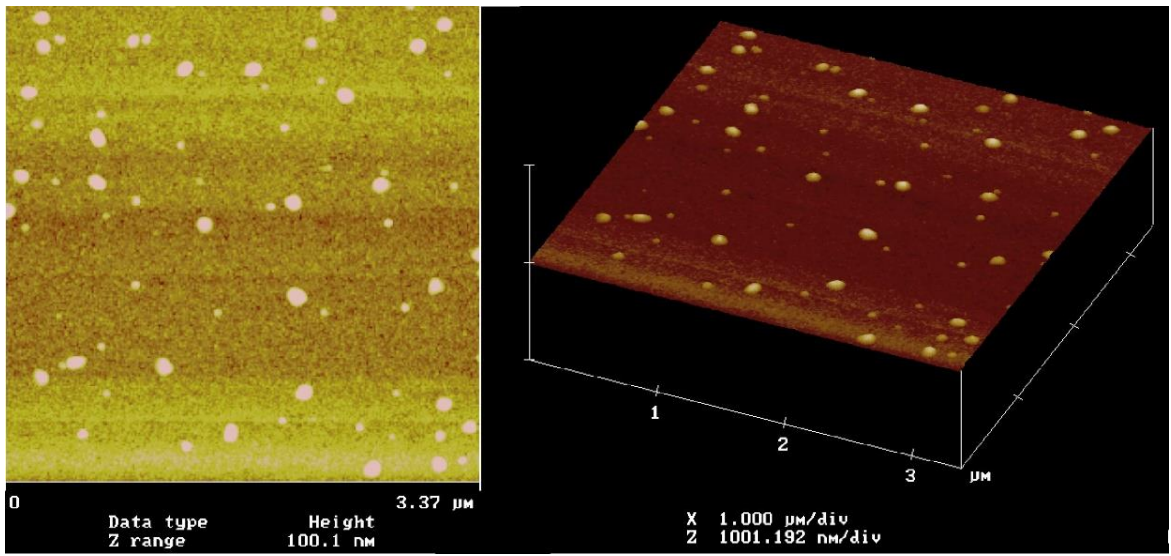

**Figure 2.17**

AFM images of $NbSe_3$ surface after exposure to $CF_4$ plasma etching (30 sccm, 30 mTorr, 45 W) for 120 s.  40 nm of sample thickness was removed by the etch.  Prior to etching, the sample surface exhibited dots with density of ~ 5.8 dots/$\mu m^2$.  After the etch a similar density of bumps (up to 40 $\mu m^2$ tall) was observed on the surface.  In the flat areas away from the bumps, 95% of surface points span a height range of approximately 7 nm.

Dots present on a freshly grown whisker can, and should, be removed prior to plasma etching by heating the whisker in vacuum (further discussed in section 2.6).  This eliminates masking by dots during etching and reduces additional surface roughness.



In the flat areas of the sample surface, away from the bumps in **Figure 2.17**, 95% of surface points span a height range of approximately 7 nm (compared to 1 nm on untreated samples) after exposure to the $CF_4$ etch where approximately 40 nm of $NbSe_3$ was removed. Whiskers etched with $CF_4$ plasma were contacted with silver and graphite paints, and these tests revealed that etched surfaces exhibited contact resistances comparable to the surfaces of fresh crystals. $NbSe_3$ samples thinned by a $CF_4$ plasma etch can thus successfully be contacted electrically.

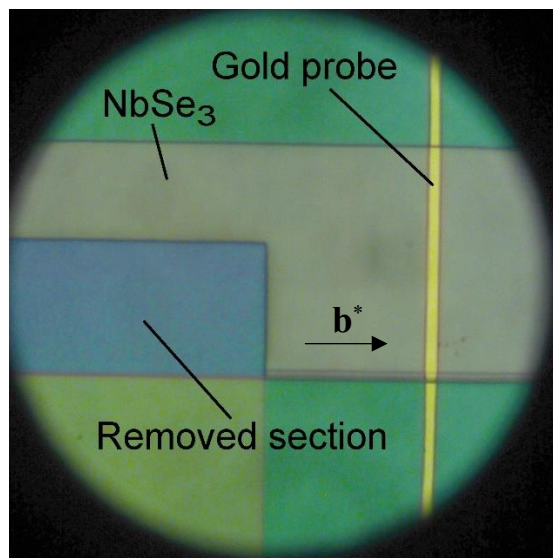

**Figure 2.18**
$NbSe_3$ whisker from which a section was removed by a $CF_4$ etch.

**Figure 2.18** shows a $NbSe_3$ crystal that was partially masked with photoresist, etched with $CF_4$ plasma to remove the unmasked part of the whisker completely, and then stripped of photoresist. When observed under an optical microscope, vertical edges of the crystal formed by etching (i.e. the edges bordering the area where the portion of the sample was removed) appear sharp and very similar to the edges formed by crystal growth (i.e. the top edge of the



NbSe$_3$ in the figure). Edges formed by removing NbSe$_3$ by SF$_6$ RIE appear to be of similar quality. This was not the case for crystals patterned using ion-milling where milled edges were observed to be darker and more pronounced on micrographs, suggesting significantly rougher than naturally-grown crystal walls (see the discussion on ion milling later in this section, and **Figure 2.22**).

## O$_2$ Plasma Etch

Standard processing techniques involving optical and e-beam lithography regularly require samples to undergo quick resist descumming by O$_2$ plasma RIE. When processing NbSe$_3$ whiskers it is important to be aware of effects of O$_2$ plasma etch on the NbSe$_3$ surface.

Exposing whiskers to prolonged O$_2$ plasma etch, even at relatively high powers (90 W), does not seem to degrade whisker appearance as observed under optical microscope. The O$_2$ plasma etches NbSe$_3$ crystals extremely slowly if at all. No change in crystal thickness was detected during post-etch inspection using a profilometer (with vertical resolution of about 10 nm) when whiskers were exposed to an O$_2$ plasma etch for 5 minutes at 90 W.

Contact resistances of contacts formed on whisker surfaces previously exposed to O$_2$ plasma (30 sccm, 30 mTorr, 90 W) for 90 s were found to be larger typically by a factor of 10 to 100 compared to contact resistances formed and measured on fresh whiskers. In addition, the contacts were extremely noisy. O$_2$ plasma has a strong effect on whisker surface conductivity, possibly by causing surface oxidation. The altered resistive surface layer can be removed by a short exposure to SF$_6$ or CF$_4$ plasma, as discussed earlier, to gain access to the conductive bulk of the whisker, if needed.



Mantel *et al.* [94] reported that they used $O_2$ plasma to clean (descum) resist traces from the whisker surface before contacting it with gold probes. We later learned[101] that they were successful in contacting very few crystals using this procedure, and that the yield of robust, low-resistance, low-noise contacts was very low. Typically, a few poor contacts would become sufficiently conductive and stable only after the samples were biased with relatively high voltage pulses. Considering our tests that showed that $O_2$ plasma diminishes conductivity of the crystal surface, we can now offer a plausible explanation for irregular nature of their contacts. Sample processing that was used prior to evaporation of metal probes likely oxidized the crystal surface and/or left insulating residues. High voltages most likely forced dielectric breakdown of the oxide/residue, sometimes establishing sufficient electrical contact between the metal probe and the bulk of the crystal.

Consequently, while the $O_2$ plasma etch can be used to descum thin resist residues from $NbSe_3$ whiskers, it should always be followed by a short $SF_6$ or $CF_4$ plasma etch prior to metal contact deposition to achieve low-resistance, stable contacts as discussed in section 2.7.

## Ion Milling

We explored another dry-etching technique: ion-milling of $NbSe_3$ by argon ions. The Veeco Microtech Ion Milling System available at the CNF uses argon ions from a gridded, broad-beam ion source to ion-mill material without the use of reactive chemistry. In contrast to reactive ion etching, ion milling dislodges clusters of atoms from material surface using kinetic energy of non-reactive accelerated ions of argon.



**Figure 2.19** and **2.20** show AFM images corresponding to two different regions on the surface of a NbSe$_3$ whisker which had 200 nm of the material removed from the surface by ion milling. During the process the substrate was rotating in plane while tilted at a 45° angle to the incident beam of ions. Milling samples at large angles combined with sample rotation is known to result in more isotropic etching that produces smoother sample surfaces.

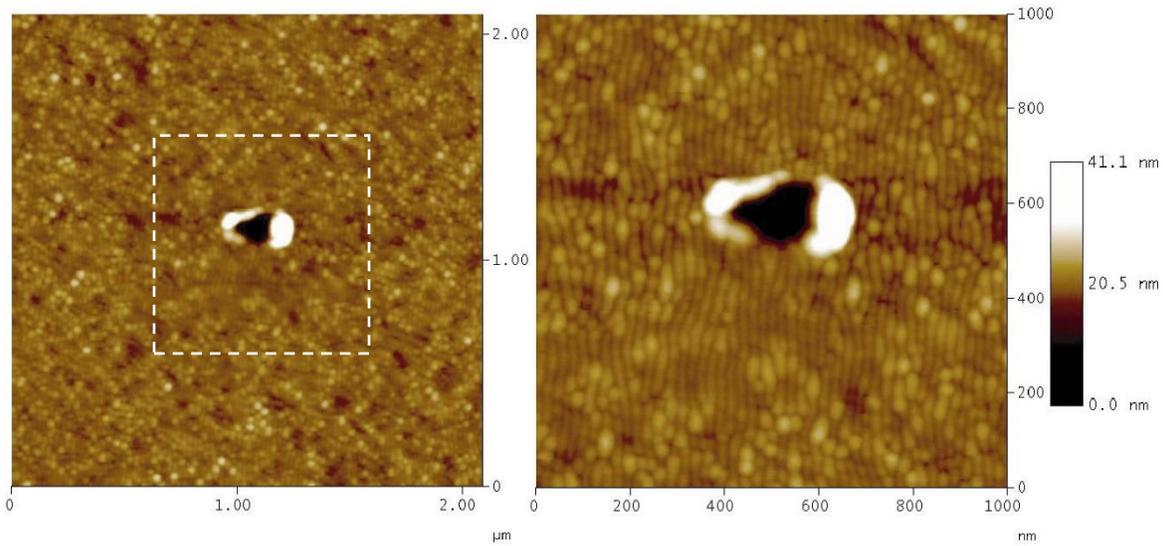

**Figure 2.19**

AFM images of an ion-milled NbSe$_3$ whisker mounted on a Si wafer with a drop of isopropyl alcohol. The dashed rectangle is the area imaged at higher magnification and displayed on the right. The whisker was ion-milled with Ar ions for 4 minutes (in cycles of milling for 15 s followed by 45 s intervals with ion beam turned off to allow the sample to cool off), and a thickness of more than 200 nm of NbSe$_3$ was milled away. Ion-milling parameters were as follows: substrate tilt angle relative to the incident ion beam was 45°, base pressure $10^{-5}$ torr, beam voltage V = 500 V, I = 70 mA, current density 0.535 mA/cm$^2$, suppressor voltage 200 V, discharge voltage 40 V, beam current I =70 mA, with neutralizer reading above 70.

Both sets of images show that the milled surface has an uneven topography. At higher magnifications one can observe structures resembling quasi-ordered wave fronts with well-oriented ridges and valleys. In addition, **Figure 2.19** shows a crater-like structure on the



surface. During imaging the AFM tip was scanned along the long (**b**[*]) axis of the whisker corresponding to the horizontal axis of both figures.

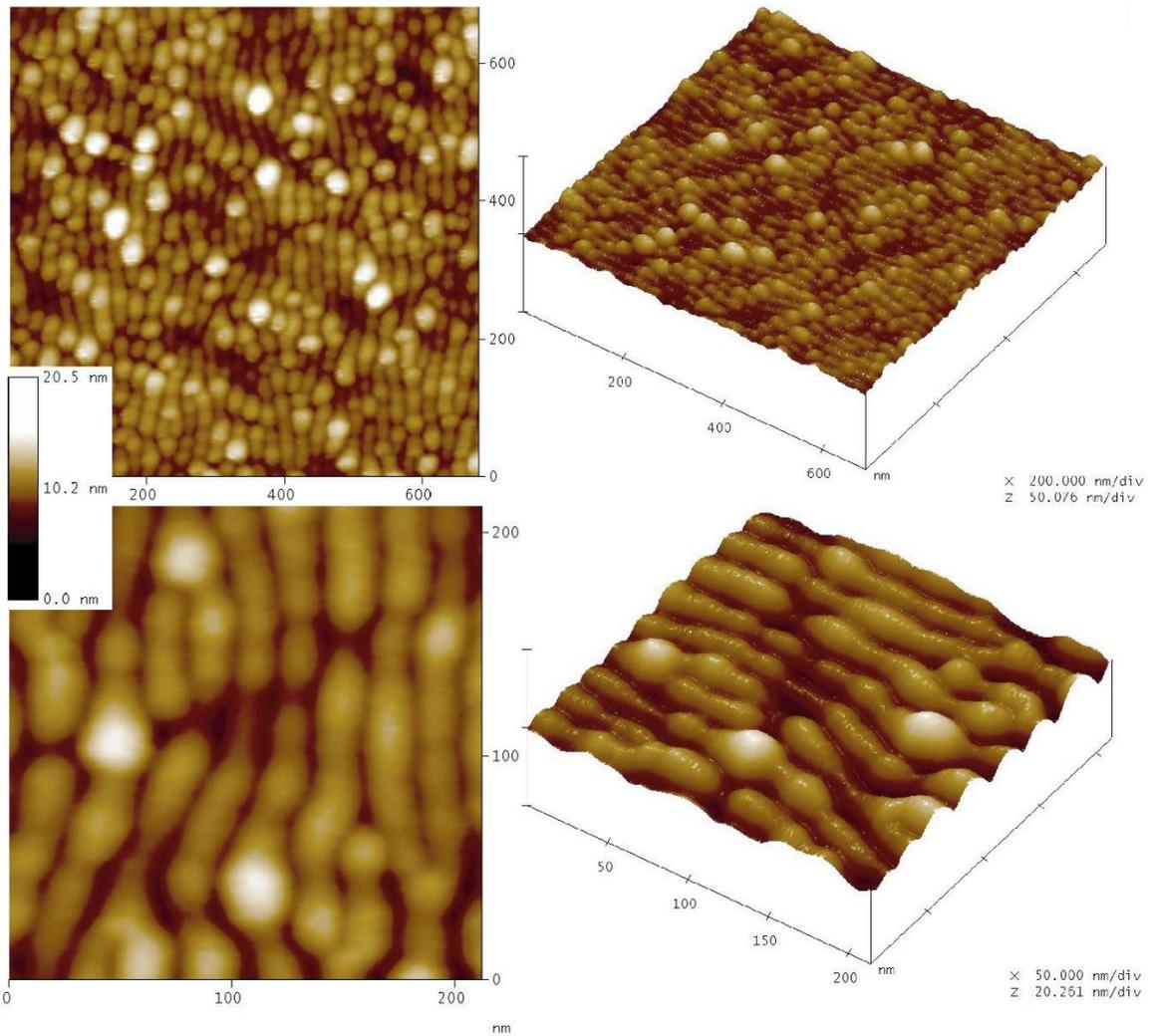

**Figure 2.20**

AFM images of the sample in **Figure 2.19** acquired on the milled surface away from the crater-like defect.

**Figure 2.21** shows that the orientation of ridges and valleys does not change as the scanning direction of the AFM tip is changed from being parallel to being perpendicular to **b**[*]



eliminating a possibility that these structures are artifacts associated with scanning

directionality.

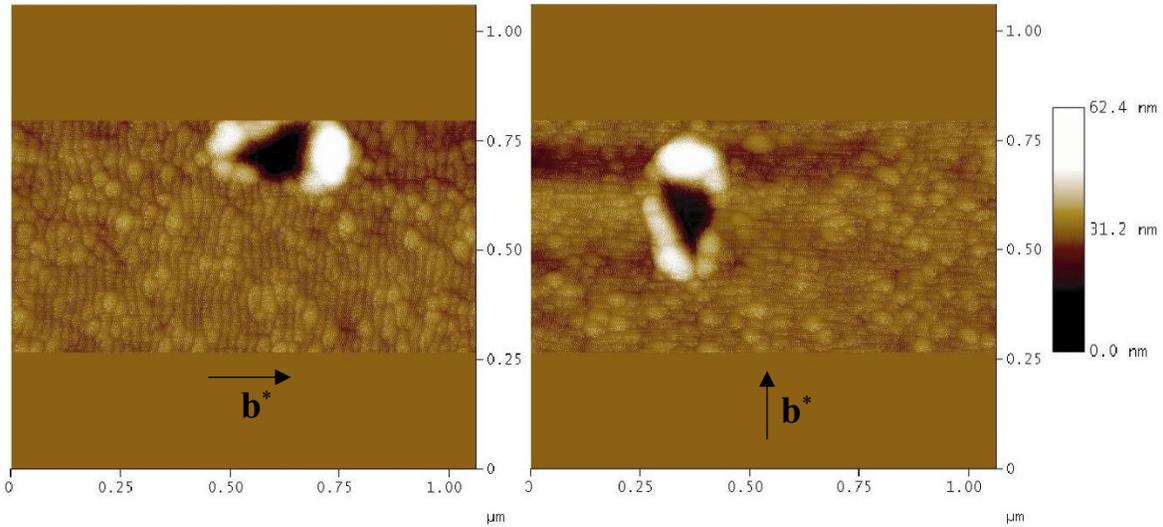

**Figure 2.21**

AFM images of the sample in **Figure 2.19** obtained with the scan axis parallel (perpendicular) to **b**$^*$ long axis of the whisker in the left (right) image (i.e. the sample was rotated by 90° between the two scans; the scan axis is parallel to the horizontal axis of both images).

The images reveal several interesting features:

1. Ridges and valleys run perpendicular to the b$^*$ axis of the crystal.

2. The wave periodicity (ridge to ridge distance) is typically between 19-25 nm.

3. Surface extrema points span a height of ~13 nm (highest peak to lowest valley) with 95 % of the surface data points spanning 5.8 nm, i.e. most of the surface is within ± 2.9 nm around the 50% bearing height$^*$. Even though the ridges and valleys have a definite overall orientation, they are not perfectly ordered locally.  The ridge direction wanders, and the height varies along the ridge.

---

$^*$ 50 % bearing height is defined as a position of a plane that crosses the surface so that 50% of the surface points are above the plane and 50 % are below it.

4.  At certain places a ridge can split into two branches, or two ridges can collapse/merge into one.  A ridge can suddenly terminate, or a new one can emerge between two neighboring ridges.  This morphology resembles line dislocations in ordered two-dimensional structures.

5.  Crater observed in **Figure 2.19** and **2.21** has an elongated shape with the following dimensions: width of ~ 350 nm along $\mathbf{b}^*$, and ~190 nm perpendicular to $\mathbf{b}^*$ direction.  The height of the crater spans ~ 50 nm.  It is not clear if the crater is an isolated case of such a structure on the surface, or if the surface is peppered with more of them (the largest area around the crater we scanned gives a rough upper bound on crater density of no more than one crater per 22 $\mu m^2$).

Unlike cracks and usual grain boundary edges (surface steps) on the crystal that run along $\mathbf{b}^*$, the ridges and valleys observed here run perpendicular to the $\mathbf{b}^*$ direction.  Because of the sample rotation during milling combined with the angled beam irradiation, the preferred orientation of the ridges and valleys cannot be associated with the direction of the incident Ar ions.  As mentioned, we ruled out AFM scanning direction as a cause of the observed directionality.  Also, the periodicity and the local disorder of the structures remains the same when the size of the scanned area is changed.  Furthermore, the observed local disorder (i.e. "line-dislocations") is typically associated with structural disorder and is not observed in periodic phenomena associated with resonant tip oscillations or light fringes that are sometimes responsible for artificial periodic noise often observed on AFM images.  All this



points to the observed quasi-ordered structures as being independent of the imaging technique, and instead as being a real representation of the post-ion-mill surface morphology of $NbSe_3$.

The crater-like structure observed on the set of images is most likely associated with surface dots observed on fresh $NbSe_3$ whiskers. Lateral dimensions and the height spanned by the crater agree well with typical dot sizes we have observed on other crystals (see section 2.6). The tall rim of the crater is likely a trace of a dot left behind as a result of rupturing or outgassing of the dot material (probably Se) that could occur due to heating of the substrate during ion milling.

**Figure 2.22** shows the difference between natural whisker edges (formed during crystal growth) versus whisker edges that were formed by milling away part of the whisker unmasked by patterned resist. The milling in this case was done at 0° tilt angle (i.e. ions impacted the crystal at normal incidence to the substrate) as to minimize the rounding of the edges. The milled edges appear more pronounced probably either due to non-uniformly collimated ion beam that can round the edges during milling or due to edge roughness, suggesting that ion milling, unlike $CF_4$ and $SF_6$ RIE process, produces edges inferior to the ones formed during crystal growth. This difference may result because ion milling degrades the edges of the resist mask more strongly than does RIE. Etching in the ion milling process is accomplished using large kinetic energy of ions, which equally impact both the unmasked part of the whisker and the resist mask (and its edges). In RIE, the ions have a lower kinetic energy but ideally react more selectively with the sample material than with the resist, allowing the etching of $NbSe_3$ to be maximized relative to that of the resist. During RIE,



degradation of the protective resist mask and its edges is minimized allowing the RIE process to produce sharper edge walls on the whisker as compared to that produced by the ion mill process.

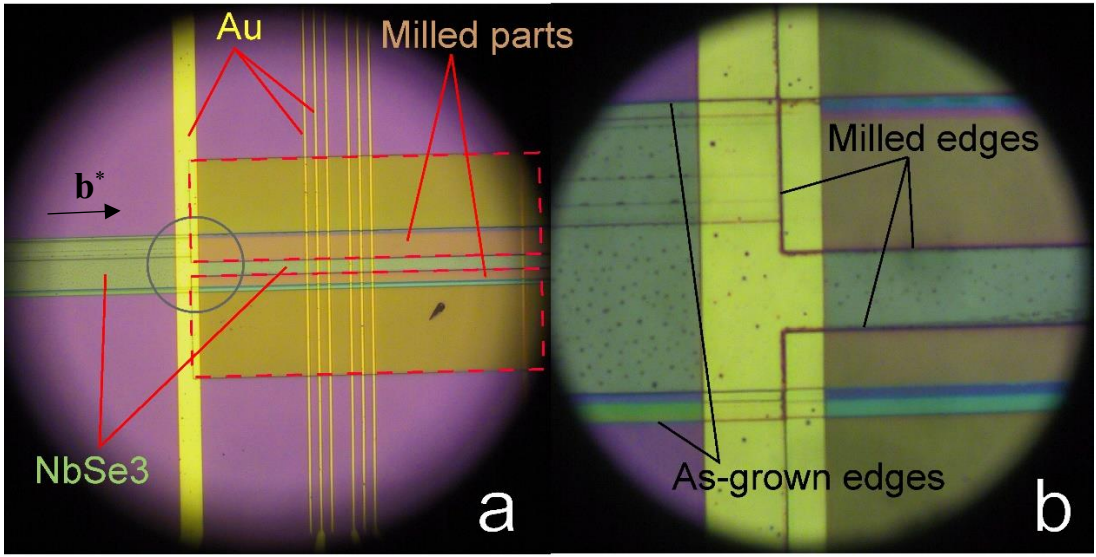

**Figure 2.22**

Micrograph of NbSe₃ whisker with selected parts of the whisker removed by ion milling. (a) Steps on the surface of NbSe₃ are visible on left side of the whisker. On the right side the sections of the whisker plagued with steps were completely removed by milling the areas (in dashed red rectangles) unprotected by resist. Gold probes were patterned subsequently to contact the sample. (b) Magnified view of the area circled in (a). Edges of the whisker formed by milling appear darker and more pronounced then the whisker edges formed during crystal growth. The milling was performed with the ion beam at normal incidence to the sample to minimize the rounding of the edges.

**Table 2.1** compares and summarizes various aspects of NbSe₃ surface quality as produced by different dry etch processes discussed in this section and compares these treated samples to as-grown, unprocessed crystals.



**Table 2.1  A comparison of dry etch effects on NbSe₃**

| Treatment ▶ Effect ▼ | Untreated (as-grown crystals) | SF₆ RIE | CF₄ RIE | O₂ RIE | Ar Ion Milling |
|---|---|---|---|---|---|
| **Etch rate** | N/A | Fast (100-300 nm/min), hard to control thickness | Slow (10-30 nm/min), easier to control thickness | Negligible | Moderate (50-60 nm/min) |
| **Post-etch surface conductivity** | Conductive | Conductive | Conductive | Resistive | Not tested (likely conductive) |
| **Z-range of surface (away from dots)** | 2.4 nm | 51 nm | 14 nm | Not measured | 13 nm |
| **Height span of 95% of surface (away from dots)** | 1.0 nm | 31 nm | 7 nm | Not measured | 6 nm |
| **Effectiveness in removing dots** | Dots often present, 30 - 200 nm tall | Removes dots | Ineffective in removing dots. 40-nm-tall dot remnants. | Removes dots | Crater-like dot remnants (?) ~50 nm tall |
| **Post-treatment surface topography and lateral ordering** | Isotropic, no discernable order | Honeycomb-like structures, mostly disordered | Isotropic, no discernable order | Not measured | Quasi-ordered ridges perpendicular to $\mathbf{b}^*$ with line-dislocation disorder. |
| **Surface structure lateral size (typical)** | 50 - 150 nm | 100 - 500 nm | 20 - 60 nm | Not measured | 19 - 25 nm ridge to ridge periodicity. 190-350 nm wide crater |
| **Appearance of etched edges** | N/A | Comparable to as-grown edges | Comparable to as-grown grown edges | N/A | Noticeably rougher or rounder edges |
| **Treatment purpose or advantage** | Conductive and smooth surfaces | Complete removal of unmasked parts of whiskers i.e. for lateral patterning | Thinning of whiskers. Vertical patterning. Conductive and smooth surfaces. | Descumming resist. Must follow by SF₆ or CF₄ plasma to remove resistive layer | Potentially a good way to thin crystals |
| **Treatment disadvantage or cautions** | Difficult to isolate crystals of specific thickness | Too fast for thinning crystals. Rough surfaces. | Remove dots prior to CF₄ etch to produce smooth surfaces. | Leaves surfaces non-conductive. Etching of NbSe₃ is ineffective. | Heating. Rough or rounded milled edges. |



## Revealing Subsurface Grain Boundaries

**Figure 2.22** is an example of how one can use a simple microfabrication technique to remove whisker segments with *visible* steps on the surface. As discussed earlier, grain boundaries running along $\mathbf{b}^*$ are often present in NbSe$_3$, both on the surface and in the bulk of the whisker as shown in **Figure 2.2**. Whiskers selected for measurements with seemingly "perfect" crystals surfaces can be harboring steps and grain boundaries hidden below the surface.

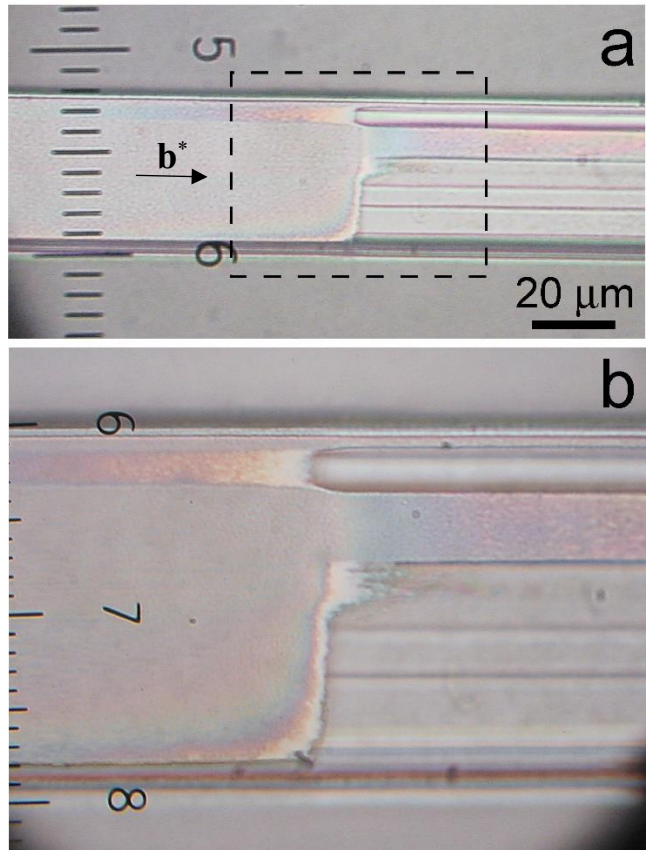

**Figure 2.23**

(a) Low and (b) high magnification optical micrographs of a NbSe$_3$ whisker exposed to a plasma etch test. The end of the whisker on the right was etched by SF$_6$ plasma revealing subsurface grain boundaries parallel to $\mathbf{b}^*$. The left side was covered by a microscope glass slide to shield it during etching.



Etching can reveal these hidden grain boundaries. The optical micrograph in **Figure 2.23** shows a NbSe$_3$ whisker that has been exposed to SF$_6$ plasma so that its right side has been etched while its surface on the left side was masked and protected from the etch by a glass microscope slide. The unetched surface on the left appears mostly uniform and flat, while the etched surface on the right clearly reveals sub-surface grain boundaries running along **b**$^*$direction.

For comparison, **Figure 2.24** shows optical micrographs of a NbSe$_3$ whisker for which etching did not reveal any additional grain boundaries below the surface. Here the microscope slide masking the left end of the whisker was held some distance above the whisker producing a soft masking transition. The right side of the whisker has been completely etched away. The resulting right edge where the ribbon "disappears" is gradual, extending a few hundred microns as indicated by the subtle color change along the crystal. This is an example of a very rare single crystal NbSe$_3$ with no observable grain boundaries in the cross-section. This quick etch procedure is likely the simplest and the most convincing test that can visually confirm through an optical microscope that the bulk of a selected whisker is composed of a single crystal.



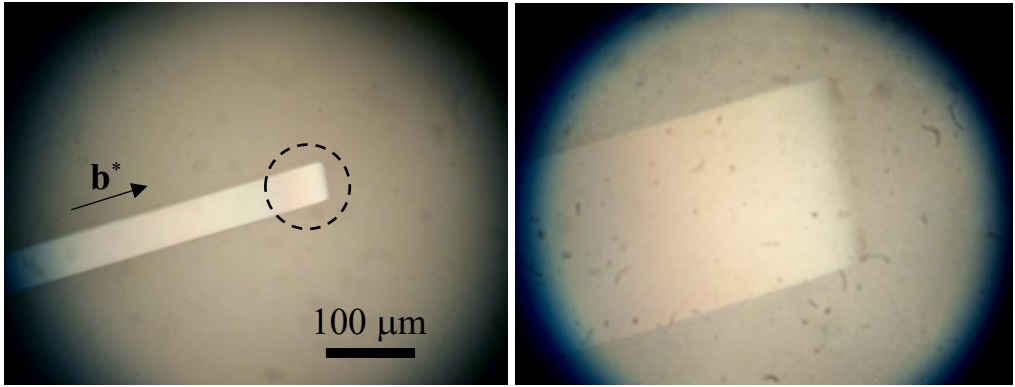

**Figure 2.24**

NbSe₃ crystal with a uniform cross section shown at two different magnifications under optical microscope (the right image is a magnified region of the dashed circle in the left image). The right side of the ribbon was completely etched away by plasma, while the left side was masked and protected by a microscope glass slide held some distance above the whisker. This produced a gradual thickness change in the whisker from its full thickness on the left side to crystal "disappearing" on the right end. This thickness gradient samples the full cross-section of the whisker revealing that there are no observable grain boundaries present in the bulk of the whisker.

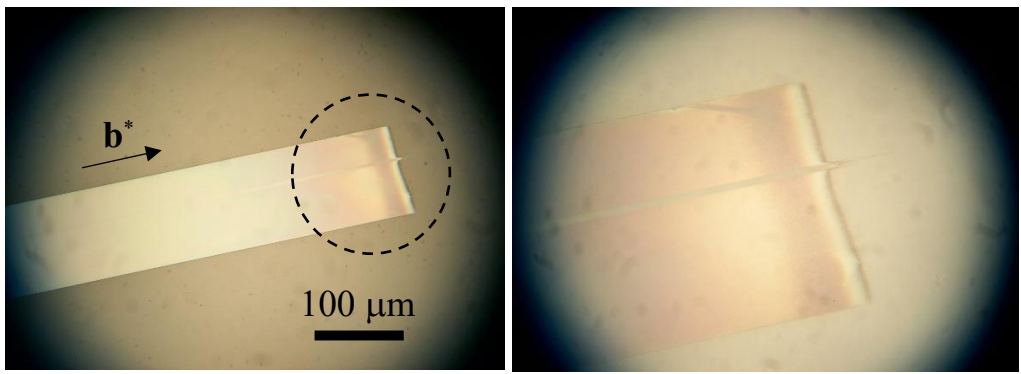

**Figure 2.25**

Another NbSe₃ ribbon shown at two different magnifications (again the right image is a magnified region of the dashed circle in the left image) and masked during plasma etch similarly as the ribbon in **Figure 2.24**. Here the region of the ribbon with thickness gradient near the ribbon's right end reveals a grain boundary in the whisker cross section that is not visible on the surface of the unetched (masked) far left end of the whisker.



Another crystal, shown in **Figure 2.25,** has a step that is not discernable on the unetched part of the surface but becomes visible and more pronounced along the thickness gradient toward the thin end of the whisker.

While the presence of such grain boundaries could be inferred from involved X-ray diffraction and CDW transport measurements, a quick etch test can be a very simple but effective sample diagnostic that can save time in experiment preparation. For example, prior to preparing $NbSe_3$ samples for transport measurements, a small length of selected whiskers can be cut off the sample ends and etch-tested to expose subsurface morphology. Samples with revealed grain boundary steps can be identified and eliminated from experiments before lengthy sample preparation and measurements are undertaken. Alternatively, if CDW transport measurements of a sample shows signature hints of non-uniform sub-surface morphology, the sample whisker can be etched and examined for confirmation after the experiment, i.e. the etch tests can be utilized only when transport results are indicative of multi-step low-quality sample to either exclude the sample data form the analysis or to at least account for expected effects of grain boundary presence. Furthermore, since etch tests can provide direct information about grain boundary spatial distribution in the whisker, one could intentionally select "stepped" samples to investigate and correlate their transport signatures with the corresponding grain boundaries in the bulk. Again, one can avoid destroying the actual samples intended for measurement by cutting off only small longitudinal sections of whiskers for etch testing. A simple etch test can be a versatile and time-saving investigative procedure for assessing crystal quality during sample preparation.



## 2.6 Surface Dots

Early work by Maher *et al.*[35,89] and Adelman *et al.*[90] with microfabricated electrical probes has shown that a more reliable and less perturbing way to form electrical contacts with $NbSe_3$ whiskers is by fabricating them on a substrate and then placing the whisker on top of them, rather than depositing the contacts on top of whiskers. Even so, the reliability with which high quality contacts can be formed has been relatively poor. Many carefully selected whiskers of good crystalline quality never make it to the experiment and are wasted because of inability to adequately contact them.

The quality of the electrical contact formed between a whisker and a metal pad on the substrate depends on the physical contact between the whisker and the metal. Any kind of debris or roughness on the whisker or the metal surface may jeopardize an attempt to properly establish electrical contact between the two.

We have looked more closely at whisker surfaces to address this problem. Images of freshly grown $NbSe_3$ crystals obtained by optical microscopy often show crystal surfaces plagued by "dots" as shown on **Figure 2.26**. Closer inspection with SEM in **Figure 2.27** and AFM in **Figure 2.28** shows that dots are "bumps" rather than "pits", resembling water droplets on a hydrophobic surface.



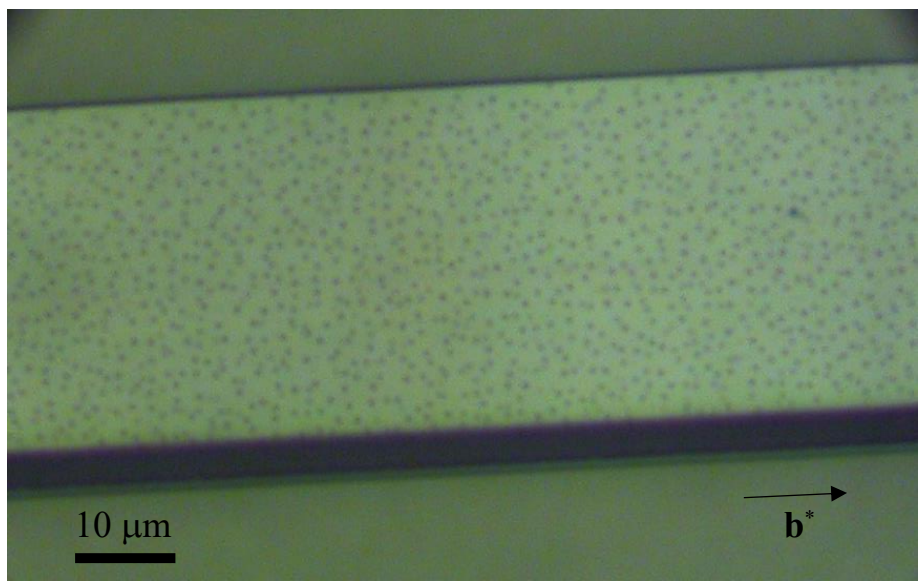

**Figure 2.26**

Surface of a NbSe₃ whisker as seen under an optical microscope.  The surface is covered by "dots" that are often observed on freshly grown whiskers.

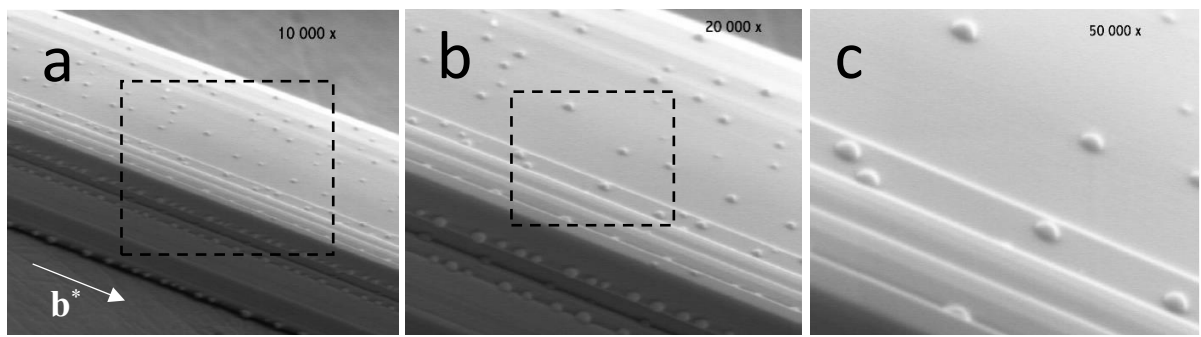

**Figure 2.27**

As-grown NbSe₃ whisker observed by SEM.  Dashed rectangles are the areas observed under a higher magnification and displayed in the image to the right.  The surface of the whisker is covered by droplet-shaped "dots".



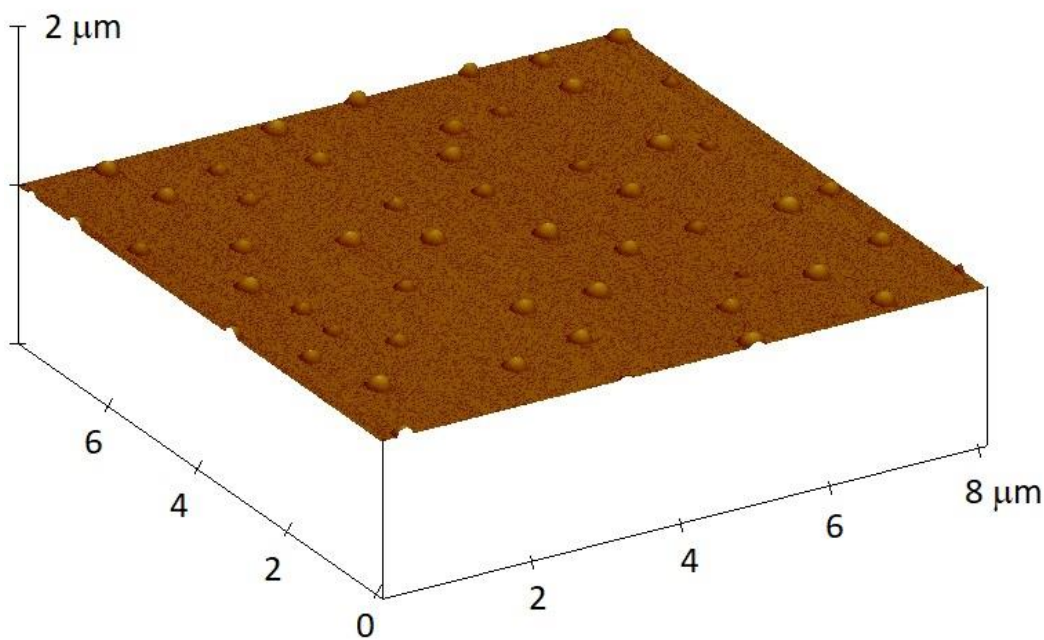

**Figure 2.28**

AFM image of a freshly grown NbSe$_3$ showing droplet-like dots on the surface. Dot density and size can vary widely from sample to sample and from batch to batch. Inspection of several samples yielded density range of 0.7 dots/$\mu$m$^2$ - 3.4 dots/$\mu$m$^2$, dot height 30 nm - 120 nm, and lateral diameter 80 nm - 350 nm.

In most cases dots are a nuisance, as when they hinder proper electrical contact between metal probes and the whisker. Sometimes, however, they can aid in obtaining additional information about crystal quality. Dots are most often large enough to be visible under optical microscope, while small cracks and step edges on the crystal are not always discernable. When cracks and step edges are present on a surface, dots tend to "line-up" along them. Figure 2.27 **(a)** and **(b)** show examples of dot arrays along step edges in the lower (shaded) parts of the figures. **Figure 2.29 (a)** is an SEM photo of a sample which shows an array of dots aligned with the **b**$^*$ axis on the crystal. The higher-magnification images **(b)** and **(c)** reveal that the dot array follows a crack in the crystal surface.



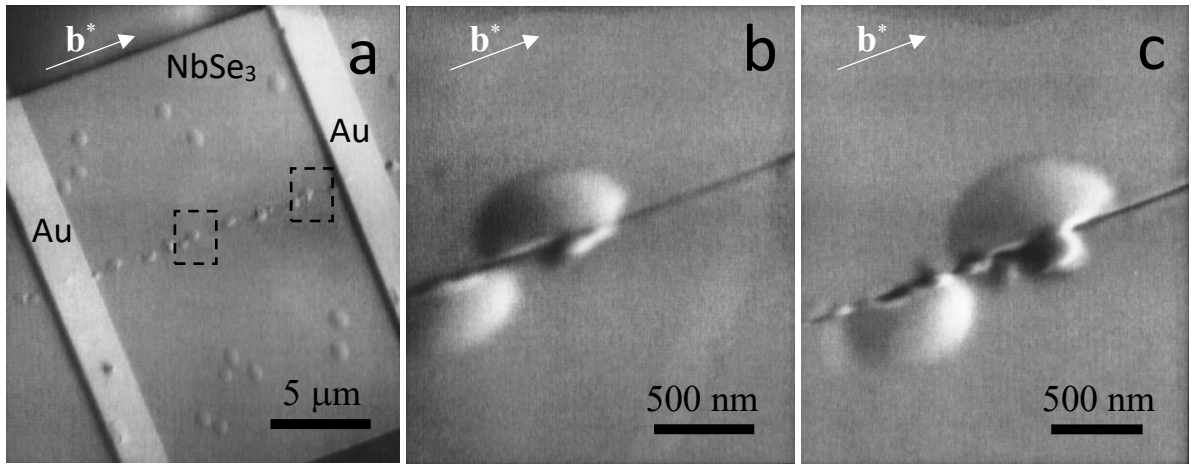

**Figure 2.29**

SEM images of a NbSe₃ sample with gold probes patterned over the whisker. Dashed rectangles in (a) are magnified areas shown in (b) and (c). Image (a) shows an array of dots lined up along the **b**\* direction along the sample. (b) and (c) reveal a crack in the whisker along the array of dots.

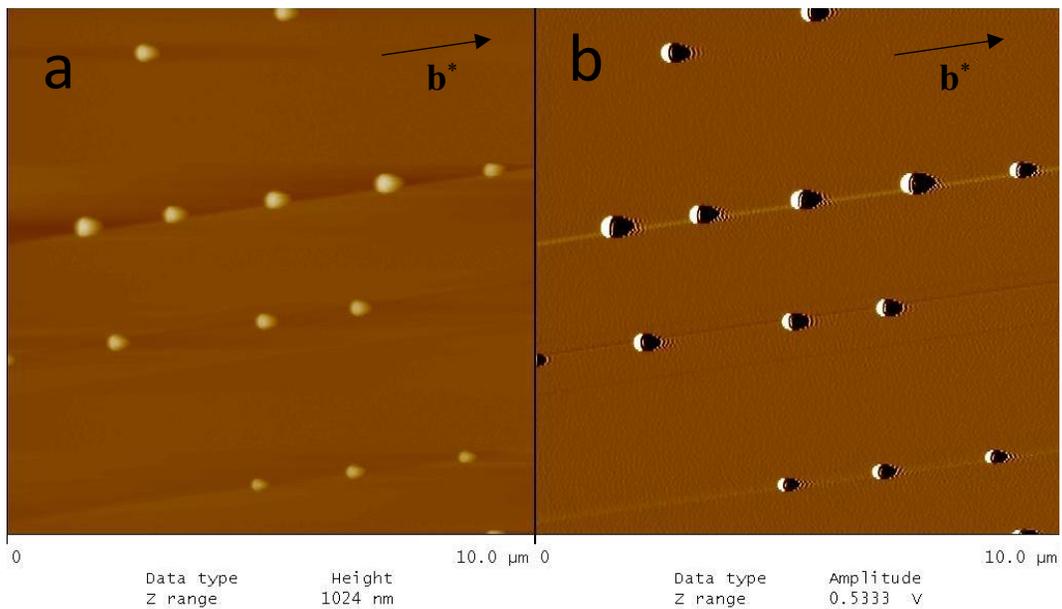

**Figure 2.30**

(a) Amplitude (height) AFM scan of a NbSe₃ surface, and (b) its corresponding phase scan. Dots line up on top of longitudinal grain boundaries.



**Figure 2.30** is an AFM image of a crystal with step edges running along $\mathbf{b}^*$ axis. Again, it can be clearly seen that dots preferentially form along the step edge grain boundaries. The subtle step edges on this sample were not resolvable under optical microscope, while the ordered dot "trains" along the whisker's long axis were.

Dot ordering can, thus, be a good indicator that the crystal has an "imperfect" surface when quickly inspected by an optical microscope.

## Origin of Dots

NbSe$_3$ crystals are grown by placing elemental Nb and Se on one end of a long sealed quartz ampoule. The ampoule is evacuated prior to sealing. During growth, a temperature gradient is applied so that the end containing the reacted Nb and Se is warmer than the other end. Any species with finite vapor pressure will then be transported from the warm to the cooler end, where they can nucleate and form crystals. Adding excess selenium over the stoichiometric value facilitates this vapor transport process. At ambient pressure, the boiling temperature of Se is 685° C,[102] and crystal growth is typically performed at temperatures in this range. As a result, during growth the tube is filled with Se vapor. The tube is cooled to room temperature under a temperature gradient so that the excess Se condenses and eventually solidifies at the cold end.

The most likely origin of the observed droplet-like dots on the surface is condensation of Se during cool-down. The dots observed on a surface of a freshly grown crystal are thus expected to be Se or possibly some form of Se oxide. This is consistent with the data in **Figure 2.9** and our discussion in section 2.2 on heating of NbSe$_3$ samples.



## Composition of Dots

Scanning Auger Nanoprobe (PHI-670) and Electron Microprobe (EPMA tool JOEL 8900) analysis tools were used to characterize the atomic composition of dots on surfaces of $NbSe_3$ whiskers. Preliminary measurements by both tools revealed that the content of Se atoms is slightly higher while the content of Nb atoms is slightly lower on the dot as compared on the crystal surface away from the dots. More precise quantitative analysis was, unfortunately, not possible when using these tools due to difficulties associated with sample probing. Dots were observed to disappear (most likely via evaporation) under focused electron beams in vacuum environment on timescale of tens of seconds, a shorter time scale than required by the tools to acquire a sufficient data sample. Nevertheless, the indication of increased Se levels on dot surfaces agrees with the idea that dots could be condensed Se in some form.

## Dots as Conductive Bottlenecks

Dots interfere with the formation of low resistance contacts between $NbSe_3$ whiskers and metal pads on substrates. Dot-covered whiskers can usually be successfully contacted, suggesting that the dots are sufficiently conductive. This is consistent with their composition being selenium, since both Se and some of its oxides are good conductors. However, the contact resistances are often large and the resulting contacts noisy suggesting that the dots represent conductive bottlenecks by reducing contact area between the metal probe and the whisker. Since the current density through the dots may be high, local heating may be significant. Furthermore, local heating can cause uneven thermal expansion and stress between $NbSe_3$ whisker and the contact probe and result in relative "crawl" between the materials which can be detrimental to electrical contacts. "Breaking" of contacts is indeed



commonly observed during measurements at current densities that are small compared to the ones that bulk whiskers and the metal pads can sustain without appreciable heating.

## Removing Dots from Fresh NbSe₃ Whiskers

**SF₆ Plasma.**  Our tests discussed in section 2.5 revealed that $NbSe_3$ surfaces etched by RIE in $SF_6$ plasma lack dots.  These surfaces, however, are not smooth and are characterized by honeycomb-like pits and ridges.  Our inspection of samples etched by $SF_6$ plasma revealed that the typical surface roughness, arrangement, and density of pits and ridges do not coincide with the arrangement and density of dots before the etch.  Dots observed before the etch are sporadically distributed across the surface compared to the distribution of honeycomb-like pits, i.e. dot density << pit density.  In one sample dot density before the etch was 0.50-0.75 dots/$\mu$m$^2$, two orders of magnitude smaller than the density of pits (50-60 pits/$\mu$m$^2$) observed on the surface after the etch.  Post-etch honeycomb-like structures are thus most likely not related to dots that are potentially present on the sample before exposure to the $SF_6$ plasma etch (at least not in a direct way where dots act as micro masks).

Dots can be removed from whisker surfaces by a weak $SF_6$ plasma etch.  **Figure 2.31** shows micrographs of two different segments of a single whisker.  The whisker surface was uniformly covered by dots before the $SF_6$ etch.  The right segment shown in (b) was exposed to the etch while the left segment in (a) was masked from etching by a microscope slide.  The dots are absent on the etched part of the whisker.



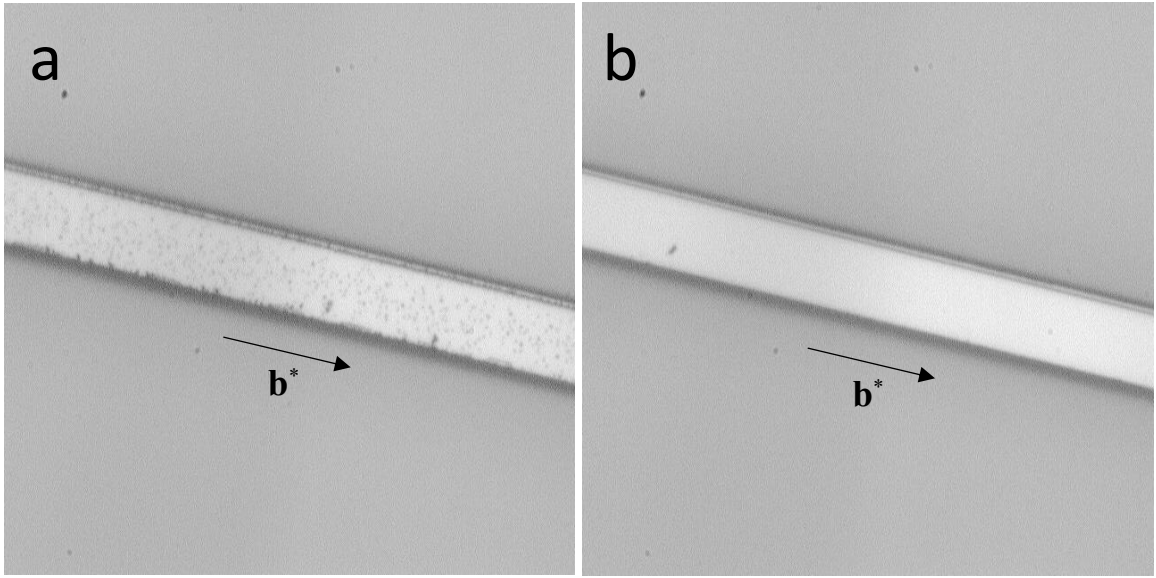

**Figure 2.31**

(a) A segment of a NbSe$_3$ whisker that was protected from the RIE SF$_6$ plasma etch by a microscope glass slide. (b) Another segment of the same whisker that was exposed to the plasma etch for 90 s (20 sccm, 0.15 W/cm$^2$).

**CF$_4$ Plasma.** We have not performed enough testing to conclude whether removing dots by CF$_4$ plasma can be effective. In section 2.5 we have shown that a CF$_4$ plasma etch of short duration does not remove dots completely (etch rate of the dot material was never determined). For thinning whiskers, we recommend the use of CF$_4$ plasma etch because it leaves NbSe$_3$ surface smooth and conductive, but dots should always be removed by some other effective method (heating in vacuum, for instance) prior to the CF$_4$ etch process.

**O2 Plasma.** A whisker with dots was etched in O$_2$ plasma (30 sccm, 30 mTorr) for 90 s at a relatively high-power setting of 90 W. After the etch, the inspection of the crystal under optical microscope revealed that the dots were removed. It is not clear whether dots were removed by O$_2$ plasma or they outgassed due to heating in vacuum during the etch (high



power etching can heat substrates). We do not recommend utilizing $O_2$ plasma for removal of surface dots since the exposure compromises the whisker's surface conductivity and hinders subsequent electrical contacting of whiskers (unless the exposure to $O_2$ plasma is followed by either $SF_6$ or $CF_4$ plasma which can remove the resistive layer formed on the surface - see section 2.5 and **Table 2.1**).

**Heating in Vacuum.** We found that the most effective and the least aggressive way to remove dots is by moderately heating whiskers in vacuum. **Figure 2.32** is an optical micrograph of a crystal before and after heating in vacuum showing that heating removes dots and leaves the surface clean. Our tests revealed that heating in air ambient also removes dots, however heating in vacuum does not compromise the surface conductivity (and our ability to subsequently contact whiskers) while heating in air does (see section 2.2).

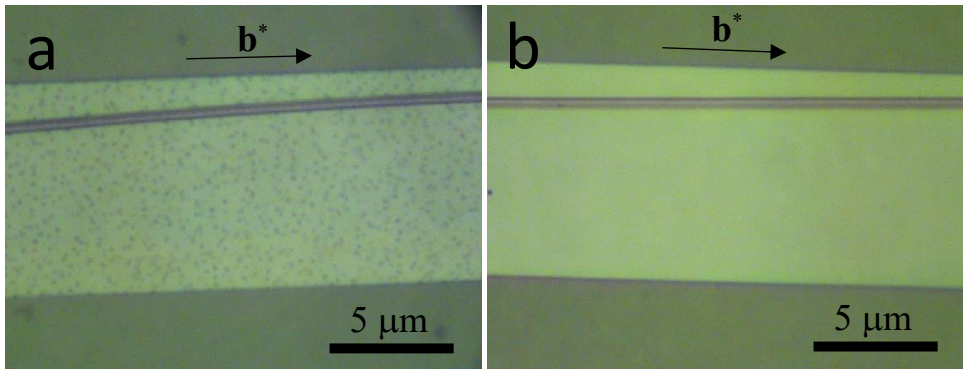

**Figure 2.32**

NbSe$_3$ whisker (a) before and (b) after heating to 300° C at vacuum pressure of $1.3 \times 10^{-6}$ torr for two hours. The sample was left to cool in vacuum overnight.

Heating in vacuum environment should thus be the preferred method for removing surface dots from NbSe$_3$ samples.



## Dots Induced by Processing

**Figure 2.33** shows a sample that was heated in air ambient for several hours[*] and then exposed to extensive processing while covered by resist. Surface of NbSe$_3$ shows large dots that were initially not visible under optical microscope (if they had been present initially, they were smaller than what could be resolved by an optical microscope). In many of our tests we observed that whiskers acquired pronounced dots during standard processing steps.

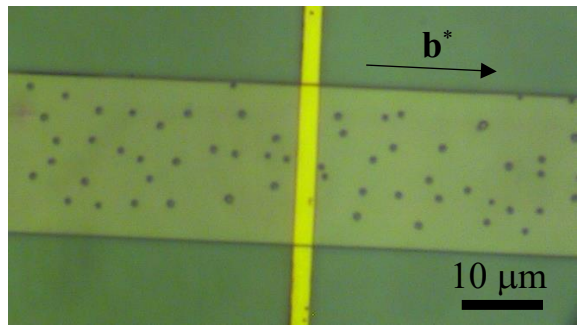

**Figure 2.33**

Micrograph of a NbSe$_3$ whisker with visibly pronounced dots on its surface. The dots appeared after the sample was subjected to extensive processing in an early attempt to pattern gold probes (shown in yellow) over the whisker. The processing involved heating the sample to 240° C for several hours in an air-ambient oven to cure polyimide (used as a glue holding the whisker to the substrate). A bilayer resist stack (copolymer + PMMA) was then applied over the sample. The resist was cured by heating the sample on a hot plate in air ambient to 170° C three times for 15 minutes each time. The resist mask was patterned using e-beam lithography and developed to open the trenches for subsequent gold evaporation. Before the evaporation the trenches were exposed to a sequence of three short plasma etching steps (O$_2$ + CF$_4$ plasma followed by SF$_6$ plasma followed by O$_2$ plasma). After gold evaporation, the resists were removed chemically by a wet lift-off process.

---

[*] These tests were performed in the initial stages of our investigation before we realized that heating in air ambient compromises whisker surface conductivity. After this realization all polyimide curing was performed in a high vacuum oven, and any subsequent exposure of crystals to heat in air ambient (unavoidable in some processing steps) was kept to a minimum.



Due to time restrictions, we were not able to determine with great certainty the exact causes that induce the dots associated with processing, but a closer look at seven processed samples and individual processing steps that varied slightly from sample to sample points to processing that involves whiskers *heated in air ambient for a prolonged periods of time while being covered by e-beam resists* (copolymer and/or PMMA). Longer heat treatments and higher temperatures seem to lead to more pronounced dots. For example, **Figure 2.34** and **Figure 2.35** show whiskers that were both exposed to this type of processing at some point during sample fabrication. While the former shows small dots on the surface (also more clearly seen on the SEM image shown in **Figure 2.39** in section 2.7), the latter lacks dots (at least ones large enough to be visible under an optical microscope). A notable difference in processing is that the whisker in **Figure 2.35** was exposed to a less aggressive heating treatment (15 min at 114° C) as compared to the sample in **Figure 2.34** (140 min at 135° C).

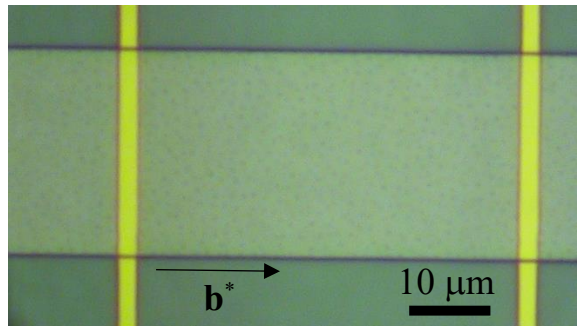

**Figure 2.34**

During processing to pattern gold probes (shown in yellow) this sample was exposed to heating in air ambient for a total of 140 minutes at 135° C. Very small "processing dots" are visible on the surface under optical microscope.



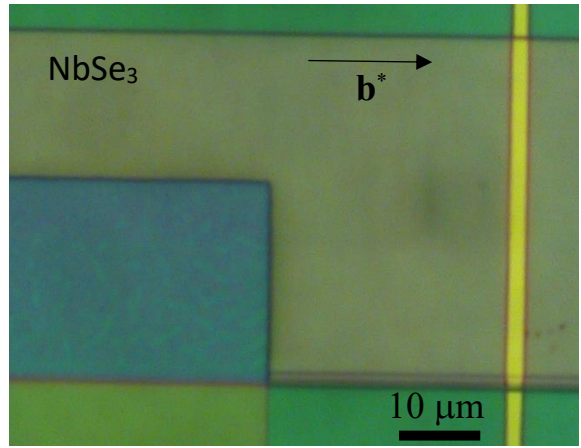

**Figure 2.35**

During processing to pattern gold probes (shown in yellow) over a NbSe₃ whisker previously patterned by RIE, this sample was heated in air ambient for 15 minutes at 114° C while coated with resist.  No "processing dots" are discernable on the microscope image of the sample.

"Processing dots" are not easily removed by common resist stripping agents (methylene chloride and acetone), and dots usually remain on the whisker upon process completion.   The chemical composition of these dots has not been probed.  "Processing dots" were sometimes, although more rarely, observed on samples that were exposed to air-ambient heating, and then *after this initial heating* were also exposed to other processing steps such as exposure to resists and other chemicals, etching, and heating during etching.   We've seen in section 2.2 that heating to moderate temperatures alone (without exposure to other processing) does not produce dots on a crystal surface, but heating in air ambient may leave oxide layers or residues (including dot residue) on the surface.

Our best explanation for dots that appear during processing is that remnants of incompletely removed dots that are present after sample growth are trapped under resist.  This dot residue



then reacts with the resist during prolonged heating, resulting in a residue that cannot be removed by resist stripping agents.

There was some indication that a whisker coated by resist and heated in vacuum is less likely to acquire dots than a covered whisker heated in air. Original dots or dot residue could become volatile during the heating process forming bubbles in an uncured resist (our tests discussed in section 2.2 indicate that Se or Se oxide on whisker surfaces outgases easily at elevated temperatures). The bubbles are more likely to escape from the wet resist if a sample is heated in a vacuum environment then in an environment with room pressure.

Dots never appeared on crystals that were processed using optical lithography. This is likely because prolonged air-ambient heating of crystals in contact with optical resists is usually not necessary (Shipley resists used at CNF for optical lithography are typically baked at 115° C for only 2 minutes prior to resist exposure).

Additional, more specific, tests are needed to identify the exact cause of "processing dots". As mentioned before, we suggest that all sample heating be done in vacuum environment when possible.[*] When this is not possible, steps that require heating in air ambient should be kept to a minimum (both in temperature and duration).

---

[*] Heating in air ambient can be eliminated if one has access to hotplates with a vacuum environment. We did not have access to such a hotplate at CNF, so baking of a copolymer + PMMA resist stack on our samples was performed on a hotplate in air ambient for 15 minutes, followed by heating in an evacuated oven purged by backpressure of $N_2$ at 165° C for 90 minutes to complete resist curing. The use of a hotplate immediately after the resist is spin-coated over the substrate is required to set the resist quickly so that it cures uniformly and without cracks. After a short bake on a hotplate, the curing process can be completed in a vacuum oven when needed.



## 2.7 Electrical Contacts

Having good electrical contacts on a NbSe$_3$ sample is essential for probing its collective transport properties. Stable contacts result in higher signal-to-noise ratio in measurements. Robust contacts do not fail in mid-experiment and eliminate the need to prepare new samples and repeat measurements. A reliable procedure to produce good contacts prevents wasting rare single crystal NbSe$_3$ whiskers. Here we define more explicitly what we mean by "good" contacts. The definition is arguably somewhat arbitrary, but it is based on pragmatic considerations relevant to our experiments.

### What Are "Good" Contacts?

In order to tailor a fabrication process that produces nominal electrical contacts we first need to establish and define figures of merit for ideal contacts. We can then anticipate factors that may be obstacles in the fabrication process, understand them, and find ways to control them to achieve desired effects.

An *ideal* electrical contact between a whisker and a metal probe would:

- have low enough resistance to allow large current flow between the metal probe and the whisker (current densities on the order of 1 mA/$\mu$m$^2$) without it being a significant source of local ohmic heating;

- have high enough resistance to minimize perturbation of local electric field in the whisker and prevent current shunting of the whisker through the contact metal;



- withstand material stresses due to multiple thermal cycling from room temperature down to cryogenic temperatures;

- be stable to yield good signal-to-noise ratio in measurements (i.e. insignificant source of noise);

- be produced by a reliable fabrication process (i.e. a process yield of 100 %).

In practice fabricated contacts are almost never ideal, and their quality can vary substantially. We will categorize fabricated contacts according to their quality in a crude way as follows:

- "Good" contacts produce low noise levels with typical voltage fluctuations due to contact noise below ~ 0.01 mV when the sample is biased by a typical DC current required in measurements. These contacts have low resistance comparable to whisker resistance at low temperature, are not a significant source of ohmic heating, and can withstand multiple thermal cycling;

- "Fair" contacts are a substantial source of noise with signal fluctuations up to 1 mV at typical DC biases. The contacts have a resistance comparable, or slightly higher, than whisker resistance at low temperature and are not reliable (often break upon cooling);

- "Bad" contacts are broken contacts or contacts with resistances in excess of 100 kΩ. The contacts are strong sources of noise with signal fluctuations on the order of 100 mV or higher when the sample is DC biased.



We will use this simple classification to evaluate the success of a fabrication processes by tracking how many "good", "fair" and "bad" contacts it produces. For some measurements, e.g., DC *I-V* measurements, "fair" contacts are sufficient to give reliable results. But for more sensitive measurements of, e.g., voltage fluctuations ("narrow band" and "broad band" noise) associated with CDW motion, "good" contacts, which more stringently resemble the "ideal", are necessary.

## Existing Ways to Electrically Contact Crystals

A typical CDW transport measurement configuration involves four contact probes applied to the whisker: two biasing (current) probes and two measurement signal (voltage) probes. Often it is desirable to have more than four probes. For instance, to measure the spatial configuration of CDW current along a whisker, one may apply a current bias through two (current) probes spaced far apart along the whisker, and then measure signals between neighboring pairs of inner (voltage) probes along different segments of the whisker located between the current probes.[36] In order to obtain a good spatial resolution of the measurement, signal probes need to be spaced closely and should have small width dimension along the whisker's $\mathbf{b}^*$ direction. Furthermore, a narrow width of the probe is desired in order to minimize perturbation of the measurement by the probes themselves which shunt the whisker locally along the contact area. Attaching such contact probes to the whiskers is not trivial because the probes require microscopic fabrication precision and alignment to whiskers which are fragile microscopic objects. Here we review some of the older contact-making techniques and describe novel techniques we developed and optimized.



**Conductive Paint Applied Manually.** Some of the early attempts to electrically contact NbSe₃ whiskers involved simple techniques such as manually applying small blobs of conductive (carbon or silver) paint or conductive epoxies over a whisker on a substrate. This can be done using a tip of a home-made "poke tool" consisting of a thin, sharpened wire (tungsten can be "sharpened" to a very fine tip by a torch heat) attached to a handle. The conductive paint on the substrate is then further connected to fine electrical wires or connectors linked to instruments. These techniques are fast and simple and are still used today to make quick electrical connections for less sensitive measurements. We used them to contact whiskers in the tests described in the earlier sections. The main disadvantages of such contacts are that 1) they often produce noisy and non-reliable contacts that break easily upon cooling, and 2) they are messy and often quite large compared to whisker's width and height dimensions. After years of practice the narrowest contact probes the author is able to produce are typically 50-100 μm wide. Recall that most high-quality NbSe₃ whiskers are typically 1-10 μm wide and < 1 μm thick. As mentioned, large contacts put limitations on studying physics on small length scales and can significantly perturb electric fields in the whisker, i.e. current flowing through the whisker can shunt the part of the whisker that is in contact with a conductive probe.[103]

**Extruded Indium Wires.** Another contacting technique involves gently pressing indium wires over the whiskers. The wires are created by extruding soft indium metal through a small pinhole in a machined piece of metal and can be made as small as 25-50 μm in diameter.[104] Although smaller than contacts produced by applying conductive paint, such



probes are still relatively large, and it is difficult to control placement and positioning of wires over the whisker. Pressing the wires against the whisker can also damage the fragile crystals.

**Probes-on-Bottom (POB).** More-sophisticated methods used to contact whiskers involved evaporation of indium probes through a shadow mask over NbSe$_3$ whiskers[34] or over substrates on which whiskers can be subsequently positioned.[87] The latter method had then evolved to use more standardized fabrication techniques and involves photo-lithographic patterning of gold (or chrome topped with gold) probe arrays on alumina substrates.[36,90] With this method, the whisker is carefully mounted post-fabrication on top of the probes, and then a small drop of a solvent (ethyl acetate) containing dissolved polymer (ethyl cellulose) is dispensed over it. When the solvent evaporates, the polymer residue encapsulates the whisker securing the whisker against the probes. **Figure 2.36 (a)** is a sketch showing a side view illustration of a sample prepared using this method. **Figure 2.36 (b)** shows a top-view micrograph of an actual NbSe$_3$ whisker on gold probes prepared in this way.



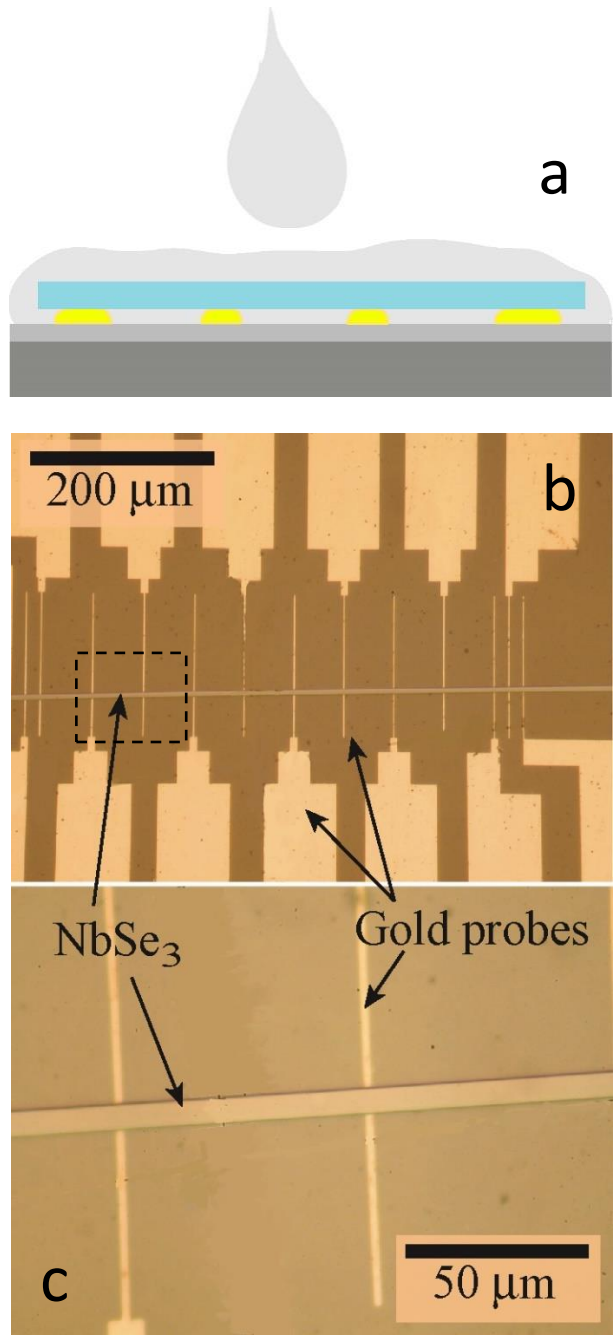

**Figure 2.36**

(a) A side-view illustration of a sample in a "probes-on-bottom" (POB) configuration. It shows a whisker (in blue) laying over four metallic probes (in yellow) fabricated on a substrate and encapsulated by a polymer (in light gray). (b) Top-view micrograph of a NbSe$_3$ sample in a POB configuration encapsulated and affixed to the substrate by a (transparent) polymer. (c) The area in the dashed rectangle in (b) shown at a higher magnification.



This "probes-on-bottom" (POB) modernized approach proved very successful. It is relatively simple since the contact probes are fabricated on a substrate without delicate whiskers being present during fabrication. The microfabrication techniques allow precise control of a geometrical configuration of probes and a small probe size that can be made as narrow as one micrometer with optical lithography. These advantages allowed Adelman and Lemay *et al.*[36,90] to directly measure spatial CDW current distribution along the whisker that is associated with the CDW current conversion and phase slip in the proximity of current contacts.

When choosing a suitable substrate for a process, it is important to minimize the mismatch between the thermal expansion coefficients of the substrate and the whisker. This minimizes stress between different materials and ensures that there is very little thermal "crawling" of the whisker on the substrate upon cooling, and that the whisker remains in good physical and electrical contact with the metal probes beneath it. The alumina substrates originally used performed relatively well in this regard. They are, however, expensive, impossible to cleave, hard to cut, and often come in shapes poorly suited for mounting in fabrication equipment typically made to accommodate round standardized silicon wafers. To overcome these difficulties, we have adapted the existing POB approach to silicon wafers. The wafers are thermally oxidized before fabrication to obtain insulating substrates. The probes are then patterned on a substrate similarly to what was done in the work of Adelman and Lemay *et al.*[36,90] A schematic and an outline of the POB process on an oxidized silicon wafer is given in **Figure 2.37** and in section 2.8, and the full details of the recipe are given in the Appendix C, Recipe 1.



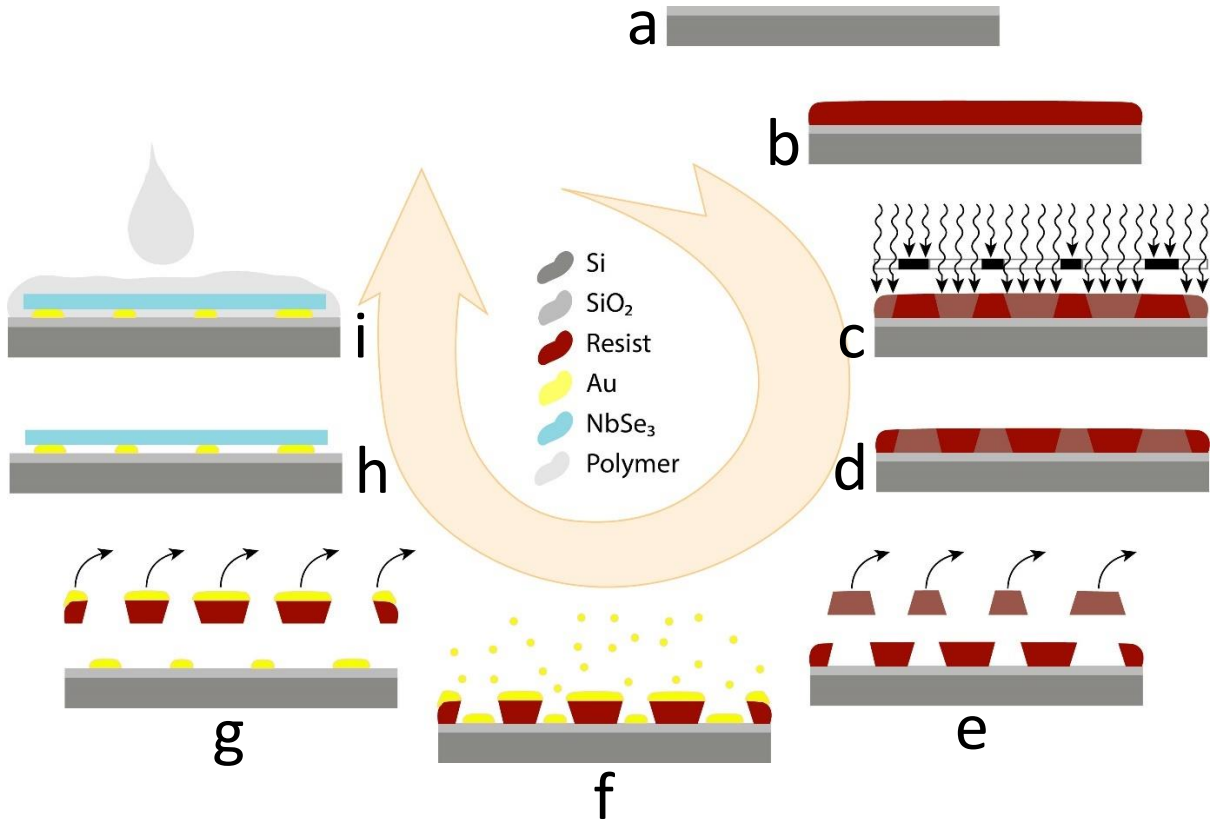

**Figure 2.37**

Fabrication process for the "probes-on-bottom" (POB) method. (a) Start with a thermally oxidized silicon substrate. (b)-(e) Use optical lithography to pattern a resist mask for electrical contacts. (f) Evaporate Au probes on the substrate through the resist mask. (g) Au probes remain on the substrate after a mask lift-off. (h) Position a NbSe$_3$ whisker over the probes. (i) Apply a drop of polymer dissolved in a solvent. When the solvent evaporates, the remaining polymer encapsulates the whisker pressing it against the probes and substrate. Further technical details of the POB microfabrication recipe are contained in Appendix C, Recipe 1.

Although a large improvement over manual techniques, the POB contacting method is far from ideal. A whisker must be manually positioned and aligned over the probes. Even after the solvent-containing polymer is applied, the whisker often does not form good electrical contacts with all the probes, likely due to polymer getting between the probes and the whisker. In addition, any microscopic imperfection or debris on either the probe or the whisker (such as dots on whisker surface discussed in section 2.6) may hinder electrical



contact between the two. Typically, about one third of contacts exhibit unacceptable noise levels and/or contact resistance. A drop of solvent can be used to re-dissolve the existing polymer and re-glue the whisker to the probes in hope to improve the electrical contacts. Most often the improvement cannot be made, and the crystal itself is almost impossible to clean and reuse without damaging it.

Another disadvantage of the POB method is that whiskers end up bending over the profile of the probes patterned on top of the substrate. An array of probes is like a row of railroad ties. When the whisker is laid over them it can deform because the probes have finite thickness (typically ~100 nm – comparable to a typical whisker thickness). We have not thoroughly investigated what effect the elastic straining of the crystal may have on CDW transport in such a scenario, however unnecessary bending of crystals should be minimized in order to avoid plastic deformations and its underlying crystal disorder which can substantially change CDW transport properties.

Because its Fermi surface is partially ungapped, $NbSe_3$ remains a good conductor at low temperatures. As a result, even modestly large contact resistances can pose as conduction bottlenecks where most ohmic heating occurs. Typical contact resistances obtained using the POB technique with 2-$\mu$m-wide probes are on the order of a few k$\Omega$, which is one to two orders of magnitude larger than the resistance of a whisker of typical size. Furthermore, contact resistance increases as the probe or sample width decreases, so POB methods are limited if one wants to study physics at yet smaller (sub-micrometer) dimensions.



In order to overcome these limitations and improve on previous work, and encouraged by enormous capabilities of modern technology in microfabrication processing, we set out to develop more ideal electrical contacts with procedures that at the same time allow direct patterning of NbSe$_3$ samples.

## Developing Probes-on-Top (POT) Method

We chose a "probes-on-top" (POT) configuration where lithographically defined metal probes are evaporated over the whisker that had previously been affixed to the substrate. Some basic forms of the POT approaches have successfully been used before,[89] but in this early work the potential of microfabrication was not fully exploited to gain benefits of patterning NbSe$_3$ crystals and to access small scales in sample preparation that can easily be achieved by the currently available technology. This is in part because at the time not much was known about the effects of microfabrication processing on NbSe$_3$ whiskers (i.e. heating in air or in contact with resists and other common agents, exposure to different reactive plasma species, etc.). Latyshev *et al.*,[91] Mantel *et al.*,[93,94] and others[95,97] later pushed the boundaries with their POT contacting methods by bringing NbSe$_3$ into the cleanroom and applying microfabrication techniques directly to the crystals, but their methods have shortcomings (some of which we discussed in section 2.5), and/or their published work lacks sufficient technical detail to reproduce their procedures reliably.

In the following we investigate factors that affect the quality of electrical contacts when metal probes are evaporated over a whisker. These include whisker surface preparation and contact-probe step coverage over the whisker edge. We show how to minimize the effects of the



impeding factors and develop an optimized POT process that to our knowledge yields the most reliable electrical contacts at the whisker-probe interface to date. The outline of the POT fabrication process is given in section 2.8, which also includes additional recipes we developed that can be used in conjunction with the POT recipe to permanently affix $NbSe_3$ whiskers to substrates, and thin and/or pattern them if needed. More complete technical details of the POT fabrication recipe are listed in Appendix C, Recipe 4.

**Surface Preparation for Probe Deposition.** As we described in sections 2.2 and 2.5, heating of whiskers in air ambient or their exposure to $O_2$ plasma compromises the ability to electrically contact them. Unfortunately, both processes are quite common steps in standard clean room recipes when patterning thin films. We have shown that the poorly conducting surface layer can be removed from the whisker by $CF_4$ plasma etching with controllable slow etch rates. We show that executing this step immediately prior to evaporation of metal probes onto the whisker is crucial for obtaining good electrical contacts between the probes and the whisker when the whisker had been previously exposed to $O_2$ plasma and/or heating in air. This pre-evaporation etch step takes place after patterning of a resist mask through which the metal is to be evaporated onto the whisker. The $SF_6$ or $CF_4$ plasma then reaches the whisker though the trenches in the resist mask.

We have attempted to electrically contact twenty-three $NbSe_3$ whiskers by evaporating multiple Au probes on each (a thin Ti layer is also evaporated prior to Au deposition to serve as an adhesion layer – a common practice when evaporating Au). Some of the whiskers had been exposed to heating in air and/or $O_2$ plasma prior to contact evaporation. We varied several parameters during the process development and the following general trends emerged:



- The longer the crystals were exposed to heating in air ambient and/or to $O_2$ plasma, the lower the yield of "good" contacts.

- The larger the number of processing steps that exposed the whisker to chemicals and heat, and the longer it took to complete them, the lower the yield of "good" contacts.

- The longer the exposure to $CF_4$ plasma prior to metal evaporation, the higher the yield of "good" contacts.

It was clear that the duration of the $CF_4$ etch had a particularly strong impact on the contact outcome

**Figure 2.38** shows the results of etch duration tests performed on six $NbSe_3$ whiskers. We subsequently evaporated Ti/Au probes over the samples and each whisker ended up with a number of probes that ranged between 13 and 16. The exact processing recipe applied to the six samples prior to the $CF_4$ etch varied slightly from sample to sample, but the exposure to heat and $O_2$ plasma was comparable for each. The figure shows that longer $CF_4$ etching results in higher yields of "fair" and "good" contacts. The exposure of $NbSe_3$ surface to 40 seconds of etching resulted in all contacts of "good" quality after probe patterning.



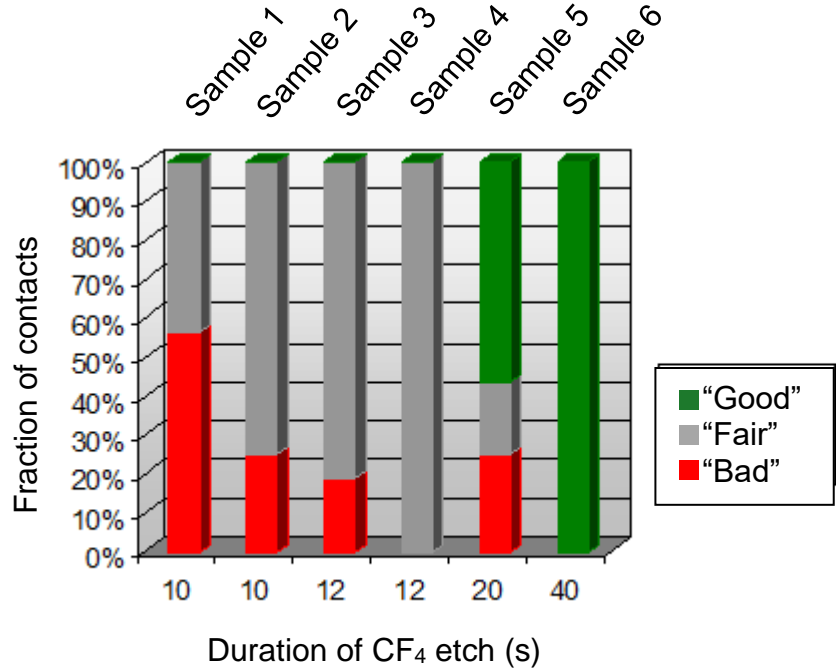

**Figure 2.38**

The effect of CF$_4$ plasma etch duration on the success to contact whiskers by POT method. Each histogram represents one of six whiskers whose surface was exposed to the etch. Subsequently, each whisker had multiple probes evaporated on top of it ranging between 13 to 16 probes per whisker. Terms "good", "fair", and "bad" contacts are defined earlier in this section. Used CF$_4$ plasma etch parameters: gas flow 30 sccm, pressure 30 mTorr, power 20 W.

The advantage of contacting whiskers by POT method (as compared to the POB method) is

that the top of the whisker can easily be prepared by this type of etching to obtain clean and

conductive contacting surfaces.

As we mentioned, the CF$_4$ etching is performed through a patterned resist mask. The etch

removes a thin layer of the unmasked segment of the whisker which results in slightly

modified geometry of the whisker surface. One should thus choose the shortest etch time that

still yields good contacts. After metal evaporation, the resulting probes are slightly imbedded



into the whisker. This can be beneficial because the probes lock the whisker into place on a substrate and help minimize thermal "crawl" of the whisker during cooldown or during local ohmic heating in measurement which, in turn, minimizes breakage of contacts. This will be discussed in more detail on a concrete example in section 2.9. The effects of this slightly modified surface geometry of the crystal on transport were not systematically studied here, however the samples prepared in this way did not display deteriorated CDW transport characteristics, i.e. *I-V* characteristics, mode-locking, threshold fields, etc. were comparable to the ones observed in high quality unprocessed samples.

**Contact Probe Step Coverage.** It is important to be aware of another potential difficulty when trying to contact the whiskers by POT technique. In order to ensure good electrical contact between the metal probe and the whisker, metal probes must be continuous across the whisker edge, i.e. the step coverage by the evaporated metal across the whisker edge has to be sufficient. A general rule-of-thumb to obtain continuity of evaporated metal across a vertical step edge is that the metal thickness to step height ratio should be at least 1:1. It is possible to obtain good metal continuity over the step with smaller metal-to-whisker thickness ratios but the results may not always be consistent.

Whisker thickness can become a limiting factor for a POT method. Thick whiskers require that a thick layer of metal be evaporated into patterned resist trenches. In order to perform the last processing step properly (lift-off of a resist mask), it is recommended that the resist mask be at least 2-3 times thicker than the evaporated metal. For instance, a 0.7 μm thick crystal optimally requires > 0.7 μm thick metal probe layer which in turns requires > 1.4 μm thick resist mask. When large-feature probes (>1 μm wide) are desired, so that optical lithography



can be used to define trenches in the resist mask, this is not a problem, since optical resists can easily be prepared this thick. If submicron probe widths are desired one would typically employ the use of electron-beam lithography (EBL). Defining narrow trenches in EBL resists thicker than 1 μm can be problematic because: 1) very thick layers of standard EBL resist stacks like copolymer/PMMA bi-layers used in lift-off processing crack during curing, and 2) resist dosing for e-beam exposure in thick resist is complicated by the lateral- and back-scattering of electrons which results in poor definition of fine (submicron) lines in thick resist. Thus for submicron definition of probes the POT process restricts the use of whiskers to thinner crystals, or one can resort to thinning of whiskers by etching to reduce their thickness.

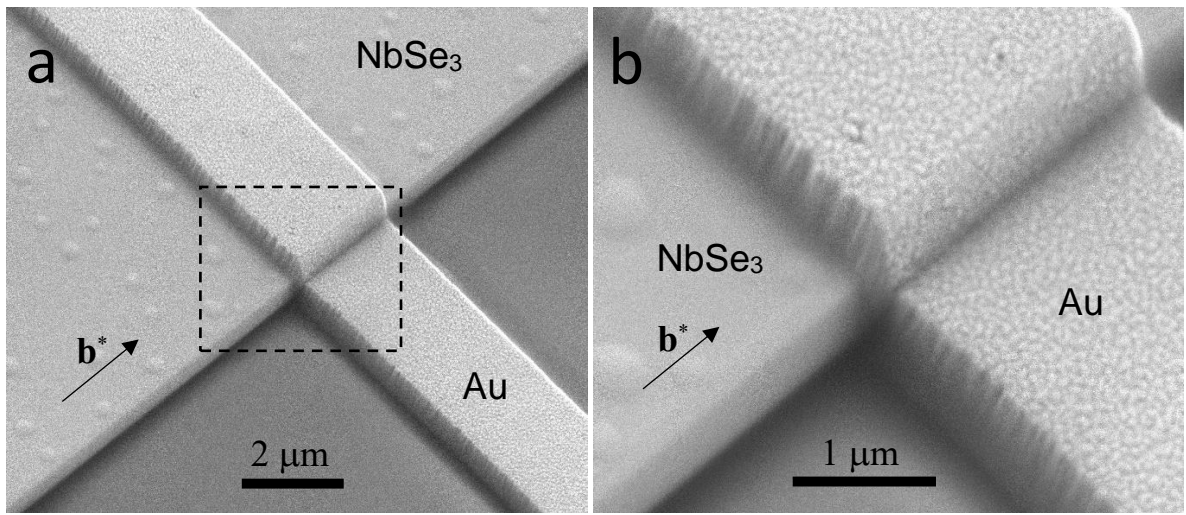

**Figure 2.39**

(a) SEM image of a sample showing step coverage by a metal probe over a whisker edge. 376-nm-thick whisker with 455-nm-thick gold probe deposited over it. Ratio 455 nm / 376 nm = 1.2 > 1 ensures sufficient step coverage by metal and electrical continuity of the probe. The dashed rectangle is the area magnified and displayed in (b). The probe is partially imbedded into the whisker surface as discussed in the main text.



**Figure 2.39** shows sample Gold12a used in experiments described in chapter 3. Gold probes patterned over the crystal are electrically continuous over the step edge. Although the probes implemented here were approximately 2 μm wide, the recipe we developed employs EBL in the fabrication process and can produce probes of submicron width on thin whiskers.

## 2.8  Fabrication Recipes

Tests and considerations discussed in this chapter and techniques we developed helped us tailor several useful recipes for preparing NbSe$_3$ samples. Individual recipes can be stringed together as needed to create complex sample structures. For example, to create a patterned NbSe$_3$ crystal on a silicon substrate with electrical contacts fabricated using POT method use Recipes 2, 3, and 4. The following is a list of recipes with a brief description of steps in each with technical details containing specific fabrication parameters given in Appendix C.

### Recipe 1: Contacting Whiskers by Probes-on-Bottom (POB) Method

- Pattern Au (with a thin Ti adhesion layer) probes on a thermally oxidized Si or alumina substrate using optical lithography.

- Position a whisker over the probes.

- Dispense a drop of a polymer dissolved in a solvent over the whisker resting on the probes. When the solvent evaporates the whisker will be fixed to the probes and substrate.



## Recipe 2: Affixing Whiskers to Substrate

- Spin polyimide on a Si substrate.

- While polyimide is still wet, position a whisker on the substrate and cure in a vacuum oven.

- Etch away excess polyimide that is not covered by the whisker by $O_2$ + $CF_4$ plasma if needed (this can make whisker surface too resistive for contacting, so a pure $CF_4$ etch should later be used prior to evaporating metal probes).

## Recipe 3: Thinning or Patterning Whiskers

- Use electron beam lithography to pattern a resist mask on a whisker.

- As needed thin or completely etch away exposed parts of the whisker by $CF_4$ plasma.

## Recipe 4: Contacting Whiskers by Probes-on-Top (POT) Method

- Use electron beam lithography to pattern a bilayer resist mask on the whisker. The resist mask should have trenches where Au probes will be evaporated over the whisker.

- Use $CF_4$ plasma etch to remove insulating layer from the unmasked part of the whisker.

- Evaporate Au (after a thin Ti adhesion layer) on the substrate and perform lift-off to remove the resist mask coated with the excess metal leaving behind defined metal probes over the whisker.



## Recipe 5: Electroplating (Current) Probes into a Whisker Cross-Section

- Pattern gold probes to be used as voltage probes in measurements on a Si substrate by POB method.

- Position a whisker over the probes.

- Spin polyimide over the structure to encapsulate the whisker on top of the voltage probes.

- By using contact lithography, pattern a resist mask with trenches where electroplated gold probes will form in a later step.

- Use a combination $O_2$ and $CF_4$ plasma to etch polyimide from the bottom of the trenches, thus exposing the whisker surface in the trenches.

- Use $CF_4$ etch to completely etch away the whisker material in the trenches forming gaps in the whisker cross-section.

- Evaporate seed Au layer for electroplating, then lift off the resist mask to remove excess seed metal.

- Electroplate Au over the seed layer. This will fill whisker gaps with Au so that probes are imbedded into the whisker cross section (i.e. the length of the whisker will be interrupted by the probes that cross it).



## 2.9    An Example of Sample Preparation

The culmination of all the tests performed on NbSe$_3$ that we presented in this chapter is a development of a complex fabrication process to prepare a robust measurement sample.  The process sequentially combines multiple recipes listed in section 2.8 to affix a NbSe$_3$ whisker to an oxidized silicon substrate (Appendix C, Recipe 2), thin a length segment of a whisker to a desired thickness (Appendix C, Recipe 3), and then using POT method successfully contact with gold probes both the thinned segment and the unetched (control) segment of the whisker (Appendix C, Recipe 4).  The full process is illustrated in

**Figure** 2.40 and the SEM images of the fabricated sample Gold12 are shown in **Figure 2.41**. The thickness of the segment of the whisker shown on the right side of the main panel was reduced from the original thickness of 376 nm to 152 nm by CF$_4$ etch.



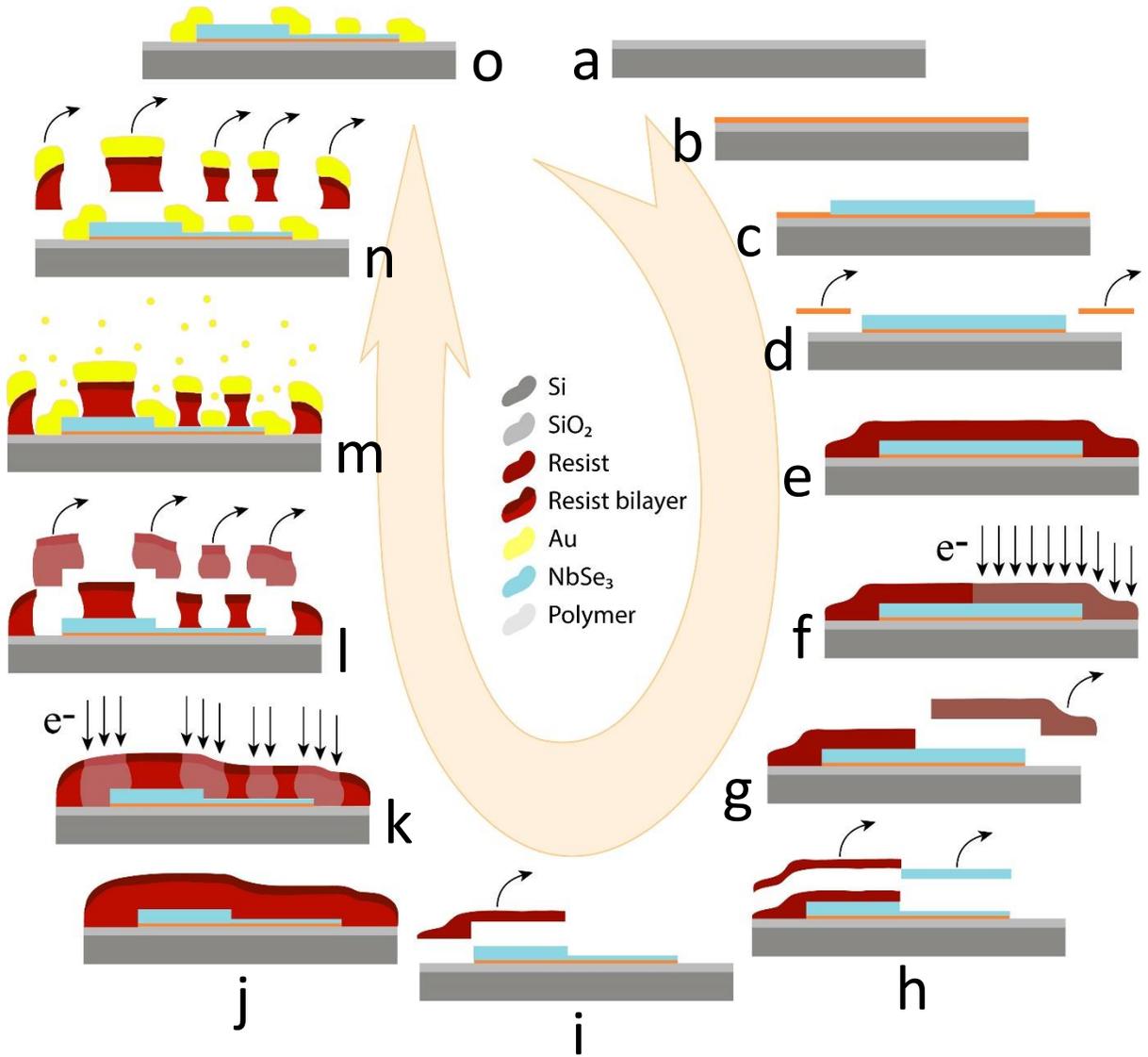

**Figure 2.40**

Fabrication process to thin a segment of a NbSe$_3$ whisker and then apply "probes-on-top" (POT) method to apply electrical contacts. (a) Start with a thermally oxidized silicon substrate. (b) Spin-coat the substrate with polyimide. (c) Position a whisker on the wet polyimide and cure in vacuum oven. (d) Remove excess polyimide by O$_2$ + CF$_4$ plasma. (e)-(g) Use EBL to pattern a resist mask that will leave a segment of the whisker unprotected. (h) Thin the unmasked part of the whisker by CF$_4$ plasma. (i) Strip the resist mask. (j)-(l) Use EBL to pattern a bilayer resist mask for metal probes. (m) Remove resistive layer from the whisker surface by CF$_4$ plasma, then evaporate Au. (n) Perform lift-off. (o) Defined Au probes remain on the whisker and substrate after the mask lift-off. Further technical details of the full process are contained in Appendix C, Recipes 2, 3, and 4.



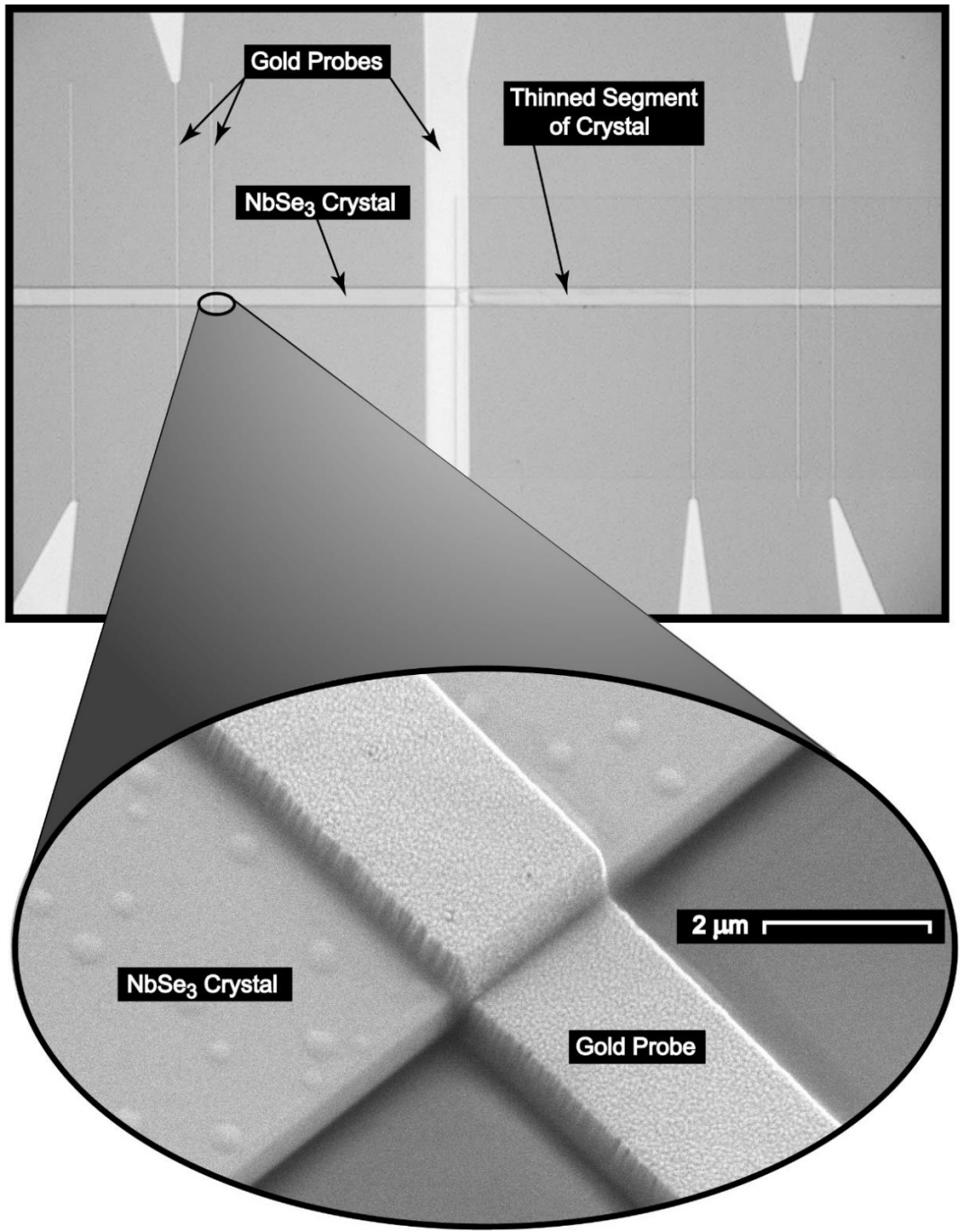

**Figure 2.41**

SEM images of sample Gold12. The unetched segment (sample Gold12a) of the NbSe$_3$ whisker is on the left side of the main image. As-grown thickness of the crystal is 376 nm. The segment on the right (sample Gold12b) has been thinned from 376 nm to 152 nm by CF$_4$ plasma etch. The expanded inset shows step coverage of the gold probe over the crystal edge of the thicker (Gold12a) segment. "Processing dots" are visible on the whisker surface.



During cool-down from room temperature to 14 K all thirteen fabricated gold probes retained "good" electrical contact with the whisker. All contacts also survived full thermal cycling to room temperature and back to 14 K. The contacts all remained stable during three months of data acquisition. This is in contrast to samples with contacts prepared by more traditional techniques like POB or conductive paint techniques, for which approximately two-thirds of contacts typically remain in "good" condition after cooldown, with a typical contact lifetime of 1-2 weeks before contact stability of many probes degrades beyond useful (this is often linked to the total current density and the local heating produced at the contacts in the experiment, as we discussed earlier).

Loss of electrical contacts due to thermally induced stresses between a whisker, substrate, and metal probe during thermal cycling or due to local ohmic heating is a common mode of contact failure. At low temperatures ohmic heating is pronounced at the conductivity bottlenecks of the circuit which are typically at the contact between a whisker and a current probe. This is especially true for high-resistance silver paint or POB contacts. At high current densities the local heating creates "hot spots" that often destroy good electrical connections. The POT method used on Gold12 sample produced electrical contacts with resistance of ~20 $\Omega$ per 2 $\mu$m × 25.6 $\mu$m contact area, compared to a few k$\Omega$ typical for contacts fabricated by POB technique with similar contact area and for silver paint contacts with significantly larger contact areas. Low resistance contacts minimize "hot spots" which, in addition to the "locked-into-crystal" configuration of POT probes, is a likely reason why the sample Gold12 outlived all the samples we measured with contacts prepared using earlier, more traditional methods.



The inset of **Figure 2.42** displays two data sets for single particle resistance, $R_S$, vs. temperature, $T$, one for the thick and one the thin segment of sample Gold12. Each data set shows formation of two CDWs at the expected transition temperatures. The data was taken by biasing each segment with a sufficiently small current (10 μA) to keep the CDWs pinned below $E_T$ so that all the current is carried by the single particle channel of the whisker. $R_S$ was measured across a whisker length between voltage probes spaced by $L_v = 200$ μm. The data was taken in a 4-probe configuration for each segment to ensure contact resistances were excluded. For easier comparison, in the main panel of the figure we have scaled the $R_S$ data for the etched segment (Gold12b) by the thickness ratio $t_b/t_a = 376$ nm / 152 nm = 0.40 and plotted it with the $R_S$ data for the non-etched segment (Gold12a). The curves overlap very well showing no anomalies in the $R_S$ vs. $T$ profile for the etched segment and indicating that etching does not change single particle resistivity even when the thickness of a whisker is reduced by more than a factor of two. Collective transport measurements for both segments are presented in Chapter 3. This method of preparing samples, although more involved, has many great advantages over the traditional techniques used in the past.



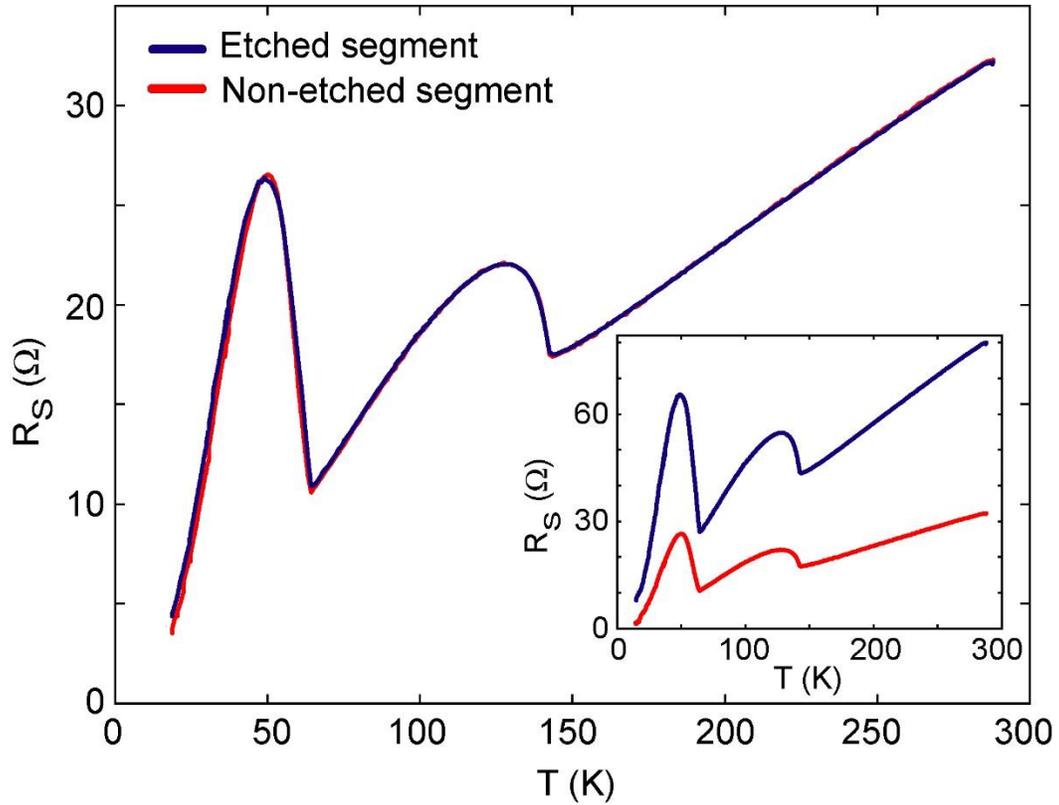

**Figure 2.42**

Inset of the figure shows two $R_S$ vs. $T$ data sets for sample Gold12: in blue for the etched (Gold12b) and in red for the non-etched (Gold12a) NbSe$_3$ segment. Each data set was taken in a 4-probe current-biased configuration between voltage probes separated by $L_v$= 200 μm. 4-probe configuration ensured that contact resistances were excluded. Main panel shows the same data sets but with resistance data for the etched segment scaled by the thickness ratio $t_b/t_a$ = 152 nm / 376 nm = 0.40 and plotted against the data for the non-etched segment.

## 2.10  Highlights of Developed Sample-Preparation Methods

Here we summarize the merits of the new sample preparation methods presented in this

chapter by listing the advantages over the existing techniques. For completeness we also

point out their downsides.



**Optimized Dry Etch Methods for NbSe₃.** We identified both ion-mill (physical) and RIE (predominately chemical) etching parameters for NbSe$_3$. For RIE we explored a wide parameter space that includes gas specie, pressure, flow rate, power, and etch rate, and we presented useful sets of parameters in Appendix B. We have identified which etch procedures compromise crystal surface conductivity and which leave surfaces sufficiently conductive for subsequent contacting with electrical probes. We have evaluated post-etch surface roughness for various plasma etch species. The optimized dry-etch techniques allow us to precisely pattern NbSe$_3$ crystals into various geometries with lateral and vertical control.

**Enabled Control over Crystal Thickness Parameter.** By implementing slow reactive ion etching with established etch rates of NbSe$_3$ one can custom tailor the final thickness of a sample crystal. This contrasts with the old practice of manually cleaving thick crystals into thinner ones which does not allow a precise control over the final crystal thickness. In practice the etch-produced surface roughness puts a lower limit on the thickness of a usable sample to approximately 50 nm. In comparison, as-grown crystals of such small size are extremely fragile and difficult or impossible to transfer from a growth batch to a substrate with tweezers. The etching techniques in combination with lithography can now produce many segments of a single crystal each of a different thickness as needed.

**Revealing Grain Boundaries within the Crystal Bulk.** A smooth crystal surface, as observed under an optical microscope, is often a poor indicator of the quality of the crystal order in the bulk of NbSe$_3$. We have shown that etching techniques can reveal grain boundaries or cracks associated with them that are hidden beneath the surface in the bulk of



the crystal.  The technique can be used before or after experiments to investigate crystal quality.

**Surface Dots.**  We have identified heating and dry etching techniques to remove dots that are often observed on freshly grown whisker surfaces and which can hinder electrical contacting of crystals.  We have shown that ordered arrays or trains of dots can be a tell-tale sign of steps associated with grain boundaries or cracks on the whisker surface which are too small to otherwise be discernable under optical microscope.

**High Yield of Robust and Stable Electrical Contacts.**  We have developed a process to apply metal probes over $NbSe_3$ whiskers that produce long-lasting electrical contacts with "good" characteristics as defined in section 2.7.  Contacts can survive multiple thermal cycling from room to low (< 20 K) temperatures.  The stability and robustness of contacts produced by the new POT technique in part comes from the geometry of the contact (the probes are "locked" into the crystal), in part due to low resistance of fabricated contacts, and in part due to a layer of polyamide "glue" between the whisker and the substrate.  Improved stability and robustness of contacts is a major achievement of the new POT method which we expect will significantly improve sample preparation in the future.  The technique also outperforms the earlier, more traditional methods in terms of contact yield and reproducibility.

**Improved Temperature Stability and Uniformity During Measurements.**  By gluing a whisker to a substrate by polyimide, we ensure that along its bottom side the whisker is thermally connected to the substrate.  Although polyimide is far from being an ideal thermal conductor, a glued whisker is better thermalized than a whisker in vacuum and partially in



contact with the substrate.  With POT method, metal probes are evaporated onto the whisker which ensures that the whole area where a probe crosses the whisker is in a good thermal contact with it.  Good thermal anchoring dissipates contact "hot spots" produced by ohmic heating, provides uniform cooling of the whisker, and minimizes the occurrence of thermal gradients and instabilities along the whisker during measurements.

**Minimizing Crystal Strain.**  Another advantage of the POT technique is that it avoids whisker-on-rail-road-ties configuration which is characteristic of samples with POB contacts. In the POT case the probes are patterned over the whisker which lies flat on the substrate which completely avoids crystal bending (and any structural disorder like microcracks that bending may induce) and prevents the perturbation of intrinsic CDW transport that this may cause.

**Flexibility of Patterning Crystals and Probes with Submicron Precision.**  Identifying and optimizing dry etching techniques for NbSe$_3$ allows us to precisely pattern the crystals into various sizes and geometries with lateral and vertical control. Furthermore, utilizing electron beam lithography allows submicron pattern and alignment precision for small-feature needs in the future (e.g. to probe CDW physics on sub-micron length-scales) that can easily be adapted to individual crystal requirements.  Since no two NbSe$_3$ whiskers are alike in terms of physical dimensions and grain boundary structure, pattern designs must be tailored to each whisker individually. It is thus important to have flexibility in producing different probe and crystal patterns with ease.  Electron beam lithography provides this flexibility through software pattern control, so that pattern changes can often be performed "on the fly" (in contrast to standard optical lithography which typically requires a separate lengthy process to



produce a physically separate optical mask each time pattern changes need to be implemented).  Furthermore, EBL can achieve excellent alignment of patterns along whisker's **b**$^*$ axis, with much better precision than with capabilities of optical lithography.

**Process Compatibility and Ease of Integration with Standard Fabrication Techniques.**

Individual recipes we developed (Recipes 1-5 in Appendix C) are well characterized and make the integration of the whole sample preparation process easy and flexible.  For example, in the recipe sequence Recipe 2-4 used to produce sample Gold12, one can repeat Recipe 3 several times to produce a sample with several whisker segments each with a different crystal thickness; or Recipe 3 can be entirely omitted in the sequence when thinning of the whisker is not desired.  The use of polyimide in Recipe 2 attaches the whisker permanently to the substrate so that the whisker is not prone to move or float away when soaked in various solvents routinely used in microfabrication processing.  A whisker mounted in this way can be used in multi-step lithography and other fabrication processes.  The use of silicon wafers instead of alumina as substrates allows us to easily integrate $NbSe_3$ processing with other standard processing techniques, since most microfabrication tools are configured for standardized silicon wafers as substrates. Moreover, our recipes incorporate standard and common fabrication techniques and can be utilized in most decently equipped microfabrication facilities.

**Cost, Technical Expertise, and Sample Preparation Time.**  A full process to electrically contact a sample using POT method is more expensive, lengthy to complete, and requires more technical expertise as compared to traditional techniques because of its higher complexity and utilization of sophisticated microfabrication tools.  Given the very high



success rate of producing working samples and long-lived electrical contacts our process is a more efficient option in the long run than the older techniques. It does however require higher skill and a longer training period for a person performing fabrication. Multiple samples can be processed in the same fabrication run to save time.

**Modified Crystal Cross-Section and Current Shunting through Probes.** In order to get robust and stable electrical connections in the POT recipe the probes are slightly etched into the bulk of the crystal. This modification of crystal geometry could affect the CDW transport, especially in thinner crystals where a fraction of the modified crystal thickness may be appreciable. In addition, low resistance contacts could perturb intrinsic CDW transport properties by modifying electric fields inside the whisker, i.e. a metal probe contacting a whisker locally shunts it. Sample Gold12, with metal probes imbedded in two whisker segments each of different thickness, displayed relatively clean CDW signatures like differential resistance curves with well-defined threshold fields, nominal mode-locking characteristics, and very sharp NBN peaks in the slow branch. We did, however, observe somewhat lower-quality mode-locking characteristic in the thinner segment of the whisker where the fractional depth of the imbedded probe into the whisker cross-section was larger, and thus expected to have a stronger perturbing effect. These perturbations can be minimized by fabricating narrower probes or by patterning the whisker geometry to turn a part of the whisker itself into transverse probes as was done in the work of Mantel *et al.*[94] as shown in **Figure 2.11**. The ends of these probes, which are far away from the active area of the whisker where the collective transport is being investigated, can subsequently be contacted by metal probes using our POT method to further connect the whisker to the measurement apparatus.



**Thick Crystals Difficult to Contact.**  The POT contacting process is optimized for whiskers with thickness up to approximately 0.4 μm.  Thicker crystals are more difficult to contact because they require a thicker layer of evaporated metal (gold) to obtain a good step coverage of the probe over the crystal. A thicker probe requires a thicker patterned resist mask in order to adequately perform lift-off of metal after deposition. Typical thickness of e-beam lithography resists thus puts an upper limit on whisker thicknesses that we can process with POT method.  To overcome this limitation one can first use the approach of Mantel *et al.*[94] where narrow transverse probes are carved directly out of NbSe$_3$ ribbon by dry etching.  The end of the probe away from the active part of the whisker with collective transport can then be thinned by etching to below 0.4 μm, and then further contacted by metal probes using the POT technique.

The methods and recipes developed and described in Chapter 2 enable the use of multistep lithography on NbSe$_3$ whiskers.  We have successfully integrated sample preparation of three-dimensional NbSe$_3$ whiskers of varying geometries and sizes with standard clean-room processing techniques and equipment originally developed for thin film technology.  The use of nanofabrication techniques should further enable the future exploration of sub-micrometer physics of CDWs in NbSe$_3$.



# 3 Dynamics of Temporally-Ordered Collective Creep in NbSe₃

As discussed in chapter 1, CDW collective transport in the high temperature regime ($2T_P/3 < T < T_P$) is relatively well understood with FLR-based models accounting for most experimentally observed features, while in contrast, the transport at low temperatures ($T < 2T_P/3$) is still poorly explained. When Lemay *et al.*[55] accidentally observed narrow band noise (NBN) oscillations (a signature of collective dynamics) in NbSe₃ samples in the lower temperature regime between $E_T$ and $E_T^*$ where a CDW also exhibits creep-like behavior, it became even more obvious that the collective dynamics in this regime were complex and not well understood. To date on the CDW collective transport phase diagram (a map of the evolution of $j_C(E)$ with temperature), temporally-ordered collective creep remains the most puzzling feature, and no model correctly predicts it.

Experimentally, it has been well established that at low temperatures there exists a distinct branch of the phase diagram bound by two distinct thresholds, $E_T$ and $E_T^*$, where the $j_C(E)$ relation has a very different functional form from what is observed below $E_T$ (where $j_C$ is essentially zero) and above $E_T^*$ (where the fast dynamics resembles the high-temperature collective sliding state). This intermediate "slow branch" at low temperature links pinned CDW state below $E_T$ to sliding CDW state above $E_T^*$ as shown in **Figure 1.6**. It is quite astonishing that no model to date correctly predicts a whole branch of the transport phase diagram.



Most of the chapter 3 and 4 address dynamics in the slow branch of NbSe$_3$ in an attempt to qualitatively and quantitatively understand this mysterious transport regime and illuminate the underpinnings of the dynamics that create rich transport behavior that is observed experimentally. The extended goal of the work presented in the remaining pages of this dissertation is to unify our experimental findings and those of others into a consistent picture of a collective transport phase diagram for CDW (and related) systems.

A starting point in experimentally mapping a transport phase diagram is a simple current-voltage (*I-V*) measurement which reveals a functional form of the velocity-force relation. As discussed earlier, CDW current $I_C$ in the slow branch ($E_T < E < E_T^*$) of the NbSe$_3$ phase diagram is up to nine orders of magnitude smaller than the total current $I_{tot}$ flowing through the whisker due to large parallel conduction by single particles: $I_C << I_{tot} = I_S + I_C$ where $I_S$ is the single particle current. Thus, one cannot use *I-V* data obtained in a conventional way to extract collective current vs. voltage relation ($I_C$-*V*). When Lemay *et al.*[55] discovered coherent oscillations in the slow branch of the NbSe$_3$ phase diagram, it became possible to map CDW current density $j_C(E,T)$ by measuring the frequency of coherent oscillations, $f_{NBN}$, at different electric fields $E$ and temperatures $T$, and calculating $j_C(E,T)$ from $f_{NBN}$ by using **(1.7)**. It also became obvious that the collective dynamics in this regime was not well understood, since collective behavior was not expected to persist in the regime where CDW exhibits creep. As noted, such a behavior had not been predicted by any theoretical model to date. While the incoherent glassy dynamics below $E_T$ and coherent sliding above $E_T^*$ are relatively well understood, explaining the behavior between the two thresholds remains a challenge.



We will show that the slow branch in NbSe$_3$ is characterized by distinct dynamics that allows creep to coexist with temporally-ordered motion. One of the striking findings of the work by Lemay *et al.*[55] was that in *undoped* NbSe$_3$ samples a simple quantitative analysis of $j_C(E,T)$ data in the slow branch suggested that a small *local* length-scale associated with sample volume per residual impurity plays a crucial role in the dynamics of *collective* creep. Our initial intent was to expand on this work and use crystals, pure and doped with different types of impurities as well as different doping concentrations as controlled parameters, to investigate the effects of disorder and pinning on low-temperature dynamics in NbSe$_3$. At the heart of the experiments presented in this chapter is a measurement of narrow-band noise (NBN) in high quality NbSe$_3$ crystals. The results presented shine light on the mechanisms of the temporally-ordered collective creep by identifying and quantifying relevant physical parameters (length and energy scales) that define its dynamics, and further by characterizing the transition from coherent creep to coherent sliding at $E_T^*$.

## 3.1 Sample Preparation and Experimental Setup

Aside from documenting the experimental procedure part of this chapter, this section aims to provide an "instruction manual" for a potential future graduate student listing enough detail so that the measurements can easily be reproduced.

### Pure and Doped NbSe$_3$ Crystals

As discussed earlier only a small fraction of whiskers in a typical NbSe$_3$ batch (we estimate less than one in a thousand) are high quality single crystals such as the one shown in **Figure**



**2.24**.  To prepare samples we have carefully selected and separated such whiskers from the batch under optical microscope using tools with tips of microscopic dimensions[*].  Finding single crystal whiskers is somewhat an art form, and is almost always a tedious and slow process, but it is an extremely important step in sample preparation.  Samples composed of multiple crystal domains are unlikely to carry a homogeneous current density through the entire crystal cross section which precludes observation of clean intrinsic signatures of collective transport such as a single well-defined NBN peak.  In whiskers with steps in the cross-section such as the ones shown in **Figure 2.5 (a)** and **(b)**, and **Figure 2.23,** finite-size effects can produce different current densities in different parts of the cross-section causing CDW to shear between the regions.[36-38,40,105]  Since $j_C \propto f_{NBN}$, (see **(1.7)**), multiple domains with different current densities can produce multiple, wide, and overlapping NBN peaks of coherent oscillations in the frequency domain of a measured voltage signal.  In addition, CDW shear produces increased levels of $1/f$-like broad band noise (BBN)[37] which can completely obscure NBN peaks in the spectrum especially at low frequencies.  Through careful sample selection and collective-transport test measurements discussed later in this section, we have singled out samples with clear CDW signatures.  All samples considered in our experiments to determine $j_C(E,T)$ in the slow branch of NbSe$_3$ displayed single sharp NBN peaks characteristic of uniform transport in high-quality single crystal whiskers.

NbSe$_3$ samples were selected from crystal batches that were grown pure (undoped) or in presence of elemental tantalum (Ta) or titanium (Ti) to purposely introduce impurities in the crystals.  In the periodic table Ta is in group V B, just below Nb, and its valance shell has the same electronic configuration as Nb, i.e. Ta is an isoelectronic impurity. As a result, Ta impurities are expected to replace niobium atoms in NbSe$_3$ lattice without causing a substantial distortion. Ti is in IV B group and is a non-isoelectronic impurity and as such is expected to produce stronger disorder effects. Consequently, Ta and Ti impurities are expected to have different effects on CDW pinning. The work by McCarten *et al.*[19] showed that both impurity types in NbSe$_3$ produce weak pinning within the context of the FLR model, but that Ti has much stronger effect on $E_T$ threshold and residual resistance ratio ($RRR$) than Ta impurities. For example, at impurity concentration of $n_i \approx 100$ ppm, Ti doping produces roughly 40 times larger $E_T$'s, and roughly *25* times smaller $RRR$'s than does Ta-doping at a comparable doping concentration.

---

[*] Fine spring scissors and tweezers (Dumont #5 Fine Forceps, item No. 11254-20) available from Fine Science Tools, and homemade "poke tools" with torch-sharpened tungsten tips

From our experience we note that the process of selecting whiskers without visible steps from doped crystal batches was noticeably more time-consuming than from pure batches. Since impurities introduce additional disorder in the lattice, one can expect that highly doped crystal batches will contain scarce numbers of whiskers devoid of steps in the cross-section. This is indeed what we observe in practice.

## Characterizing Impurity Concentration, $n_i$

Impurity concentration $n_i$ of each crystal batch was inferred from bulk values of residual resistance ratio $RRR$. $RRR$ is defined by



$$RRR = \frac{R_S(T = 300K)}{R_S(T = 4.2K)} \qquad \textbf{(3.1)}$$

where $R_S$ is a measured single particle resistance of a whisker segment at a given temperature. Measurements were performed on a few (at least two or three) whiskers selected from a batch, and the appropriate $RRR$ value for the batch was then obtained by averaging the measured values. With decrease of temperature, the phonon population in a crystal freezes out and at $T$ = 4.2 K the single particle conduction is essentially limited by disorder scattering, while at $T$ = 300 K disorder scattering provides a negligible contribution. Measured value of $RRR$ thus provides information about the extent of disorder present in the crystal and is related to impurity concentration as follows

$$\frac{1}{RRR} = \frac{1}{r_0} + \frac{n_i}{b_i} \qquad \textbf{(3.2)}$$

where $r_0$ is a contribution from residual defects other than impurities, and $b_i$ is an empirically determined coefficient that depends on impurity type. For the $RRR$ measurements we have chosen whiskers with no (or very few) steps in the whisker cross-section. In such samples of moderately high crystalline quality the main contribution to disorder in the crystal is associated with impurities rather than the non-impurity related structural defects. In fact, this is the case even in undoped samples where a small but finite number of spurious impurities outweighs the effects of other structural disorder.[42] The $1/r_0$ term is thus much smaller than the impurity contribution $n_i/b_i$ in **(3.2)** and can be neglected. Impurity concentration $n_i$ can then be approximated by



$$n_i \approx \frac{b_i}{RRR} \, . \tag{3.3}$$

$b_i$ coefficients for Ta and Ti were empirically determined by McCarten *et al.*[19] and were found to be $b_{Ta} \approx 3 \times 10^{20}$ cm$^{-3}$ and $b_{Ti} \approx 10^{19}$ cm$^{-3}$ respectively to within a factor of two[*]. Since the exact impurity varieties in the undoped crystals are not known, we used $b_{Ti}$, a coefficient for a non-isoelectronic impurity, to roughly estimate the $n_i$ in undoped crystals. This is justified because among all impurities the non-isoelectronic types dominate in lowering the *RRR* value in undoped crystals.

To measure *RRR*, whiskers were mounted onto a lab-made dipstick probe that can be submerged into liquid helium to measure $R_S$ at 4.2 K. The end of the probe contains alumina substrates with patterned four parallel gold electrodes suitable for a 4-probe measurement configuration. NbSe$_3$ whiskers are placed over the electrodes and secured to the substrate with a drop of polymer dissolved in a solvent as described in section 2.7 on contacting the whiskers by the "probes-on-bottom" method. To ensure good electrical contact between the electrodes and the whisker, we also applied a trace of silver paint over the whisker before applying the polymer coat. A 4-probe measurement configuration with current injected through the two outer (current) electrodes and voltage drop measured across the whisker segment between the two inner (voltage) electrodes ensured at both temperatures that the measured value of $R_S$ is a resistance of solely the whisker segment between the voltage electrodes, with resistance of electrical contacts and wires excluded. At 4.2 K collective CDW motion in NbSe$_3$ exhibits very large threshold fields so that a relatively small bias current of 10 µA in a sample is carried by uncondensed single particles, rather than the CDW channel (i.e. CDW is in a pinned state), ensuring that the voltage we measure indeed probes single particle resistance $R_S$.

---

[*] $1/r_0$ is especially insignificant correction to **(3.2)** when the magnitude of the uncertainty in $b_i$ coefficients is considered.

**Table 3.1** shows a list of crystal batches characterized to estimate their impurity concentration. All samples used in subsequent measurements of NBN in the slow branch (listed in **Table 3.2**) were selected from these batches.

**Table 3.1  NbSe$_3$ growth batches and doping**

| Batch | Doping type | *RRR* | $n_i = b_i/RRR$ ($10^{17}$ cm$^{-3}$) | Samples used in NBN measurements (see **Table 3.2**) |
|---|---|---|---|---|
| 1 | Ta | 40 | 75 | George 7 |
| 2 | Ta | 75 | 40 | George 6 |
| 3 | Ta | 110* | 27 | George 4, George 20 |
| 4 | Ti | 15 | 6.7 | Rudy 1 |
| 5 | Ti | 30 | 3.3 | Rudy 7 |
| 6 | Pure | 180** | 0.55 | Nely 3f |
| 7 | Pure | 200 | 0.50 | Gold 12a, Gold 12b |
| 8 | Pure | 400 | 0.25 | Serge |

*Measurements of $E_T$ dependence on thickness at 77 K performed on five crystals from Batch 3 indicate that the batch may have a somewhat lower *RRR*, between 60-90. This range is obtained when the independently measured values of $E_T$ from samples in this batch are compared to data in **Figure 3.2** from reference [19].

**RRR* measurements were performed on thin crystals where finite size effects must be accounted for. The value reported here is an estimated bulk RRR value obtained by adjusting the measured *RRR* to account for *RRR* dependence on sample thickness by comparing with data in **Figure 3.1** from reference[19]. The actual measured *RRR* value was lower than the bulk estimate reported here.

McCarten *et al.*[19] showed (see **Figure 3.1**) that finite-size effects play an important role in ordinary size crystals and 1/*RRR* decreases with crystal thickness $t$ and approaches a thickness independent, bulk value in thick crystals (typically above a few micrometers for undoped crystals).



The $E_T$ dependence on thickness shown in **Figure 3.2** was associated with crossover from three-dimensional to two-dimensional weak pinning when the bulk transverse CDW correlation length becomes limited by the crystal thickness.[*]

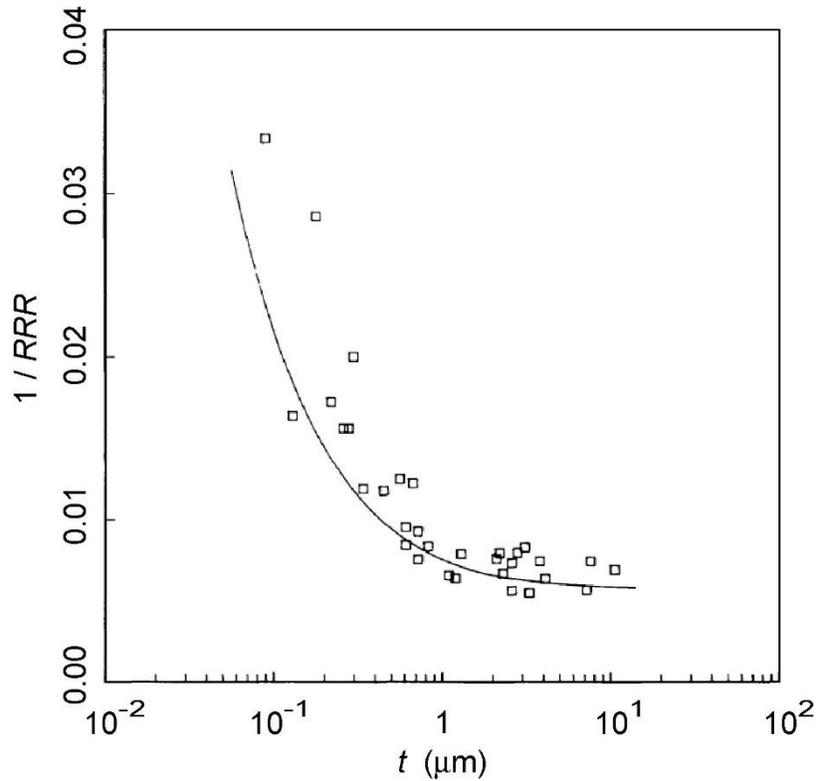

**Figure 3.1**

1/*RRR* vs. crystal thickness *t* for undoped NbSe₃ crystals. The solid line is a fit by Fuch's theory of surface scattering, assuming 100% diffuse scattering and a transverse single-particle mean free path of 0.7 μm. Obtained from reference [19].

---

[*] Typical widths and lengths of ordinary NbSe₃ whiskers are larger than length scales where finite-size-effects limit behavior for these dimensions, so mainly sample thickness must be accounted for.



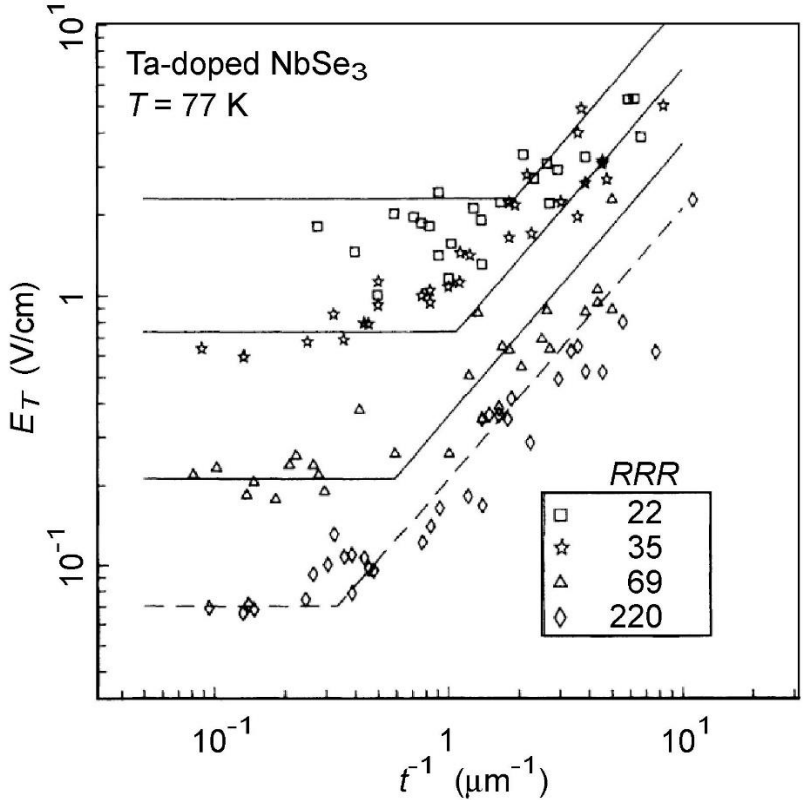

**Figure 3.2**

Threshold electric field $E_T$ at 77 K vs. $1/t$ where $t$ is crystal thickness of samples from an undoped batch (diamond symbol) and three Ta-doped batches (square, star, and triangle symbols). The legend provides bulk *RRR* values of samples measured. Obtained from reference [19]. The solid lines represent two-dimensional (small $t$ limit) and three-dimensional (large $t$ limit) weak pinning fits to equations described in the same reference.

To properly extract bulk *RRR* values, the measurements must be performed on thick whiskers where *RRR* value is approximately independent of $t$. Our measurements suggested that the *RRR* dependence on thickness is less pronounced in more doped crystals, nevertheless the variation in thickness between the chosen whiskers will introduce some uncertainty in *RRR* values reported in **Table 3.1**. In order to minimize this uncertainty, we made every effort to select thicker whiskers for the *RRR* measurements. *RRR* values reported are based on whiskers with thicknesses in a range of 0.85 μm – 2.0 μm except for Batch 6 (see ** in the



table caption).  We note that crystals with $t > 2$ μm and with no steps in the cross-section are rare and extremely hard to find in a batch, especially for doped batches.  The data in **Figure 3.1** indicate that crystal-to-crystal variation of bulk $RRR$ values (i.e. for $t$ approximately larger than 1 μm in thickness independent regime) is about 20%.  We thus estimate at least this much uncertainty in our reported values of bulk $RRR$.

Due to several factors that can introduce errors in $RRR$ measurements, as well as the large uncertainty in coefficients $b_i$, $n_i$ values in **Table 3.1** should be primarily used as a relative measure of doping levels between different batches while considering the uncertainties discussed.

## Samples and Electrical Contacts

**Table 3.2** summarizes the samples used in the transport experiments we present in this chapter as well as their doping, size, and electrical contact considerations.  Samples consist of NbSe$_3$ whiskers obtained from batches listed in **Table 3.1**, packaged onto silicon or alumina substrates and contacted with 2-μm-wide gold probes by either "Probes-On-Bottom" (POB) or "Probes-On-Top" (POT) methods discussed in section 2.7.  Specific recipes and procedures are listed in the last column of **Table 3.2**.  $L_i$, $L_v$, $L_{v1}$ and $L_{v2}$ are distances shown in **Figure 3.3**.  They define current- and voltage-probe configuration in the measurements.  All the samples were selected by choosing whiskers that had no visible steps on its top surface when viewed under an optical microscope.  The bottom side of the samples Gold12a, Gold12b, and Nely3f was additionally inspected by flipping each whisker over gently on a substrate, and it was confirmed that there were not visible steps on the underside of these samples.



Samples Gold12a and Gold12b are two contiguous segments of a single whisker (Gold12) attached to a Si substrate (Appendix C, Recipe 2). The segment Gold12a remained at its as-grown thickness of 0.376 um, while the segment Gold12b was thinned from 0.376 μm to 0.152 μm by $CF_4$ etch (Appendix C, Recipe 3). Electrical contacts were prepared with the POT method (Appendix C, Recipe 4) by evaporating Au (and thin Ti adhesion layer) over both segments as described. The individual recipe outlines were presented in section 2.8. Section 2.9 includes additional details of Gold12 sample preparation and characterization including images of the fabricated sample in **Figure 2.41** and a comparison of single particle resistance vs. temperature measurements for both segments in **Figure 2.42**. SEM inspection of the sample revealed that no steps or cracks were visible on the crystal surface on unetched segment Gold12a. The etched segment Gold12b revealed no additional grain boundaries indicating that the whole whisker is likely a single crystal $NbSe_3$.



**Table 3.2  Samples**

| Sample | Doping/RRR | Thickness $t$ (μm) | Width $w$ (μm) | $L_i/L_v$ (μm) | $L_{v1}/L_{v2}$ (μm) | Method to electrically contact samples |
|---|---|---|---|---|---|---|
| George7 | Ta/40 | 0.800 | 2.0 | 690/210 | 230/250 | POB, Recipe 1 on alumina substrate |
| George6 | Ta/75 | 0.276 | 1.4 | 710/210 | 250/250 | POB, Recipe 1 on alumina substrate |
| George4 | Ta/110 | 0.290 | 10.0 | 690/210 | 310/110 | POB, Recipe 1 on alumina substrate |
| George20 | Ta/110 | 1.31 | 1.5 | 630/70 | 70/490 | POB, Recipe 1 on alumina substrate |
| Rudy1 | Ti/15 | 1.0 | 3.8 | 320/70 | 140/110 | POB, Recipe 1 on alumina substrate |
| Rudy7 | Ti/30 | 0.16 | 24.8 | 710 / 140 | 320/250 | POB, Recipe 1 on alumina substrate |
| Nely3f | Pure/180 | 0.50 | 36.4 | 1840/50 | 860/930 | POT, Recipes 2-4 on thermally oxidized Si |
| Gold12a[*] | Pure/200 | 0.376 | 25.6 | 900/200 | 350/350 | POT, Recipes 2-4 on thermally oxidized Si |
| Gold12b[*] | Pure/200 | 0.152 | 25.6 | 900/200 | 350/350 | POT, Recipes 2-4 on thermally oxidized Si |
| Serge[**] | Pure/400 | 0.7 | 2.4 | - / - | - / - | POB, Recipe 1 on alumina substrate |

List of samples in which the clear NBN signature was observed in slow branch and their geometrical considerations. POB and POT stand for "Probes-On-Bottom" and "Probes-On-Top" method, respectively, discussed in section 2.7.  Recipes are outlined in section 2.8 and given in Appendix C.

[*]Gold12a and Gold12b are two different segments of a single NbSe$_3$ whisker.  They are listed as two separate samples because Gold12b was thinned down from whisker's original thickness by RIE and its transport characteristics such as $E_T$ and $E_T^*$ differ from characteristics of Gold12a.

[**]Sample Serge is from reference [55].



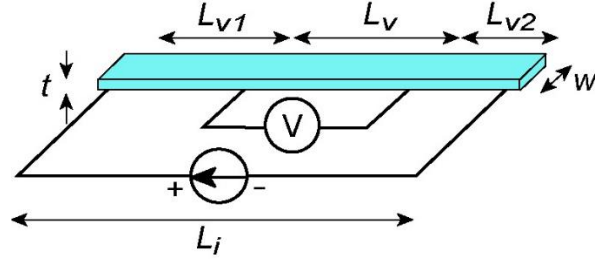

**Figure 3.3**

Four-probe measurement configuration with labeled geometrical considerations and a sample shown in blue.

The list of samples in **Table 3.2** is a selected fraction of approximately 40 samples of NbSe$_3$ that were prepared and investigated to some extent. Samples not listed here include the ones on which measurement procedures could not be pursued to completion because: 1) the quality of electrical contacts degraded significantly upon cooldown due to the partially reliable "probes-on-bottom" contacting method; 2) the samples did not have differential resistance vs. current and/or mode-locking characteristics consistent with single-crystal whiskers (most often due to multiple crystal domains in the cross-section hidden under the crystal surface); and 3) samples had large 1/f-noise floor at low frequencies so that it was not possible to resolve NBN peaks of coherent oscillations in the slow branch.

Each electrically contacted whisker was characterized to determine its thickness $t$ as follows:

$$t = \frac{\rho_S L_v}{w R_S} \qquad (3.4)$$

where $\rho_S$ = 1.85 $\Omega\mu$m is resistivity of NbSe$_3$ at room temperature,[19] $w$ is the width of the whisker measured under optical microscope with uncertainty of ±0.2 μm, and $R_S$ is the



resistance of the whisker segment of length $L_v$ at room temperature. $R_S$ was determined using a 4-probe configuration as shown in **Figure 3.3** by current biasing the whisker through two outer probes and measuring the voltage across the two inner probes that are $L_v$ distance apart. This ensured that the value of $R_S=V/I$ reflects the resistance of the whisker segment only, excluding the resistance of the wires and electrical contacts. For each given whisker, resistances of many whisker segments of different lengths $L_v$ were measured (10-20 of them) and an average resistance per unit length $R_S/L_v$ was determined for use in calculating $t$. The uncertainty in $t$ due to both the uncertainty in $w$ and the uncertainty in $R_S/L_v$ is indicated by the number of significant digits in the value of $t$ listed in **Table 3.2**, i.e. the last significant digit listed is the only uncertain digit. To cross-check the reliability of this method to determine $t$, we measured the thickness of a few samples by a step profilometer (KLA Tencore profilometer available at Cornell Nanofabrication Facility). We found both measurements to be in a reasonably good agreement. For instance, a resistance measurement for sample Gold12a yielded $t = 0.376$ μm while the profilometer measured $t = 0.39 \pm 0.01$ μm.

## Collective Transport Checks

Samples were cooled in a Laybold-Heraeus GMBH closed-cycle helium-gas cryostat shown in **Figure 3.4**. It has two cooling stages encased in a vacuum can. The sample is mounted on the second stage and during cooling the can is evacuated to 5 mTorr by a mechanical pump to prevent ice build-up on samples. Gold-platted "pogo" pins soldered to external wire leads make contact to patterned gold pads on the sample chip. The temperature of the sample cold stage is controlled by a LakeShore DRC-93C temperature controller that drives a Kapton-encapsulated resistive heater wrapped around the sample cold stage. Two diodes thermally



connected to the cold stage sense temperature and provide feedback to the controller. The controller sends temperature data to the computer via GPIB interface. Temperature variation with time depends on the operating temperature but is typically below ±0.05 K. Due to temperature gradient over the cold finger, diode temperature can be slightly different from the actual sample temperature, and this can introduce a systematic error in temperature readings. This error is minimized by placing one of the diodes very close to the sample (within 5 mm) and connecting it well thermally to the copper block that holds the sample. The cooldown takes place after the sample is mounted onto a cold stage, and electrical connections between the crystal and external wiring are checked and established.



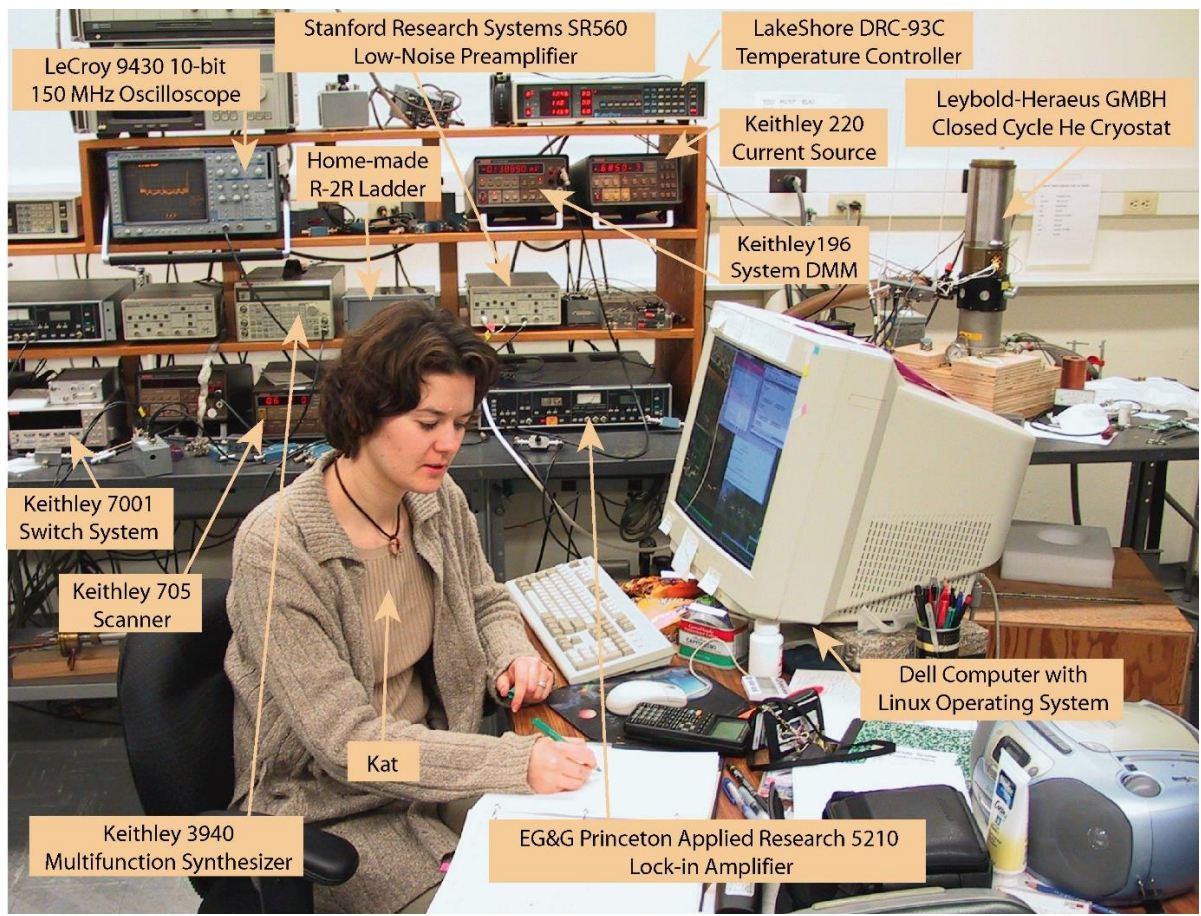

**Figure 3.4**
Laboratory equipment in the group of Prof. Robert Thorne. The author is shown performing coherent oscillation measurements on a NbSe$_3$ sample located in the cryostat. Basement of Clark Hall, Cornell University, 2003.

In addition to initial inspection under an optical microscope, the quality of all samples in **Table 3.2** was evaluated by several measurements that reveal CDW collective transport signatures. In addition to single particle resistance vs. temperature data to observe two CDW transitions as shown in **Figure 2.42** for sample Gold12, this also includes measuring *I-V* and differential resistance vs. bias current curves at various temperatures to observe collective transport thresholds at $I_T$ and $I_T^*$ (corresponding to $E_T$ and $E_T^*$) as in **Figure 1.5**.



**Figure 3.5** shows differential resistance vs. electric field data measured at various temperatures below $T_{P2}$ in the Ti-doped sample Rudy7. A similar behavior was observed in the other samples. *In the upper temperature range* the steep drop in differential resistance corresponds to electric field $|E| = E_T$ where CDW depins and begins to slide. With decrease of temperature CDW conduction for $|E| > E_T$ begins to freeze developing a shoulder which with further decrease of temperature evolves into the second characteristic threshold $E_T^*$. *In the low temperature range* the field at the ends of the hysteresis loop roughly corresponds to $E_T$ while the sharp drop in differential resistance indicates $E_T^*$. This low-field resistance hysteresis is associated with phase slip near current contacts which affects CDW strain between the contacts and will be discussed further in the next chapter. In some samples at low temperatures a second hysteresis was observed at $E_T^*$ (as seen in **Figure 1.5 (b)**). Both hysteresis loops are sensitive to bias sweep rate at which data is taken.

**Figure 3.6** shows dependence of $E_T$ and $E_T^*$ on temperature extracted from data in **Figure 3.5**. Similar behavior was observed in the other samples.



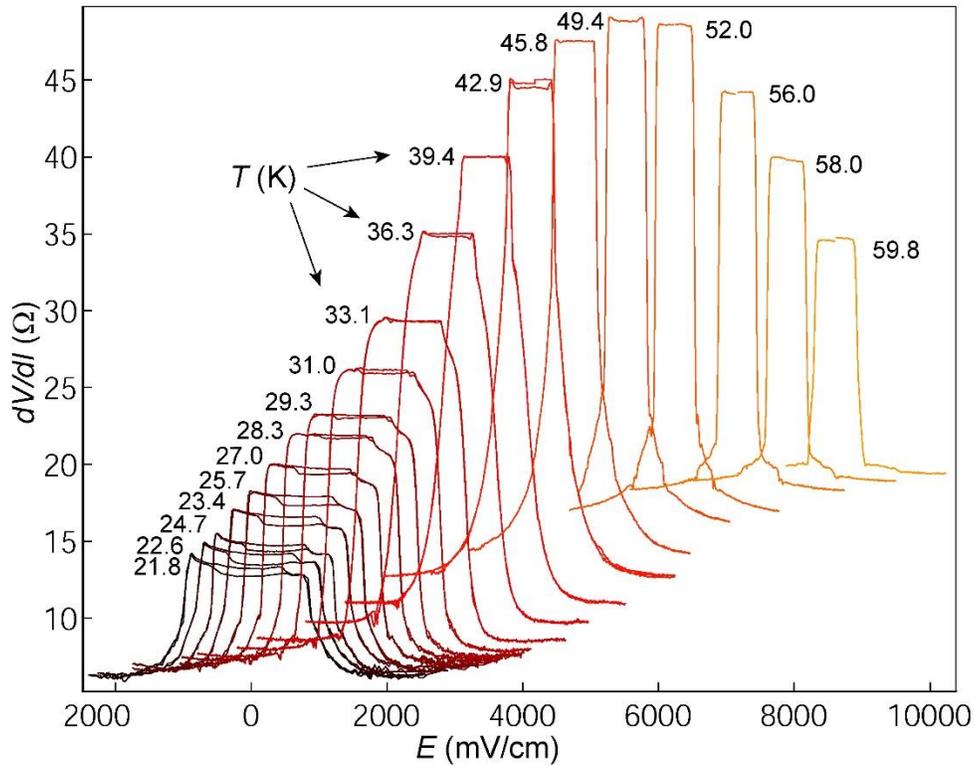

**Figure 3.5**

Differential resistance vs. electric field measured at different temperatures for the Ti-doped sample Rudy7. Data sets are offset horizontally so that the offset of each curve is proportional to its temperature i.e. horizontal offset = $T \times$ (9 mV/cm). A vertical offset of each curve is zero. Differential resistance is measured in a current-biased configuration by using a lock-in amplifier. Bias current was then mapped to the corresponding electric field by separately measuring _I_-_V_ data. At low temperatures current bias was swept in both directions revealing the low-field hysteresis observed for $|E| < E_T$.



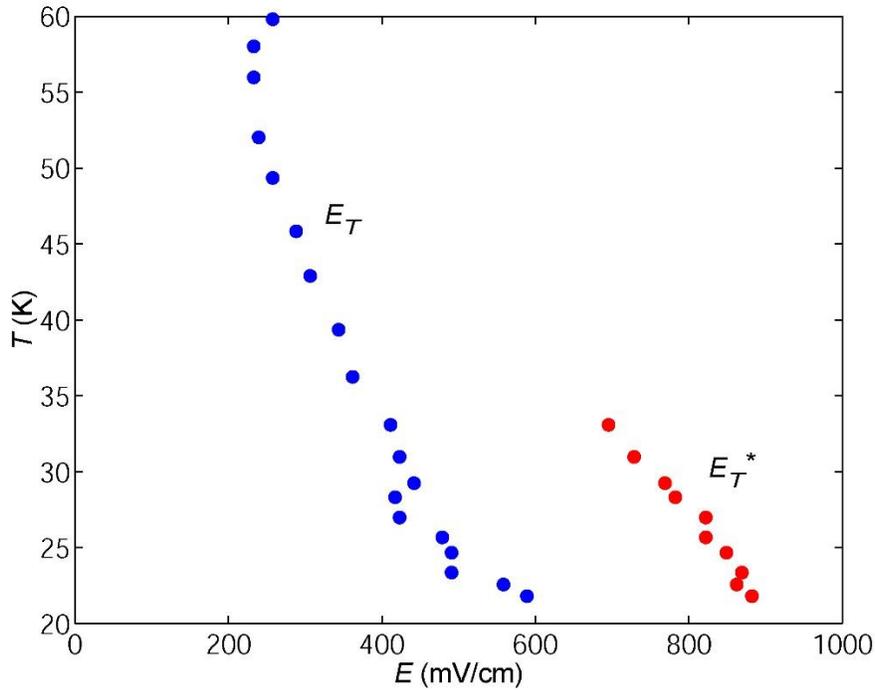

**Figure 3.6**

Evolution of $E_T$ and $E_T^*$ with temperature observed in sample Rudy7.

Irregularities in *I-V* and *dV/dI* vs. *I* data can reveal imperfections such as multiple crystal

domains or steps in the crystal cross-section that may not be visible under an optical

microscope. **Figure 3.7** with data in **(a)** at 40.8 K shows two rounded shoulders above $I_T$ that

in **(b)** at 25.6 K develop into two switching thresholds i.e. two $E_T^*$ thresholds are observed

instead of one in each data set. This behavior is consistent with a presence of a step on the

whisker which divides the cross section into two regions of different thickness. Since $E_T^*$

depends on thickness, CDW sliding begins at different thresholds in the two regions.[37,38,40,105]

This is in contrast with data of **Figure 3.5** and **1.5** from samples with no steps in the cross-



section and with nominal transport characteristics. Samples with signatures such as the ones shown in **Figure 3.7** were typically not pursued in NBN measurements since we sought to base our analysis on NbSe$_3$ crystals of highest quality. By selectively excluding such whiskers we increased our likelihood of probing true intrinsic transport behavior of single crystals rather than the behavior complicated by artifacts of imperfect crystal geometries. For completeness and comparison, in section 3.2 on results, we briefly discuss qualitative features of the noise spectra we observed in a few whiskers plagued with steps, multiple crystal domains and other imperfections.

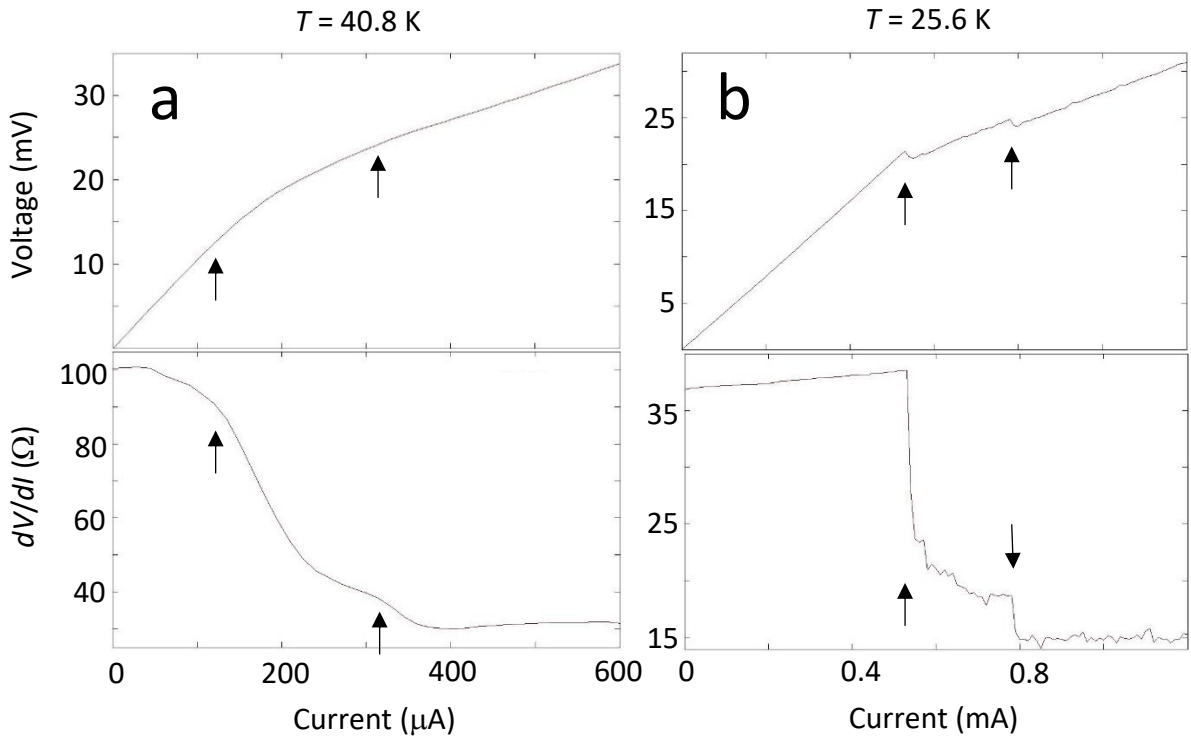

**Figure 3.7**

Voltage vs. current and $dV/dI$ vs. current data for a Ta-doped sample at two different temperatures show at least two $I_T^*$ thresholds. In (a) data obtained at $T = 40.8$ K shows two shoulders begin to develop, and in (b) at $T = 25.6$ K at least two clearly developed switching thresholds can be observed.



To further probe coherence of the CDW motion in several samples listed in **Table 3.2**, we have also investigated mode-locking of CDW motion to an external AC drive in the sliding regime of CDW transport. We have observed good quality mode-locking characteristics consistent with those observed in high-quality single crystals of $NbSe_3$.

All these experimental checks were performed to ensure that our measurements are performed on samples that can clearly reveal *intrinsic* CDW transport behavior that is not complicated by extraneous factors.

## Narrow-Band-Noise Measurement Configuration

To measure $j_C(E)$ in the slow branch of $NbSe_3$ phase diagram where $j_S \gg j_C$ we exploit the fact that CDW velocity of a DC-biased sample has a component periodic in time, and that the frequency, $f_{NBN}$, of this component is proportional to the average CDW velocity, and thus to $j_C$. The frequency of the periodic component can be isolated by performing spectral noise measurements i.e. by Fourier-transforming time domain voltage signal into the frequency domain. By extracting $f_{NBN}$ from a time-domain voltage trace measured at a bias field $E$ and at temperature $T$, we can map out $j_C(E,T)$ relation via **(1.7)**. Note that the temporal order of the CDW motion in this regime is a prerequisite for successfully extracting $j_C$ from these measurements, and we expect to observe NBN only in samples of exceptional quality in which temporal order can persist over large, *collective* length-scales.



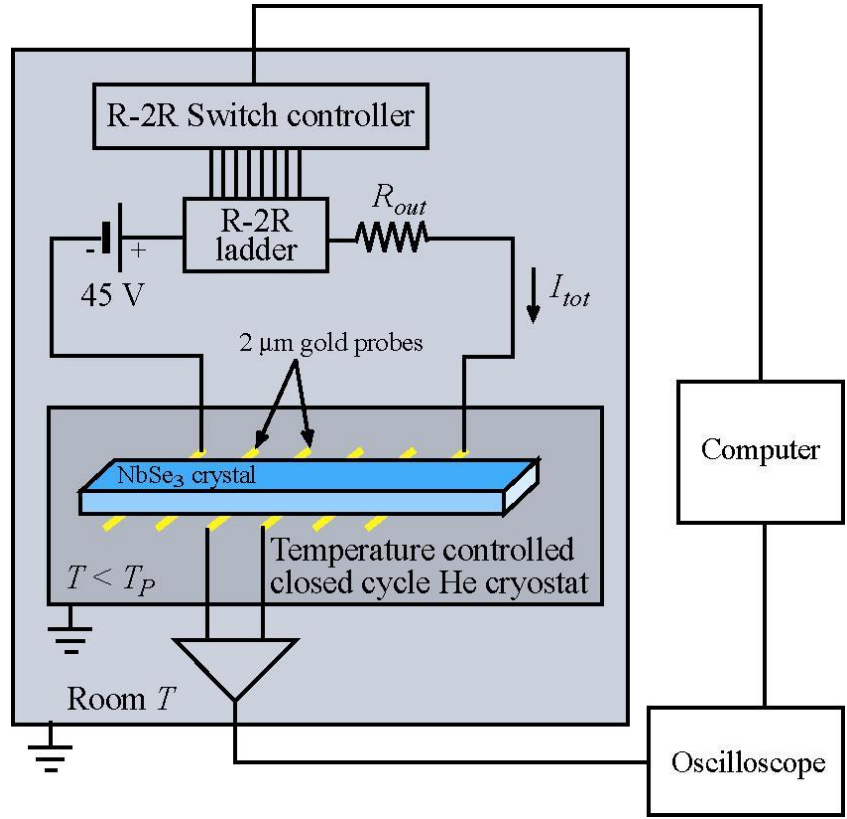

**Figure 3.8**

Narrow-band-noise measurement configuration.

A four-probe measurement configuration and the experimental setup are shown in **Figure 3.8**. A whisker segment of length $L_i$ is biased through two probes by a DC current $I_{tot}$. The current in the sample is carried by the CDW and single particle channels $I_{tot} = I_C + I_S$. A corresponding voltage signal was measured over a smaller sample segment of length $L_v$ between voltage probes and was amplified by a Stanford Research Systems SR-560 low-noise amplifier with a typical gain between $10^2$-$10^4$. The amplified signal was input into the LeCroy 9430 oscilloscope where a voltage trace $V(t)$ (8 s - 30 s long, and typically a sample of 2,000-10,000 data points), was fast-Fourier-transformed (FFT) into the frequency domain



and averaged 3-30 times (to increase signal-to-noise ratio) to obtain a power spectral density plot $P(f)$ as follows:

$$V(t) \rightarrow FFT\big(V(t)\big) \rightarrow V(f) \rightarrow V(f)^2 \propto P(f) \qquad \textbf{(3.5)}$$

$P(f)$ data was then downloaded to the computer via GPIB interface where the background noise signal $V(t, I_{tot}=0)$, measured separately, was subtracted in quadrature within our data analysis program. By repeating these measurements at different bias currents $I_{tot}$ we obtain a set of spectra $P(f, I_{tot})$. Finally to map $P(f, I_{tot})$ to $P(f, E)$ we use the $V(I_{tot})$ and $E(I_{tot}) = V(I_{tot})/L_V$. $V(I_{tot})$ is measured separately by replacing the amplifier and the oscilloscope in the circuit by Keithley 196 System Digital Multi Meter.

A typical set of spectra $P(f, E)$ is shown in **Figure 3.9**. Narrow-band noise associated with the periodic component of the CDW motion appears as a peak in each $P(f)$ spectrum at a frequency $f_{NBN}$ which for a full set of spectra provides $f_{NBN}(E)$ relation. At last to obtain temperature dependence, the measurements are simply repeated at different temperatures, and $f_{NBN}(E, T)$ is converted to $j_C(E, T)$ using **(1.7)**.

Special care must be taken to isolate the system from the external noise. Our set-up was optimized for a bandwidth between 0.5 Hz – 50 kHz. In order to minimize the contribution from the power grid 60 Hz noise and the harmonics, all non-passive components in direct contact with the sample circuit were battery powered. Current biasing of the sample is accomplished by using a home-made $R$-$2R$ ladder in series with an external resistor $R_{out} \gg R_{sample}$ (See **Figure 3.8**). An $R$-$2R$ ladder[106] is a voltage-dividing network of resistors and switches connected to a 45 V battery source. By opening and closing specific switches one



can control the current output of the *R*-2*R* ladder that biases the sample. A computer controlled Keithley 705 Scanner connected to the ladder orchestrates to obtain a desired current bias. Voltage signal from the sample is filtered with a band-pass filter (with adjustable bandwidth) built into the low-noise amplifier which is also battery powered. In addition, all the home-made parts of the equipment were properly shielded, which further reduces stray noise pick-up. All these precautions were necessary in order to isolate the CDW coherent oscillation signal from the external noise.

## 3.2 Results

### Coherent Oscillations in the Low Temperature Slow Branch of NbSe$_3$

A typical NBN data set obtained as discussed in the previous section is shown in **Figure 3.9**. It shows averaged power spectral density as a function of frequency at different electric fields measured on sample Gold12a at $T = 22.55$ K. Sharp NBN fundamental peaks are clearly visible above the background noise as well as weaker first and second harmonics at several fields. No peaks are observed for $E < E_T$. Peaks first emerge just above $E_T$ at low frequencies and move to higher frequencies as we increase $E$ field through the slow branch. At fields above $E_T^*$ (not shown in **Figure 3.9**), we observe increase in the background broad band noise level, and NBN peaks abruptly jump to frequencies above the bandwidth optimized in our experimental set-up (> 50 kHz). The values of $E_T$ shown in **Figure 3.9**, as well as those of $E_T^*$, are independently determined from differential resistance measurements as in **Figure 3.5**.

We have observed slow-branch coherent oscillations with similar behavior in a total of twelve different NbSe$_3$ samples, but the quantitative analysis discussed in the following sections is



based on the investigation of the ten high-quality samples listed in **Table 3.2**. These samples for a given driving field exhibit single fundamental NBN peak, as well as clear collective signatures described in section 3.1 ensuring that there is no significant filamentary conduction throughout the whisker cross section.

Slow branch NBN peaks were observed in the temperature range between 15 K - 32 K. Above this range $f_{NBN}$ quickly exceeds 50 kHz as we increase driving field, and our measurement becomes limited by the experimental setup. Below ~ 18 K temperature stability of the measurement becomes compromised, both due to cooling capability of our cryostat and due to ohmic heating of the sample. $E_T$ increases with decrease of temperature, and at low temperatures large current densities (often on the order of 1 mA/$\mu$m$^2$) are required to access the slow branch regime. The presence of ohmic heating can be confirmed by non-linearities in the $I$-$V$ curves below $E_T$ (i.e. below $I_T$) where only single particle ohmic conduction is expected. The reduced temperature stability and low resistivity of NbSe$_3$ samples at low temperatures produce small signal-to-noise ratios in measured voltage signals. As a result, below 15 K it was difficult or impossible to resolve NBN peaks from the background noise. The same also affects the measurement of $dV/dI$ with bias current, and $E_T$ and $E_T^*$ become difficult to discern below this temperature.



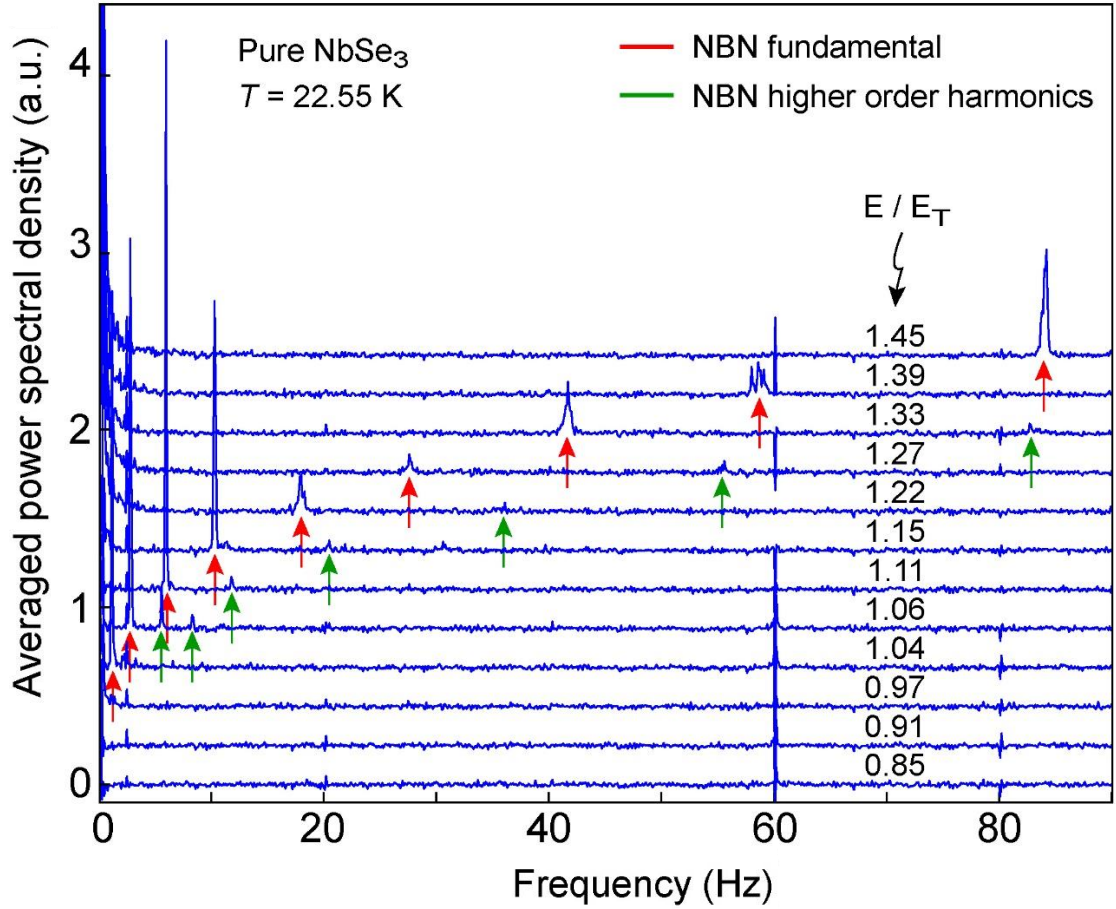

**Figure 3.9**

Coherent oscillations in slow branch of NbSe$_3$. Averaged power spectral density $P$ as a function of frequency and electric field at $T = 22.55$ K for sample Gold12a. Spectra at different $E$ fields are offset vertically. For a given $E > E_T$ a sharp NBN peak is visible at $f_{NBN}$ (red arrows). Some higher-order harmonics of the peak (green arrows) can also be discerned at $2f_{NBN}$ and $3f_{NBN}$.

We have observed NBN peaks with narrow spectral widths in all samples in **Table 3.2**. The

highest amplitude and narrowest NBN peaks for a given sample were observed at slow

average CDW velocities, typically below few hundred Å/s (or at $f_{NBN}$ up to a few tens of Hz),



and as low as 5 Å/s ($f_{NBN}$ = 0.4 Hz or $v_C$ ≈ 2 cm/year)[*]. While the CDW moves at extremely

slow, creep-like velocities in this regime, it remarkably retains temporal order that results in

coherent oscillations. We have observed that the NBN peak's width and amplitude change

along the slow branch often in a qualitatively similar way from sample to sample. While the

largest peak amplitudes were observed for biases just above $E_T$, our preliminary analysis

showed that the quality factors $Q$ of the peaks were the largest near the midpoint of the slow

branch at $E = (E_T^* - E_T)/2$ as shown in **Figure 3.10.** Here $Q = f_{NBN} / \Delta f_{NBN}$, where $\Delta f_{NBN}$ is

defined as the peak full width at half maximum (FWHM). The figure further shows that the

highest-$Q$ peaks are observed in pure samples, with the highest $Q$-values for isoelectronically-

doped (Ta) samples slightly lower, and with the values in samples with non-isoelectronic Ti

impurities significantly lower. The largest recorded quality factors in pure, Ta-, and Ti-doped

samples are 425, 340, and 45 respectively. These observations suggest that the spectral

widths and the "noise" in coherent oscillations may be affected by the impurity content. This

should be explored in the future. It is remarkable that the motion in this regime was observed

to be coherent even in crystals with the highest doping concentrations and with the most-

perturbing, non-isoelectronic Ti impurities.

---

[*] For an interesting order-of-magnitude comparison note that land masses that formed the Himalayan mountain range or the tectonic plates that form North American fault line off the coast of California (also systems with quenched disorder exhibiting collective motion) move at ~10 cm/year.



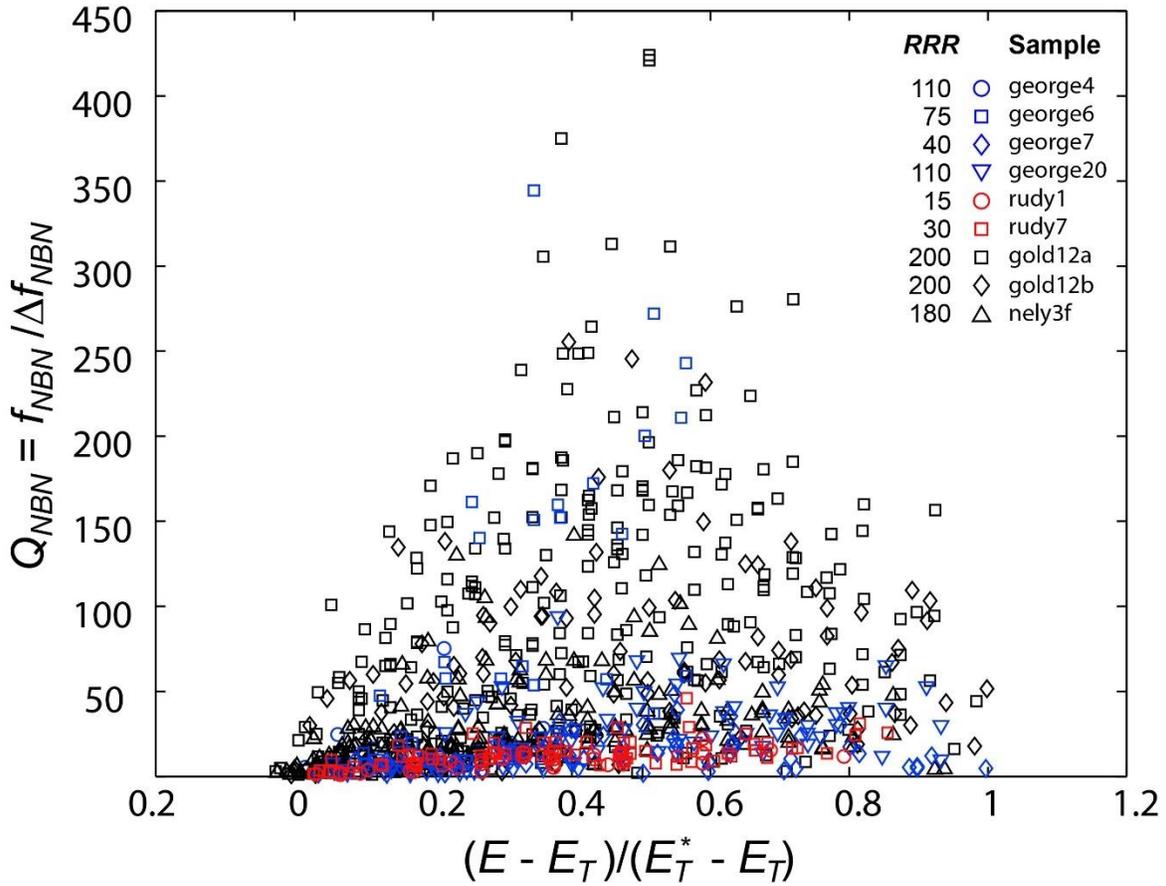

**Figure 3.10**

Quality factors of coherent oscillations observed in the low-temperature slow branch creep regime of pure and Ta- and Ta-doped samples in **Table 3.2**. Here $Q = f_{NBN}/\Delta f_{NBN}$, where $\Delta f_{NBN}$ is defined as the peak full width at half maximum (FWHM).

We observed that the NBN peaks in slow branch are metastable on timescale of many minutes or hours and can jump to slightly different frequencies while the bias current through the sample is held constant. In addition, when a NBN measurement is repeated after re-setting the bias current from zero to $I_{tot}$, the observed peak can have a slightly different $f_{NBN}$ value. The spread of the frequency values typically falls within a range of a few $\Delta f_{NBN}$. This behavior in slow branch is consistent with the idea that for a given drive, a distribution of



pinning centers can produce metastable configurations of the CDW phase within FLR volumes, all very close in energy,[22] and that the mechanism of slow branch transport involves large collective length-scales.

Our differential resistance measurements show that $E_T$ observed at high temperatures smoothly evolves with a decrease of temperature towards the field where coherent oscillations first appear in the slow branch. See, for example, **Figure 3.5**. Furthermore, the onset of NBN peaks with electric field in the slow branch is accompanied by a sudden increase in BBN level just as is the onset of collective CDW motion at $E_T$ in the high temperature regime. BBN level increases further at $E_T{}^*$ when the slow motion switches to high-velocity sliding. The increase of BBN at both thresholds corresponding to $I_T$ and $I_T{}^*$ can be seen in **Figure 3.11** for sample Gold12b at $T$=22.45 K. For comparison, **Figure 3.12** shows increase of BBN level in a high temperature regime at $T$=42.60 K which occurs at the collective threshold $E_T$.



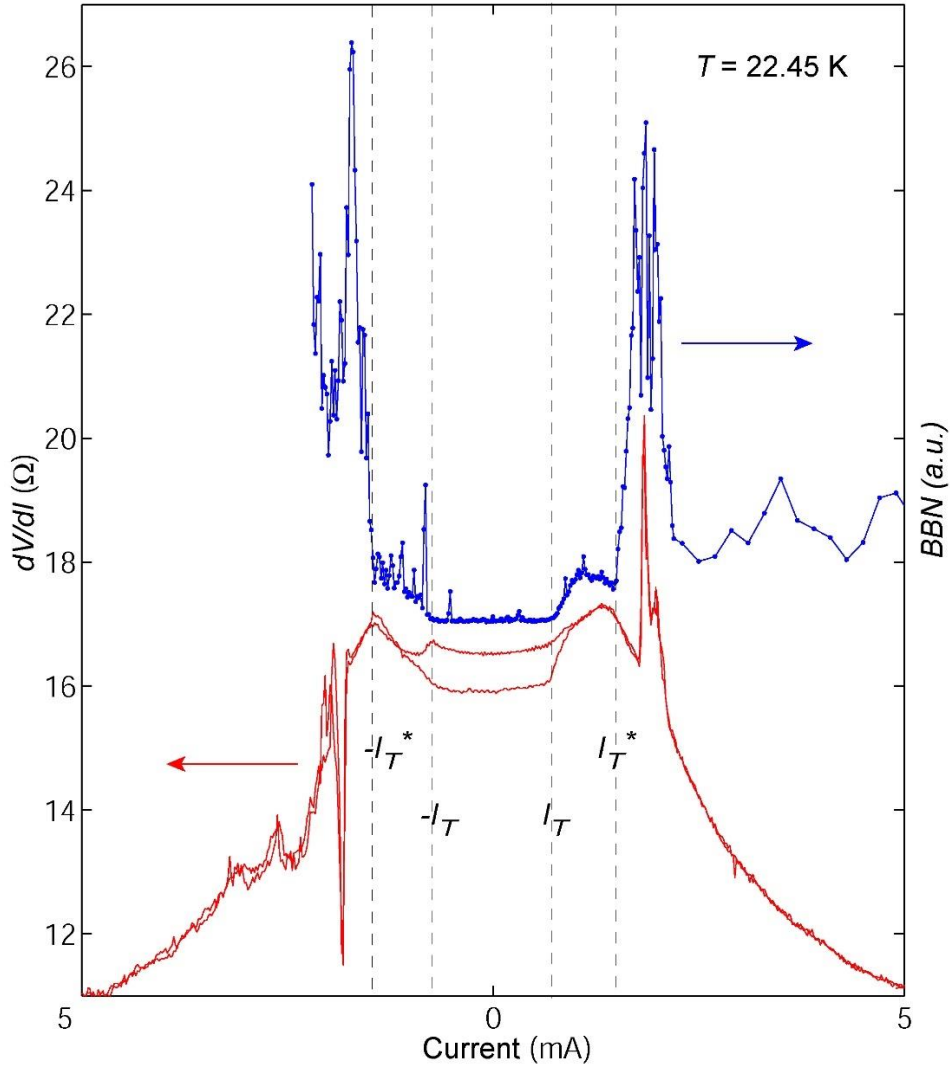

**Figure 3.11**

BBN level data (blue) superimposed on *dV/dI* (red) plot for sample Gold12b at *T* = 22.45 K. The data shows that both $|I_T|$ and $|I_T^*|$ (corresponding to $|E_T|$ and $|E_T^*|$) are accompanied by increase of BBN level. Some ohmic heating was present (as indicated by upward curvature of *dV/dI* data below $|I_T^*|$). The spike and dip in the *dV/dI* data, observed just above $I_T^*$ and below $-I_T^*$ respectively, are artifacts associated with a non-optimized measurement set-up with the lock-in amplifier.



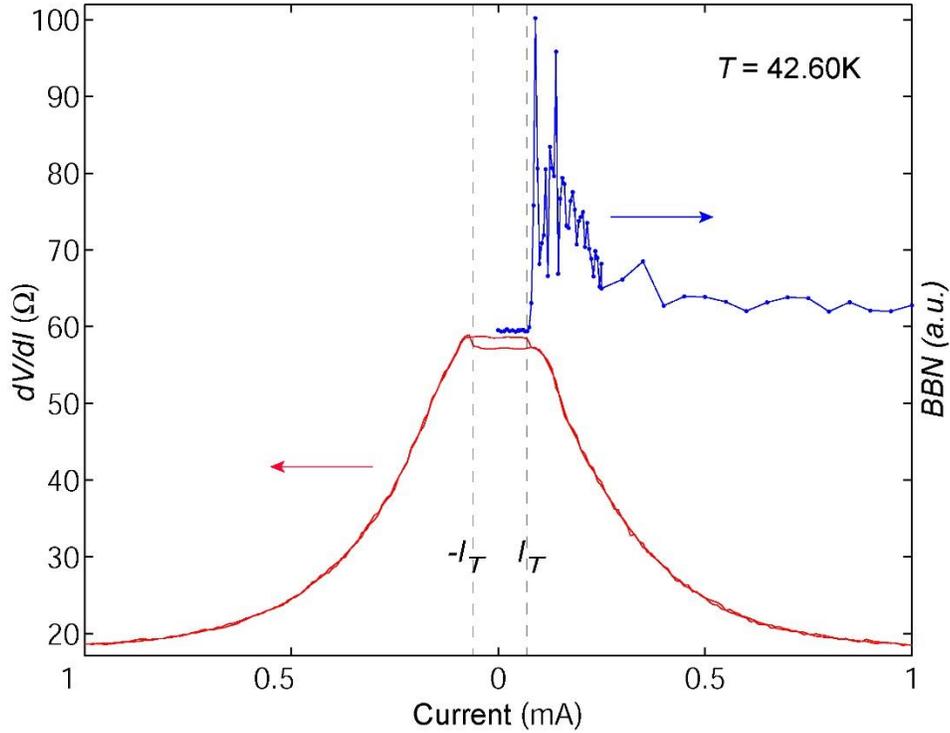

**Figure 3.12**

BBN level superimposed on *dV/dI* plot at sample Gold12b in the upper temperature regime at *T* = 42.60 K. Data shows a single characteristic collective threshold at |*I*$_T$| and a corresponding increase in BBN level at that threshold.

For completeness we next discuss qualitative observations from numerous other samples that we partially investigated but have omitted from **Table 3.2**. In these samples, electronic measurements and transport tests typically revealed signatures consistent with multiple crystal domains in the whisker. Since inspection under optical microscope did not reveal steps on the whisker surface, it is likely the whiskers had non-uniform subsurface morphology like the one shown in **Figure 2.23**. The observations on these samples can be summarized as follows:

- In several samples we observed more than one NBN peak in slow branch, at frequencies not corresponding to harmonics or subharmonics of the fundamental



frequency. This is consistent with earlier observations in whiskers with multiple crystal domains or steps in the cross-section.[37] Such samples tend to exhibit rounded thresholds (in case of many smaller domains) or multiple distinct thresholds (in case of a few domains in a whisker) in *I-V* curves. We have observed this multi-step/multi-threshold correlation in at least two samples. One sample which had two distinct switching $E_T^*$ thresholds in the *I-V* curve, also exhibited two fundamental NBN peaks in slow branch.

- Numerous other whiskers did not exhibit any visible NBN peaks in the slow branch. Instead the data showed large low-frequency BBN noise floor above $E_T$ that further increased above $E_T^*$ as shown in **Figure 3.13**. Maher *et al.*[37] showed that a dominant source of BBN in NbSe$_3$ whiskers with large BBN power is CDW phase slip occurring along thickness step(s) in the whisker cross-section. The large noise floor in the slow branch of **Figure 3.13** may be obscuring any signal from coherent oscillations, which are expected to have complicated structure and may not be well defined in such samples. These BBN spectra show that even in these samples a distinct behavior exists in the slow branch since the increase of low frequency noise at both $E_T$ and $E_T^*$ indicates that the dynamics significantly changes at both thresholds.



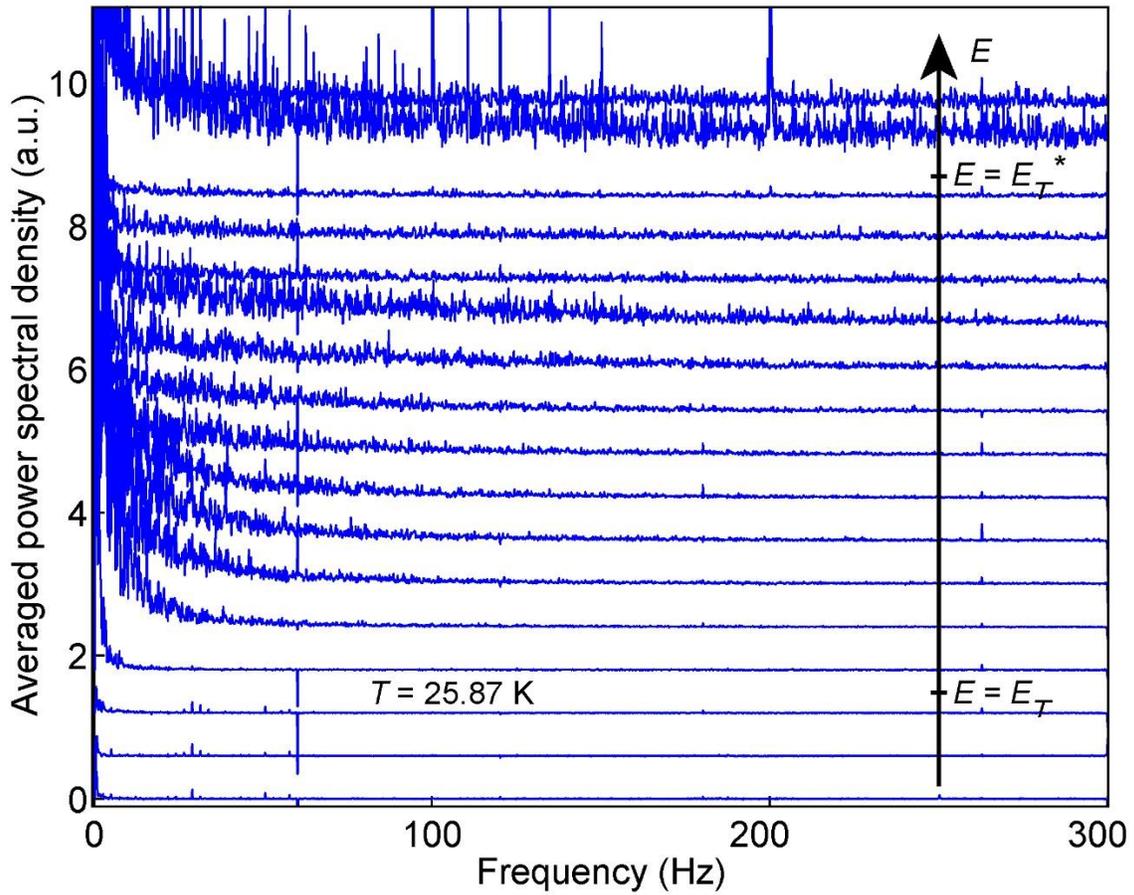

**Figure 3.13**

Noise spectra through slow branch in sample George18. Spectra at different $E$ fields are offset vertically. No NBN peaks are visible but we observe discernible increase in $1/f$–like noise first at $E_T$ and then further at $E_T^*$. Here $E_T$ and $E_T^*$ are identified from independent $dV/dI$ measurements.

In summary, our spectroscopic and differential resistance measurements reveal an onset of collective CDW motion at low $T$ that commences at a well-defined threshold field. $E_T$ identified as collective pinning threshold at high temperatures evolves smoothly towards this threshold field at low temperatures. This onset of motion at low temperature is accompanied by an increase in BBN, also observed at onset of collective motion $E_T$ at high temperatures.



This points to $E_T$ as being the true threshold of CDW collective transport across the transport phase diagram. Observation of coherent oscillations (demonstrating that the motion is temporally-ordered over large length scales) and a well-defined threshold for collective motion clearly indicate that non-local, collective dynamics characterized by large FLR length scales (~micrometers in NbSe$_3$) is an important ingredient of the low temperature dynamics. Some characteristics of the CDW transport in the slow branch such as extremely slow velocities (sub-Hz to kHz frequencies of coherent oscillations) and quality factors up to a few hundred are in contrast with observed characteristics of high temperature collective behavior where CDW velocities above $E_T$ correspond to orders-of-magnitude larger frequencies (from MHz to hundreds of MHz), and NBN peaks with quality factors that can exceed thousands or even tens of thousands.[54] Collective effects clearly play an important role in defining dynamics in the low temperature regime, but these differences, and the fact that FLR model completely fails to predict two thresholds in this regime, indicate that the "ingredients" leading to collective dynamics are not the only important ones in this regime. Note also that the notion of "collective coherent creep" is quite remarkable since creep motion in other systems typically originates from stochastic processes (i.e. thermal disorder), incoherent by nature.

## Modified Anderson-Kim Form of $j_C(E, T)$

The $f_{NBN}(E, T)$ relation obtained from many sets of spectra such as the one in **Figure 3.9** is shown in **Figure 3.14** for both segments of the pure sample Gold12, and **Figure 3.15** shows data representative of **(a)** Ta- and **(b)** Ti-doped samples. Similar data sets where obtained for all samples in **Table 3.2**.



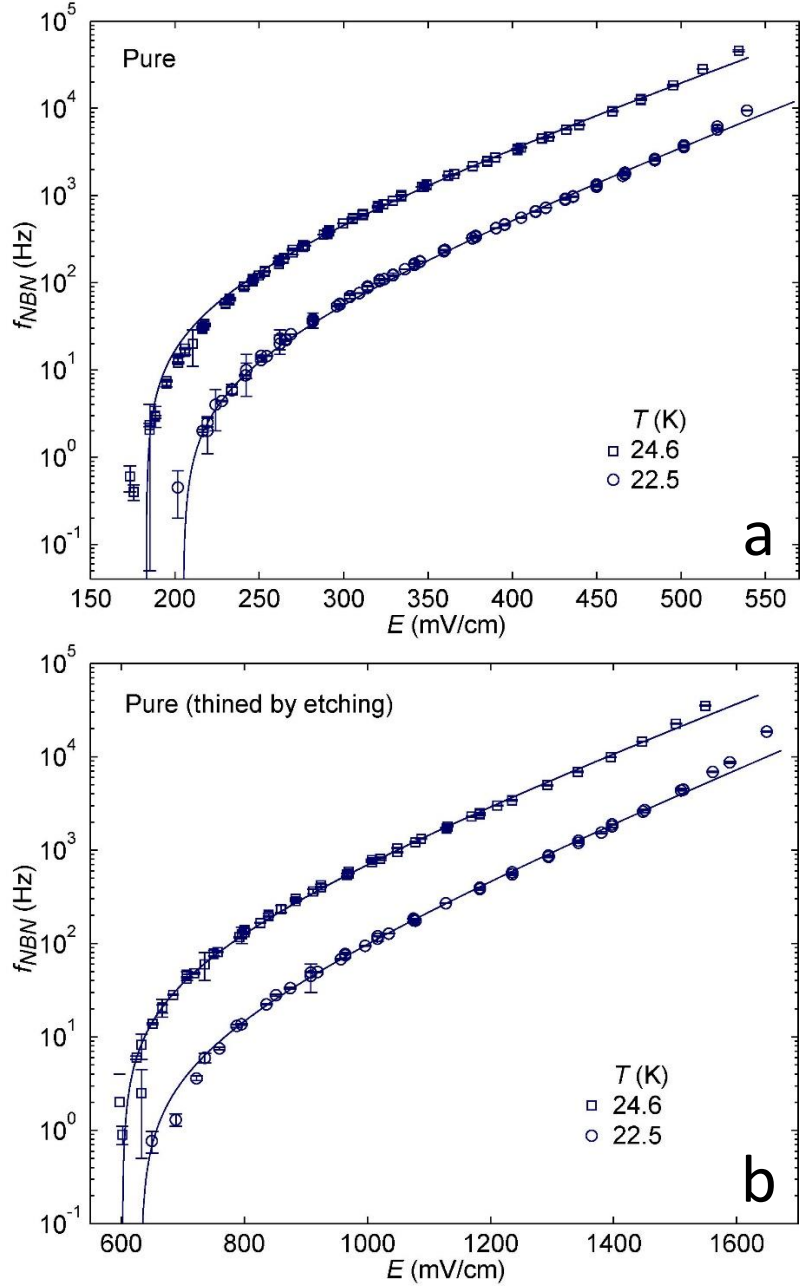

**Figure 3.14**

$f_{NBN}$ measured as a function of electric field $E$ and temperatures $T$ for pure samples (a) Gold12a and (b) Gold12b. Data are fitted by **(3.6)** where $f_{NBN}$ relates to CDW current through $f_{NBN} = j_C/(en_C\lambda_C)$. Fitting lines span the whole slow branch at each temperature, i.e. they originate at $E_T$ and terminate at $E_T^*$. $E_T$ and $E_T^*$ were independently measured in $dV/dI$ measurements. $E_T$ enters into the fitting scheme through a reduced field $E-E_T$ in **(3.6)**. Error-bars shown are measured $\Delta f_{NBN}$ (FWHM for each NBN peak).



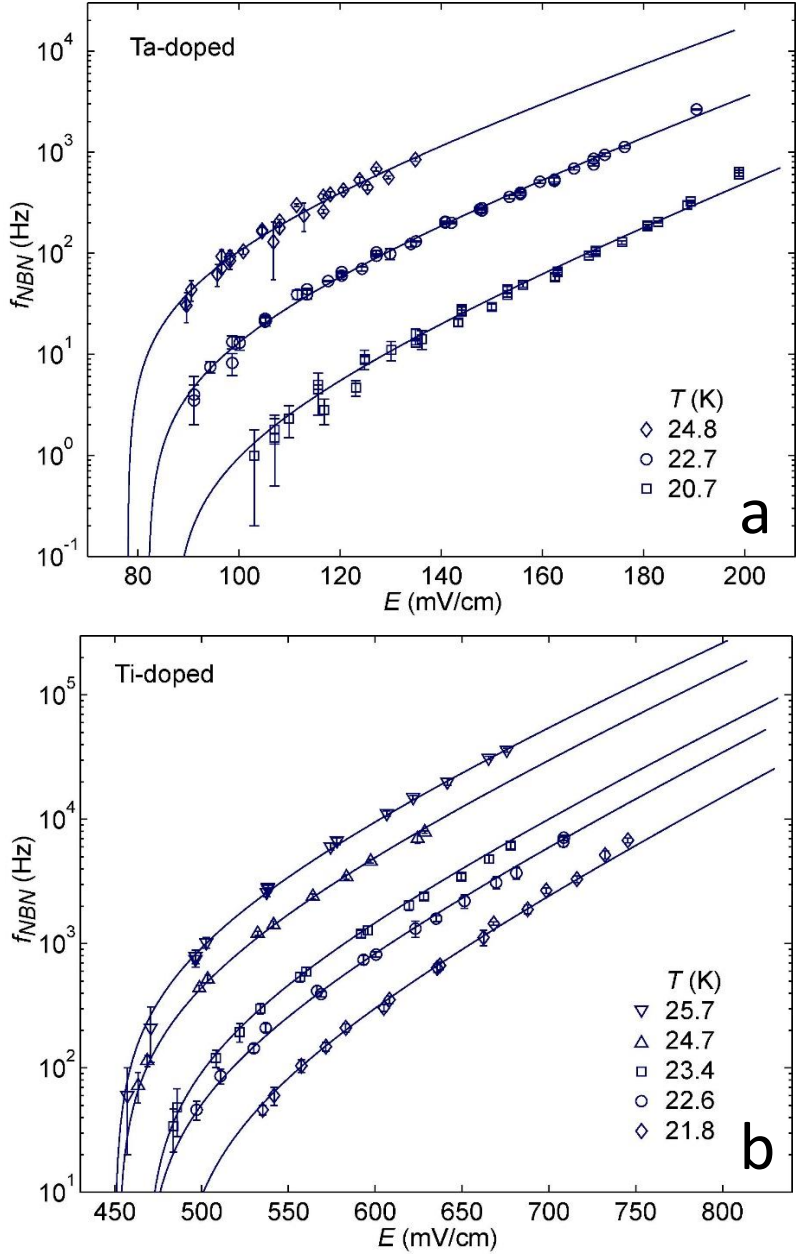

**Figure 3.15**

$f_{NBN}$ measured as a function of electric field $E$ and temperatures $T$ for (a) the Ta-doped sample George20 and (b) the Ti-doped sample Rudy7. The same data fitting procedure is used as in **Figure 3.14**. Error-bars shown are measured $\Delta f_{NBN}$.



The data map out current - field (or velocity - force) relation of CDW transport for $E_T < E < E_T^*$ through $j_C(E, T) = en_C v_C(E, T) = en_C \lambda_C f_{NBN}(E, T)$ (equation **(1.7)** which cannot be obtained by standard $I$-$V$ measurements. All the data can be well fitted by the form:

$$j_C(E,T) = \sigma_0 (E - E_T) e^{\frac{-[T_0 - \alpha(E - E_T)]}{T}}$$ **(3.6)**

where $\sigma_0$, $T_0$, and $\alpha$ are fitting parameters, and $E_T$ is the threshold field measured independently by differential resistance measurements. The entire data set for one sample, including data subsets measured at different temperatures (covering range 20 K – 26 K), was fitted together at the same time so that each sample yielded one set of three fitting parameters. While we were able to observe NBN peaks and collect data at temperatures as low as 15 K, parasitic heating skews $j_c(E)$ data in our measurement near or below ~18 K and causes deviations from the fit below this temperature. The heating below 18 K was independently confirmed by nonlinearities in single particle conductance for $E < E_T$ in differential resistance measurements as discussed earlier. We have thus purposely excluded data obtained in this low temperature range from the fits.

Equation **(3.6)** resembles the Anderson-Kim formula for thermal creep of flux lattices in superconductors[107,108] where the electric field $E$ is replaced by reduced field $E$-$E_T$ to account for the onset of conduction at a finite threshold $E_T$ observed in our system. Flux-lattice creep is thermally assisted and mechanisms that cause it are stochastic in nature. In the creep regime of the moving flux lattice, collective behavior does not generally prevail, and the motion of the flux lines over large length-scales is not correlated. In that context the Anderson-Kim form describes incoherent creep and it is thus not obvious why the form



should also apply to temporally-ordered creep-like motion in which collective dynamics must play an important role. One would expect that a functional form for coherent CDW transport should somehow account for collective effects. Our modified Anderson-Kim (MAK) form captures this through $E_T$, a threshold field independently measured and associated with collective pinning. We emphasize, however, that this form is not derived from an analytical model that considers specifics of a CDW system. It is merely a phenomenological form inspired by a model for (incoherent) creep that successfully describes our data while the existing CDW models do not.

Next, we more closely look at the fitting parameters and examine their possible physical interpretation.

**Fitting Parameters.** Flux-lattice creep in superconductors is a consequence of local thermally activated barrier hopping aided by an applied external drive. Our hypothesis is that in the CDW case the motion in the slow branch is also limited by creep-like microscopic dynamics on a length-scale much smaller than the length-scale responsible for collective dynamics. To understand the (microscopic) dynamics that allows persistence of temporally-ordered collective creep, one may address the following obvious questions:

1. What limits the motion above the collective pinning threshold forcing CDW to creep at extremely low velocities (i.e. what are the relevant barriers)?

2. What are the typical length- and energy-scales associated with the microscopic mechanism that allows CDW to advance forward in creep regime?



3. How can collective dynamic behavior that is responsible for coherent oscillations (NBN) coexist with the creep dynamics and not be destroyed in the process?

Here we address these questions. The MAK functional form in **(3.6)** fitted to our data can be interpreted as follows:

- $E$-$E_T$ factor accounts for the collective dynamics with characteristic length-scale of $L_\phi$ that determines size of FLR domains. Collective motion begins at the collective pinning threshold $E_T$ and is driven by the reduced electric field $E$-$E_T$.

- Exponential factor describes local barrier hopping that occurs on length-scale $L_{barr} < L_\phi$. $k_B T_0$ corresponds to energy of a microscopic barrier effectively reduced by an amount $k_B \alpha (E$-$E_T)$ due to electric field in excess of $E_T$ that drives the CDW transport.

For each sample, **Table 3.3** lists the values for fitting parameters $\sigma_0$, $T_0$, and $\alpha$, as well as the sample doping concentration $n_i$, threshold fields $E_T$ and $E_T^*$ at a given temperature, and a barrier concentration $n_{barr}$ predicted from the fit through $\alpha$ in **(3.7)** and **(3.8)** and using $j_C / f_{NBN}$ $= e n_C \lambda_C = 0.4018$ Am$^{-2}$Hz$^{-1}$ with the parameter values given in **Table 1.1**.

From the fits we find $T_0$ to vary little across the samples, with no obvious correlation to doping type or concentration, and sample thickness, with the average $T_0 = 487 \pm 49$ K with data taken at several temperatures for most samples. This gives $k_B T_0 \approx 1.2 \Delta_C$ with $\Delta_C$ given in **Table 1.1**. $\Delta_C$ is the energy-scale for both quasiparticle excitation above the CDW gap and for theoretically predicted strong pinning barriers in a CDW.[21] Let us consider the case of $T_0$ being associated with quasiparticle excitation and CDW-quasiparticle interaction as being



responsible for the observed thermally activated dynamics in this regime. Quasiparticles interact with CDW and can screen CDW fluctuations. As we decrease $T$, quasiparticles freeze below the gap and this screening is reduced. In fully gapped CDW materials, de-screening can produce observable effects like stiffening of CDW elastic constants which causes $E_T$ to decrease and vanish in $T \to 0$ limit. In NbSe$_3$ single particle carriers from the un-gapped parts of the Fermi surface are present even at low temperatures. In fact, as we lower the temperature the single particle conductivity rises, which aids screening. This is consistent with the observed increase of $E_T$ as we decrease $T$ in NbSe$_3$ (see for example **Figure 3.6**). De-screening is thus not expected to occur in NbSe$_3$. This suggest that a more likely scenario is that $T_0$ is a characteristic energy scale associated with strong pinning barriers rather than with the energy associated with de-screening.

**Table 3.3 Fit parameters**

| Sample | Doping/ $RRR$ | $n_i/10^{16}$ (cm$^{-3}$) | $E_T/E_T^*$ (at given $T$) (mV/cm)/(mV/cm) | $\sigma_0/10^9$ ($\Omega^{-1}$m$^{-1}$) | $T_0$ (K) | $\alpha$ (Kcm/ mV) | $n_{barr}/10^{16}$ (cm$^{-3}$) |
|---|---|---|---|---|---|---|---|
| George7 | Ta / 40 | 750 | 215/359 (24.70 K) | 0.177 | 430.7 | 0.5140 | 0.5656 |
| George6 | Ta / 75 | 400 | 295/901 (24.60 K) | 60.76 | 591.4 | 0.1318 | 2.206 |
| George4 | Ta / 110 | 270 | 60/165 (24.61 K) | 26.42 | 495.1 | 0.8361 | 0.3477 |
| George20 | Ta / 110 | 270 | 78/198 (24.84 K) | 2.704 | 490.5 | 0.8454 | 0.3439 |
| Rudy1 | Ti / 15 | 67 | 305/599 (23.82 K) | 8.209 | 551.7 | 0.2926 | 0.9930 |
| Rudy7 | Ti / 30 | 33 | 452/814 (24.67 K) | 0.754 | 436.0 | 0.3170 | 0.9174 |
| Nely3f | Pure / 180 | 5.6 | 200/659 (23.82 K) | 0.1365 | 455.2 | 0.2016 | 1.443 |
| Gold12a | Pure / 200 | 5.0 | 183/540 (24.58 K) | 0.188 | 446.1 | 0.3396 | 0.8562 |
| Gold12b | Pure / 200 | 5.0 | 602/1635 (24.58 K) | 0.4025 | 494.4 | 0.1249 | 2.326 |
| Serge[*] | Pure / 400 | 2.5 | 200/796 (22.80 K) | 0.35 | 478 | 0.136 | 2.451 |

[*]For the sample Serge, the value for $T_0$ reported in reference [55] is slightly different than the value listed in this table. This is a matter of different definitions for $T_0$. The fit parameter reported in their work (here we relabel it as $T_0{}^{55}$) is related to our $T_0$ through $T_0 = T_0{}^{55} - \alpha E_T$. With this relation between the parameters the functional form of the fit to the data reported in their work is the same as the MAK form here.



Fitting parameter $\sigma_0$ spans a range of more than two orders of magnitude ($0.14\times10^9 - 61\times10^9$ $\Omega^{-1}m^{-1}$) with no obvious dependence on impurity type or concentration. These values of conductivity can be compared to the ones observed in high-velocity sliding regime ($E >> E_T^*$) where CDW conductivity is approximately constant and conduction is essentially linear in field[109] with $f_{NBN}$ typically observed in 0.1 – 10 GHz range. We can estimate that in this regime, for typical driving fields on the order of characteristic fields $E_T^*$, the conductivity values of the observed fit parameter $\sigma_0$ would correspond to washboard frequency range $\sigma_0 E_T^*/(en_C\lambda_C) \sim 10 - 10,000$ GHz, higher but close to the $f_{NBN}$ range typically observed. In the context of our MAK form interpretation, the energy $k_B\alpha(E-E_T)$ by which the local barrier is reduced due to $E-E_T$, can be written in terms of an effective volume per barrier $V_{barr}$ as follows:

$$k_B\alpha(E-E_T) = en_C V_{barr}\lambda_C(E-E_T) \qquad (3.7)$$

so that $V_{barr}$ and the barrier concentration $n_{barr}$ can be extracted from the fitting parameter $\alpha$:

$$V_{barr} = 1/n_{barr} = \frac{k_B\alpha}{en_C\lambda_C} \qquad (3.8)$$

Earlier we suggested that $T_0$ describes energy scale associated with strong pinning. We did not, however, address what these barriers can physically be attributed to and how they limit the motion in slow branch. Based on two highly pure samples Lemay *et al.*[55] found that $n_{barr}$ agrees with the residual impurity concentration $n_i$ previously measured in undoped samples. A natural conclusion was that impurities act as microscopic barriers limiting the CDW phase advance and causing CDW to creep. In **Figure 3.16** we explore the correlation between $n_{barr}$,



obtained from the fit parameter $\alpha$, and $n_i$ which is independently determined by *RRR* measurements. Our data from both pure and doped samples (solid symbols) show that $n_{barr}$ varies over a range of less than one order of magnitude ($0.35 \times 10^{16}$ cm$^{-3}$ – $2.5 \times 10^{16}$ cm$^{-3}$) with no obvious dependence on $n_i$, while $n_i$ spans much larger range ($2.5 \times 10^{16}$ cm$^{-3}$ – $750 \times 10^{16}$ cm$^{-3}$). For comparison on the same graph we plot a dashed line corresponding to $n_{barr} = n_i$, the trend our data would follow if the barrier concentration matched the impurity concentration in samples. This illustrates a gap between $n_{barr}$ and $n_i$ which grows as $n_i$ increases in doped samples. Only as $n_i$ decreases to the values corresponding to residual impurity concentrations observed in pure samples, $n_i$ and $n_{barr}$ become comparable, in agreement with what Lemay *et al*.[55] observed.



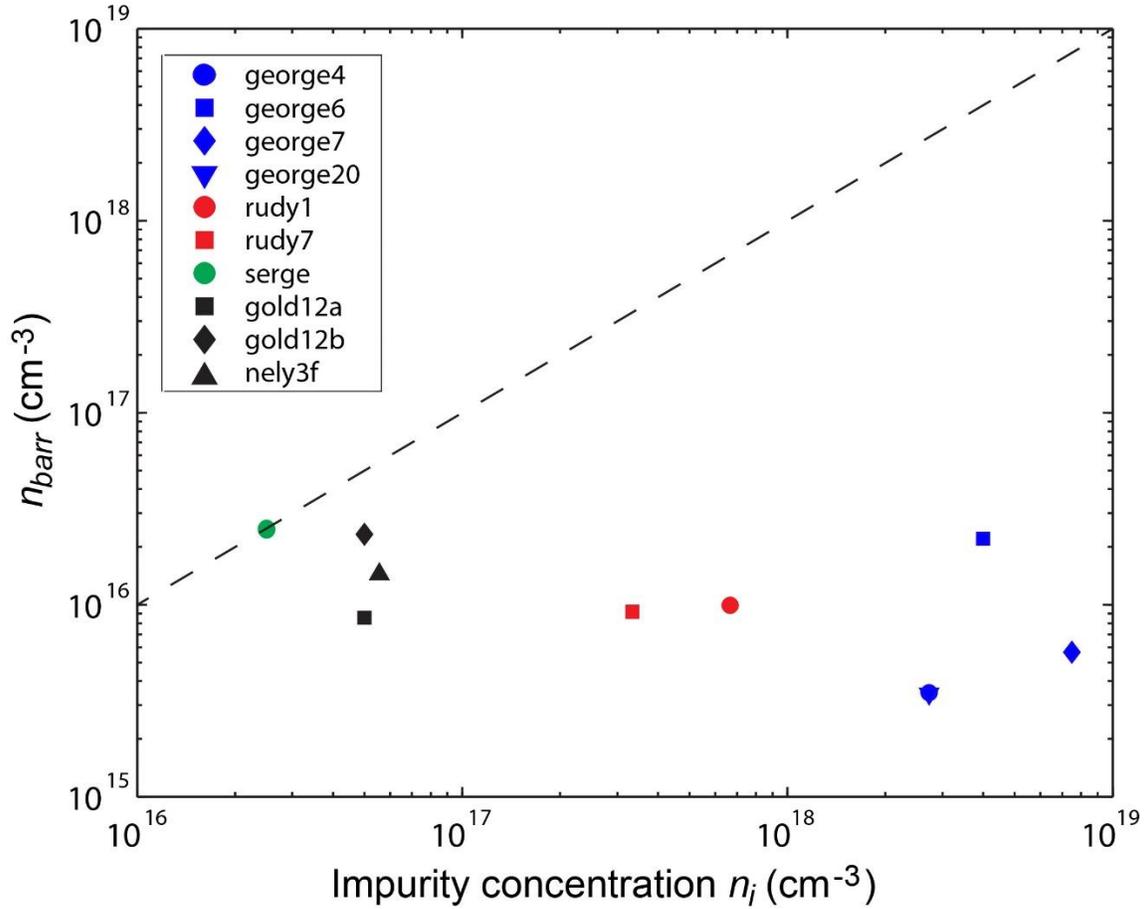

**Figure 3.16**

Correlation between barrier concentration $n_{barr}$ extracted from the fit parameter $\alpha$ and independently measured impurity concentration $n_i$ for each sample. Dashed line corresponds to $n_{barr} = n_i$ which data is expected to follow if the impurity concentration in our samples matched the barrier density (i.e. one-impurity one-barrier scenario).

Values of $n_{barr}$ extracted from our data confirm that volume per barrier $V_{barr}$ corresponds to a length scale that is orders of magnitude smaller than the length scale responsible for collective dynamics. This is consistent with our assertion that in addition to collective dynamics, local dynamics must also play an important role in the regime of temporally-ordered collective creep.



The effects of disorder present due to doping are clearly reflected in our data:

- In Ti-doped (non-isoelectronic impurity) samples, the highest measured quality factors are significantly lower ($Q = 45$) than in Ta-doped ($Q = 340$) or in pure samples ($Q = 425$).

- The values of $E_T$ and $E_T^*$ in $dV/dI$ measurements were consistent with expected values for a given level of doping and sample thickness.

Yet, the number of barriers that limit motion in the slow branch does not seem to depend on changing impurity concentration. When the impurity concentration is changed by up to a factor of 300, the values of $n_{barr}$ vary by less than one order of magnitude in a seemingly uncorrelated way. The average value of $n_{barr}$ is, however, comparable to the spurious impurity concentration measured in pure samples. Why this is the case is not clear at this point. If impurities are not the local barriers, what else can cause strong local pinning? Or could the barriers still be associated with impurities that are somehow "screened", and this screening is more pronounced in more doped samples, so that only an effective number of impurities can be "felt" by CDW (this is consistent with the fact that $n_{barr}$ is always smaller or equal to $n_i$, but never greater)? These are speculative questions, and both further experimental investigation and modeling efforts are needed to precisely answer them. What is clear is that:

- the functional form of the velocity-field relation does not change with doping level or type;

- both local and collective dynamics play an important role;



- the intricacies of the dynamics on the microscopic scale are more complex than the one-impurity-one-barrier scenario originally suggested by Lemay *et al.*[55]

The following analysis may further inform some of the questions about $n_{barr}$.

**Figure 3.17** shows that the quantity $E_T{}^*$-$E_T$ was found to scale roughly linearly with $n_{barr}$. Note that $E_T{}^*$ and $E_T$ are determined from differential resistance measurements (not from NBN data) and $E_T{}^*$ is not contained in the MAK form, so that $n_{barr}$ cannot be correlated to $E_T{}^*$-$E_T$ through the fit, i.e. $E_T{}^*$-$E_T$ is independent of the analysis that produced parameter $n_{barr}$. Plots of $E_T$ vs. $n_{barr}$ and $E_T{}^*$ vs. $n_{barr}$ (not shown) indicate some correlation that could also be linear but the scatter is large compared to $E_T{}^*$-$E_T$ vs. $n_{barr}$ plot shown in **Figure 3.17**.



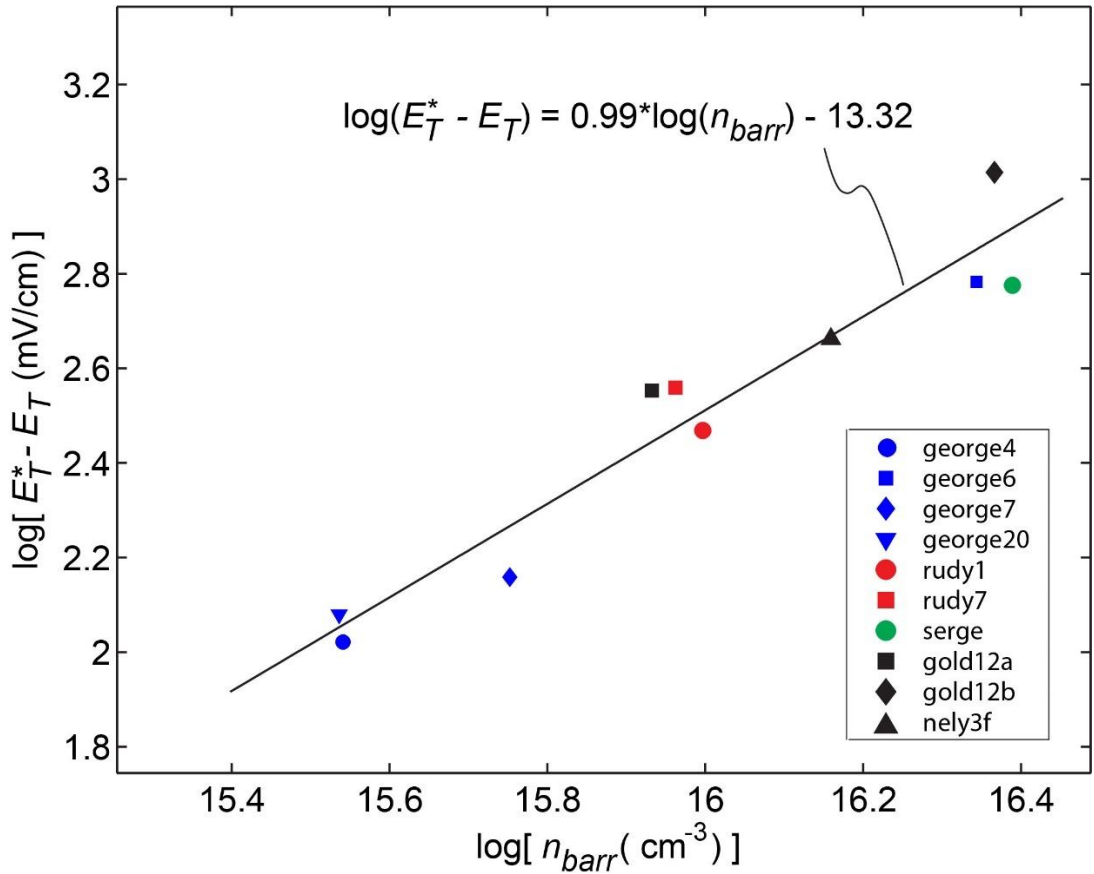

**Figure 3.17**

The slope of 0.99 of the line fitted to data on this log-log plot shows that $E_T^*$-$E_T$ has close to linear dependence on $n_{barr}$. Each data point was taken at a slightly different temperature in a narrow range between 23.8 K – 24.8 K except sample "serge" taken at 22.8 K.

Note that each data point here corresponds to a slightly different temperature since each was taken during a different sample cool-down. Quantity $E_T^*$-$E_T$ varies very little with temperature in this range (see, for example, **Figure 3.6**), so temperature variation from sample to sample is not expected to contribute significantly to the data scatter in **Figure 3.17**. $E_T$ and $E_T^*$ also depend on thickness which explains why for example Ta-doped samples (blue symbols) span almost the whole range of $E_T^*$-$E_T$ in the graph.



Since $E_T$ and $E_T{}^*$ are known to depend on both impurity concentration, $n_i$, and crystal thickness, $t$, and we observed that $n_{barr}$ does not correlate with $n_i$, dependence of $E_T{}^*$-$E_T$ on $n_{barr}$ may suggest that $n_{barr}$ may also be affected by finite size effects. This does not seem to be the case as can be seen from data for samples George20 and George4 in **Table 3.2**. While both have the same $n_i$ (i.e. both came from the same batch) and roughly the same $E_T{}^*$-$E_T$ and $n_{barr}$, the thicknesses of the two samples is significantly different (1.31 μm and 0.290 μm). Our data from a limited number of samples is not sufficient to draw definite conclusions, but $n_{barr}$ dependence on sample thickness could be further investigated more systematically in the future.

The fact that $n_i$ does not correspond to, nor scale with $n_{barr}$ can make one question the validity of interpretation of $n_{barr}$ extracted from the fitting parameter $\alpha$ in **(3.7)** and **(3.8)**. Yet the correlation between $E_T{}^*$-$E_T$ and $n_{barr}$ in **Figure 3.17** suggests that $n_{barr}$ is a meaningful parameter that determines the size of the slow branch in the low temperature regime of CDW transport. $E_T{}^*$ and $E_T$ are thresholds for CDW motion, and they must reflect the barriers that impede the motion. Thus, correlation between $E_T{}^*$-$E_T$ and $n_{barr}$ suggests that it is not unreasonable to associate the fitting parameter $\alpha$ with some sort of barriers that determine dynamics of CDW motion in this regime. In theoretical models[20,22,110] local strong pining is predicted to produce a threshold $E^{strong}$ that would depend linearly on barrier concentration. If the "reduced" threshold $E_T{}^*$-$E_T$ can be associated with $E^{strong}$, this then further supports the idea that local strong pinning barriers limit the motion in slow branch, but these strong pinning centers are not directly correlated to intentionally added impurities (since $n_{barr}$ does



not correlate with the sample impurity concentration). The question remains what these barriers represent physically.

## Universal Graph

Equation **(3.6)** can be rewritten in $y = xe^x$ form, where

$$x = \frac{\alpha(E - E_T)}{T} \tag{3.9}$$

$$y = \frac{\alpha}{\sigma_0 T} e^{\frac{T_0}{T}} j_C \tag{3.10}$$

**Figure 3.18** shows all $j_C(E, T)$ data sets: 25 $j_C(E)$ sets measured at different temperatures between 18 K – 27 K and totaling more than 700 data points each corresponding to one observed NBN peak obtained from the nine samples listed in **Table 3.2** and transformed as in **(3.9)** and **(3.10)** using fitting parameters $\sigma_0$, $\alpha$, and $T_0$, and measured values $j_C$, $E_T$ and $T$, and plotted as $y$ vs. $x$. Agreement between all the data sets and the modified Anderson-Kim from expressed as $y = xe^x$ line on the graph is striking. Excellent agreement is observed over at least five orders of magnitude in $y$, independent of temperature, doping type and level, and crystal thickness, strongly confirming that this functional form is at least a very good approximation of the true form that describes dynamics in this regime.



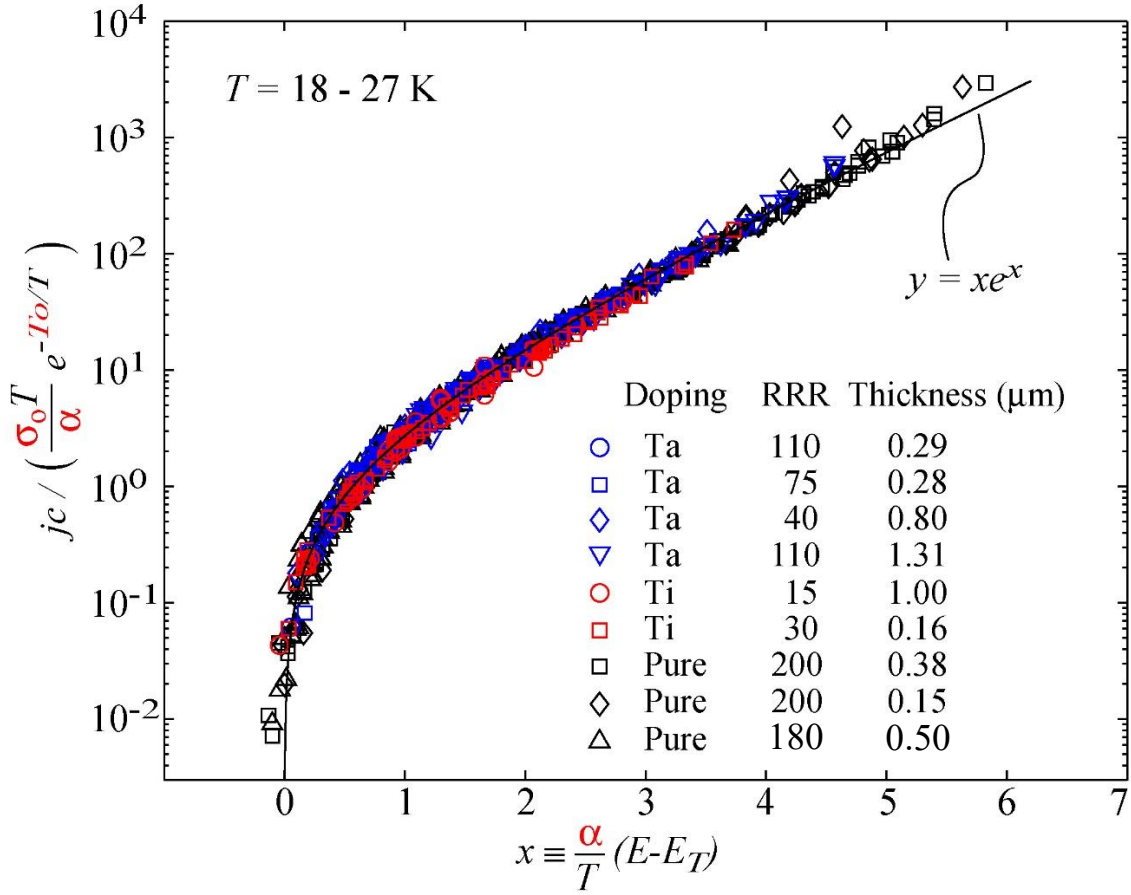

**Figure 3.18**

$j_C(E)$ data for all the temperatures measured from all nine samples listed in **Table 3.2** transformed as in **(3.9)** and **(3.10)** and plotted as $y$ vs. $x$. Solid line represents modified MAK form expressed as $y = xe^x$.

In addition, MAK form has been well fitted to data between $E_T$ and $E_T^*$ in fully gapped CDW conductors "blue bronze" ($K_{0.3}MoO_3$)[111,112] and orthorhombic-TaS$_3$[113] confirming that fully gaped CDW conductors and partially gapped NbSe$_3$ share strikingly similar behavior in all regimes of the CDW transport phase diagram. This further supports our assertion that single particle screening in partially gapped NbSe$_3$ or de-screening in fully gapped materials has little to do with the dynamics that produces this velocity-field relation at low temperature.



## Transition from Slow to Fast Branch in NbSe$_3$

In this section we examine the nature of transition from CDW creep to sliding at $E_T^*$. **Figure 3.19** shows a plot of $f_{NBN}^* \equiv f_{NBN}(E_T^*)$ vs. $1/T$. When it was possible, the value of $f_{NBN}^*$ was directly measured from the NBN peak near $E_T^*$ from plots like **Figure 3.9** (by increasing bias from below to very near $E_T^*$ and reading off a frequency of the corresponding NBN peak). When the NBN peaks at $E_T^*$ were not resolvable above the noise background in the spectra, or when $f_{NBN}^*$ was above the frequency detectable by our setup, the $f_{NBN}^*$ was deduced from the $f_{NBN}(E)$ data by extrapolating the MAK form to independently measured $E_T^*$ field. (i.e. since we plotted our fit lines in **Figure 3.14** and **Figure 3.15** to terminate at $E_T^*$, the values of $f_{NBN}(E_T^*)$ can simply be read off from the fitted graphs).



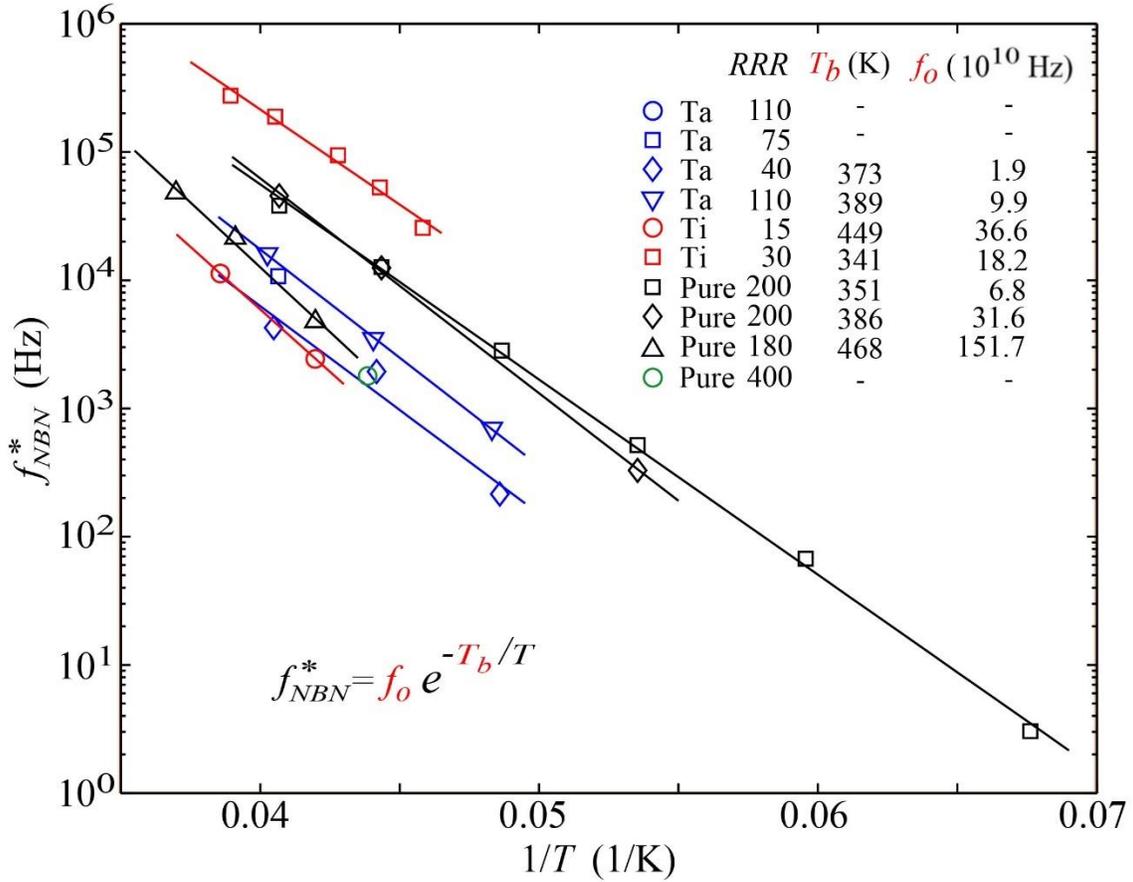

**Figure 3.19**

Temperature dependence of narrow band noise frequency at $E = E_T^*$. Data sets for different samples are fitted by **(3.11)** with extracted fit parameters $T_b$ and $f_0$ listed in the figure. Data point with the green circle symbol is from reference[55].

$f_{NBN}^* \propto v_C(E_T^*) \equiv v_C^*$ corresponds to the highest CDW velocity in the slow branch, just before the switch to the fast branch sliding state. Line fits to data in **Figure 3.19** show that the "threshold" frequency $f_{NBN}^*$ is thermally activated:

$$f_{NBN}^* = f_0 e^{-T_b/T} \tag{3.11}$$

with fitting parameter $f_0$ spanning $2 \times 10^{10}$ Hz – $152 \times 10^{10}$ Hz depending on the sample, and $T_b$ in a range of 341 K – 468 K (see inset of **Figure 3.19**). In the following we show that this



thermally activated behavior is consistent with the MAK form assuming the difference between the two thresholds, $E_T^*$-$E_T$, is independent of temperature. To see this, we use MAK form **(3.6)** and evaluate $j_C$ at $E = E_T^*$:

$$j_C^* \equiv j_C(E_T^*) = \sigma_0(E_T^* - E_T)e^{\frac{-[T_0 - \alpha(E_T^* - E_T)]}{T}} = \sigma_0(E_T^* - E_T)e^{\frac{-T_b}{T}} \tag{3.12}$$

where

$$T_b \equiv T_0 - \alpha(E_T^* - E_T) \tag{3.13}$$

Then, if $E_T^*$-$E_T$ is independent of $T$, it follows from **(3.12)** that $j_C^* \propto e^{(-T_b/T)}$ which is consistent with **(3.11)** and behavior observed in **Figure 3.19**. In the temperature range where $E_T$ and $E_T^*$ can both be clearly identified from our independent differential resistance measurements we have indeed observed $E_T^*$-$E_T$ to be approximately constant with temperature (e.g. see **Figure 3.6**).

Since $T_0$ is roughly constant for all samples, and $1/\alpha$ ($\propto n_{barr}$) is proportional to $E_T^*$-$E_T$ (see **Figure 3.17**), the MAK form then predicts from **(3.12)** and **(3.13)** that $T_b$ should be the same for all samples. We indeed observe a narrow spread in $T_b$ values extracted from the fits to data in **Figure 3.19** or visually by looking at the slope (corresponding to $T_b$ on this semi-log plot) of the fitted lines which are all nearly parallel with the average value of $T_b = 394 \pm 44$ K. This value is in remarkable agreement with $\Delta_C$, i.e. $k_B T_b = (0.97 \pm 0.11)\Delta_C$ where $2\Delta_C = 70$ meV is CDW gap energy of lower Peierls transition in NbSe$_3$ given in **Table 1.1**.

We emphasize that the thermally activated behavior of $f_{NBN}^* \propto j_C(E_T^*)$ (i.e. each data set falls on a *straight* line in **Figure 3.19**) and constant value of $T_b$ obtained from the fits (i.e. the



straight lines are *parallel*) are consistent with the MAK form but *independent* of the agreement observed between MAK form and $f_{NBN}(E)$ data in **Figure 3.14**, **3.15**, and **3.18**. $f_{NBN}^{*}$ data points plotted in **Figure 3.19** are values of $f_{NBN}$ obtained at *independently* determined $E_T^{*}$ field measured in differential resistance measurements, i.e. note that in the MAK form $j_C$ (and thus $f_{NBN}$) is a function of $E_T$ but not of $E_T^{*}$, and that at a given temperature different samples have different value of $E_T^{*}$ dictated by independent parameters such as sample doping, and thickness.

In the above discussion we argue that two independent sets of measurements support the idea that MAK form consistently describes CDW transport dynamics in the low temperature creep regime:

1. $j_C(E)$ data in the slow branch (**Figure 3.14**, **3.15**, and **3.18**) shows remarkable agreement with the MAK form.

2. At different temperatures we independently measure $E_T^{*}$ by differential resistance measurements. $E_T^{*}$ depends on independent and extrinsic parameters like doping level and thickness. We measure NBN peaks at field $E_T^{*}$ to obtain $f_{NBN}^{*}$. The plot of $f_{NBN}^{*}$ vs. $1/T$ data shows thermally activated $f_{NBN}^{*}$ and constant $T_b$ for all samples.

3. When we use MAK form in point 1) to compute $f_{NBN}(E_T^{*})$ we see that MAK form predicts the observed thermally activated behavior in point 2).

**Figure 3.19** suggests that a threshold (or a critical) frequency or velocity (rather than a threshold field) is a relevant parameter for the transition from CDW creep to CDW sliding, i.e. $f_{NBN}^{*}$ and $v_C^{*}$ have well-defined, thermally activated behavior with barriers of energy scale



$\Delta_C$, while $E_T{}^*$ is weakly dependent on temperature. Larkin and Brazowski[23,110] (also see reference [22]) predicted that a barrier for soliton generation at strong pinning centers is $\sim \Delta_C$ with the activation frequency $\sim 10^{10}$ Hz, values in agreement with the observed values of our fit parameters $T_b$ and $f_0$. This activation frequency is on the order of the CDW washboard frequency observed in high-velocity sliding regime at high temperatures where motion is defined by FLR elastic dynamics. Based on this we can interpret the transition from the slow to the fast CDW transport branch as follows: at a critical velocity $v_C{}^*$ associated with $f_{NBN}{}^*$ local barriers become ineffective in limiting the creep motion, and CDW switches from creep to coherent sliding characterized by collective elastic dynamics.

## 3.3   Questions, Further Directions, and Suggested Improvements

### Anomalous Behavior in Several NbSe$_3$ Samples

We wish to mention a few instances of anomalous transport behavior in NbSe$_3$ samples, unlike the behavior we have described so far and what is typically reported in literature. The behavior was observed in a total of three different NbSe$_3$ samples all which came from an isolated growth batch. It is not clear if something unusual happened during the crystal growth process, and we can offer no explanation for the observed behavior, but we think it is worth recording here since we are not aware that anyone reported it previously.

The anomaly is the following: As one decreases temperature below $T_{P2}$, instead of observing a single $E_T{}^*$ threshold emerge in differential resistance measurements, one observes that several (up to three) $E_T{}^*$-like "thresholds" appear. Unlike multiple sliding thresholds observed in samples plagued by crystal non-uniformities such as multiple crystal domains or



steps in the crystal cross-section as in **Figure 3.7**, the individual "thresholds" in these samples emerge *each at a different temperature*. This is shown in **Figure 3.20 (b)**. For comparison, in **(a)** of the same figure we show an equivalent set of differential resistance curves observed in a sample with a typical behavior.

Inspection by optical microscope revealed that the samples with anomalous behavior had no steps visible on the sample surface. $R$ vs. $T$ data showed two well-defined CDW transitions occurring at expected transition temperatures, and no other unusual behavior. We note that additional "thresholds" in $dV/dI$ curves emerge in a fashion very similar to the emergence of $E_T^*$ in samples with typical behavior: each additional threshold first appears as a rounded "shoulder" that grows and becomes more sharply defined as one lowers the temperature further. The first shoulder appears above $E_T$ but at notably higher temperatures than in typical samples. Then subsequent shoulders appear one at the time at higher fields with decrease of temperature.

CDW transport in the samples of this isolated batch appears to acquire these novel features as the $k_B T$ energy scale changes, and this behavior does not in an obvious way appear to be associated with current inhomogeneity in samples with multiple crystal domains, and we thought this was worth mentioning here.



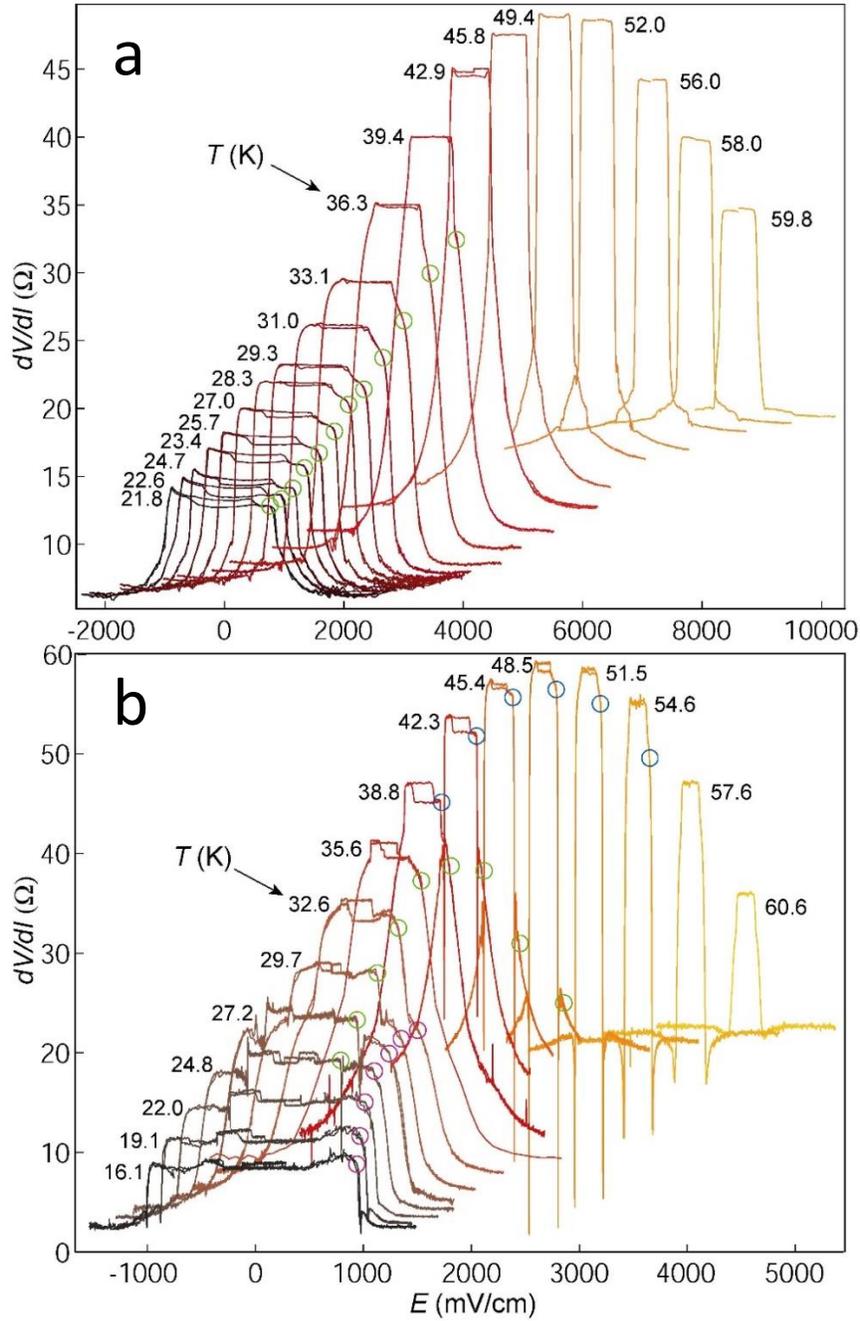

**Figure 3.20**

(a) Typical set of *dV/dI* vs. *E* curves measured at different temperatures in NbSe₃ samples (here on sample Rudy7). Green circles point to formation of $E_T^*$ threshold as temperature is decreased. (b) An equivalent set of curves measured on a sample (Nely3a) with anomalous behavior. Circles of different colors point to formations of three different $E_T^*$-like "thresholds" that become defined at different temperatures. The data sets are offset horizontally so that the offset of each curve is proportional to its temperature



Inspection by optical microscope revealed that the samples with anomalous behavior had no steps visible on the sample surface. $R$ vs. $T$ data showed two well-defined CDW transitions occurring at expected transition temperatures, and no other unusual behavior. We note that additional "thresholds" in $dV/dI$ curves emerge in a fashion very similar to the emergence of $E_T^*$ in samples with typical behavior: each additional threshold first appears as a rounded "shoulder" that grows and becomes more sharply defined as one lowers the temperature further. The first shoulder appears above $E_T$ but at notably higher temperatures than in typical samples. Then subsequent shoulders appear one at the time at higher fields with decrease of temperature.

CDW transport in the samples of this isolated batch appears to acquire these novel features as the $k_B T$ energy scale changes, and this behavior does not in an obvious way appear to be associated with current inhomogeneity in samples with multiple crystal domains, and we thought this was worth mentioning here.

## Mode-Locking to NBN in Slow Branch

Since the CDW motion in the slow branch is temporally-ordered, mode-locking of the CDW periodic component to an external AC drive should in principle be possible. At higher temperatures when an external AC drive is applied in addition to the DC drive, one observes plateaus in the $I$-$V$ curves in the sliding regime analogous to Shapiro steps in superconductors. The plateaus occur at DC drives where $f_{NBN} = f_{AC}$ and where $f_{NBN}$ is a rational multiple or fraction of $f_{AC}$, as the periodic component of the CDW motion locks to the drive. Simple $I$-$V$ measurements however cannot be used in the low temperature slow branch regime to detect



mode-locking since the CDW current $I_C$ is masked by orders of magnitude larger single particle current $I_S$ and cannot be directly resolved. Our attempts to measure subtle changes in differential resistance using standard lock-in amplifier techniques, as is typically done in mode-locking measurements in a high temperature sliding branch, were not successful in resolving this effect in slow branch either. We propose a simple experiment by which one is more likely to observe the mode-locking phenomenon in this regime of CDW transport. One could perform measurements to generate spectral plots as in **Figure 3.9** and observe NBN peaks at different (DC) electric fields in the presence of an external AC drive field. On such a plot the peak of an external AC drive should appear at a fixed frequency $f_{AC}$. As we measure noise spectra at progressively larger DC fields, and as the frequency of the NBN peak, $f_{NBN}$, approaches $f_{AC}$, we should observe that the external AC peak "attracts and locks" CDW NBN peak at $f_{AC}$. As we increase DC drive further, we should observe that the CDW NBN peak "reluctantly detaches" from the AC peak as it moves towards higher frequencies. By characterizing this "momentary halting" of NBN peak at $f_{AC}$ while sweeping DC drive one could extract coherence characteristics of the CDW creep in slow branch. This is an analogous measurement scheme to obtain Shapiro steps, however in this case we would be locking extremely slow CDW creep to external AC drive at frequencies as low as few Hz. We suggest that this is a promising technique, and we expect that locking of NBN peaks to external AC peaks in the frequency domain should be easily observable.

## Quality Factors and Harmonic Content of NBN in Slow Branch

Our preliminary inquiry into $Q$-factors of NBN peaks revealed how doping affects coherence of the motion and how the coherence evolves with external drive in slow branch (see **Figure**



**3.10).** One could also examine how quality factors depend on crystal segment size/volume. Dependence on length, width, and thickness of the whisker could probe relevant coherence lengths and finite size effects in this regime of the transport. Detailed examination of harmonic content of NBN could further illuminate microscopic dynamics and CDW interaction with local pinning centers[21] in the slow branch.

## Doped vs. Undoped Samples

Many questions remain in trying to understand the exact role impurities play in microscopic dynamics of CDW transport at low temperatures. Why is $n_{barr}$ parameter obtained from the fit comparable in both doped and undoped samples even though the effects of doping are clearly visible in other measurements? Doping level and type has a strong effect on values of $E_T$ and $E_T^*$, and on quality factors of coherent oscillations, but it does not appear to influence fit parameters of $j_C(E)$ in an obvious way. Our data (**Figure 3.17**) suggests that $n_{barr}$ obtained from the fit is a parameter relevant in the dynamics of CDW transport at low temperatures, but that correlating one barrier with one impurity is not valid. $n_{barr}$ extracted from pure and doped samples does however correspond to the spurious impurity concentration observed in undoped samples.

## Observing Temporally-Ordered Motion in Slow Branch of Other Materials

The data in **Figure 3.9**, **3.14**, **3.15**, and **3.18** in this chapter clearly illustrate the temporal order of the motion and the form of the $j_C(E)$ relation in in the slow branch of NbSe$_3$. The question



remains how universal this behavior is among other CDW systems and further in DW systems in general.

As we mentioned before, MAK functional form of the velocity-field relation between two thresholds has been experimentally confirmed in two fully gapped CDW materials[111-113] testifying to the ubiquity of this behavior across systems. We stress that evidence of *temporally-ordered motion between $E_T$ and $E_T$\** has not been observed in any CDW system other than $NbSe_3$ to date. This is not entirely surprising considering that no other CDW system equals $NbSe_3$ in terms of crystalline quality of the synthesized material. Crystal inclusions and disorder in other widely studied CDW materials such as "blue bronze" or $TaS_3$ produce significantly less homogeneous CDW current distribution in these crystals. Resulting quality factor of the periodic CDW motion is thus not expected to be high and NBN peaks are expected to be more difficult to observe in slow branch. Nevertheless, experimentally confirming the coherent, temporally-ordered nature of the CDW motion in this regime in another CDW system would be an important milestone for building a complete transport phase diagram for this class of systems. We point out that washboard oscillations in the creep regime of collective transport have also been observed in flux lattice vortices of type-II superconductors (another class of systems with collective transport in periodic media with quenched disorder), in two different systems: $NbSe_2$[114] and high-$T_C$ superconductor $Bi_2Sr_2CaCu_2O_y$[115].



## Need for New Models

Even though **(3.6)** describes temporally-ordered collective creep data remarkably well, it is a somewhat arbitrarily chosen functional form. New modeling efforts are needed to uncover the true form of $j_C(E)$ relation (which should be very close to the MAK form) in this regime and build a complete microscopic picture of temporally-ordered collective creep. The theory should address the following questions:

- What are the relevant barriers, and how do they relate to impurity type and concentration?

- What is the nature of the transition at $v_T^* \equiv v(E_T^*)$?

- How is the behavior modified by decrease of quasiparticle density with decrease of temperature (de-screening of DW) in fully gapped density wave materials?

- How general is this local-collective picture for driven disordered systems?

## 3.4   Conclusions

The low temperature dynamics in $E_T < E < E_T^*$ regime of NbSe$_3$ is well described by the modified Anderson-Kim functional form given by equation **(3.6)** in both doped and pure samples. Fitting parameters $T_0$ and $\alpha$ ($\propto 1/n_{barr}$) point to energy- and length-scales much smaller than the ones responsible for collective dynamics. Nevertheless, CDW collective behavior prevails and signatures like coherent oscillations and onset of motion at collective threshold $E_T$ demonstrate that large collective length- and energy-scales remain important in



the regime where CDW exhibits creep-like behavior. Quality factors of coherent oscillations – metrics of motion's temporal order– are typically lower in doped than in undoped samples, and more so in samples doped with more perturbing non-isoelectronic Ti impurities, than in ones doped with Ta. This indicates that impurities do affect microscopic dynamics, and/or that doped samples are more likely to suffer from structural problems and result in compromised crystals displaying non-ideal CDW behavior.

The scale for density of barriers that limit the creep is given by $n_{barr} \sim 10^{16}$ cm$^{-3}$. This value agrees with a concentration of residual impurities in undoped samples, but $n_{barr}$ remains unchanged with increase of impurity concentration $n_i$ in doped samples. This implies that the simple one-impurity-one-barrier picture does not hold as previously thought. Nevertheless, $1/n_{barr}$ sets an important new length scale that characterizes creep, is independent of doping, and is orders of magnitude smaller than the one associated with collective dynamics. In addition, the energy scale of the fitting parameter $T_0$ in the MAK fits agree well with the energy scale predicted for local strong pinning centers suggesting that the local dynamics in addition to the collective one (and/or their interplay) set the behavior of these systems.

Threshold frequency $f_{NBN}{}^*$ (or threshold velocity $v_C{}^*$), not $E_T{}^*$, is the relevant parameter that defines transition from creep to sliding state at low temperatures. $f_{NBN}{}^*$ is thermally activated, and the agreement of our fitting parameter values with theoretical predictions involving strong pinning centers suggests that transition from creep to sliding occurs when a CDW reaches sufficient velocities with activation frequencies that render strong pinning barriers ineffective in retarding the motion in the slow branch. This then culminates in an abrupt and often



hysteretic transition to sliding. Above the transition the conduction is nearly ohmic, and the dynamics resembles high temperature sliding regime governed by elastic collective dynamics.

Before concluding we elaborate on the label *temporally-ordered collective creep* as one to suitably describe the CDW motion in the slow branch. One was unlikely to encounter this terminology in literature before Lemay *et al.*[55] discovered oscillations in the slow branch of NbSe$_3$ and showed that they correspond to the traditional washboard oscillations observed in driven CDW systems. Coherent oscillations or NBN confirm that the motion is *temporally ordered*. The motion is a result of *collective* dynamics because: 1) the onset of motion with drive where first NBN peaks appear is characterized by a well-defined threshold field that can be identified with the *collective* threshold $E_T$, and 2) observation of NBN in our measurements would not be possible if the motion was not correlated over very large sample volumes corresponding to length scales dominated by *collective* dynamics. And finally, the motion is extremely slow, temperature activated, described by a functional form almost identical to a from that describes creep in a different system, and by fitting parameters suggesting that strong local pinning and local dynamics are important – all signatures of *creep*.

Our observations in NbSe$_3$ show that both local and collective pinning is important in defining the low temperature CDW behavior. Since de-screening of CDW with decrease of temperature does not occur in partially-gapped NbSe$_3$, and yet NbSe$_3$ exhibits most of the characteristic signatures of low temperature dynamics common to fully-gapped materials, we conclude that models based on de-screening that predict behavior observed in fully-gapped materials cannot be fully correct. Our results and conclusions strongly suggest that present theoretical understanding of low temperature CDW transport is incomplete. New modeling



efforts that will provide a satisfactory picture in agreement with experimental observations in these systems are needed.



# 4 Relaxation of CDW Polarization in the Low Temperature Regime

## 4.1 Introduction

The experiment presented in this chapter is a result of very preliminary measurements and is in that respect an incomplete "work-in-progress" that is meant to be pursued in the future in more detail. Our objective was to design an experiment to probe slow relaxation dynamics that is mediated by the slow-branch CDW creep discussed in the previous chapter. To induce the relaxation, we utilize switching current pulses that polarize the CDW strain profile (i.e. induce displacement of CDW phase between current contacts).

A rich impurity landscape in CDW systems interacts with a CDW and through pinning can maintain macroscopic displacements of the CDW phase along the sample for long times. After a biasing current in a sample is switched off, the CDW can remain polarized on a time scale of hours or longer in NbSe$_3$.[116] In the upper temperature regime, these metastable states[117-119] with long relaxation times have been successfully probed through x-ray scattering from crystal lattice that couples to the CDW wavevector[120-123] and through transport relaxation measurements[83,116,124,125] in response to switching current pulses. The experiments in this chapter are similar to the latter, but here we attempt to probe the dynamics in the *low temperature regime* of the lower ($T_{P2}$) CDW transition in NbSe$_3$ with hope to gain additional insights about the dynamics responsible for the temporally-ordered collective creep. The



time domain measurements presented here are complementary to the experiments presented in chapter 3.

As discussed earlier, in the low temperature regime of NbSe$_3$, the CDW is pinned below $E_T$ and in the slow branch ($E_T < E < E_T{}^*$) a CDW current $I_C$ is orders of magnitude smaller than the single particle current $I_S$, and its contribution to the total current $I_{tot}$ in this part of $I$-$V$ curve can be neglected, i.e. for all practical purposes the applied current flowing through a whisker for $E < E_T{}^*$ is $I_S$. When we measure a time trace of a voltage signal $V(t)$ over a whisker segment of length $L_v$, we are probing a local single particle resistance $R_S(t)$ averaged over that segment:

$$R_S(t) = V(t)/I_S \approx V(t)/I_{tot} \qquad (4.1)$$

Single particles are coupled to the CDW via Coulomb interaction, and they screen CDW fluctuations on a timescale much shorter than seconds (i.e. on the timescale of a measurement in this experiment single particles respond to the CDW instantaneously). Thus, since single particles "dress and follow" the CDW to screen its strain in a polarized configuration, we are indirectly probing changes in local CDW strain with time by measuring $V(t) \propto R_S(t)$. This in turn allows us to track CDW relaxation. However, $R_S$ is only sensitive to changes in CDW polarization when the changing strain profile in the probed sample segment can produce a net change in average local single particle resistivity. In a standard four-probe measurement configuration (see **Figure 3.3**), this is easily accomplished when the voltage probes are asymmetrically placed with respect to the center point between the current contacts. When an electric field is applied across a whisker between the current contacts, the CDW responds by



deforming its phase to stretch near one current contact and compress near the other resulting in a polarized configuration, as shown in **Figure 4.1**, and a CDW strain between the contacts. The produced strain profile across the sample is approximately linear (except very near the current contacts)[35,66,126,127] and is antisymmetric around the sample mid-point between the current contacts. If the voltage probes are symmetrically placed around the mid-point, the strain profile between them averages to zero because the strain has one sign to the left, and the opposite sign to the right of the midpoint. This makes symmetrically placed voltage probes insensitive to changes in CDW polarization, while probes offset from the center can provide the necessary asymmetry.

When $E > E_T$, CDW phase slip near the current contacts facilitates CDW-to-single-particle current conversion:[35,36,83,124,126-130] phase fronts are inserted near one current contact and removed at the other as shown in **Figure 4.1 (c)**. The phase-slip process is mainly localized near current contacts where the absolute strain is the largest.

As a result of these boundary conditions and the strain induced by CDW polarization, in the low temperature regime of NbSe$_3$ (below approximately 35 K), single particle resistance exhibits hysteretic behavior for for $|E| < E_T$ in differential resistance vs. current measurements shown in **Figure 3.5**. Spatially resolved measurements by Lemay *et al*.[66] showed that the vertical size of the hysteresis loop is the largest when a pair of voltage probes are closest to one of the current probes (as expected).



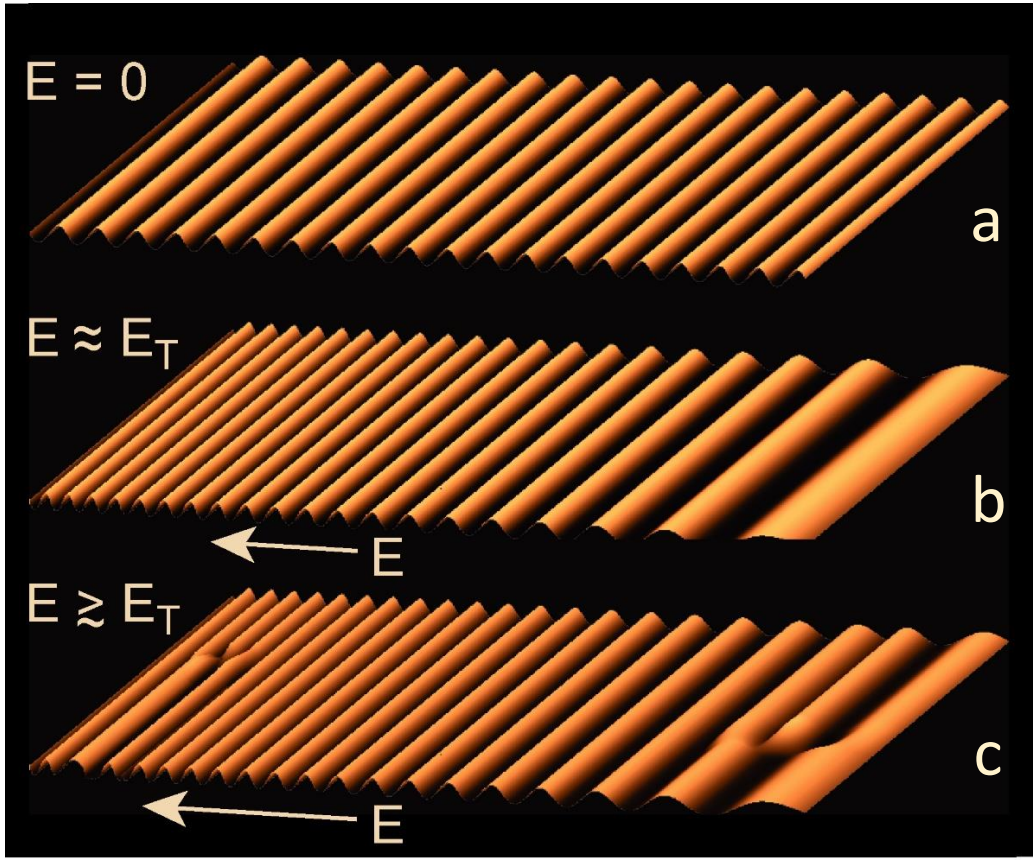

**Figure 4.1**

Polarization of macroscopic CDW phase by electric field, and formation of phase dislocations (phase slip) near current contacts above $E_T$.

In the low temperature regime, this low-field resistance hysteresis can be used to clearly identify the *slow branch* in NbSe$_3$ where temporally-coherent creep is observed. In $dV/dI$ vs. $I$ curves, the slow branch is bound by the end of the hysteresis loop, which Lemay *et al.*[66] showed can be identified with $I_T$, and by a sharp, sometimes hysteretic, drop in resistance at $I_T^*$ as shown in **Figure 1.5**.



## 4.2 Experimental Method

Our measurements of transient voltage response to bipolar switching current pulses were performed in the low temperature regime at $T = 22.5$ K where the slow and the fast branch of CDW transport can be easily identified. We used a standard four-probe measurement configuration in **Figure 3.3** where sample current bias was provided by Keithley 220 Current Source, and the voltage signal was measured directly by a Keithley 196 System Digital Multi Meter. Data collection and experiment control were facilitated via a GPIB interface to a desktop computer. We did not attempt to optimize the circuit for fast measurements since our aim was to probe the slow dynamics of the creep regime.

The biasing current pulse sequence is shown in **Figure 4.2 (a)** and proceeds as follows: First we apply a current $|I_A| > I_T^*$ corresponding to point A to force the CDW phase into a polarized configuration between current probes illustrated in **Figure 4.1 (c)**. The CDW reaches sliding steady state in the fast branch during this polarization pulse which lasts 2 or 3 seconds. It can be verified from our data (based on our measurements of CDW relaxation times after biasing samples above $I_T^*$) that the CDW can reach the steady state on these time scales. The fast dynamics above $I_T^*$ should also aid the CDW to reach the steady state fast compared to these time scales. Note that typical washboard frequencies observed above $E_T^*$ in the experiments of chapter 3 at these temperatures are at least ~1 kHz, and more typically above many 10s of kHz.

Next, we step the current to the opposite direction, though $I = 0$, to a finite current bias of the opposite sign (point B in **Figure 4.2 (a)**). We measure a voltage trace, $V(t)$, at a rate of 3-4 data points per second while maintaining the current constant at B for several minutes. This



measurement cycle is then repeated many times, but each time the final current bias is slightly increased (points C, D, E, F and G). $V(t)$ data is continuously collected while we step the current through the whole sequence in **Figure 4.2 (a)**. We have measured $V(t)$ traces across sample segments at both ends of the sample (near each current contact) to probe the dynamics while $R_S$ is increasing or decreasing through the low-field resistance hysteresis loop as shown in **Figure 4.2 (b)** and **(c)**. The insets in **(b)** and **(c)** show the corresponding current and voltage probe configurations for each case. The offset position of the voltage probes from the midpoint between current probes ensures that the measurement is sensitive to changes in CDW strain as discussed earlier.



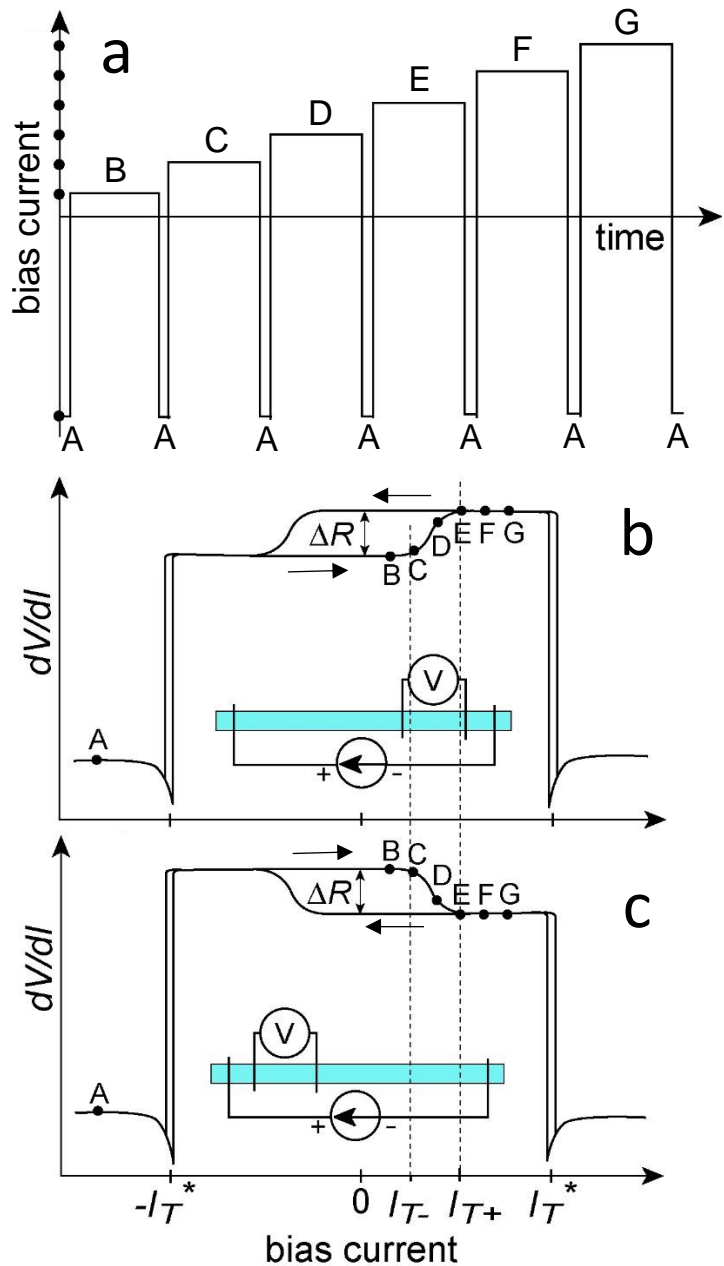

**Figure 4.2**

(a) A schematic of a bias current sequence vs. time applied to the sample showing specific bias points. Plateaus at points A are all either 2 or 3 seconds long, while plateaus at points B, C, D, E, F, and G last 2 or 3 minutes. (b) and (c) show the corresponding bias points on $dV/dI$ vs. $I_{tot}$ diagram. Insets in (b) and (c) show the position of current and voltage probes on the sample for a given bias current polarity that produces the corresponding orientation of the low-field resistance hysteresis.



## 4.3   Data and Results

**Four Distinct Regimes of Relaxation at Low Temperatures**

**Figure 4.3, 4.4,** and **4.5** show three $V(t)$ data sets measured on samples Gold12a (whisker thickness 376 nm) and Gold12b (thinned by plasma etching to 152 nm) from Table 3.2. Schematics showing current and voltage probe configurations corresponding to each data set are shown as insets in part **(a)** of each figure. Part **(a)** of each figure shows a long voltage trace $V(t)$ continuously measured during applied current sequence shown in **Figure 4.2 (a)**. The abrupt voltage switching to below $V = 0$ (vertical lines extending to the $x$-axis) corresponds to current switching to point A. The staircase "plateaus" are transient responses corresponding to points B, C, D, E, etc. Part **(b)** of each figure shows the same data, but each transient "stair" is offset by the starting voltage in that "stair", i.e. each transient measured after the current switch here starts evolving from zero volts on the graph (this is most clearly seen in **Figure 4.4 (b)**). This allows us to clearly compare transient amplitudes between point B, C, D, E, etc. Part **(c)** of each figure shows the data in **(b)** but with each transient superimposed on the same time axis where zero minutes on the graph corresponds to the time just after the current switch from point A. The same color is assigned to corresponding traces in parts **(b)** and **(c)**.

The values of currents $I_{T-}$, $I_{T+}$, and $I_T^*$ were determined in separate measurements of differential resistance vs. bias current for each sample segment. $I_{T-}$ is the current where the low-field resistance hysteresis loop begins to close during the measurement sweep, $I_{T+}$ is the current where it completely closes, and $I_T^*$ is the second characteristic threshold where differential resistance steeply drops as shown in **Figure 4.2 (b)** and **(c)**. The effects of bias



sweep rate on $I_{T+}$ are discussed in the next section. $I_{T-}$, $I_{T+}$, and $I_T^*$ are labeled in part **(b)** of each data set figure, and their values are listed in each figure's caption. The captions also contain distances between contact probes $L_{v1}$, $L_v$, and $L_{v2}$ as defined in **Figure 3.3** which are also listed in **Table 4.1** for each whisker segment.



**Figure 4.3**

(a) Voltage trace $V(t)$ measured during the applied current sequence shown in **Figure 4.2 (a)** on sample Gold12a. The inset shows the corresponding low-field hysteresis case and biasing configuration with $L_{v1} = 150$ µm, $L_v = 50$ µm, and $L_{v2} = 350$ µm. Current plateaus B, C, D, E, etc. are stepped in steps of 50 µA from 0 to 2000 µA. Each polarizing current pulse (point A) has amplitude of -3000 µA and lasts 3 s. $I_{T-} = 519$ µA, $I_{T+} = 931$ µA, and $I_{T*} = 1750$ µA. (b) The same data of the "staircase" voltage trace in (a) but each transient "stair" is offset by the starting voltage in that "stair", i.e. each transient measured after the current switch starts evolving from zero volts on the graph. (c) Same data as in (b) but with each transient superimposed on the same time axis where zero minutes on the graph corresponds to the time just after the current switch from point A. The same color is assigned to corresponding voltage traces in (b) and in (c).



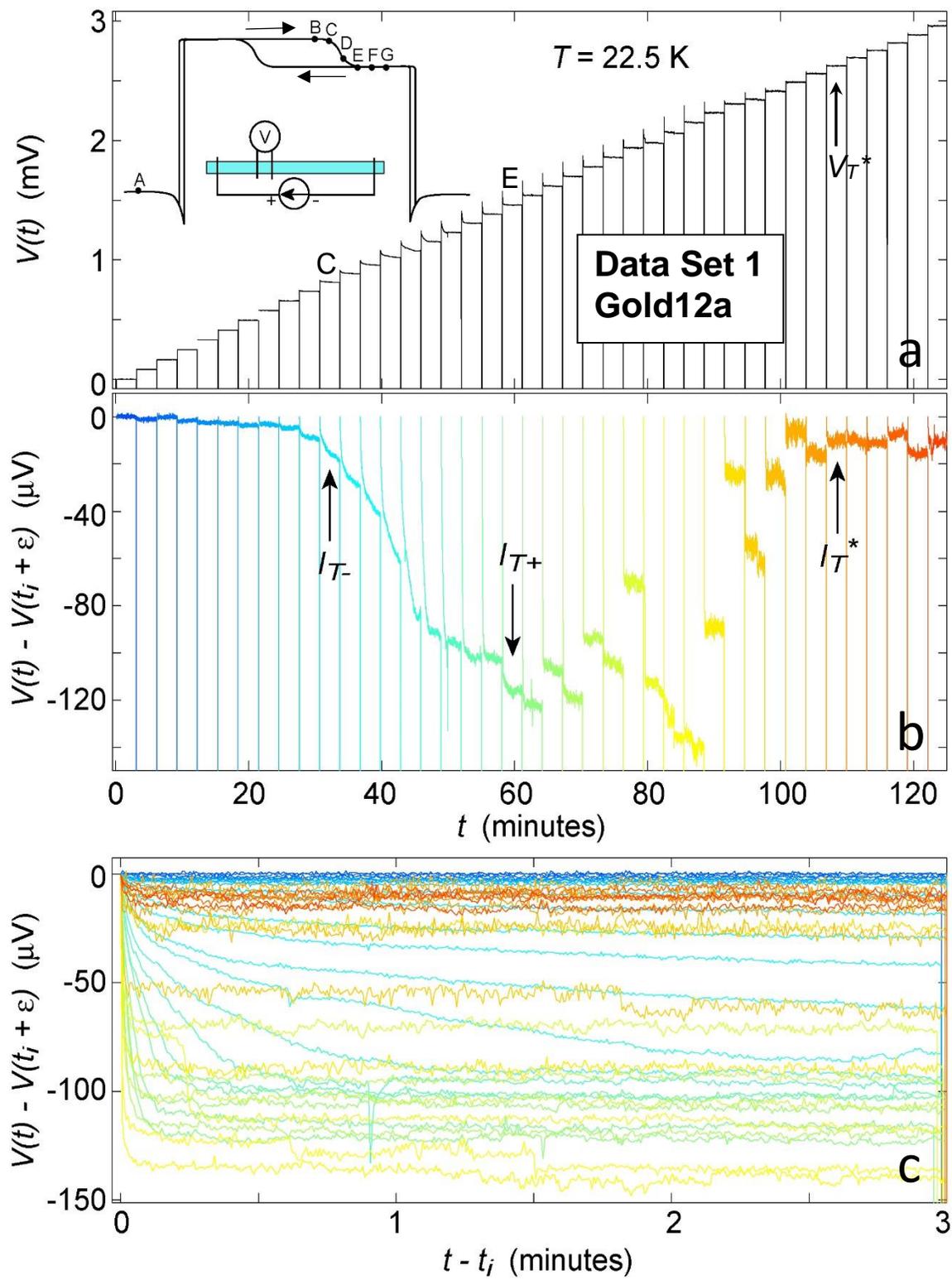



**Figure 4.4**

(a), (b), and (c) $V(t)$ measured on sample Gold12b (thinned part of the whisker) and plotted in the same way as the data in **Figure 4.3** . The inset of (a) shows the corresponding low-field hysteresis case and biasing configuration with $L_{v1} = 50$ μm, $L_v = 150$ μm, and $L_{v2} = 350$ μm. Current plateaus B, C, D, E, etc. are stepped in steps of 50 μA from 400 μA to 1700 μA. Polarizing pulses (point A) had amplitude of -3000 μA and lasted 3 sec. $I_{T-} = 775$ μA, $I_{T+} = 981$ μA, and $I_T^* = 1594$ μA. The same color is assigned to corresponding voltage traces in (b) and in (c).



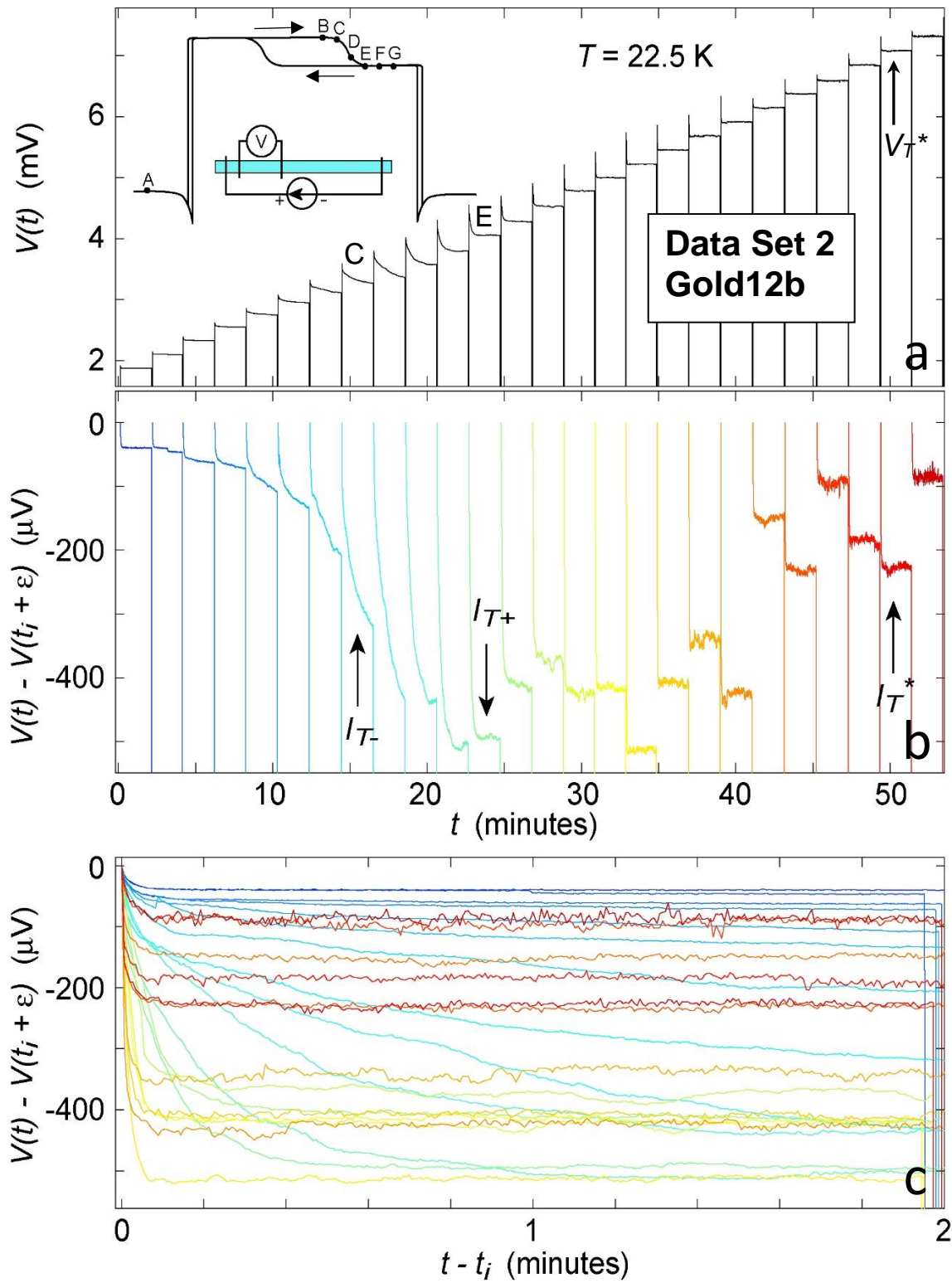



**Figure 4.5**
(a), (b), and (c) $V(t)$ measured on sample Gold12a (thick part of the whisker) and plotted in the same way as traces in **Figure 4.3** for the same sample but in the opposite half of the sample, where the polarity of CDW polarization is expected to be opposite to those in **Figure 4.3** and **4.4**. In this case voltage transients *increase* instead of decaying. The inset of (a) shows the corresponding low-field hysteresis case and biasing configuration with $L_{v1} = 350$ μm, $L_v = 150$ μm, and $L_{v2} = 50$ μm. Current plateaus B, C, D, E, etc. are stepped from 400 to 1300 μA in steps of 50 μA. Polarizing pulses (point A) had amplitude of -2000 μA and lasted 2 s. $I_{T-} = 510$ μA, $I_{T+} = 939$ μA, and $I_T{}^* = 1748$ μA (note that bias current was stepped only to 1300 μA, so that $I_T{}^*$ is outside of the graph's range). The same color is assigned to corresponding voltage traces in (b) and in (c).



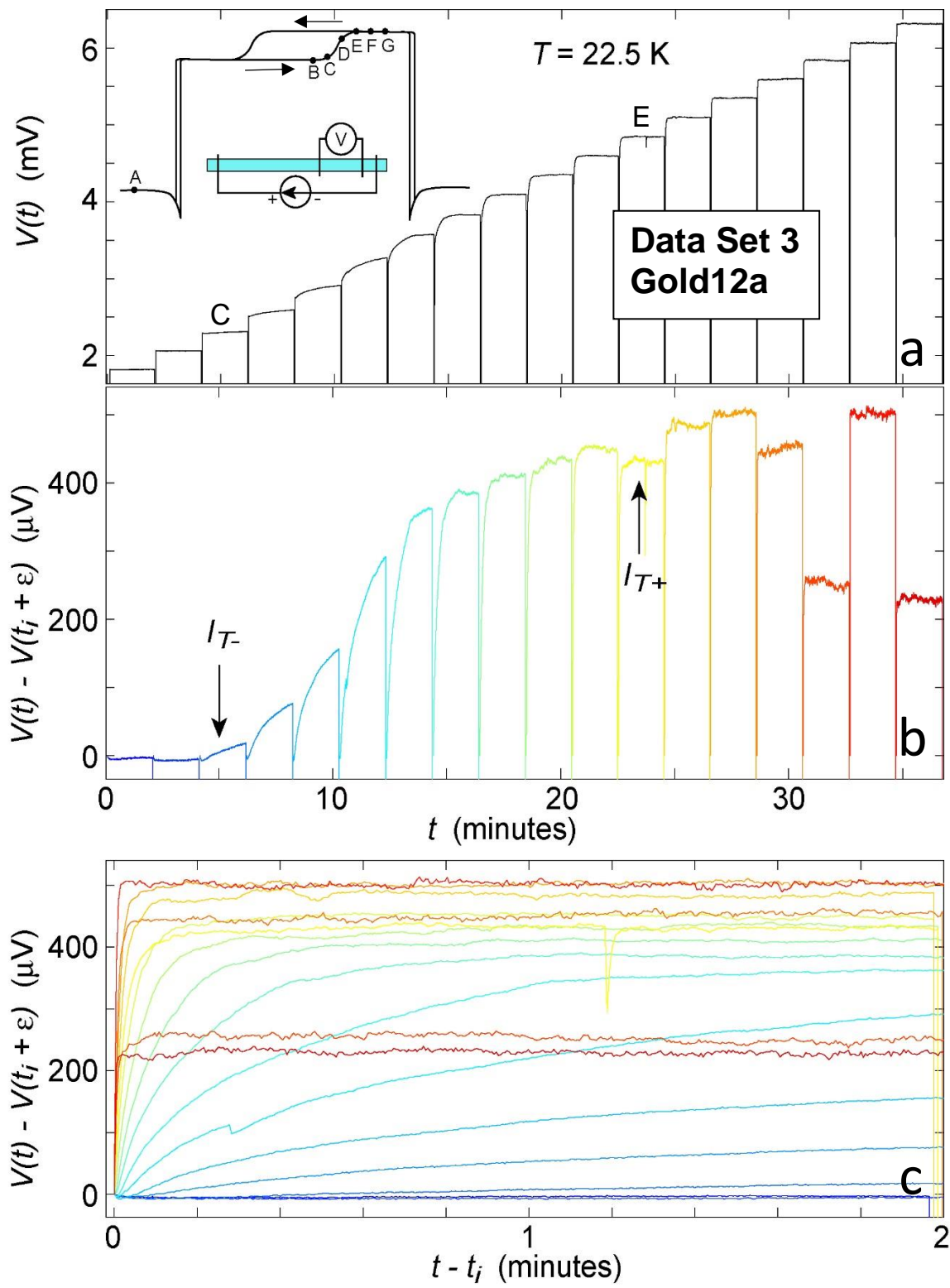



**Figure 4.3** shows a data set of sample Gold12a (non-etched segment of the whisker) corresponding to a biasing configuration shown in **Figure 4.2 (c)**. Note that here single particle resistance is expected to decrease as the bias current is stepped through points B, C, D, E, F, and G. The data show that in this biasing configuration the voltage signals decay with time for most bias currents, except for $I < I_{T-}$ where very little or no transient response was observed on timescale of minutes. As we step the current to near $I_{T-}$, we begin observing very slow relaxation with time scales exceeding our measurement time of three minutes (lightest blue curves). Above $I_{T-}$ the evolution of transients becomes very clear and the amplitude of transients, $\Delta V$ (defined as the maximum amplitude swing of the transient in the time observed), increases smoothly until bias current reaches $I_{T+}$. Between $I_{T-} < I < I_{T+}$ relaxation speeds up from decay times of many minutes (measurement is limited by the 2-minute dwell time at each current bias) to several seconds at $I_{T+}$. Above $I_{T+}$ decay times quickly become short, below 1 s and beyond the measurement sampling rate of 0.36 s per data point for this data set. As we step the bias current above $I_{T+}$, the amplitude of transients $\Delta V$ seems to evolve in a seemingly disordered, jumpy fashion, with a general trend of decreasing as bias current approaches $I_T^*$. Above $I_T^*$ the decay remains fast, but $\Delta V$ seems to stabilize close to the values observed below $I_{T-}$.

The same qualitative behavior is observed from data on the thinned segment of the whisker (sample Gold12b) as shown in **Figure 4.4.** The data again correspond to a measurement configuration shown in **Figure 4.2 (c)** (i.e. probing the same polarity of the CDW polarization), but with different values of $L_{v1}$, $L_v$, and $L_{v2}$. The most noticeable qualitative



difference between the two data sets is that in the latter, a more noticeable decay was observed for $I < I_{T-}$ (note that in the latter set we did not measure transients for I < 400 μA).

For a measurement configuration in **Figure 4.2 (b)** which probes the opposite polarity of the CDW polarization (with voltage probes near the opposite half of the sample), single particle resistance is expected to *increases* as the bias current is increased through points B, C, D, E, etc. Transient $V(t)$ data of sample Gold12a taken in this configuration is shown in **Figure 4.5**. In this case the measured voltage signal *increases* with time at most bias currents. The qualitative evolution of transients $V(t)$ at different bias points B, C, D, E, etc., relative to bias points $I_{T-}$ and $I_{T+}$ is again consistent with what is observed in the preceding two data sets, except that in this case, as expected, $V(t)$ *increases* with time after each current switch confirming that the measured transients are indeed related to changes in local single particle resistivity coupled to CDW dynamics.

When no pulses of opposite polarity were applied between different bias points (i.e. current was stepped through B, C, D, E, F, G without reversing bias to A), we observed no transients on timescale of minutes with our setup. Comparatively very little or no transients were observed when we applied polarization pulses in a way that current switch never swung through $I = 0$. That is, we had to induce a *reversal in bias current direction* in order to observe significant transients on timescale of minutes, consistent with previously observed pulse-sign memory effects.[116,125]

To focus more closely on the transient evolution between $I_{T-} < I < I_{T+}$ we have re-plotted the data set in **Figure 4.3** so that the panel in **(c)** is displayed with the *x*-axis on a log plot in **Figure 4.6** (note that the data appears truncated at short times because the data point



corresponding to zero minutes cannot be plotted on this log scale). Plotted in this way, the data reveals the speed-up in relaxation for biases $I_{T-} < I < I_{T+}$. It also shows that the relaxation has a non-simple form. Re-plotting of data on a standard semi-log and log-log plots (not shown) revealed that the relaxation does not follow simple exponential or power law decays. It appears that more than one relaxation mechanism, and more than one characteristic relaxation time, may be in play in this regime. We defer the more detailed analysis and attempts to fit the data with various functional forms to a future investigation as it is obvious after these preliminary measurements that these experiments need to be repeated to 1) expand trace measurement times to access longer relaxation times, and 2) increase sampling rates to access timescales much shorter than a few seconds. One could then try fitting different functional forms that were observed e.g. the stretched exponential form in the x-ray scattering experiments probing CDW relaxation in the high temperature regime (see for example references [120,123]). Also see the experiments probing relaxation by transient transport measurements similar to ours by Gill,[116] Fleming,[125] and Adelman *et al*.[83]



**Figure 4.6**

Same data as in **Figure 4.3** but with panel **(c)** replotted with the *x*-axis on a log plot.  Note that in this plot, the first data point for each trace, corresponding to zero minutes, cannot be displayed, i.e. log(0).  As a result, the data is truncated, and it appears that each trace begins the decay at a different voltage on the *y*-axis.  This is because zero minutes on this graph corresponds to a point in space that is to the left of the page (i.e. at -∞).



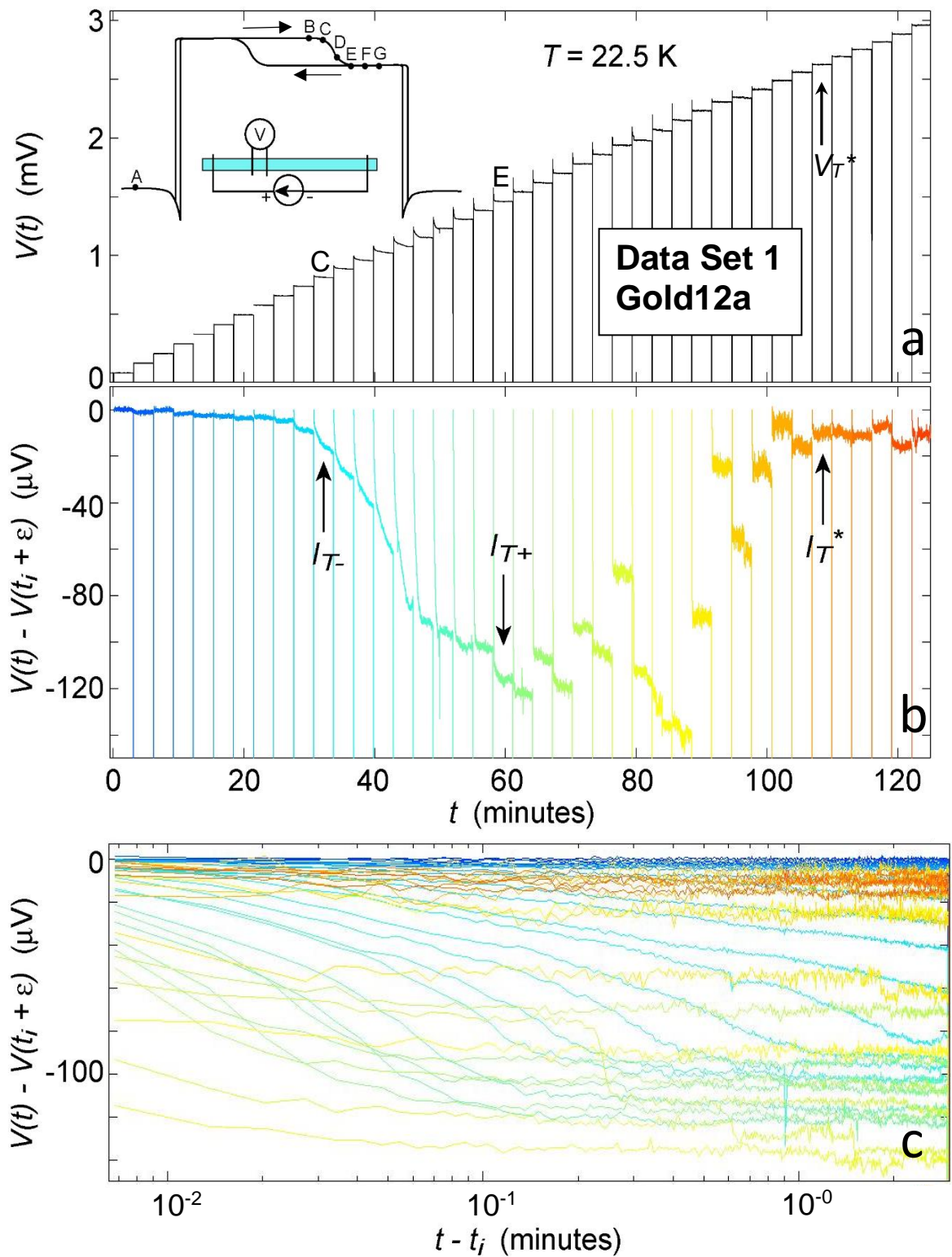



**Table 4.1** lists relevant parameters for each of the three measured data sets $V(t)$. $\Delta R$ is the height of the resistance hysteresis loop in differential resistance vs. current curves (measured separately) as shown in **Figure 4.2 (b)** and **(c)**. The values of $\Delta R$ listed in the table are negative if $R_S$ decreases as we sweep through points B, C, D, E, F, and G in **Figure 4.2**, and as positive if $R_S$ increases. $I_{T-}$ and $I_{T+}$ are the values of currents where the hysteresis loop begins to close and where it closes, respectively, in measured sweeps of $dV/dI$ vs. $I$. $I_T^*$ is the current where the switching threshold is observed in the same measurement. The value of $\Delta V_+$ reported here is the transient amplitude of the signal obtained when the current is switched to a bias closest to $I_{T+}$, and this value is typically very close to the maximum transient amplitude observed in a data set. $L_{v1}$, $L_v$, and $L_{v2}$ specify geometry of the biasing configuration and are defined in **Figure 3.3** in chapter 3.

**Table 4.1 Parameteters of transient data sets**

| V(t) data set: | 1 | 2 | 3 |
|---|---|---|---|
| **Sample:** | **Gold12a** | **Gold12b** | **Gold12a** |
| $L_{v1}/L_v/L_{v2}$ **(μm)** | 150/50/350 | 50/150/350 | 350/150/50 |
| $\Delta R$ **(Ω)** | -0.098 | -0.243, -0.490* | 0.514 |
| $I_{T-}$ **(μA)** | 519 | 775 | 510 |
| $I_{T+}$ **(μA)** | 931 | 981 | 939 |
| $I_T^*$ **(μA)** | 1750 | 1594 | 1748 |
| $(\Delta R)(I_{T+})$ **(μV)** | -91 | -238, -481* | 483 |
| $\Delta V_+$ **(μV)** | -118 | -493 | 452 |

*Multiple sweeps of $dV/dI$ vs. $I$ revealed that in this sample $R_S$ is bi-stable for $I_{T-} > I > I_{T+}$ and produces resistance hysteresis loops that can have two different heights $\Delta R$, depending on the sweep. Here we report both values we observed.

**Table 4.1** shows that $\Delta V_+$ agrees with $(\Delta R)(I)$ for $I = I_{T+}$ in both sign and amplitude to within 26%, 2.4%, 6.6% for the sample sets 1, 2, and 3 respectively. This implies that the maximum transient amplitude for each set should be observed at $I = I_{T+}$ (as is the case), and this can be



explained as follows:  If $I_{T+}$ is a true threshold for collective transport, then no phase slip occurs for biases below $I_{T+}$.  For biases between $I_{T-}$ and $I_{T+}$ the strain grows as the CDW displaces from its initial polarized configurations in response to the driving force, but phase slip has not started yet.  At $I_{T+}$ the strain is the largest (and so is the relative change in strain) and phase slip begins.  At this point the strain is reduced by phase slip, and transient voltage swings should become smaller (which is what we observe in transients above $I_{T+}$).  Sudden change in shapes of transients near $I_{T+}$ where slow relaxation tails suddenly flatten out (and $V(t)$ signal becomes noisier, see for example **Figure 4.3 (c)** and **4.6(c)**), is where the CDW has reached steady state creep.  The bias where transients develop (at $I_{T-}$) is where creep dynamics begin to facilitate polarization change (i.e., the CDW phase starts to overcome local barriers and polarization rearrangement begins), but phase slip near contacts has not yet started.  Then, when the strain is large enough, phase slip starts and $\Delta V$ begins to shrink above $I_{T+}$.  The jumpy evolution of $\Delta V$ with increase of bias above $I_{T+}$ could be indicating that the CDW can sample a *distribution* of available metastable states while settling into a steady state creep.

During steady state coherent creep, the CDW must overcome large barriers for collective pinning which are orders of magnitude larger than local pinning barriers, and the collective transport does not start until $I_{T+}$.  But for $I_{T-} < I < I_{T+}$, the CDW phase can rearrange locally at some pinning centers (although very slowly because the microscopic processes that facilitate this are thermally and field activated resulting in very slow transient responses) before steady state motion begins.  $I_{T-}$, $I_{T+}$, and $I_T{}^*$ separate each $V(t)$ data set into four regimes with distinct



transient behaviors which are consistent with this picture (which is also consistent with the conclusions of the previous chapter), although additional work could provide further details.

We note that the relaxation times we observed of up to several seconds for biases switching to above $I_{T+}$ are much longer than what has been previously observed in NbSe$_3$ in experiments involving uni-polar or bi-polar current pulses to above collective pinning threshold $I_T$. At higher temperatures between 34 – 46 K, typical observed relaxation timescales are between 1 - 50 μs.[116,125] For the upper ($T_{P1}$) CDW, relaxation times as slow as 1 s have been observed,[120] but more typical values for temperatures between 70-110 K are from milliseconds to several hundred milliseconds.[120-123] It is remarkable that collective coherent transport can persist through dynamical processes that span such a wide range of timescales from microseconds to seconds.

## The Effects of Slow Dynamics between $I_{T-}$ and $I_{T+}$ on Low-Field Hysteresis in Differential Resistance Measurements

Here we show that the effects of very slow dynamics between $I_{T-}$ and $I_{T+}$ can easily be observed in differential resistance vs. bias current measurements when bias sweep rates are not slow, and the measurement is not probing steady state dynamics. For example, when the bias current is swept too fast, for biases between $I_{T-}$ and $I_{T+}$ the CDW does not have time to relax and reach the steady state configuration at each bias point by the time the measurement is performed. This will result in a corresponding measured value of $R_S$ that is higher or lower (depending on whether we are probing a stretched or compressed side of the polarized CDW) than the steady state value of $R_S$ for that bias. Fast current sweep rates can thus significantly distort the shape of the hysteresis loop and push the point where the loop closes, $I_{T+}$, to higher



absolute values of bias currents.  This is easily observable with typical sweep rates used to measure differential resistance curves.

**Figure 4.7 (a)** shows a part of a measured $dV/dI$ vs. $E$ curve obtained from a sample that was probed in a measurement configuration shown in the inset of part **(b)**.  The region in the red circle in **(b)** shows schematically the part of the curve displayed in **(a)**.  To obtain the data, bias current was swept to well above the switching threshold $I_T^*$ in both directions at different rates indicated by the different colors lines and with rates listed in the inset of the figure. Current was then mapped to the electric field on the $x$-axis through a separate $I$-$V$ measurement.  Experimentally to implemented different sweep rates here, for different sweeps we stepped current bias in different size increments while keeping the dwell time at each current point the same for all the sweeps (it was experimentally easier to implement it this way).  Hence there are more data points in curves corresponding to slow



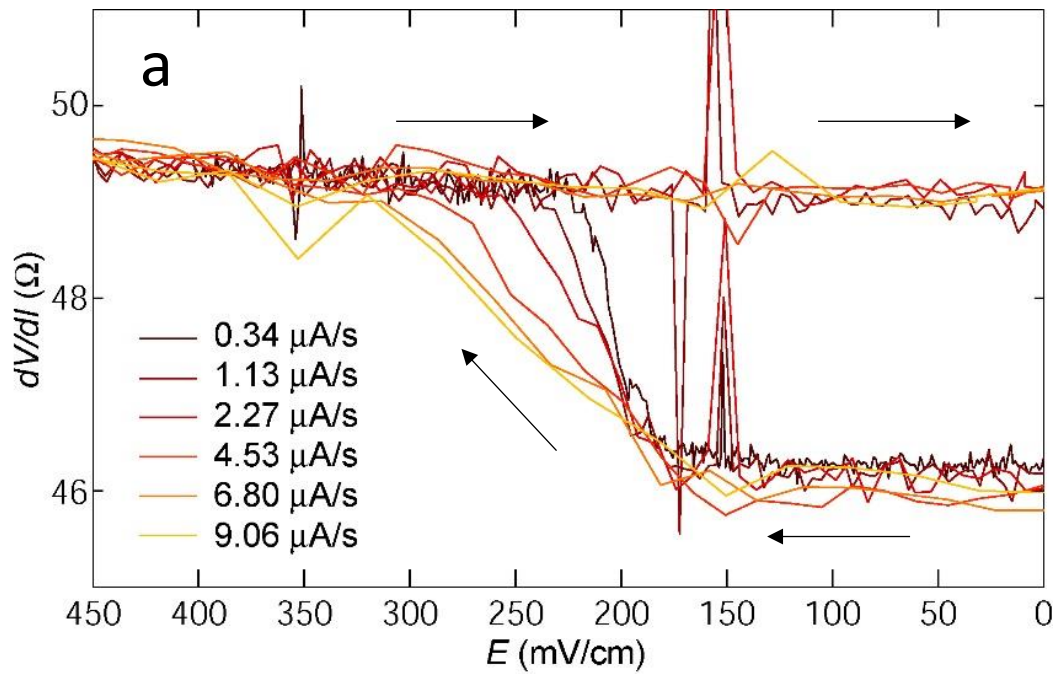

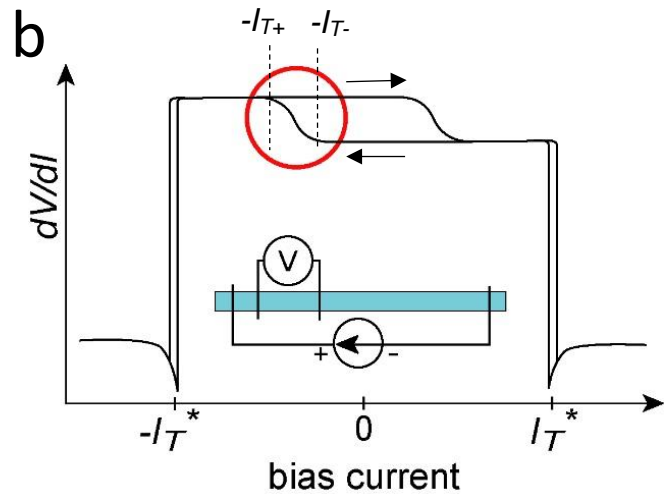

**Figure 4.7**

(a) A section of a measured *dV/dI* vs. *E* curve corresponding to the area indicated by the red circle in (b). For a fast current sweep, closing of the hysteresis loop is "delayed" until larger absolute biases are reached (the glitches in data at 150 mV/cm are artifacts of the mode-locking amplifier setup used). (b) Schematic of a *dV/dI* vs. *I* curve and a corresponding measurement configuration in which data in (a) were taken.



sweeps than in the ones with fast sweeps as can be seen in **Figure 4.7 (a)**. The sweep with the slowest rate (current stepping at 0.34 μA/s, dark brown lines) shows the hysteresis loop closing at $E_{T+}$ = 220 mV/cm while the fast sweep rate of 9.06 μA/s (light orange line) closes the loop at $E_{T+}$ = 330 mV/cm, a significantly larger absolute value of electric field.

**Figure 4.8** shows the dependence of $E_{T+}$ on current sweep rate.

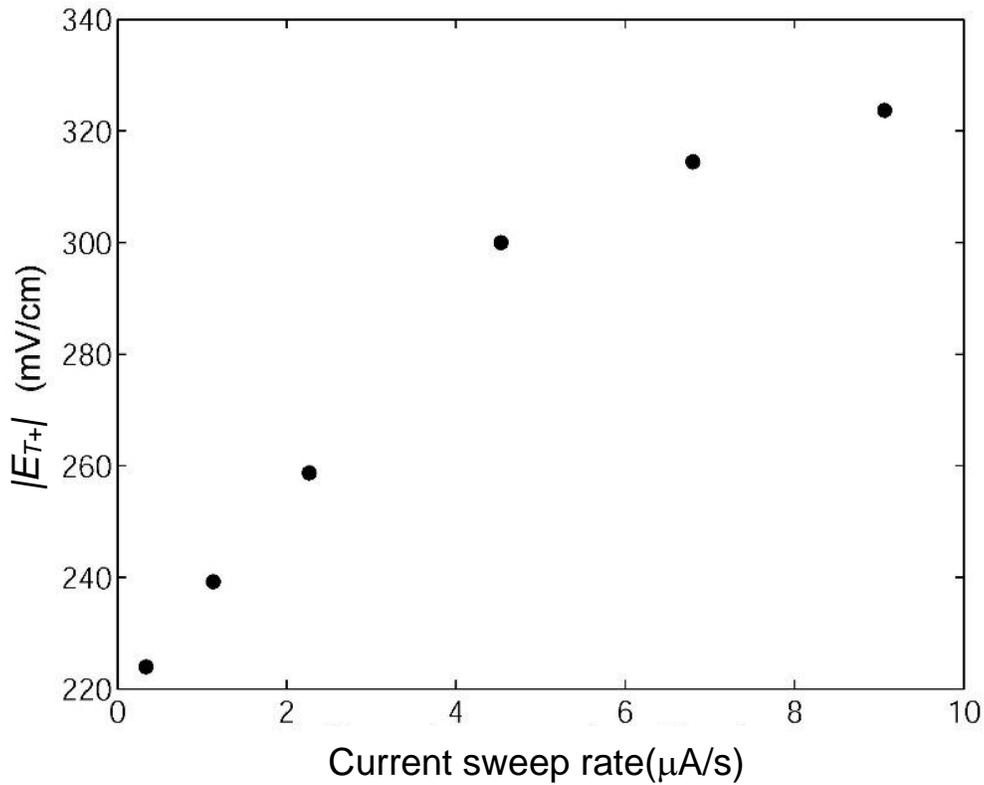

**Figure 4.8**

Electric field where low-field hysteresis loop closes, $|E_{T+}|$, in differential resistance vs. current measurements and its dependence on current sweep rate.

Lemay *et al.*[43,66] showed that $I_{T+}$, the current where the low-field hysteresis loop closes in differential resistance vs. current measurements, can be identified with the collective threshold $I_T$ and the onset of temporally-ordered collective creep observed in the low



temperature regime. But the slow dynamics between $I_{T-}$ and $I_{T+}$ can complicate $dV/dI$ vs. $I$ measurements, and as we have shown, some care must be taken to ensure the measurement is probing steady state dynamics in order to properly determine the value of $I_{T+}$. Although we were aware of this effect when we measured $dV/dI$ vs. $I$ curves corresponding to $V(t)$ data sets 1-3, and all our $dV/dI$ vs. $I$ measurements are generally performed at conservatively slow sweep rates, the full extent of the characterization presented in **Figure 4.7** and **4.8** had not yet been performed. In retrospect the sweep rates we used were likely not sufficiently slow, so that the reported values of $I_{T+}$ were slightly inflated compared to the values we would get when probing steady state behavior. We estimate that this systematic error is small and the reported values of $I_{T+}$ are at most 10% above the steady state values (i.e. the steady state value of $I_{T+}$ is within one or two neighboring V(t) bins of the bin labeled as $I_{T+}$ in part **(b)** of **Figure 4.3, 4.4**, **4.5,** and **4.6**).

## Suggestions to Extend Current Work

- Explore temperature dependence of $V(t)$ relaxation.

- Perform spatially resolved $V(t, x)$ measurements where $x$ is a distance along a sample between current contacts.

- Expand measurement times and increase sampling rates to capture longer-than-minutes and shorter-than-seconds relaxation times.

- Repeat measurements of same $V(t)$ sets to see if the observed transient amplitudes, $\Delta V$, are reproducible for a given sample, especially between $I_{T+}$.and $I_T^*$.



- Future studies should focus on looking more carefully at functional forms of relaxation and comparing them to forms observed at high temperatures.

## 4.4   Summary

We measured CDW voltage transients in the low-temperature regime of NbSe$_3$ in response to sudden reversal of bias-current direction. The transients are observed when the CDW, compressed near one end of the sample and stretched near the other, relaxes into oppositely polarized configurations induced by a sudden swing in the current drive. The initial polarization state was prepared with a fixed polarizing current $|I_A| > I_T^*$. The current was then swung through zero to some bias current in the opposite direction. The cycle was then repeated many times with the final bias progressively stepping through different values from 0 to well above $|I_T^*|$. Many detailed checks need to be completed carefully before final conclusions should be drawn from these measurements, but we can summarize some basic findings and ideas that emerged.

We observed significant memory effects (transient with large amplitudes) for currents switching to final biases between $I_{T-}$ and $I_T^*$. The largest voltage transients were observed following a switch to a bias current at or just above $I_{T+}$. As the bias current switching is stepped toward $I_T^*$, $\Delta V$ gradually decreases, but in somewhat disordered and jumpy manner, towards $\Delta V$ values observed when the bias is switched to currents below $I_{T-}$.

The relaxation following current switching to biases between $I_{T-}$ and $I_{T+}$ is extremely slow, and changes from a time scale of minutes or longer at $I_{T-}$ to a few seconds at $I_{T+}$. Current



switching to biases above $I_{T+}$ produces relaxation times that quickly become shorter than a fraction of a second. The slow relaxation dynamics for $I_{T-} < I < I_{T+}$ can be attributed to the slow settling of the CDW into the creep state where the dynamics are orders of magnitude slower than in the sliding state. At $I_{T+}$ the observed relaxation times of a few seconds are consistent with the dynamical time scale at the onset of temporally-ordered collective creep in the slow branch at frequencies of ~0.5 Hz reported for these samples in chapter 3 (see **Figure 3.14**). This indicates that the same dynamics is responsible for the changes in the CDW transient response near $I_{T+}$ and for the onset of temporally-ordered creep. The picture that emerges from these time domain experiments supports the claim that $I_{T+}$ can be identified as the true threshold below which no phase slip and steady state collective transport occur, i.e. $I_{T+} = I_T$.

The effects of the sluggish dynamics for $I_{T-} < I < I_{T+}$ can be observed in differential resistance vs. bias current measurements. The current where the low-field hysteresis closes, $I_{T+}$, can be observed to depend on bias sweep rate. In the low temperature regime, these long relaxation times (many minutes to seconds) should be taken into consideration, and care should be taken when measuring differential resistance vs. current to allow the system to relax into a steady state at each bias point before performing the measurement.

The microscopic mechanisms that facilitate steady state creep above the collective threshold $I_T (= I_{T+})$ must also facilitate CDW relaxation into a new polarized state for biases between $I_{T-}$ and $I_{T+}$, i.e. relaxation in this slow regime should be limited by the same dynamics that limits the creep. Thus, the CDW displacement timescales associated with voltage transients we observed for $I_{T-} < I < I_{T+}$ must be linked to timescales to overcome barriers responsible for creep. In addition to length- and energy-scales provided by the experiments in chapter 3 that



set the dynamics for temporally-ordered coherent creep, the relaxation experiments in this chapter provide a rough timescale that can be associated with creep over local barriers on the order of many minutes to seconds.

Transient response measurements in the low temperature regime are a good approach (both in terms of simplicity and experimentally accessible timescales) to further explore microscopic dynamics of these systems in this still poorly understood regime.



# 5 New Phase Diagram of CDW Transport

In this chapter we briefly summarize our results and show how they fit within a broader context of the CDW transport by incorporating our findings into a consistent picture of the CDW transport phase diagram.

In chapter 1 through a survey of existing literature, we introduced common features of CDW transport. Experiments show that under external drive, many different CDW conductors, including the partially gapped $NbSe_3$, exhibit very similar behavior. We pointed out that a significant discord exists between experiment and theory, and that the evolution of the basic *I-V* characteristic with temperature is not well understood. Discovery of temporally-ordered collective creep in the slow branch of the $NbSe_3$ phase diagram confirmed this further, but it also provided means to study this regime of the phase diagram in more detail.

Chapter 2 addresses sample preparation techniques. Here we presented results of extensive characterization and testing that allowed us to overcome technical challenges of adopting bulk $NbSe_3$ whiskers to the existing thin-film processing technology at micro- and nano-scale. We developed techniques to thin and pattern bulk whiskers and devised simple etch-tests that can reveal structural defects in the bulk. We presented new and improved techniques to electrically contact whiskers. The information in this chapter should prove useful to anyone tackling $NbSe_3$ sample preparation with modern microfabrication techniques in the future.



Chapter 3 focuses on low temperature collective transport in the slow branch ($E_T < E < E_T^*$) of the CDW phase diagram. By measuring the frequency of coherent oscillations of the temporally-ordered creep at different electric fields, we have mapped the velocity-field relation in nine high-quality NbSe$_3$ samples, pure or doped by isoelectronic tantalum or non-isoelectronic titanium impurities. We find that for $E_T < E < E_T^*$ the velocity-field relation in all samples is well described by the modified Anderson-Kim form, showing thermally activated dependence on temperature and approximately exponential dependence on electric field. While the persistence of coherent oscillations and the onset of motion at the collective pinning threshold indicate that the collective dynamics are important, the fit parameters reveal that another, much smaller length and energy scale are relevant at low temperatures. Energy scale ~$\Delta_C$ points to a single-particle-like excitation, while the volume-per-barrier $1/n_{barr}$ << FLR domain indicates local nature of the barriers. One hypothesis is that the barriers are associated with phase advance past local strong pinning defects. Curiously, the barrier concentration obtained from fits to experiment is independent of the concentration of intentionally added impurities; in all samples, doped and pure, it roughly corresponds to the estimated residual impurity concentration in pure samples. This suggests either that the strong local defects are not associated with added impurities, or that there is some other excitation that limits dissipation in the slow branch. The velocity at which CDW switches from coherent creep to coherent sliding at $E_T^*$ is thermally activated with activation barriers ~$\Delta_C$, again pointing to a local rather than collective process.

Chapter 4 describes preliminary experiments to probe CDW relaxation associated with the dynamics in the slow branch of the NbSe$_3$ phase diagram. Long relaxation timescales of



minutes are consistent with the slow creep dynamics. We also observe signatures of metastable configurations of steady-state creep.

By incorporating the new results presented in this dissertation with the ideas based on the existing work on CDW transport, we summarize and offer the following explanation of the CDW phase diagram:

**High Temperature Regime.** At high temperatures the behavior is defined by collective elastic dynamics, and the observed properties (e.g., $E_T$, form of $I$-$V$ relation, form of AC conductivity, X-ray determined CDW structure, etc.) are qualitatively well described by the FLR model. Characteristic length scale $L_\phi$ is defined by large phase coherent domains containing many impurities ($L_\phi \sim$ micrometers in NbSe$_3$). The FLR model predicts that in absence of thermal disorder ($T = 0$ limit), the CDW is pinned below the collective pinning threshold $E_T$. In experiments at finite temperatures, thermal disorder modifies this scenario in that it allows the CDW to creep below $E_T$, and theory predicts that a few diverging barriers in the distribution produce extremely slow, glassy dynamics characterized by thermally assisted incoherent creep. However, unlike in, e.g., vortex lattices, the energy barriers for collective motion are huge in NbSe$_3$ and this incoherent thermal creep is negligible except very close to $T_P$. Above $E_T$, the CDW slides exhibiting coherent oscillations with $f_{NBN} \propto j_C \propto v_C$. These oscillations are expected if the collective dynamics are elastic. The oscillations are extremely coherent even just above $E_T$, suggesting that there is no regime where plasticity plays an important role. At high fields for $E \gg E_T$, the CDW velocity is weakly temperature and field dependent. The depinning field $E_T$ is determined by collective pinning of the FLR domains by impurities and scales with $n_i^2$ in bulk crystals, as predicted. $E_T$ dependence on temperature



is determined by elastic constants which depend on interaction of the CDW with single particle carriers. In fully gapped materials like blue bronze and $(TaSe_4)_2I$, as $T \rightarrow 0$ and quasiparticles freeze out, screening of CDW fluctuations by quasiparticles becomes less and less effective, the elastic constants stiffen, and $E_T$ is expected to decrease towards zero (as is observed experimentally). In partially gapped $NbSe_3$ (and m-$TaS_3$), the single particle density remains large and saturates at low $T$, and the single particle conductivity increases with decrease of temperature, which aids screening of CDW fluctuations and softens the elastic constants, thus causing $E_T$ to increase (again consistent with experiment) as illustrated in **Figure 5.1**. Thus, the FLR model of collective pinning and elastic dynamics provides a consistent picture of many aspects of CDW transport, especially in the high temperature regime.

**Low Temperature Regime.** However, models of CDW dynamics based upon FLR model fail completely to describe the essential qualitative behavior at modestly low ($T < 2T_P/3$) temperatures. Our experiments suggest the reason: in addition to the length and energy scale associated with large collective FLR domains and collective pinning, a second (much smaller) length and energy scale emerges at low temperatures that is more characteristic of local or single particle excitations. This energy scale $\sim \Delta_C$ is comparable to barriers predicted to be associated with local pinning by strong pinning centers that fix the CDW's phase at their position. The length scale extracted based upon an Anderson-Kim-like model is defined by volume per barrier $1/n_{barr} \ll$ FLR domain volume, suggesting local nature of the pinning centers. Thus, as the temperature is decreased, local strong pinning centers become important in shaping of dynamics. Glassy dynamics and incoherent thermal creep are still in principle



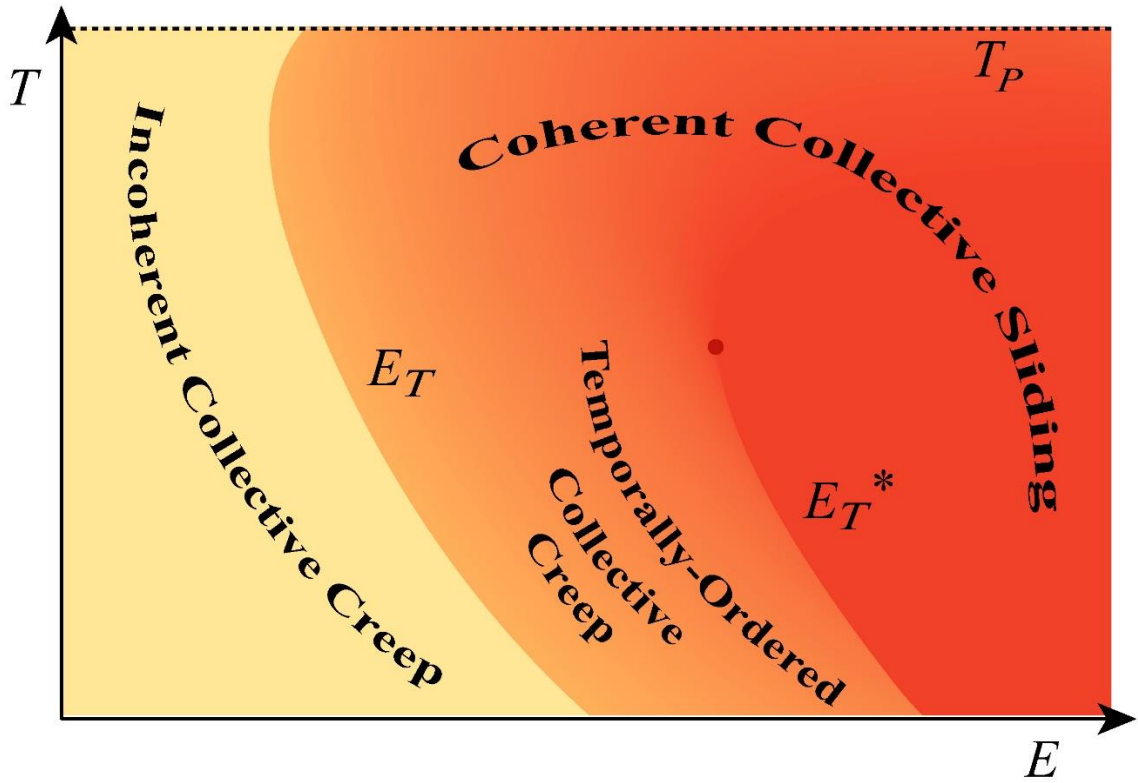

**Figure 5.1**

Phase diagram of CDW transport.

observed below collective pinning threshold $E_T$, but in practice the collective pinning barriers are so large that essentially no motion occurs. At $E_T$ the FLR domains can finally start to move, and collective motion can begin. However, at low temperatures the local strong pinning barriers limit the motion, producing creep-like dynamics: extremely slow CDW velocities with thermally activated behavior. But these local, thermally activated and presumably random events do not destroy the coherent elastic collective dynamics on large length scale. Rather, the retardation of motion by many local barriers within the large FLR domain amounts to an overall average kinetic-friction-like force. Overall, the motion is still temporally ordered, coherent oscillations persist, and we can still observe narrow-band noise



in this regime. With increase of electric field, the CDW washboard frequency, $f_{NBN}$, eventually reaches a critical frequency at which coherent creep abruptly gives way to coherent sliding at velocities/frequencies that are orders of magnitude larger. This first order dynamic transition is characterized by a thermally activated threshold CDW velocity $v_C^*$ (rather than by threshold field $E_T^*$). Above $E_T^*$ the motion is again described by collective elastic dynamics much like the motion observed in the high temperature regime above $E_T$.

One possible explanation for this transition is that at CDW velocity $v_C^*$ the washboard frequency becomes comparable to activation frequency over strong pinning barriers. At this point the retardation of motion by the barriers becomes ineffective, and at $E_T^*$ a first-order-like (discontinuous and hysteretic) dynamic transition occurs where coherent creep switches into coherent sliding. A detailed theoretical framework explaining this transition is still missing and many questions remain unresolved: Is the transition caused by purely local effect (as barrier-crossing argument suggests), or are elastic collective dynamics important? What are the potential barriers relevant to the transition? Are defects important, or is there some intrinsic excitation that is key to CDW motion (e.g., soliton excitation)?

With the lack of a comprehensive picture, we offer in **Figure 5.2** a cartoon depiction of CDW dynamics based on the simple "particle-in-a-washboard" concept. This simplistic picture is likely far from what is microscopically actually taking place, but it is useful in capturing ingredients that appear important in shaping of the phase diagram: elastic degrees of freedom, periodic potential, barriers at two length and energy scales, thermal fluctuations, disorder, and a driving force provided by the potential tilt.



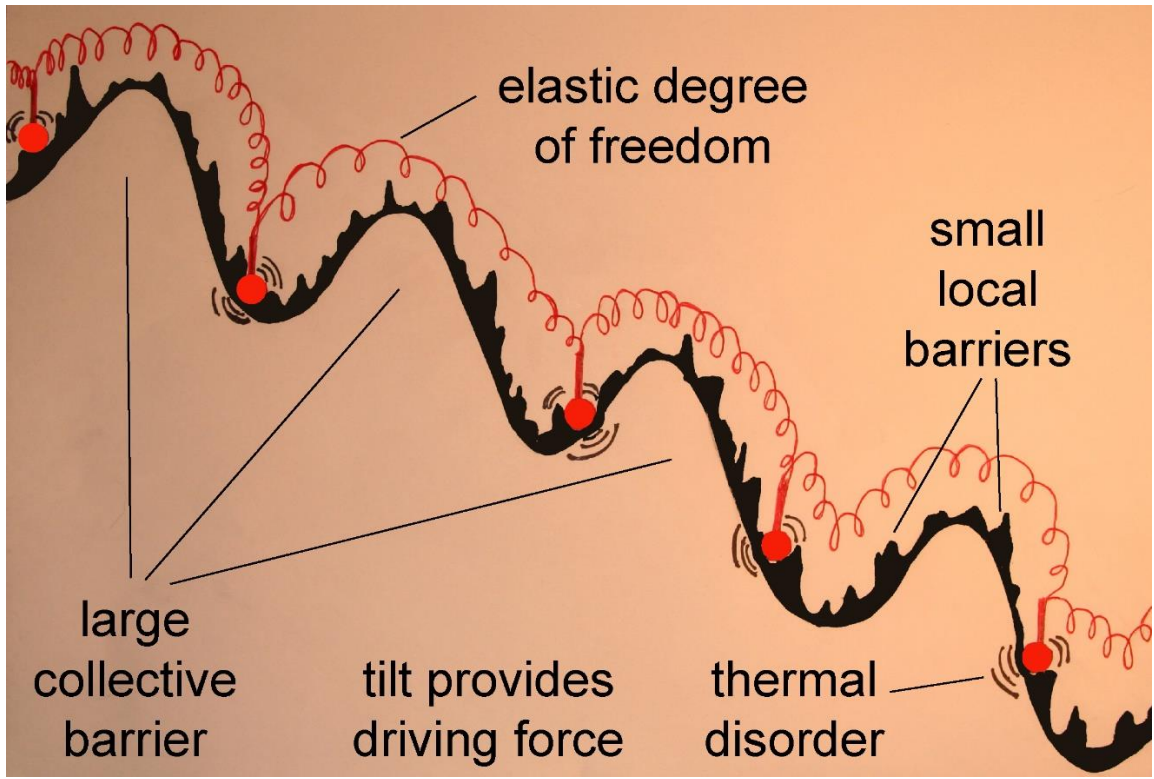

**Figure 5.2**

Cartoon of CDW dynamics within an FLR domain depicting CDW as an extended elastic medium in a periodic potential tilted by a driving force. CDW motion is limited by two types of barriers: large collective barrier posed by periodic potential extended over the length-scale of the elastic medium (within an FLR domain), and small local barriers. Thermal disorder "jiggles" CDW, and aids CDW advancement past the local barriers.

We showed that collective and local dynamics simultaneously governs transport. The dynamics on these two length/energy scales can coexist (i.e. local dynamics of creep does not destroy coherent oscillations in slow branch) because they are vastly different in size. Our findings strongly suggest that the assumption that only one length/energy scale is important is a major omission of theoretical models which could explain their failure to predict the complex CDW dynamics observed in experiments. We hope that the discussion of the CDW phase diagram we present here will aid future theoretical efforts in producing a clear



microscopic picture of dynamics that captures the common features of collective transport in these systems.



APPENDIX A

# Order Parameter Phase in CDWs and Superconductors: A Comparison

The following is a part of the A-examination work[10] developed by the author to fulfill a partial requirement for a Ph.D. degree in physics at Cornell University (see next page).






K. Cicak, K. O'Neill, and R. E. Thorne

Physics Department, Cornell University, Ithaca, NY 14853-2501, USA



**Abstract.** Both the superconducting and charge-density wave (CDW) states have complex order parameters characterized by a magnitude and phase. The special relation of the superconducting phase to the many-body number eigenstates results in the Josephson effect and many other widely studied properties. In this brief review we discuss the role of the nature of the CDW ground state and the role of the CDW phase in an analogous context.


## 1. INTRODUCTION

Although the BCS superconducting and CDW ground states describe different physical systems, the mathematical apparatus that describes them is extremely similar. Both are coherent quantum ground states, and can be explained as instabilities of the Fermi sea due to electron-phonon interactions. In both cases, theoretical analysis starts from the Frohlich Hamiltonian [1,2]:

$$H = \underbrace{\sum_{k,\sigma} \epsilon_{k,\sigma}\, c^{+}_{k,\sigma} c_{k,\sigma}}_{\text{electron term}} + \underbrace{\sum_{q} \hbar\omega_q\, b^{+}_{q} b_{q}}_{\text{phonon term}} + \underbrace{\sum_{k,q,\sigma} g_{k,q}\, c^{+}_{k+q,\sigma} c_{k,\sigma}\, (b^{+}_{-q} + b_{q})}_{\text{electron-phonon interaction}} \tag{1}$$

Where $\epsilon_{k,\sigma}$ is the (renormalized) energy of an electron with momentum $k$ and spin $\sigma$, $c^{+}_{k,\sigma}$ and $c_{k,\sigma}$ are electron creation and annihilation operators, $\hbar\omega_q$ is the energy of a phonon of momentum $q$, $b^{+}_{q}$ and $b_{q}$ are phonon creation and annihilation operators, and $g_{k,q}$ is the electron-phonon coupling constant. The last term in (1) describes the electron-phonon scattering: a phonon in mode $q$ (-$q$) is annihilated (created) when an electron is scattered from a state with momentum $k$ into a state with momentum $k+q$. The 3D vector spaces of $k$'s and $q$'s are reduced to 1D in the CDW case.

A CDW is a periodic modulation of electron charge at twice the Fermi wavevector, $2k_F$, coupled to a corresponding ion-lattice modulation ("frozen" $2k_F$-phonon). Consequently, the CDW ground state can be expressed in two equivalent representations: one in terms of $2k_F$-phonons, the other electron-hole pairs. Table 1 compares the superconducting (BCS) ground state to the corresponding CDW ground state in these representations [1,2].

**Table 1**

| Superconducting State | CDW State (2 equivalent representations) | |
|---|---|---|
| **Cooper Pairs:** Electrons over-screen ions, producing an effective attractive potential and pairing between electrons with opposite spin and momentum. These pairs condense into a macroscopically occupied state: | **$2k_F$-Phonons:** Electron-phonon scattering in 1D produces a macroscopically occupied phonon state with wavevector $2k_F$: | **Electron-Hole Pairs:** An electron of momentum k is scattered by a $2k_F$-phonon, increasing its momentum by $2k_F$ and leaving behind a hole of momentum k-$2k_F$: |
| $\| \psi_S \rangle = \prod_{k=k_1,k_2,\ldots,k_M} (u_k + e^{i\phi_S} v_k c^{+}_{k\uparrow} c^{+}_{-k\downarrow}) \| 0 \rangle$ | $\| \psi_{cdw,\,ph} \rangle = e^{-\|\alpha\|^2} \sum_{n=1,\ldots,\infty} \left( \frac{\alpha^n}{\sqrt{n!}} (b^{+}_{2k_F})^n + \frac{(\alpha^*)^n}{\sqrt{n!}} (b_{-2k_F})^n \right) \| 0 \rangle$ | $\| \psi_{cdw,\,e\text{-}h} \rangle = \prod_{k=k_1,k_2,\ldots,k_M} (v_k c^{+}_{k} + e^{-i\phi_{cdw}}\, u_k c^{+}_{k-2k_F}) \| 0 \rangle$ |
| $u_k,\, v_k$ are real and satisfy $u_k + v_k = 1$<br>$\phi_S$ = superconducting phase | $2\alpha = \Delta_o e^{i\phi_{cdw}}$, $\Delta_o$ = one-half of CDW gap<br>$\phi_{cdw}$ = CDW phase | $u_k,\, v_k$ are real and satisfy $u_k + v_k = 1$<br><br>(continued $\rightarrow$) |

| Superconducting State | CDW State (2 equivalent representations) | |
|---|---|---|
| $|\Psi_S\rangle$ has the following form: | $|\Psi_{cdw,ph}\rangle$ has the following form: | $|\Psi_{cdw,e-h}\rangle$ has the following form: |
| $|\psi_S\rangle \propto |0\rangle + e^{i\phi_S}\begin{pmatrix}\text{group of}\\\text{Cooper-pair states}\\\text{each with}\\Q_{tot}=0\end{pmatrix} +$ $e^{2i\phi_S}\begin{pmatrix}\text{group of}\\\text{Cooper-pair states}\\\text{each with}\\Q_{tot}=0\end{pmatrix} +$ $e^{3i\phi_S}\begin{pmatrix}\text{group of}\\\text{Cooper-pair states}\\\text{each with}\\Q_{tot}=0\end{pmatrix} + ...$ | $|\psi_{cdw,ph}\rangle \propto e^{i\phi_{cdw}}\begin{pmatrix}\text{state with 1}\\2k_F\text{-phonon}\\Q_{tot}=-2k_F\end{pmatrix} +$ $e^{-i\phi_{cdw}}\begin{pmatrix}\text{state with 1}\\2k_F\text{-phonon}\\Q_{tot}=-2k_F\end{pmatrix} +$ $e^{2i\phi_{cdw}}\begin{pmatrix}\text{state with 2}\\2k_F\text{-phonons}\\Q_{tot}=-4k_F\end{pmatrix} - e^{-2i\phi_{cdw}}\begin{pmatrix}\text{state with 2}\\2k_F\text{-phonons}\\Q_{tot}=-4k_F\end{pmatrix} +$ $e^{3i\phi_{cdw}}\begin{pmatrix}\text{state with 3}\\2k_F\text{-phonons}\\Q_{tot}=-6k_F\end{pmatrix} + e^{-3i\phi_{cdw}}\begin{pmatrix}\text{state with 3}\\2k_F\text{-phonons}\\Q_{tot}=-6k_F\end{pmatrix} + ...$ | $|\psi_{cdw,e-h}\rangle \propto |0\rangle + e^{i\phi_{cdw}}\begin{pmatrix}\text{group of}\\\text{N-electron states}\\\text{each with}\\Q_{tot}=-2k_F\end{pmatrix} +$ $e^{2i\phi_{cdw}}\begin{pmatrix}\text{group of}\\\text{N-electron states}\\\text{each with}\\Q_{tot}=-4k_F\end{pmatrix} +$ $e^{3i\phi_{cdw}}\begin{pmatrix}\text{group of}\\\text{N-electron states}\\\text{each with}\\Q_{tot}=-6k_F\end{pmatrix} + ...$ |
| where $Q_{tot}$ is total momentum. | where $Q_{tot}$ is total momentum. | where $Q_{tot}$ is total momentum. |
| $|\Psi_S\rangle$ = superposition of states with different Cooper-pair numbers but the same momentum. | $|\Psi_{cdw,ph}\rangle$ = superposition of states with different numbers of $2k_F$-phonons. | $|\Psi_{cdw,e-h}\rangle$ = superposition of states with the same number of electrons but different total momentum. |
| $\Rightarrow |\Psi_S\rangle$ *has an ill-defined Cooper-pair number but a well defined momentum.* $\phi_S$ *connects states with well-defined Cooper-pair numbers in* $|\Psi_S\rangle$. | $\Rightarrow |\Psi_{cdw,ph}\rangle$ *has an ill-defined phonon number and ill-defined momentum.* $\phi_{cdw}$ *connects states with well-defined phonon numbers in* $|\Psi_{cdw,ph}\rangle$. | $\Rightarrow |\Psi_{cdw,e-h}\rangle$ *has a well-defined electron number and ill-defined momentum.* $\phi_{cdw}$ *connects states with well-defined total momentum in* $|\Psi_{cdw,e-h}\rangle$. |
| In this sense $\phi_S$ and $\phi_{cdw}$ are both true quantum-mechanical phases of their corresponding states. In addition $\phi_{cdw}$, unlike $\phi_S$, has a more physical meaning: $\phi_{cdw}$ describes a CDW position relative to the lattice, and appears explicitly in the phenomenological equations of motion. | | |
| **Order Parameter:** Operator $c_{-k\uparrow} c_{k\downarrow}$ destroys a Cooper pair. Since $|\Psi_S\rangle$ is a superposition of states with different number of Cooper pairs, $$\sum_k \langle c_{-k\downarrow} c_{k\uparrow}\rangle_{|\psi_S\rangle} = e^{i\phi_S} \sum_k u_k v_k \propto \Delta_o e^{i\phi_S}$$ $\neq 0$ in general $\Rightarrow$ *good order parameter.* | **Order Parameter:** Operator $b_{2k_F}$ destroys a $2k_F$-phonon. Since $|\Psi_{cdw,ph}\rangle$ is a superposition of states with different number of $2k_F$-phonons, $$\sum_k \langle b_{2k_F}\rangle_{|\psi_{cdw,ph}\rangle} \propto \Delta_o e^{i\phi_{cdw}}$$ $\neq 0$ in general $\Rightarrow$ *good order parameter.* | **Order Parameter:** Operator $c_k c^+_{k-2k_F}$ destroys an electron-hole pair (changes momentum of a state by $2k_F$). Since $|\Psi_{cdw,e-h}\rangle$ is a superposition of states which differ in momentum by $2k_F$, $$\sum_k \langle c_k c^+_{k-2k_F}\rangle_{|\psi_{cdw,e-h}\rangle} = e^{i\phi_{cdw}}\sum_k u_k v_k \propto \Delta_o e^{i\phi_{cdw}}$$ $\neq 0$ in general $\Rightarrow$ *good order parameter.* |
| **Off-Diagonal Long Range Order (ODLRO):** The matrix of the order parameter operator has non-vanishing off-diagonal elements when represented in a basis of particle (i.e. electron-, Cooper pair-) number eigenstates. | **ODLRO:** The matrix of the order parameter operator has non-vanishing off-diagonal elements when represented in a basis of particle (i.e. $2k_F$-phonon-) number eigenstates. | **ODLRO:** The matrix of the order parameter operator has non-vanishing off-diagonal elements when represented in a basis of momentum eigenstates. |
| In superconductors, the special role of $\phi_S$ in connecting states with different number of charged particles in $|\Psi_S\rangle$ and associated ODLRO with respect to electron-number eigenstates leads to the Josephson effect (zero bias electrical current across an SNS junction). The CDW state has no ODLRO with respect to the basis of electron-number eigenstates and zero-bias charge current should not be observed in CDW-N-CDW junctions. An equivalent phenomenon for CDW systems, however, has been predicted: a zero bias momentum current across the junction [3]. | | |

# Etch Rates

**Table B.1**

| Material Etched | Plasma Type | Power (W) | Pressure (mT) | Gas Flow (sccm) | DC bias (V) | Etch Rate** (nm/min) |
|---|---|---|---|---|---|---|
| NbSe$_3$ | SF$_6$ | 10 | 15 | 20 | - | 85 ± 5 (2) |
| | | 20 | 15 | 20 | - | 110 ± 34 (11) |
| | | 20 | 30 | 30 | - | 302 ± 24 (2) |
| | | 30 | 15 | 20 | - | 137 ± 9 (3) |
| | | 45* | 15 | 20 | - | 206 ± 83 (6) |
| | CF$_4$ | 20 | 30 | 30 | ~210 | 29 ± 23 (4) |
| | | 30 | 30 | 30 | - | 12 |
| | | 45 | 30 | 30 | - | 21 |
| | O$_2$ | 90 | 30 | 30 | - | Negligible |
| | Ar Ion Mill*** | | | | | 53 |
| e-Beam Resist NEB-31 | O$_2$ | 90 | 30 | 30 | - | 125 ± 87 (4) |
| | SF$_6$ | 45 | 15 | 20 | - | 56 |
| e-Beam Resist PMMA | O$_2$ | 90 | 30 | 30 | | 594 |
| | SF$_6$ | 20 | 15 | 20 | | 43 ± 12 (5) |
| | CF$_4$ | 20 | 30 | 30 | ~210 | > 28 ± 19 (16) |
| e-Beam Resist UV-5 | O$_2$ | 45 | 30 | 30 | | 122 |
| | | 90 | 30 | 30 | | 245 |
| | SF$_6$ | 45 | 15 | 20 | | 64 |
| Polyimide PI-285 | O$_2$ | 45 | 30 | 30 | - | 118 ± 36 (4) |
| | | 90 | 30 | 30 | - | 260 |
| | SF$_6$ | 20 | 15 | 20 | - | 19 |
| | | 45 | 20 | 20 | - | 38 ± 5 (2) |
| Thermal SiO$_2$ | Ar Ion Mill*** | | | | | 36 |
| | SF$_6$ | 20 | 15 | 20 | | 9 |
| | | 45 | 15 | 20 | | 23 |
| | | 90 | 22 | 20 | | 38 |

*Make sure the whisker is well thermally connected to a substrate and perform etching in multiple short intervals (20-30 seconds each) and allow for cooling of several minutes between intervals. Otherwise, the etching at this high power may result in damaged samples as shown in **Figure 2.12.**



[**]The reported value is the mean and standard deviation of several etch rate values obtained from (a) multiple etch test runs, and/or (b) different etch rates measured on different crystal facets of a single whisker (i.e. differently oriented crystal grains in a multistep whisker sometimes etched at very different rates). The number in parenthesis indicates the total number of etch rate measurements that were averaged. Where parenthesis are missing, the etch rate reported is based on one etch test performed. Note very large uncertainties for some reported values. This is not surprising considering that the etch tests were performed over large timespan of a couple of years, and the conditions of the machine used (Applied Materials RIE system) changed over time with changes in the vacuum chamber environment due to various other materials etched in the machine, occasional repairs resulting in subtle changes in tool configuration, etc. These factors are known to influence etch rates, and it is a good practice to have the etch rates recalibrated from time to time. In addition, the expected variation in the etch rate along different crystallographic axes of a crystal (i.e. differently oriented crystal grains) also contributes to the large uncertainty.

[***]All etch runs to determine the etch rates were performed in Applied Materials reactive ion etcher at Cornell Nanofabrication Facility (CNF) except for Ar ion milling which was performed in the VEECO ion mill at CNF with the following parameters: beam voltage 500 V, beam current 70 mA, supressor voltage 200 V, discharge voltage 40 V, neutralizer current 60 mA, current density 0.481 mA/cm$^2$. The etch rates for ion milling were included for completeness and comparison purposes.



APPENDIX C

# Fabrication Recipes

Use Si wafers which have been previously thermally oxidized containing 0.35 μm - 1.0 μm of SiO$_2$ on the surface. Alumina substrates have also been used as substrates in the past with the POB procedure.

## Recipe 1: Contacting Whiskers by Probes-on-Bottom (POB) Method

### *Make Masks for Optical Lithography*

1. **Create CAD file -** Using CATS software convert a pattern file patternname.dxf into a pattern-generator readable file patternname.dat.

2. **Fabricate a mask (a reticle for optical lithography).**

### *Pattern Metal Probes on Substrate*

3. **Prime Wafer with P-20**
   Static dispense, wait 60 sec, then spin dry. Use 3-inch Si wafers with 1 micron of oxide on them.

4. **Spin Shipley 1813 Resist**
   Static dispense, 500 rpm for 5 sec for coverage, then 4000 rpm for 30 sec

5. **Bake on Hotplate** – 115 C for 2 min.

6. **Expose on 5X g-line stepper** – exposure time 1.25 sec, focus 251, mask patternname

7. **Image reversal** – Bake in YES oven on NH$_3$ (ammonia) reversal process

8. **Flood Exposure** – using HTG contact aligner for 60 sec

9. **Develop Resist** – Soak in 321 MIF for 60 sec. Check, may need extra 30 sec.

10. **Descum Resist Trenches –** Use Branson Barrel Etcher, oxygen plasma, resist descum process.



11. **Ti/Au Evaporation –** Use SC-4500 Evaporator, e-beam evaporate, 50 A of Ti, then 1200 A of Au.

12. **Resist Lift-Off** – Soak in ultrasonic bath in acetone (or Shipley 1165) for 5-30 min until lift-off is done. Rinse with isopropanol when done, before acetone dries on the wafer.

13. **Spin Protective Resist Coat** – Spin Shipley S-1813 at 4000 rpm for 30 sec.

14. **Bake on Hotplate** – 115 C for 2 min.

15. **Cleave Wafer into Substrates.**

16. **Wash Off Protective Coating** – Soak in ultrasonic bath in acetone for 5-30 min until substrates are clean. Rinse with isopropanol when done, before acetone dries on the wafer. Substrates are ready for mounting crystals.

## Mount Crystals on Substrate

17. **Mount NbSe$_3$ Whisker -** Prepare Cu foil by cutting it into 1.5 x 1.5 cm pieces and carve $3 \times 3$ mm windows on each. Using microscope, tweezers, and scissors select a stepless crystal, cut it, and drop it over the isopropanol-wetted Cu window. Isopropanol will act as a glue holding crystal ends fixed to the Cu window. Now you can invert the foil and check the other side of the crystal under the microscope for steps. If the crystal does not have steps on either side, transfer it on the patterned metal probes on the substrate.

18. **Apply Polymer Coat -** Dispense a drop of Ethyl Cellulose dissolved in Ethyl Acetate on top of the whisker resting on the probes. Let it dry. When the solvent evaporates, and if done properly, this encapsulates the whisker over the substrate and pushes it against the metal probes for a good electrical contact. Make sure that you finish all profilometer measurements before this step.
The sample is ready for cooldown.

## Recipe 2: Affixing Whiskers to Substrate with Polyimide

Use Si wafers which have been previously thermally oxidized containing 0.35 μm - 1.0 μm of SiO$_2$ on the surface.

### Apply Polyimide Glue/Insulator

1. **SPIN** to clean
   What:                              Acetone, then Methanol



| | |
|---|---|
| Spin speed: | any |
| Spin time: | 60 sec to dry |
| Thickness: | 0 |

2. **SPIN** Adhesion Promoter

| | |
|---|---|
| What: | QZ3289 : QZ3290 (1:9), freshly mixed within 3 days |
| Spin speed: | 5000 rpm |
| Spin time: | 20 sec to dry |
| Thickness: | - |

3. **BAKE** Adhesion Promoter

| | |
|---|---|
| Bake type: | Hotplate |
| Temperature: | ~50 C |
| Bake time: | 2 min |

4. **SPIN** Polyimide

| | |
|---|---|
| What: | Probromide PI-285, diluted with QZ3288 (1:1) |
| Spin speed: | 6500 rpm |
| Spin time: | 50 sec |
| Thickness: | 61 nm |

*Mount Sample*

5. **MOUNT** NbSe$_3$ sample

| | |
|---|---|
| How: | Prepare Cu foil by cutting it into 1.5 x 1.5 cm pieces and carve 3 x 3 mm windows on each. Using microscope, tweezers, and scissors select stepless crystal, cut it and drop the over the window. Add a drop of isopropanol to glue the crystal ends to the foil. Now you can invert the foil and check the other side of the crystal under the microscope for steps. Then transfer the crystal on the polyimide-covered wafer. Crystals should not have steps on either side. |
| Note: | Polyimide thinner evaporates fast so the next step (baking) should be done within few hours. |

*Cure Polyimide*

6. **BAKE** Polyimide

| | |
|---|---|
| Bake type 1: | TOL Vacuum Oven |
| Temperature 1: | 85 C |
| Bake time 1: | 30 min |
| Pressure: | < 7e-6 torr |
| | |
| Bake type 2: | TOL Vacuum Oven |
| Temperature 2 | 150 C |
| Bake time 2: | 15 min |
| | |
| Bake type 3: | TOL Vacuum Oven |
| Temperature 3: | 230 C (try not to exceed 240 C) |
| Bake time 3: | 15 min |

*Remove Excess Polyimide and/or Thin Crystal*

7. **REMOVE EXCESS POLYIMIDE**

| | |
|---|---|
| Instrument: | **Applied Materials RIE** (reserve 1 h) |
| Gases 1: | $O_2 + CF_4$ |
| Flow1 : | 30 sccm + 3 sccm |
| Pressure 1: | 20 mTorr |
| Power 1: | 90 W |
| Time1: | ~ 30sec max (less is better)  (etch rate = 24-150 nm/min) |
| RF Voltage: | -615 to -630 V |
| NOTE: | This step may be skipped to avoid oxidation of $NbSe_3$ surface (not sure how Ti/Au will stick to the substrate covered by polyimide). |

8. **THIN NbSe$_3$ (if needed to thin the whole whisker; for thinning a patterned section only, see next recipe)**

| | |
|---|---|
| Gas 2: | $CF_4$ |
| Flow 2: | 30 sccm |
| Pressure 2: | 30 mTorr |
| Power 2: | 20 W |
| Time 2: | Etch rate = $29 \pm 23$ nm/min (top oxide layer on $NbSe_3$ etches much slower) |
| RF Voltage: | -200 V |
| | Inspect with a microscope for subsurface steps on the crystal |

# Recipe 3: Thinning or Patterning Whiskers

*Pattern Mask on NbSe$_3$*

Attach a whisker to a substrate using Recipe 2 and then proceed.

1. **SPIN Resist**

| | |
|---|---|
| What: | PMMA - 495, 8 % in Anisole |
| Spin speed: | 1000 rpm |
| Spin time: | 60 sec |
| Thickness: | 900 nm |



Alternative approach: Spin faster in two layers.  Bake first layer before spinning the second.

2. **BAKE** Resist
   Bake type:            Vacuum/$N_2$ background oven
   Temperature:          135 C
   Bake time:            60 min

3. **CONVERT** Patterns
   Username:             CICAK
   CNF Directory:        [.EBMF]
   Source files:         MASK03.DXF, (or DOSE.DXF or GOLD03.dxf)
   CATS library:         MASK03_DB.CLIB (created by READFILE command)
   Clock files:          M3CLK.CCFA (clock file associated with MASK03 file)
   CFLT file:            MASK03_01.CFLT (created by DO command)
   Machine specific file: KC0108M3.ESF (corresponding to MASK03 file, year 01, month 05)
   JOB files:            KC0108M3.JOB (contains reference to KC0108M3.ESF file)
   Note:                 In JOB files D or DM = DOSE, M = MASK, G = GOLD

4. **EXPOSE** Resist
   Instrument:           **Leica/Cambridge EBMF 10.5   (reserve 2 h)**
   Files:                KC0108M3.JOB and KC0108M3.ESF
   Field size:           3.2768 mm
   Beam size:
   Beam Current:         170 nA (or 30 nA – 170 nA with substrate B)
   Dose:                 540 uC/cm$^2$ (or 320 uC/cm$^2$ with substrate B)
   Apperture:            1
   VRU:                  0

5. **DEVELOP** Resist
   Developer:            MIBK in Isopropanol (1:1)
   Time:                 60-120 sec
   Rinse:                Isopropanol

*Etch NbSe$_3$*

6. **DESCUM PMMA FROM TRENCHES**
   Instrument:           **Applied Materials RIE        (reserve 1 h)**
   Gas 1:                $O_2$ (to descum resist)
   Flow1:                30 sccm
   Pressure 1:           30 mTorr
   Power 1:              90 W
   Time1:                6-10 sec (etch rate ~ 100-600 nm/min)
   RF Voltage:           -570 V



Alternative:   Use Branson barrel etcher (1 min process, no heat)

**7. ETCH NbSe$_3$**
Gas 2:     CF$_4$
Flow 2:     30 sccm
Pressure 2:    30 mTorr
Power 2:     20 W
Time 2:     Until crystal etched to desired thickness or removed completely
        Etch rate = 29 ± 23 nm/min (top oxide layer on NbSe$_3$ etches
        much slower)
RF Voltage:    -200 V

**8. WET ETCH TO REMOVE PMMA MASK**
Soak in:     Methylene Chloride : Acetone (9:1)
Time:      30min to 1 hour
Rinse:      Acetone then Isopropanol and blow off with air gun (careful)

**9. DRY ETCH TO DESCUM PMMA MASK**
Instrument:    **Applied Materials RIE  (reserve 1 h)**
Gas 3:      O$_2$ (to descum resist)
Flow3:      30 sccm
Pressure 3:    30 mTorr
Power 3:     90 W
Time3:      6 sec (etch rate ~ 100-600 nm/min)
RF Voltage:    -570 V
Alternative:    Use Branson barrel etcher (1 min process, no heat)
NOTE:      You may try skipping this step.

# Recipe 4: Contacting Whiskers by Probes-on-Top (POT) Method

*Pattern Bi-layer Mask on NbSe$_3$ for Gold Probe Lift-off*

Attach a whisker to a substrate using Recipe 2.  If desired, pattern the whisker using
Recipe 3, and then proceed.

1. **SPIN** Resist
What:      MMA(8.5)MAA Copolymer 11% in Ethyl Lactate
Spin Speed:    3000 rmp
Spin Time:    60 sec
Thickness:    550 nm (need total 1100 nm, need second coat)

2. **BAKE** Resist
Bake type:    Vacuum/N$_2$ background oven



| | |
|---|---|
| Temperature: | 120 C |
| Bake time: | 30 min to set the resist |

3. **SPIN** Resist (second coat)

| | |
|---|---|
| What: | MMA(8.5)MAA Copolymer 11% in Ethyl Lactate |
| Spin Speed: | 3000 rmp |
| Spin Time: | 60 sec |
| Thickness: | 550 nm (have total of 1100 nm including $1^{st}$ coat) |

4. **BAKE** Resist

| | |
|---|---|
| Bake type: | Vacuum oven/$N_2$ background oven |
| Temperature: | 135 C |
| Bake time: | 60 min |

5. **SPIN** Resist

| | |
|---|---|
| What: | PMMA - 495, 5.5 % in Anisole |
| Spin speed: | 3000 rpm |
| Spin time: | 60 sec |
| Thickness: | 290 nm |

6. **BAKE** Resist

| | |
|---|---|
| Bake type: | Vacuum oven/$N_2$ background oven |
| Temperature: | 135 C |
| Bake time: | 60-90 min |

7. **EXPOSE** Resist

| | |
|---|---|
| Instrument: | **Leica/Cambridge EBMF 10.5  (reserve 2 h)** |
| Jobfile: | GOLD01.JOB |
| Field size: | |
| Pixel size: | |
| Beam Current: | 200 nA  (for 3.2 field and VRU 0 , current ( in nA) needs to be smaller than   the smallest dose (in micro C / $cm^2$) used. |
| Dose: | 250 uC/cm^2 |
| Apparture: | 1 |
| VRU: | 0 |

8. **DEVELOP** Resist

| | |
|---|---|
| Developer: | MIBK in Isopropanol (1:1) |
| Time: | 80 sec + 15 sec increments to achieve a desired line width (not more than 2 min 30 sec total) |
| Rinse: | Isopropanol then blow dry |

9. **DESCUM/CLEAN** Surface of $NbSe_3$



| Instrument: | **Applied Materials RIE**      **(reserve 1 h)** |
|---|---|
| Gas 1: | $O_2$ |
| Flow1 : | 30 sccm |
| Pressure 1: | 30 mTorr |
| Power 1: | 90W |
| Time1: | 6 sec - 10 sec  (etch rate ~ 600 nm/min) |
| RF Voltage: | -570 V (-506 V for 1st crystal) |
| Alternative: | Use Branson barrel etcher (1 min process, no heat) |
| | |
| Gas 2: | $CF_4$ (an **IMPORTANT** step to realize good electrical contacts!) |
| Flow 2: | 30 sccm |
| Pressure 2: | 30 mTorr |
| Power 2: | 20 W |
| Time 2: | 20-40 sec  (40 sec worked well but etched too deep into the crystal).  Etch rate = $29 \pm 23$ nm/min (top oxide layer on $NbSe_3$ etches much slower) |
| | RF Voltage:          -200 V (-193 V for 1st crystal) |
| NOTE: | This step should be done immediately prior to evaporation. |

*Place Gold Contacts/Pads on NbSe₃*

10. **EVAPORATE** Ti/Au

| Instrument: | **SC-4500 Evaporator** (odd hours) **(reserve 2 h)** |
|---|---|
| Metal 1: | Ti            Note: Wait until pressure drops |
| Method: | E-beam       below 2e-6 torr before evaporating. |
| Thickness: | 40-50 Angstroms |
| Rate: | 5 A/sec |
| | |
| Metal 2: | Au |
| Method: | E-beam |
| Thickness: | up to 5500 Angstroms lifts off fine (Need thickness of Au at least $1.5 \times$ thickness of step over crystal) |
| Rate: | 5 A/sec |

11. **LIFT-OFF**

| Soak: | Methylene Chloride : Acetone (~9:1) |
|---|---|
| Time: | until gold comes off = ½ to 1 hour |
| Rinse: | Acetone then Isopropanol, blow off with air gun |

*Electrical Tests*

12. **TEST CONTACTS/SAMPLE RESISTANCE**

Set current source to $1 - 10$ μA (voltage limited to 1-10 V).

Measure voltage across the crystal contacted by gold probes. R = V/I



*Crop Wafer to Desired Size*

**13. SPIN PROTECTIVE RESIST COAT**
    What:                Shipley 1813 optical resist
    Spin speed:          4000 rpm
    Spin time:           30 sec
    Thickness:           ~1.3 microns
    NOTE:                Do not bake resist

**14. CUT WAFER**
    Machine:             Fine diamond saw in TOL (Clark Hall)
    How:                 Hold wafer piece with tweezers and gently push against the
                         running blade until it cuts through the wafer.  Cut down to a 11
                         mm x 11 mm piece.
**15. WASH OFF RESIST**
    Rinse/Soak:          Acetone, then Isopropanol, then blow dry.

16. **APPLY POLYMER COAT**
    Apply a drop of Ethyl Cellulose dissolved in Ethyl Acetate on top of crystal/substrate
    and let dry.  This may be done to provide better thermal contact between the crystal
    and the substrate and to push the crystal against the substrate.  Make sure that you
    finish all profilometer measurements before this step.

17. **SAMPLE READY FOR COOL-DOWN.**

# Recipe 5: Electroplating (Current) Probes into a Whisker Cross-Section

*Make EPS and EPC Masks*

1. **Create CAD file -** Using CATS software convert files eps1.dxf (electroplating
   substrate 1) and epc1.dxf (electroplating contacts 1) into pattern-generator readable
   files eps1.dat and epc1.dat.  Make sure your epc1.dxf pattern includes temporary gold
   paths that will electrically connect all gold patterns (that require electroplating) on the
   substrate during electroplating process.  The seed gold layer on the substrate will be
   biased within an electrical circuit during the process for electroplating to occur.  These
   temporary gold paths should be scratched off the substrate before measurements.
2. **Fabricate EPS and EPC Masks (reticles for optical lithography).**

*Make EPS Substrate (pattern voltage probes)*

3. **Prime Wafer with P-20**



Static dispense, wait 60 sec, then spin dry.  Use 3-inch Si wafers with 1 micron of oxide on them.

4. **Spin Shipley 1813**
   Static dispense, 500 rpm for 5 sec for coverage, then 4000 rpm for 30 sec

5. **Bake on Hotplate** – 115 C for 2 min.

6. **Expose on 5X g-line steper** – exposure time 1.25 sec, focus 251, mask eps1

7. **Image reversal** – Bake in YES oven on $NH_3$ (ammonia) reversal process

8. **Flood Exposure** – using HTG contact aligner for 60 sec

9. **Develop Resist** – Soak in 321 MIF for 60 sec. Check, may need extra 30 sec.

10. **Descum Resist Trenches –** Use Branson Barrel Etcher, oxygen plasma, resist descum process

11. **Ti/Au Evaporation –** Use SC-4500 Evaporator, e-beam evaporate, 50 A of Ti, then 1200 A of Au

12. **Resist Lift-Off** – Soak in ultrasonic bath in acetone (or Shipley 1165) for 5-30 min until lift-off is done.  Rinse with isopropanol when done, before acetone dries on the wafer.

13. **Spin Protective Coat** – Spin Shipley S-1813 at 4000 rpm for 30 sec.

14. **Bake on Hotplate** – 115 C for 2 min.

15. **Cleave Wafer into Substrates**

16. **Wash Off Protective Coating** – Soak in ultrasonic bath in acetone for 5-30 min until substrates are clean.  Rinse with isopropanol when done, before acetone dries on the wafer.  Substrates are ready for mounting crystals.

*Mount Crystals on EPS Substrate*

17. **MOUNT NbSe₃ sample -** Prepare Cu foil by cutting it into 1.5 × 1.5 cm pieces and carve 3 × 3 mm windows on each.  Using microscope, tweezers, and scissors select stepless crystal, cut it and drop it over the isopropanol wetted Cu window.  Isopropanol will act as a glue holding crystal ends fixed to the Cu window.  Now you can invert the foil and check the other side of the crystal under the microscope for steps.  If the crystal does not have steps on either side, transfer it on the EPS substrate.



Put a drop of isopropanol on top of crystal if needed to ensure that crystal is glued to the substrate.

## *Encapsulate with Polyimide*

18. **SPIN Adhesion Promoter** - QZ3289 : QZ3290 (1:9), freshly mixed within 5 days. Spin at 5000 rpm for 20 sec to dry. Make sure that crystals are in the center of rotation as much as possible and dispense the liquid carefully not to wash off the crystal.

19. **BAKE Adhesion Promoter** – Bake on hotplate for 50 C for 2 min.

20. **SPIN Polyimide** – Spin Probromide PI-285, diluted with QZ3288 (2:1) at 3100 rpm for 50 sec. This should give polyimide thickness of 260 +/- 90 nm. Consult thickness vs. spinning parameters chart if you need a different thickness of polyimide. Polyimide should be at least ½ the thickness of the whisker to encapsulate it. More is better, but it takes longer to etch through the polyimide later and more resist mask is then depleted during the process.

21. **BAKE Polyimide** – Bake in TOL turbo pumped vacuum oven at pressures < 7e-6 torr, at 85 C for 30 min (oven setting 0.5 hours), 150 C for 15 min (oven setting 0.3 hours), 210 C for 15 min (oven setting 0.3 hours) 30 min at pressure < 7e-6 torr.

## *Pattern Trenches for Imbedded Current Contacts*

22. **Spin Resist** – Spin Shipley 1813 on top of polyimide encapsulated crystals (EPS substrate) at 4000 rpm for 30 sec.

23. **Bake Resist** – On a hotplate (in air ambient) bake the resist at 115 C for 1 min 20 sec to 2 min.

24. **Expose Resist** – Use Carl Suss Contact Aligner to expose samples for 6 sec through the EPC mask. Make sure to align the mask to the patterned alignment mark on the EPS substrate.

25. **Develop Resist –** Develop resist in 300 MIF for 1 min. Rinse with DI water.

26. **Profilometer Inspection –** To get a general idea of sample thickness, scan the profilometer stylus across the whisker covered by polyimide and resist. Do it over the part of the whisker that will not be investigated (i.e. beyond current contacts). Also measure the thickness of the resist (i.e. plateau where resist is minus the plateau where resist has been developed). It should be ~1.3 micro m for spinning parameters given in step 22.



27. **Etch Through Polyimide –** Use Applied Materials RIE to etch away polyimide from resist trenches. O2 at 30 sccm + CF4 at 3 sccm plasma, 30 mTorr, 90 W, ~464 V, for 2 min with 3 min break for cooling after every etching minute. Etch rate of polyimide is somewhere between 180-260 nm/min. Inspect under optical microscope to make sure all polyimide is gone before continuing.

28. **Etch Through NbSe₃ Crystal –** Use Applied Materials RIE to etch away NbSe₃ from trenches where Au will be imbedded. CF4, 30 sccm, 20 mTorr, 35 W, until the crystal is completely removed from trenches. The etching should be performed so that every 1 min of etching should be followed by a 3 min break for cooling (gasses can be left flowing during cooling time in vacuum). Etching time increments should never be longer than 1 min. Inspect under optical microscope after every 3 min of etching to check if the whisker is etched through. Resist mask will withstand 16 min of etching so that afterwards there is still enough resist left to obtain a good lift-off after step 29. 16 min of etching should be enough to etch through a 0.5 micro m thick whisker. Warning: this step should be done immediately prior to the seed metal evaporation step and Au electroplating (next), i.e. within 1-2 days for best results, to avoid oxidation of etched NbSe₃ walls.

29. **Seed Metal Evaporation –** Using SC 4500 evaporator, use e-beam evaporation to deposit 50 A of Ti and then 700 A of Au

30. **Resist Lift-Off –** Soak in acetone (do not use ultrasonic bath since that will shred the crystals) until lift off is done (5min – several hours). Rinse with isopropanol before acetone dries.

31. **Measure Trench Depth –** Use profilometer to roughly determine crystal thickness (The measurement typically underestimates the real thickness of the whisker because of the polyimide coat over the whisker). Also measure trench depth where Au will be electroplated. Typically, you will need to electroplate gold at least as thick as this trench depth.

*Electroplate Au into Trenches (make imbedded current contacts)*

32. **Set up Electroplating Apparatus –** Place everything into base hood and run heater to warm the electroplating solution (Microfab Au 100) to 60 C (takes about 45-60 min).

33. **Pre-Clean Samples –** Dip samples (EPS/EPC substrate made above) into a mixture of 1 part of buffered oxide solution (BOE 30:1) and 4 parts of H2O for 30 sec. Do the same with a Au-covered wafer (that you provide) that will be used in the next step. Rinse all with DI water.

34. **Remove Impurities from Electroplating Solution –** Connect the electrodes to the Au covered wafer and place into solution in electroplating tank. Run current 0.05 A, (typically give 0.40 V) for 1 min. This will electroplate some impurities along with



Au onto the wafer and remove them from the plating solution. Remove the wafer, rinse it with DI water draining it into the plating tank. Now you are ready to electroplate your samples.

35. **Electroplate Au onto Sample** – Connect electrodes to your sample (to seed layer of Au). Place into plating tank and run current 0.02 A (gives 0.29 V) until desired thickness of Au is deposited. Plating rate at 0.02 A and 0.29 V is between 45-80 nm/min. Inspect samples by optical microscope and measure plated Au thickness by profilometer.

*Preparation of Samples before Mounting into Cryostat*

36. Use a poke tool and a scribe to gently scratch off the temporary gold paths connecting electroplated gold pads on the substrate (these temporary gold paths were patterned to keep the gold patterns electrically connected during electroplating process).

37. Cleave away excess Si wafer to make the substrates suitable size to fit into cryostat.

## Notes Related to Applied Materials RIE Machine

| Dial settings: | Loading | Tuning |
| --- | --- | --- |
| $O_2$ : | 918 | 287 |
| CF4: | 926 | 287 |
| SF6: | >940 | 288 |

| Gas | Flow | Pressure | Power Setting | RF Voltage |
| --- | --- | --- | --- | --- |
| $O_2$ | 30 sccm | 30 mT | 100 W | -550 V – 577 V |
| CF4 | 30 sccm | 30 mT | 90 W | ~ -509 V |

## Notes about Heating NbSe$_3$

NbSe$_3$ oxidizes easily when heated in air environment. This causes problems when

contacting crystals. All heating of NbSe$_3$ should be performed either in vacuum or

vacuum/$N_2$ environment when possible. If heating in air (or in environment with $O_2$ present)

is unavoidable, the heating should be kept to a minimum both in duration and temperature.